\documentclass[
11pt, 
english, 
singlespacing, 
headsepline, 
]{MastersDoctoralThesis} 

\usepackage[utf8]{inputenc} 
\usepackage[T1]{fontenc} 

\usepackage{palatino} 

\usepackage[backend=bibtex,
style= ieee,
natbib=true,
url=false,
doi=false,
isbn=false,
sorting=none,
mincitenames=1,
maxcitenames=4,
maxbibnames=4
]{biblatex} 
\renewbibmacro{in:}{%
  \ifentrytype{article}{}{\printtext{\bibstring{in}\intitlepunct}}}

\usepackage[autostyle=true]{csquotes} 

\usepackage{amsbsy}
\usepackage{amsfonts}
\usepackage{amssymb}
\usepackage{amsmath}
\usepackage{color}
\usepackage{graphicx}
\usepackage{psfrag}
\usepackage{xspace}
\usepackage{stmaryrd}
\usepackage{lettrine} 

\usepackage{minitoc} 

\usepackage{dsfont}
\usepackage{numprint}
\usepackage[super]{nth}
\usepackage{tabu}

\newcommand{\bb}[0]{\begin{eqnarray}}
\newcommand{\ee}[0]{\end{eqnarray}}
\newcommand{\ket}[1]{| #1 \rangle}
\newcommand{\bra}[1]{\langle #1 |}
\newcommand{\moy}[1]{\ensuremath{\langle #1\rangle}\xspace}
\newcommand{\idop}{\mathds{1}}
\usepackage[font=small,labelfont=bf,
   justification=justified,
   format=plain]{caption}
\usepackage{indentfirst}

\usepackage{pdfpages}

\thesistitle{Thermodynamics of quantum open systems: applications to quantum optical and optomechanical devices.} 
\supervisor{Dr. Alexia \textsc{Auff\`eves}} 
\examiner{} 
\degree{Doctor of Philosophy} 
\author{Cyril \textsc{Elouard}} 
\addresses{} 

\subject{Physical Sciences} 
\university{\href{www.univ-grenoble-alpes.fr}{Univeristé Grenoble Alpes}} 
\department{} 
\group{{}{Nanophysics and Semiconductors team}} 
\faculty{\href{http://faculty.university.com}{}} 

\AtBeginDocument{
\hypersetup{pdftitle=\ttitle} 
\hypersetup{pdfauthor=\authorname} 
\hypersetup{pdfkeywords=\keywordnames} 
\hypersetup{colorlinks=false}
}

\begin{document}

\frontmatter 

\pagestyle{plain} 

\includepdf[pages=1]{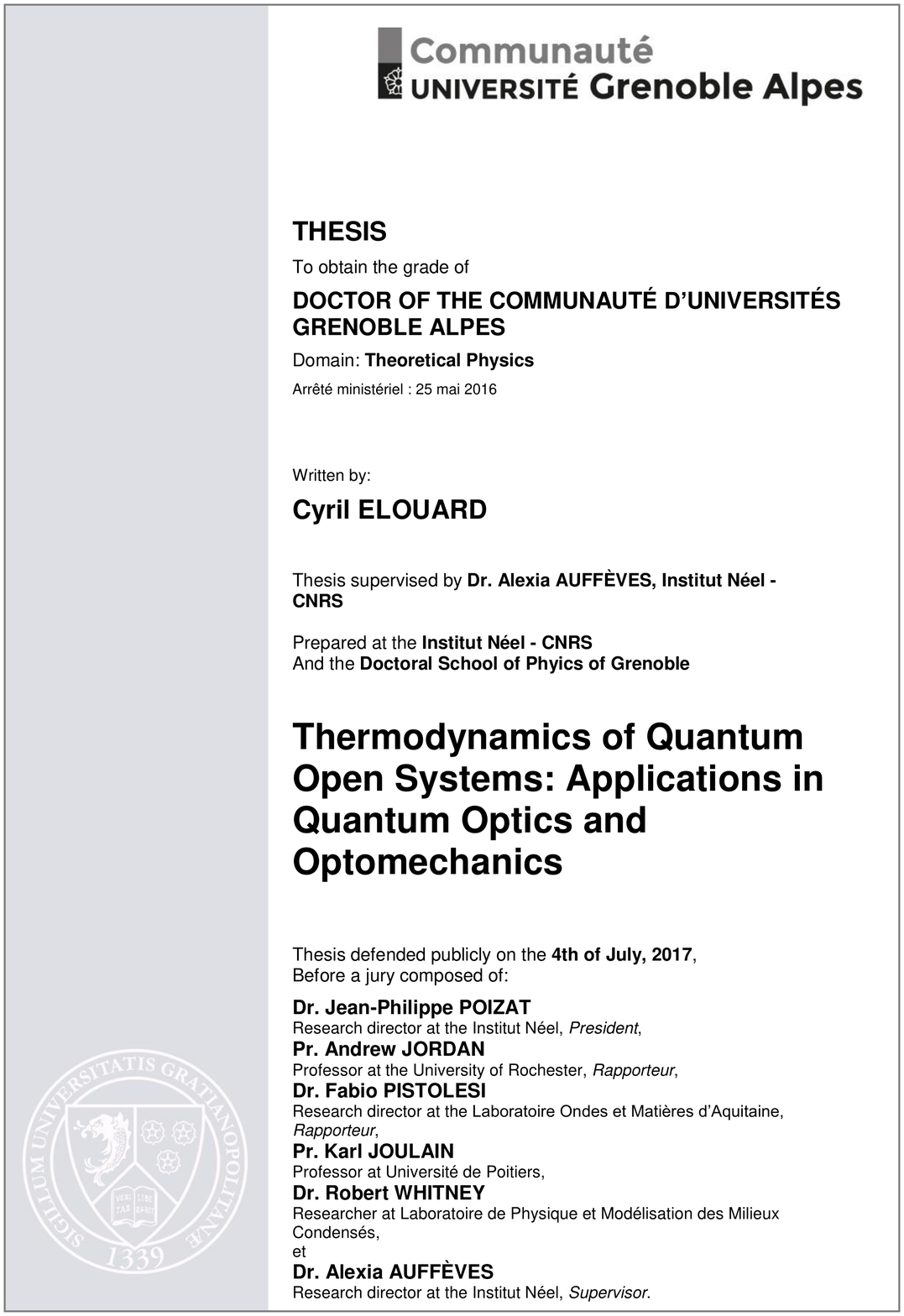}

\setcounter{tocdepth}{4}
\setcounter{minitocdepth}{4}

\dominitoc

\tableofcontents 

\listoffigures 

\listoftables 

\mainmatter 

\pagestyle{thesis} 


\addchap{Introduction}

\begin{figure}[h!]
\begin{center}
\includegraphics[width=0.7\textwidth]{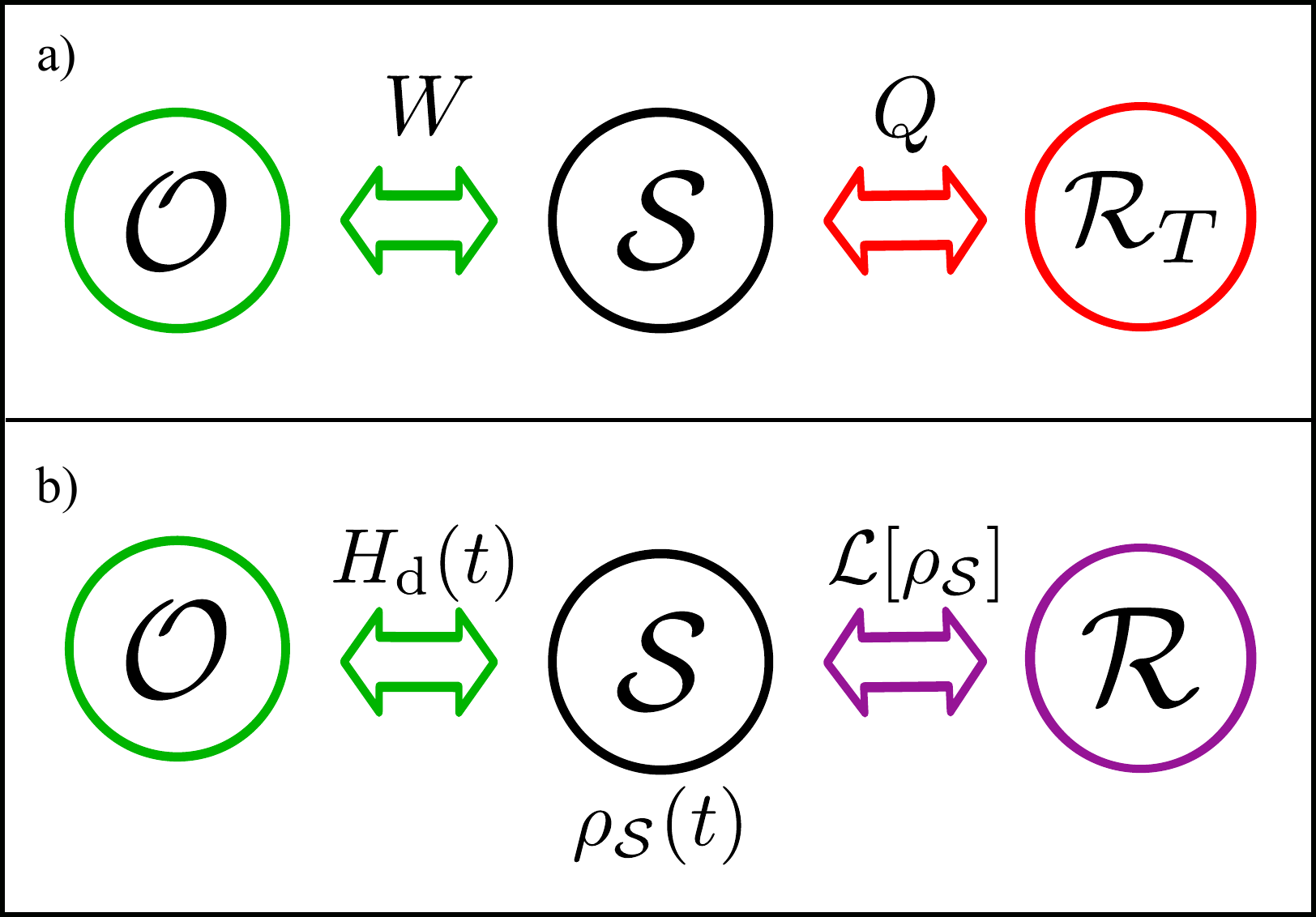}
\end{center}
\caption[Thermodynamics and quantum open systems]{\label{f0:Scenery} Thermodynamics and quantum open systems. a) Typical scenery in thermodynamics: a system ${\cal S}$ receives work $W$ from an operator $\cal O$ and heat $Q$ from a thermal reservoir ${\cal R}_T$. b) Typical situation of quantum open system: a quantum system ${\cal S}$ is driven by an operator imposing some time-dependent Hamiltonian $H_\text{d}(t)$, and to a reservoir inducing a non-unitary evolution captured by a superoperator ${\cal L}[\rho_{\cal S}]$ applied on the density operator of the system $\rho_{\cal S}(t)$.}
\end{figure}

\lettrine{T}{hermodynamics} was developed in the \nth{19} century to describe situations like the one pictured in Fig.\ref{f0:Scenery} \cite{Carnot,Kelvin,Clausius,Diu}: a system $\cal S$ (the working agent) is driven by some operator $\cal O$ and coupled to a thermal reservoir ${\cal R}_T$.  The operator interacts with $\cal S$ in a controlled/deterministic way, and provides energy called work $W$. On the other hand, the thermal reservoir interacts with the working agent in an uncontrolled/stochastic way, and the energy it provides to is referred to as heat $Q$. 

Driven by the context of industrialization in western countries, initial \mbox{motivations} to study such a situation were very concrete. A famous example is work extraction: how can we convert the disorganized heat provided by a hot thermal reservoir into useful work \cite{Carnot}, e.g. to make a train move forward? 

Studying such situations has lead to introduce the fundamental notion of irreversibility \cite{Clausius,Kelvin}: the second law expresses the arrow of time by imposing a direction to natural processes. The irreversibility is quantified by a production of entropy, which must be greater or equal to zero for an isolated system according to the second law of thermodynamics. The production of entropy solely vanishes for reversible transformations, which corresponds to quasi-static (infinitely slow) processes. The second law has crucial practical consequences as it forbids certain processes like monothermal engine cycles and sets fundamental bounds on the efficiency of thermodynamic machines. For instance, it limits the amount of work that can be extracted from a thermal reservoir during a transformation of finite duration.\\

\subsection*{Stochastic thermodynamics}

At first glance, it seems difficult to conciliate this irreversible behavior of macroscopic systems, with a description at the microscopic scale which solely involves time-symmetric physical laws. This apparent paradox was solved in the \nth{20} century owing to stochastic thermodynamics  \cite{Gibbs,Maxwell,Boltzmann}. In this framework, the reservoir exerts random forces on the system, leading to a stochastic dynamics (Brownian motion) that the operator cannot control. The system's position in phase space follows a stochastic trajectory, different for each realization of a same protocol \cite{Sekimoto10,Seifert08} and thermodynamic quantities like work, heat, entropy production are random variables that can be defined for each such trajectory. 

This stochastic behavior explains the emergence of irreversibility: Let us imagine that after applying some transformation, the operator tries to revert the dynamics followed by the system. Because of the unpredictable actions of the reservoir, reverting the protocol he has applied is not enough to make the system follow the time-reversed trajectory in phase space: $\cal S$ will not go back in general to its initial position \cite{Seifert08}. 

Owing to the central position it gives to randomness, the theory of stochastic thermodynamics provided precious insights into the thermodynamics of out-of-equilibrium systems. In particular, it enabled to prove a large class of Fluctuation Theorems (FTs), i.e. relations linking together equilibrium and out-of-equilibrium quantities \cite{Sevick08}. A famous example is Jarzynski Equality \cite{Jarzynski97} relating the variation in equilibrium free energy to the statistics of the work performed on a system driven out of thermal equilibrium.

\subsection*{Information thermodynamics}

In the meantime, thermodynamics was revisited from the information theory point of view \cite{Parrondo15}. The connection between information and thermodynamics first appeared in the resolution of Maxwell's demon paradox \cite{Maxwell,Kelvin74}, which required to discuss the physical nature of information \cite{Landauer91}. The central idea of information thermodymancis is that the working agent ${\cal S}$ can be identified with a memory, i.e. a system encoding bits of information \cite{Szilard29,Landauer61}. These investigations led to show that one bit of information can be reversibly converted into an elementary amount of energy \cite{Landauer61}. Later on, such paradigm allowed to reformulate the second law of thermodynamics so as to involve information-theoretic quantities, such that Shannon entropy \cite{Seifert12} or mutual information \cite{Lloyd89}, e.g. in the context of feedback-assisted protocols \cite{Sagawa10}.

\subsection*{Quantum (information) thermodynamics}

The beginning of the \nth{20} century also witnessed the emergence of quantum mechanics \cite{Mehra82}. The description of small systems like single atoms showed physical properties sharply contrasting with those of macroscopic systems \cite{Feynman89}.
A century later, advances of quantum and nano-technologies have urged to explore how the laws of thermodynamics translate in the quantum world, i.e. when the working agent ${\cal S}$, the reservoir $\cal R$ and/or the operator $\cal O$ are quantum systems. 

Furthermore, the quantum extension of information theory \cite{Nielsen} appeared as a promising resource for computing \cite{Shor94,Simon94} and cryptography \cite{bb84} applications. It is therefore paramount to investigate information thermodynamics with Qubits instead of classical bits of information.

The emerging field of quantum thermodynamics raises many new questions \cite{Vinjanampathy16}, e.g.:
\begin{itemize}
\item How to define thermodynamic quantities like work for quantum systems? \cite{Talkner07,Popescu14}
\item Are the laws of thermodynamics still valid for quantum systems? \cite{Brandao15}
\item What are the differences between the thermodynamics of classical and quantum systems? \cite{Campaioli17}
\item How does thermalization emerge from quantum principles? \cite{Dunjko12,Gogolin16}
\item Can we build more efficient engines/refrigerators exploiting the specific properties of quantum systems, such as coherences or entanglement? \cite{Campo14,Rossnagel14} 
\item What are the sources of irreversibility at the quantum scale? \cite{AuffevesViewPoint,Dressel16}
\item What are the entropy and energy costs of quantum tasks like quantum information processing? \cite{Bedingham16,DelRio11,Ikonen16}
\item Can thermodynamic concepts bring new insights about the foundations of quantum mechanics? \cite{Navascues14}
\end{itemize}

All these questions (and many others) have motivated a blooming research field: quantum thermodynamics, which is the framework for the present thesis.

\subsection*{The role of quantum measurement}

In this thesis, we build a theoretical framework in which quantum measurement itself is a thermodynamic process.

Indeed, it is a specificity of quantum mechanics that measurement affects the state of the quantum system under study \cite{Haroche}. This unavoidable effect plays a crucial role in many quantum protocols. Yet, the effect of quantum measurement in thermodynamics has been largely unexplored so far. Measurements have mostly been included in quantum thermodynamic protocols as practical steps where information is extracted (e.g. in Maxwell's demon protocols), and the effect of measurement considered as detrimental (just as decoherence) \cite{Lloyd97}. 

In this dissertation, we show that quantum measurement plays a key role in  thermodynamics. In particular, it has energetic and entropic footprints: As it modifies the system's state, quantum measurement induces energy variations that have to be included in energy balances, leading to new quantum terms in the first law of thermodynamics. Moreover, just as the reservoir of classical stochastic thermodynamics, quantum measurement induces randomness in the system's dynamics. Therefore, it is a source of irreversibility, which has no equivalent in the classical world.  

Building on the analogy between a measuring apparatus and a thermal reservoir, our formalism allows to explore a new area of stochastic thermodynamics in which randomness primarily comes from quantum measurement instead of Brownian motion. 

\subsection*{Testbeds from quantum optics and optomechanics}

We consider applications in which the role of $\cal S$ is played by small quantum system like a Qubit. Such situation is standard in the framework of experimental quantum technologies \cite{Dowling03}.

These devices are well described in the paradigm of open quantum systems which involves a driven quantum system (i.e. with a time-dependent Hamiltonian) coupled to a reservoir which may be a thermal bath or induce pure-dephasing (see Fig.\ref{f0:Scenery}b) \cite{Breuer}. Such a situation is very reminiscent of classical thermodynamic pictures of Fig.\ref{f0:Scenery}a \cite{Alicki79,Kosloff13}. 

Interestingly, the evolution of $\cal S$ in that context can be expressed as induced by a quantum measurement (Kraus representation \cite{Kraus83,Haroche}), which is the starting-point of the quantum trajectory interpretation of the open quantum systems' dynamics \cite{Molmer93}. These quantum trajectories are stochastic sequences of quantum states that we exploit in our framework as the analogs of the trajectories in phase space of classical stochastic thermodynamics.

\begin{figure}[h!]
\begin{center}
\includegraphics[width=0.6\textwidth]{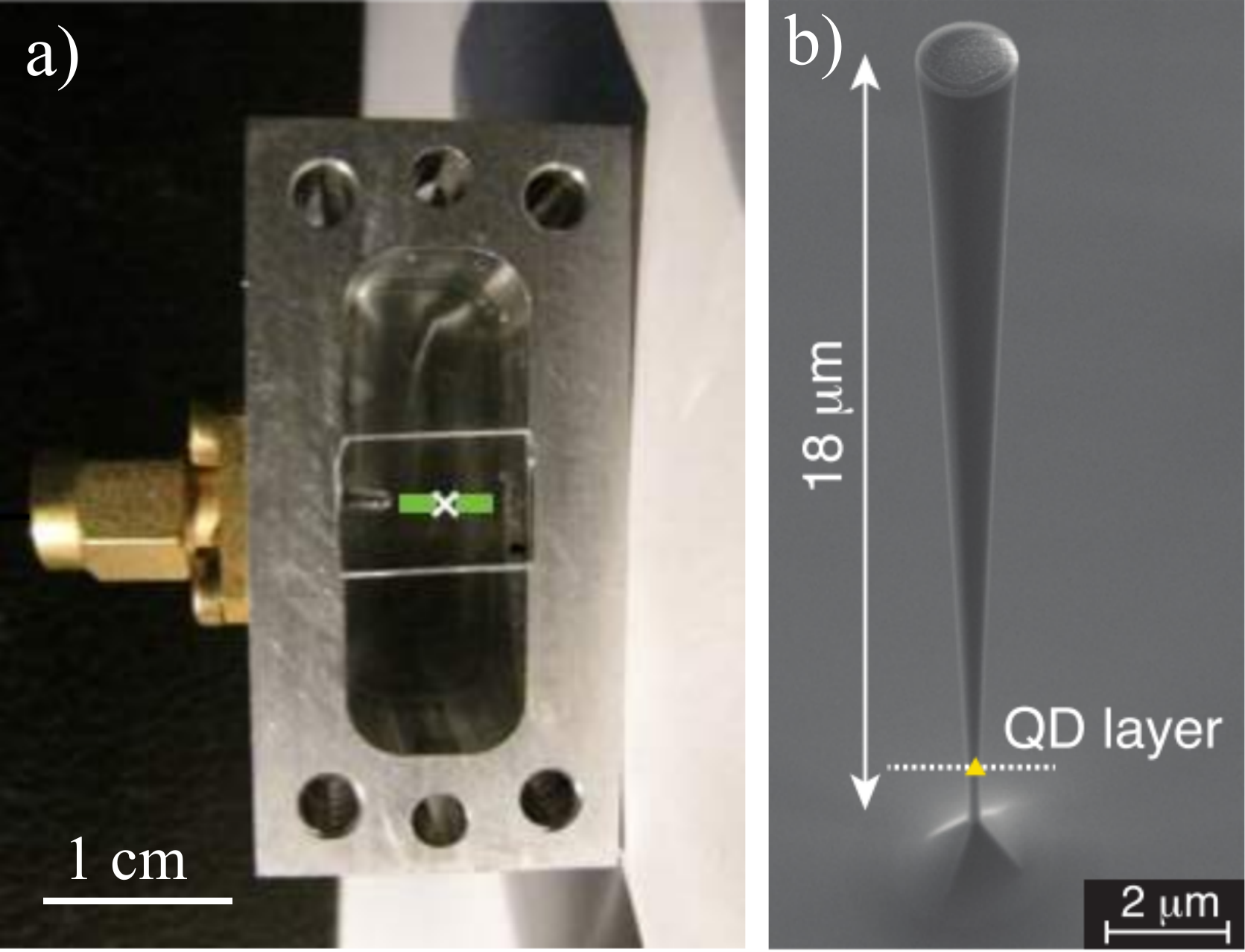}
\end{center}
\caption[Experimental platforms]{Experimental platforms. a) Superconducting circuit implementing a quantum electrodynamics experiment, involving a Qubit dispersively coupled to a cavity. Picture from the Quantum Electronics Group of Ecole Normale Sup\'erieure ( Paris, France). b) Hybrid optomechanical setup featuring a quantum dot (Qubit) embedded in a vibrating nanowire. Picture from the NanoPhysics and SemiConductors team of the the Institut N\'eel - CNRS and CEA (Grenoble, France).\label{f0:Impl}}
\end{figure}

As a testbed for our formalism, we will especially focus on two experimental platforms, involving superconducting circuits \cite{Blais04}, or optomechanical systems \cite{Marquardt09,Meystre13} (see Fig.\ref{f0:Impl}) and which allow the manipulation of single quantum objects. We specifically look at two situations, which correspond to different ways of performing work available to quantum systems. The first situation is the textbook case of quantum optics: a Qubit coupled to a coherent drive inducing Rabi oscillations \cite{CCT}. The second situation is a so-called hybrid optomechanical system, involving a Qubit parametrically coupled to a mechanical oscillator which periodically shifts its transition frequency \cite{Arcizet11,Yeo13}. In both cases, the role of the reservoir is played by the electro-magnetic environment of the Qubit.


\subsection*{Outline of the thesis}

In Chapter \ref{Chapter1}, we develop a thermodynamic description of projective quantum measurement. We investigate the energy exchanges occurring during the measurement process and identify a key quantity of our formalism: the quantum heat. We also show that the measurement is a source of irreversibility and we derive the expression of the entropy produced. As a proof of principle, we propose an engine extracting energy from the quantum fluctuations induced by the measurement instead of a the thermal fluctuations induced by a heat bath.

In Chapter \ref{Chapter2}, we extend our framework to more general types of measurements, which fully capture the thermodynamics of a driven quantum open system. We derive out-of-equilibrium thermodynamics results such as  Fluctuation Theorems. We study two types of reservoir: a thermal bath and a reservoir inducing pure-dephasing.

In Chapter \ref{Chapter3}, we investigate a fundamental situation of quantum optics: the fluorescence of a two-level atom. We exploit our formalism to provide an accurate thermodynamic interpretation of two descriptions of the system's dynamics employed in the literature: the optical Bloch equations and the Floquet master equation. We evidence the link between these two approaches and the coarse-graining time step used to describe the dynamics.

In Chapter \ref{Chapter4}, we present a hybrid optomechanical device and show that it is a promising platform to investigate quantum thermodynamics. This device involves a mechanical oscillator strongly coupled to a dissipative artificial atom. We prove that the oscillator can be used as a battery providing some work to the Qubit. Reciprocally, when the Qubit is driven, we investigate the dynamics of the hybrid system on long time-scales, and show that the Qubit behaves like a reservoir inducing noise on the oscillator. 

\chapter{Measurement Based Thermodynamics}
\label{Chapter1}

\setcounter{mtc}{2}
\minitoc



\begin{figure}[h!]
\begin{center}
\includegraphics[width=0.6\textwidth]{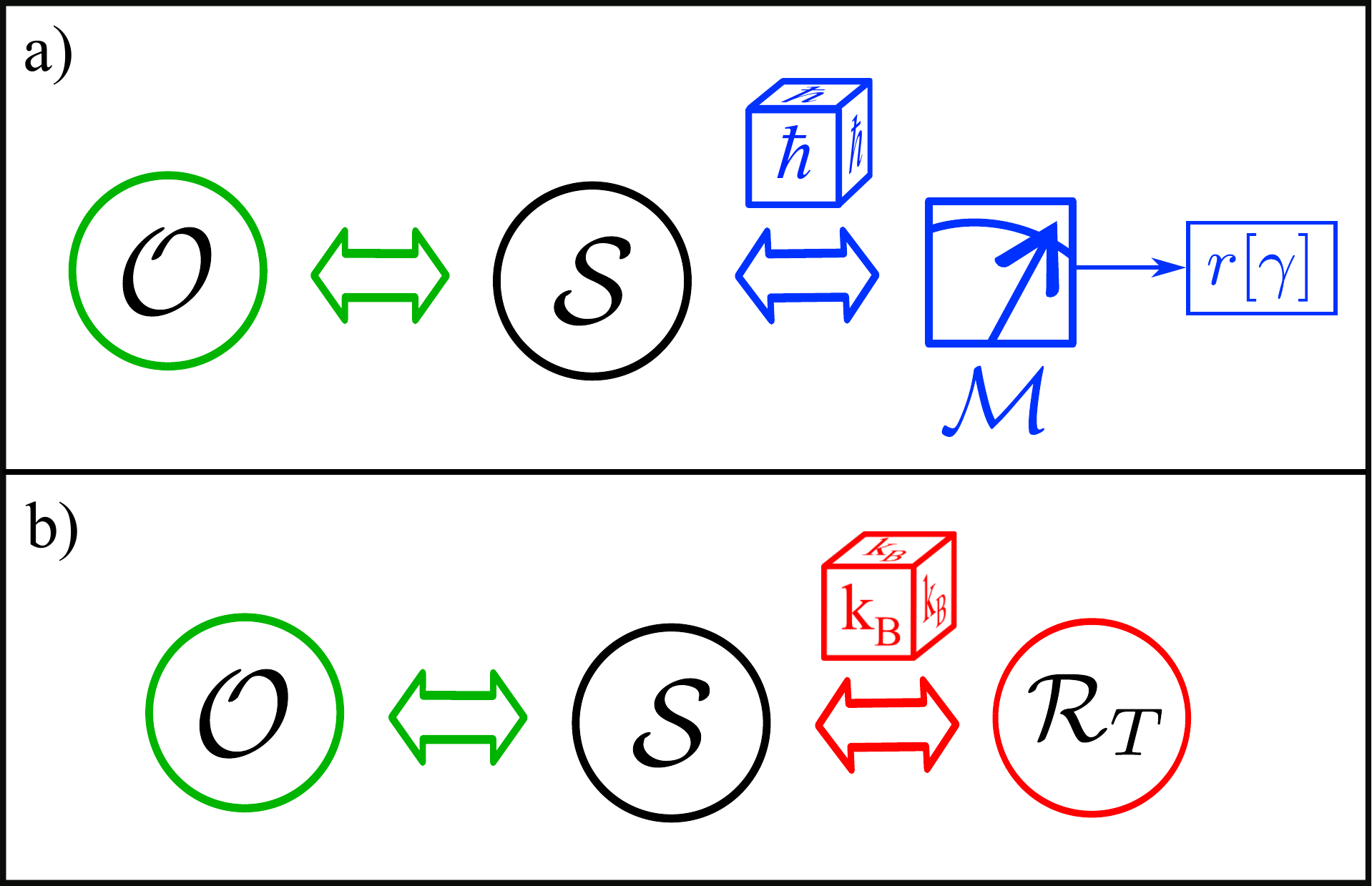}
\end{center}
\caption[Measurement-based thermodynamics]{Measurement-based thermodynamics. a) Situation under study: a quantum system $\cal S$ is driven by an operator $\cal O$ inducing a time-dependent term $H_\text{d}(t)$ in its Hamiltonian. A measuring apparatus $\cal M$ performs projective measurement of observable $\hat M$ of the system at discrete times. Due to the measurement, the evolution of $\cal S$ is stochastic, which is symbolised by the $\hbar$-dice. b) Typical situation of classical stochastic thermodynamics: the driven system $\cal S$ has a stochastic dynamics (sketched by the $k_\text{B}$-dice) because of the random forces applied by the reservoir.\label{f:Scenery}}
\end{figure}

\lettrine{W}{hat} is the thermodynamic effect of a quantum measurement? In order to answer this question, we devote this first chapter to the study of the situation depicted in Fig.\ref{f:Scenery}a. A quantum system ${\cal S}$ with Hilbert space ${\cal H}_\text{S}$ (dimension $d_{\cal S}$) interacts with an operator ${\cal O}$ and a measuring apparatus ${\cal M}$, which are both assumed to be classical systems\footnote{In particular, they are not described by quantum states belonging to a Hilbert space.}.

The operator ${\cal O}$ induces a time-dependent component $H_\text{d}(t)$ to the system's Hamiltonian. We denote $H_{\cal S}(t) = H_{{\cal S},0} + H_\text{d}(t)$ the total (time-dependent) Hamiltonian of the system, with $H_{{\cal S},0}$ the Hamiltonian of the bare system. In analogy with classical mechanics, we say that ${\cal S}$ is driven by the operator ${\cal O}$.

The measuring apparatus ${\cal M}$ is used to perform projective measurements of an observable $\hat M$ of the system, whose eigenstates are denoted $\{\ket{m_k}\}_{1\leq k \leq d_{\cal S}}$. In the line of Copenhagen interpretation of quantum mechanics, we assume that the interaction of ${\cal S}$ with ${\cal M}$ satisfies the measurement postulate: ${\cal S}$ is instantaneously projected onto one of the eigenstates $\ket{m_k}$ corresponding to the measurement outcome. The selection of the measurement outcome is a random process whose law of probability is given by Born's rule \cite{Born26}.

The situation under study is reminiscent of the canonical situation of classical stochastic thermodynamics \cite{Sekimoto10,Seifert08} which is depicted on Fig.\ref{f:Scenery}b: a driven system coupled to a thermal reservoir ${\cal R}_T$ at temperature $T$. The corresponding dynamics can be described by the Langevin equation, which takes for instance the following form for a particle in one dimension \cite{Seifert08}~:
\bb
\dot x = \mu F(x) + \zeta,
\ee
where $x$ is the particle position, $\mu$ the mobility of the particle, $F(x)$ a systematic position-dependent force. The term $\zeta$ stochastic variable of correlation $\moy{\zeta(t)\zeta(t')} \propto \delta(t-t')$ capturing the random actions of the thermal reservoir on the particle's dynamics.

 Just as the thermal reservoir, the measuring apparatus ${\cal M}$ is a source of stochasticity in the system's dynamics. The origin of the randomness in the two situations is very different: Brownian motion (pictured with a $k_\text{B}$ dice in Fig.\ref{f:Scenery}) in the classical case versus wavefunction collapse (pictured with a $\hbar$ dice in Fig.\ref{f:Scenery}) in the case under study. Still, this analogy raises a fundamental question: can we find an equivalent of the laws of thermodynamics in the absence of thermal bath, when stochasticity solely comes from quantum measurement ? The formalism exploited in this chapter is published in \cite{Elouard17}.

\section{Stochastic trajectories}

We aim to study a thermodynamic transformation of the system ${\cal S}$ between time $t_i$ and $t_f$. In the situation of Fig.\ref{f:Scenery}, a transformation is a protocol that is perfectly defined by (i) the initial state of ${\cal S}$, (ii) the driving Hamiltonian $H_\text{d}(t)$ for $t\in[t_i,t_f]$ and (iii) the instants $\{t_n\}_{1\leq n \leq N}$ at which $\cal M$ interacts with $\cal S$.

In presence of the measuring apparatus ${\cal M}$, the dynamics of the system is not unitary. If the measurement outcomes are not read, the evolution of the system along the transformation is captured by a Completely Positive Trace-Preserving (CPTP) map acting on its density operator \cite{Haroche}. 

In this dissertation, we consider that the measurement outcomes are read. As a consequence, in each realization of the transformation, the state of $\cal S$ follows a given stochastic trajectory $\gamma$. More precisely, the system evolves in a unitary way ruled by the Schr\"odinger Equation, except at times $\{t_n\}_{1\leq n \leq N}$ where it is randomly projected onto an eigenstate of $\hat M$ determine by the obtained measurement outcome $r_\gamma(t_n)$.
Therefore, the trajectory $\gamma$ is a sequence of pure quantum states conditioned on the measurement record $r[\gamma] = \{r_\gamma(t_n)\}_{1\leq n \leq N}$. Provided the system is at time $t_i$ in a known pure state, this trajectory can be fully reconstructed from the knowledge of $H_{\cal S}(t)$ and the measurement record $r[\gamma]$.

Such trajectory can be considered as the direct analog of the classical trajectory defined in standard stochastic thermodynamics \cite{Sekimoto10,Seifert08}. The two differences are (i) the space in which the trajectories evolve (Hilbert space versus phase space) and (ii) the origin of stochasticity (quantum measurement versus Brownian motion). \\

In this thesis, the state of the system along trajectory $\gamma$ will be denoted $\ket{\psi_\gamma(t)}$. The quantities associated to a single realization of a thermodynamic protocol will be indicated with brackets $[\gamma]$, e.g. $r[\gamma]$ or $W[\gamma]$, and the quantities evolving along a trajectory with a subscript $\gamma$, e.g. $r_\gamma(t)$ or $U_\gamma(t)$. The average over the set of all possible trajectories will be denoted $\moy{\cdot}_\gamma$.


\section{First law}

\subsection{Internal energy}
\label{s:InternalEnergy}

In classical thermodynamics, a given internal energy is associated with each possible state of the studied system \cite{Diu}. This quantity satisfies a conservation principle: it does not vary for an isolated system and may only change due to exchange with the surroundings under the form of heat and work. Can we find an equivalent for quantum systems ?

In this thesis, we define the internal energy of the system $\cal S$ at time $t$ along trajectory $\gamma$ as:
\bb
U_\gamma(t) \underset{\text{def.}}{=} \bra{\psi_\gamma(t)} H_{\cal S}(t) \ket{\psi_\gamma(t)}. \label{d:U}
\ee

Such definition is consistent with the quantity used in the thermodynamic framework based on quantum open systems theory \cite{Alicki79}, quantum information theory \cite{Vinjanampathy16} or resource theory \cite{Brandao13}. In these contexts, the internal energy is computed from the trace of the system's Hamiltonian multiplied by the density operator. For a pure state, the density operator takes the form $\ket{\psi_\gamma(t)}\bra{\psi_\gamma(t)}$ such that one recovers definition \eqref{d:U}.
 
Definition \eqref{d:U} is also in agreement with the definition used in the context of sochastic thermodynamics where the internal energy at time $t$ is identified with the outcome of a projective measurement of the system's Hamiltonian $H_{\cal S}(t)$. In pioneer studies, such prescription has allowed to find a quantum equivalent of many important thermodynamic results such that Jarzynski equality \cite{Esposito09,Campisi11}. When the protocol under study includes projective energy measurements, this definition coincide with ours. Indeed, right after a projective measurement at time $t$, the system's state is one of the energy eigenstates. 

\sloppy
Furthermore, our definition also provides an extension with respect to the mea\-su\-rement-based approach: it allows to attribute an internal energy to any quantum state, even to a superposition of energy eigenstates, at any time of a thermodynamic transformation. In particular, the internal energy now takes a definite value after and before a quantum measurement. Such an extension is crucial to analyse the thermodynamic consequences of the measurement process itself.
\fussy

Note that the internal energy $U_\gamma(t)$ cannot in general be obtained as the outcome of a single measurement performed on the system. Yet, it can be reconstructed provided the system's Hamiltonian $H_{\cal S}(t)$ and the trajectory $\gamma$ are known. We underline that reconstructing quantities is often required in stochastic thermodynamics, e.g. heat and work exchanged along classical trajectories are inferred from the record of the system’s evolution.\\

$U_\gamma(t)$ is conserved for an isolated quantum system. In such situation, the system's Hamiltonian is time-independent (no drive), and no measurement is performed. Therefore, the system's evolution is ruled by the Schr\"odinger equation: 
\bb
\dfrac{d \ket{\psi_\gamma(t)}}{dt} = -\dfrac{i}{\hbar} H_{\cal S}  \ket{\psi_\gamma(t)}. \label{eq:SE}
\ee 
\noindent The variation of $U_\gamma(t)$ is then given by:
\bb
\dot U_\gamma(t) &=& \dfrac{d \bra{\psi_\gamma(t)}}{dt} H_{\cal S} \ket{\psi_\gamma(t)} + \text{h.c.}\nonumber\\
&=&   \dfrac{i}{\hbar} \bra{\psi_\gamma(t)} H_{\cal S}^2 \ket{\psi_\gamma(t)} + \text{h.c.}\nonumber\\
&=& 0,
\ee
\noindent where \text{h.c.} refers to the hermitic conjugate of previous term.\\

In the two next sections, we present two ways by which system $\cal S$ can be opened. Each corresponds to an interaction with one of the two classical entities introduced above. The operator $\cal O$ induces a controlled variation of energy through the time-dependence in $H_{\cal S}$: it therefore provides work to $\cal S$. The measuring apparatus $\cal M$ induces stochastic and therefore uncontrolled energy changes which are identified to heat.  

\subsection{Work injected by the drive}

When the system is driven such that $H_{\cal S}$ depends on time, the internal energy is not conserved anymore. The increment of work injected by the drive between $t$ and $t+dt$ is defined as:
\bb
 \delta W_\gamma(t) \underset{\text{def.}}{=}  dt \moy{\dot H_{\cal S}(t)}_{\psi_\gamma(t)},\label{d:dW}
\ee 

\noindent where $\moy{\cdot}_\psi$ denotes the expectation value in state $\ket{\psi}$ and $\dot H_{\cal S}(t)$ is the time-derivative of the total system's Hamiltonian \footnote{It is frequent to express the dynamics of a driven system in a time-dependent frame (e.g. the frame rotating at the laser frequency used in quantum optics, or the interaction picture). The transformation from the lab frame to another frame generally affects the way the system Hamiltonian depends on time. We stress that the operator $H_{\cal S}(t)$ has to be written in the Schr\"odinger picture, and in the lab frame before the time-derivative in Eq.\eqref{d:dW} is computed.}. $\delta W_\gamma(t)$ vanishes if $H_{\cal S}$ is time independent (no drive). The total work $W[\gamma]$ exchanged along trajectory $\gamma$ is obtained by summing up all the work increments. In the case where ${\cal S}$ is solely coupled to the operator performing the drive, we check that: $\dot U_\gamma(t) = \delta W_\gamma(t)/dt$.

\subsection{Energy change induced by measurement: quantum heat}
\label{s:DefQq}

We now focus on the effect of a projective measurement performed by ${\cal M}$ at time $t_n$, $n\in \llbracket 1,N\rrbracket$. We denote time $t_n^+$ the time just after the measurement. The measurement is considered to be instantaneous, namely $t_n^+-t_n$ is much smaller than any other time-scale of the system's dynamics, and $H_{\cal S}$ is considered as constant during the measurement. 

The state of the system just before the measurement is $\ket{\psi_\gamma(t_n)}$, of internal energy $U_\gamma(t_n) = \moy{H_{\cal S}}_{\psi_\gamma(t_n)}$. Right after the measurement, the system is in one of the eigenstates $\ket{m_{r_\gamma(t_n)}}$ of $\hat M$ depending on the stochastic measurement outcome $r_\gamma(t_n)$. This eigenstate has internal energy $U_{r_\gamma(t_n)} = \moy{H_{\cal S}(t_n)}_{m_{r_\gamma(t_n)}}$ which in general differs from $U_\gamma(t_n)$. 

In this dissertation, we call quantum heat increment $\delta Q_{\text{q},\gamma}(t_n)$ the variation of internal energy induced by such a measurement process. Namely, the increment of quantum heat at time $t_n$ when outcome ${r_\gamma(t_n)}$ is obtained is:
\bb
\delta Q_{\text{q},\gamma}(t_n) \underset{\text{def.}}{=}  U_{r_\gamma(t_n)} -U_\gamma(t_n).\label{d:Qq}
\ee
The existence of this quantity has been noticed \cite{Manzano15,Brandner15,Abdelkhalek16} and tracked \cite{Alonso16,Yi17} in various articles under different names. In this thesis we chose to call it ``quantum heat'' which aknowledges several properties of this key quantity presented in this chapter and the next one. In particular:
\begin{itemize}
\item The existence of this quantity is specific to the quantum world. Indeed, it is a byproduct of the effect of quantum measurement on the system's state, a.k.a. wave function collapse, which transforms a pure state $\ket{\psi_\gamma(t_n)}$ into another pure state $\ket{m_{r_\gamma(t_n)}}$. This internal energy variation vanishes if the states $\ket{\psi_\gamma(t_n)}$ and $\ket{m_{r_\gamma(t_n)}}$ have the same internal energy. This happens in particular when $\ket{\psi_\gamma(t_n)}$ is one of the eigenstates of $\hat M$ such that measurement has no effect. Reciprocally, a non-zero internal energy change requires $\ket{\psi_\gamma(t_n)}$ to be a superposition of eigenstates of $\hat M$.
\item The energy change is stochastic and therefore uncontrolled. As shown in next section, the measurement process is irreversible and associated with an entropy production.
\item A non zero quantum heat increment at time $t_n$ can be found even if the system's Hamiltonian is time independent, i.e. even if the system is not driven.
\end{itemize}

When considering a single realization of a protocol involving measurements, the sum of all the quantum heat increments will be denoted $Q_\text{q}[\gamma]$. 

\subsection{Summary}

When considering a protocol involving some drive $H_\text{d}(t)$ and projective measurements of $\hat M$ at discrete times $\{t_n\}_{0\leq n \leq N}$, with $t_i = t_0$ and $t_f = t_N$, we can express the first law of thermodynamics for a single realization of the protocol, yielding trajectory $\gamma$:
\bb
\Delta U[\gamma] = U_\gamma(t_f)-U_\gamma(t_i) = W[\gamma] + Q_\text{q}[\gamma] \label{d:1stLaw},
\ee
\noindent The work and heat exchanged along trajectory $\gamma$ fulfil: 
\bb
W[\gamma] &=& \int_{t_i}^{t_f} dt \moy{\dot H_\text{d}(t)}_{\psi_\gamma(t)}\\
Q_\text{q}[\gamma] &=& \sum_{n = 1}^N \left(U_{r(t_n)} - U_\gamma(t_n) \right),
\ee

\noindent with $U_{r(t_n)} = \bra{m_{r(t_n)}}H_{\cal S}(t)\ket{m_{r(t_n)}}$.

\section{Origin of the quantum heat}
\label{s:OriginQq}

Up to now, we have considered the measuring apparatus ${\cal M}$ as a black box. As evidenced above, such description implies that a measuring device may exchange energy with the system which is measured. One can then wonder where this energy comes from, what physical system provides this amount of energy during the measurement process. 

The answer actually depends on the specific situation under study. It is enlightening to analyse two examples illustrating two possible microscopic origins for the quantum heat. In both cases, the system is a Qubit of Hamiltonian $H_0 = \hbar\omega_0 \ket{e}\bra{e}$, where $\ket{e}$ (resp. $\ket{g}$) refers to its excited (resp. ground) energy eigenstate. The Qubit is initially in its ground state $\ket{g}$. The protocol consists of only two steps: the preparation of state of the Qubit, immediately followed by a projective measurement of observable $\hat M$. Consequently, the system follows a two-state trajectory $\gamma$ fully characterized by a single measurement outcome $r[\gamma] \equiv r_\gamma(t_1)$.

We represent a Qubit state as a point on the surface of the Bloch sphere. A general state is characterized by a unit vector $\textbf{n} = (\theta,\phi)$, written as a function of the colatitude $\theta$ and the longitude $\phi$ (see Fig.\ref{f:ExQED}a):

\bb
\ket{\psi_\text{q}(\theta,\phi)} \underset{\text{def.}}{=} \cos(\theta/2)e^{i\phi/2}\ket{e} + \sin(\theta/2)e^{-i\phi/2}\ket{g} \label{d:PsiThetaPhi}.
\ee

\subsection{Example 1: measurement of the energy of a Qubit}
\label{s:ExQED1}

\begin{figure}[h!]
\vspace{0.2cm}

\begin{center}
\includegraphics[width=0.7\textwidth]{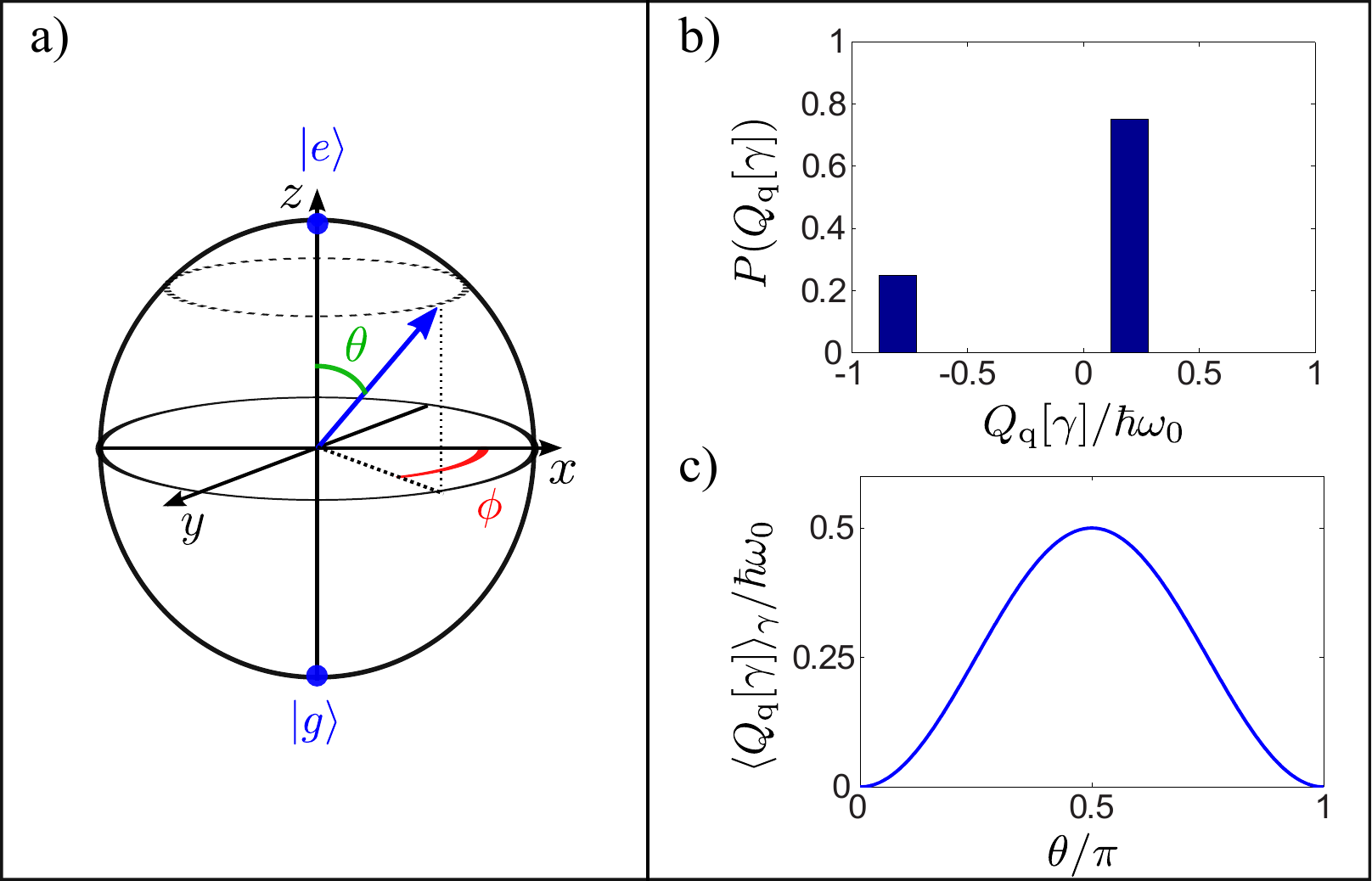}
\end{center}
\caption[Readout of a Qubit]{Readout of a Qubit. a) A pure state of a Qubit can be represented as a vector pointing at the surface of the Bloch sphere, and can therefore be characterized by the colatitude $\theta$ and the longitude $\phi$. b) Normalized probability distribution $P(Q_\text{q}[\gamma])$ of the quantum heat during the readout of a Qubit prepared in state $\ket{+_\theta}$ with $\theta = \pi/3$. c) Average quantum heat received by the Qubit during the measurement of $\sigma_\theta$. The qubit is initially in state $\ket{g}$.\label{f:ExQED} }
\end{figure}

We first consider the following protocol:

\begin{enumerate}
\item\textbf{Preparation}: A Qubit of Hamiltonian $H_0 = \hbar\omega_0\ket{e}\bra{e}$, initially in its ground state $\ket{g}$ (internal energy $0$), is prepared in the state 
\bb
\ket{+_\theta} \equiv \ket{\psi_\text{q}(\theta,0)} = \cos(\theta/2)\ket{e}+\sin(\theta/2)\ket{g}.
\ee
\noindent of internal energy $\hbar\omega_0 \cos^2(\theta/2)$ via a coupling to a resonant drive which induces a rotation in the Bloch sphere around the $y$-axes of an angle $\pi-\theta$. For a drive in a coherent state of amplitude much larger than $1$, the coupling can efficiently be modelled by a time-dependent Hamiltonian term in the Qubit Hamiltonian \cite{CCT} $H_y(t) = i(\hbar g/2) (e^{i\omega_0 t}\sigma_- - e^{-i\omega_0 t}\sigma_+)$, with $\sigma_- = \ket{g}\bra{e} = \sigma_+^\dagger$. After a time of interaction $t_{\pi-\theta} = (\pi-\theta)/g$, the Qubit is in state $\ket{+_\theta}$\footnote{This state is not an eigenstate of $H_0$ and therefore evolves in time. This free evolution of the Qubit is a rotation around $z$-axis which conserves the internal energy}. This step is deterministic and is associated with a work cost provided by the drive that can be computed from Eq.\eqref{d:dW}: 
\bb
W_\text{prep.} =\cos^2(\theta/2)\hbar\omega_0.
\ee
\item\textbf{Measurement}: Then the observable $\hat M = \sigma_z = \ket{e}\bra{e}-\ket{g}\bra{g}$ of the Qubit is measured. This measurement has two possible outcomes $r[\gamma]\in\{1,-1\}$. With probability $p_1 = \cos^2(\theta/2)$ (resp. $p_{-1} = \sin^2(\theta/2)$), outcome $1$ (resp. $-1$) is found, such that the Qubit is projected onto $\ket{e}$ (resp. $\ket{g}$). During this step, no work is performed, and the increment of quantum heat (which also corresponds to the total quantum heat because there is only one time step in the trajectory) reads:

\bb
Q_{\text{q}}[\gamma] = \left\{\begin{array}{c}
\sin(\theta/2)^2\hbar\omega_0, \quad r[\gamma]= 1 \\ 
-\cos(\theta/2)^2\hbar\omega_0, \quad r[\gamma]= -1
\end{array} \right.
\ee

The probability distribution of quantum heat along such a protocol is plotted in Fig.\ref{f:ExQED}b for $\theta = \pi/3$. As there are only two different trajectories (corresponding to the two measurement outcomes), the distribution contains two peaks except for $\theta = 0$ or $\theta = \pi$. These two cases correspond to a Qubit actually prepared in $\ket{e}$ or $\ket{g}$, such that measurement has no effect and the quantum heat distribution is delta-peaked at zero.

\end{enumerate}

Note that in this example, the average of the quantum heat $\langle \delta Q_{\text{q},\gamma} \rangle_\gamma$ over the stochastic trajectories is $0$. This a consequence of the fact that the measured observable commutes with the Hamiltonian. A non-zero average quantum heat may be obtained by measuring an observable which does not commute with the Hamiltonian, a situation which is studied in the next section.

The origin of the quantum heat in this example can be made explicit by quantizing the drive used for the preparation. We now consider as the quantum system under study the joint drive-Qubit system of Hamiltonian $H_\text{qd} = H_0 + \hbar\omega_\text{d}a^\dagger a + (\hbar g/2)(a \sigma_+ + a^\dagger \sigma_-)$, where $a$ is the lowering operator of the drive mode and $\omega_\text{d}$ its frequency. Initially, this bipartite system is in state $\ket{g}\otimes\ket{\alpha}$, where $\ket{\alpha}$ is the coherent state of complex amplitude $\alpha$\footnote{With the chosen convention, $\alpha$ has to be a pure imaginary number to induce the correct rotation around the $y$-axis.}. At time $t_{\pi-\theta}$, the system is in an entangled state 
\bb
\ket{\Psi_{\pi-\theta}} = \cos(\theta/2)\ket{e,\alpha_e} + \sin(\theta/2)\ket{g,\alpha_g}.
\ee
As the total Hamiltonian conserves the number of excitations in the total system, states $\ket{\alpha_{e/g}}$ fulfill $\bra{\alpha_e} a^\dagger a \ket{\alpha_e} =  \bra{\alpha_g} a^\dagger a \ket{\alpha_g} -1$. 

The classical behavior of the drive is obtained if the initial coherent state contains a large number of photons, i.e. $\vert\alpha\vert^2 \gg 1$. In this limit, the two states of the drive involved in the entangled state are almost undistinguishable $\vert\bra{\alpha_e}\alpha_g\rangle\vert^2 \sim 1$. The Qubit and the driving field are weakly entangled, and the reduced Qubit state can be approximated by $\text{Tr}_\text{d}\{\ket{\Psi_{\pi/2}}\bra{\Psi_{\pi/2}}\} \simeq \ket{+_\theta}\bra{+_\theta}$, which justifies the semi-classical description used before.

We now consider that the measurement of $\sigma_z\otimes \idop_\text{d}$ is performed on the total entangled state $\ket{\Psi_{\pi-\theta}}$ instead of the Qubit's state $\ket{+_\theta}$. We find that the total energy of the joint system is conserved between before and after the measurement. However, a quantum of energy that was delocalized between the Qubit and the drive mode in state $\ket{\Psi_{\pi-\theta}}$ has been relocalized either in the Qubit (if $\ket{e,\alpha_e}$ is obtained), or in the drive (if $\ket{g,\alpha_g}$ is obtained). The quantum heat found when the drive is treated classically accounts for a such a stochastic energy relocalization. Therefore, in this example, the quantum heat corresponds to energy provided by the drive. \\

\subsection{Example 2: measurement of a Qubit in a tilted basis.}
\label{s:ExQED2}

In this second example, we consider that no preparation is applied such that the Qubit remains in state $\ket{g}$. Moreover, the measured observable is now $\sigma_\theta = \ket{+_\theta}\bra{+_\theta}-\ket{-_\theta}\bra{-_\theta}$. We introduce $\ket{-_\theta} = \ket{\psi_\text{q}(\pi-\theta,\pi)}$ satisfying $\bra{+_\theta} -_\theta \rangle = 0$. As in the previous example, the quantum heat distribution carries two peaks. Interestingly, as long as $\theta$ is not $0$ or $\pi$, the quantum heat is non zero on average (see Fig.\ref{f:ExQED}c). The measurement in this case induces a net exchange of energy, not only fluctuations. Moreover, the average quantum heat is always positive: the measurement process provides energy to the system. In Section \ref{s:MPE}, we exploit this mechanism to propose an engine converting quantum heat into work.\\

Once again, a microscopic origin of the quantum heat can be derived from a more precise description of the entities interacting with the Qubit. Contrary to the previous example, the Qubit is not coupled nor correlated to any device but ${\cal M}$, and therefore the ``hidden'' energy source must be contained in the measuring apparatus. In order to find it out, we consider a practical implementation of the measurement of $\sigma_\theta$ which is based on the following protocol, implemented routinely in cavity QED or circuit QED setups:
\begin{enumerate}
\item A laser pulse inducing a counter-clockwise rotation of angle $\theta$ around the $y$-axis is sent on the Qubit. This pulse is used to map the states $\ket{\pm_\theta}$ onto states $\ket{e}$ and $\ket{g}$.
\item A projective measurement of $\sigma_z$ is performed.
\item A second laser pulse is used to map back the states $\ket{e}$ and $\ket{g}$ onto $\ket{+_\theta}$ and $\ket{+_\theta}$ (clockwise rotation of angle $\theta$ around the $y$-axis).
\end{enumerate}

Let us now analyze the energy exchanges. During step $1$ and $3$, the pulses deterministically exchange energy with Qubit (work). At the end of step $1$, the Qubit is in the state $\ket{\psi_\text{q}(\pi-\theta,0)}$ which is a superposition of $\ket{e}$ and $\ket{g}$. Consequently in step $2$ when $\sigma_z$ is measured, there is a stochastic energy change under the form of quantum heat. Steps $1+2$ are reminiscent of to the protocol treated in the previous section: we can therefore conclude that the pulse used in step $1$ also provides the energy the Qubit receives in step $2$.  

Finally, the total quantum heat term received by the Qubit during the measurement of $\sigma_\theta$ can be split into three terms accounting for energy exchanged with the laser pulses.

\subsection{Discussion}

In these two examples, we have been able to analyze which entity provides energy to the system Qubit during a projective measurement process. The two possible origins found in these examples are actually the only two possibilities: either the energy is provided by a classical system still correlated with the quantum system under study, either the energy is provided by sources used to implement the measurement process (in other words by the measuring apparatus itself).

In both examples, inferring the origin of the quantum heat has required to change our description of the situation under study. In the first case we have modified the boundaries of the system. In the second case, we have split the measurement process into three processes of different nature and we have introduced different subparts of the measuring apparatus. As a result, the form of the energy exchanges have changed, and what was first labeled as quantum heat has been transformed into an energy transfer inside a bipartite quantum system (example $1$), or partially converted into work (example $2$). 

We stress that this issue is not specific to quantum thermodynamics and is already present in classical thermodynamics when trying to define heat. Indeed, the splitting of internal energy variations between work and heat in classical protocols depends on the boundaries of the system, and on a convention about which process is controlled by the operator or not. The apparent randomness and the uncontrolled nature of a classical heat exchange follows from a lack of information about the actual microscopic dynamics, and heat is an effective quantity dependent on the chosen level of description.

Eventually, once the boundaries of the system and the description of the measuring apparatuses available to the operator are fixed, the quantum heat appears as a necessary byproduct of measurement postulate. Note that the status of this postulate is still debated nowadays, including with the philosophical point of view (See for instance Refs.\cite{Auffeves16,Zurek15,Auffeves15}). Ultimately, these different perspectives reflect the various interpretations of quantum mechanics, which coexist without altering the efficiency of the theory. 

In the remaining parts of this dissertation, we adopt the operational point of view presented in Section \ref{s:DefQq} and do not systematically question the microscopic origin of the quantum heat.

\section{Entropy produced during a measurement}
\label{s:EntropyMeas}
Because of the random nature of quantum measurement, a production of entropy can be associated to a measurement process. In this section, we explain how to compute this entropy production in the simple case of a protocol only involving the preparation of an initial state and a single projective measurement. More details and extensions are studied in Chapter \ref{Chapter2}.

As in Section \ref{s:OriginQq}, we focus on a protocol generating a two-point trajectory. The quantum system ${\cal S}$ is initially in a pure state $\ket{\psi_i} = \ket{\psi_\gamma(t_i)}$. This state might be the same for every realization of the process, or might be drawn with probability $p_i$ from a mixture $\rho_i$ of eigenstates of observable $\hat A$:
\bb
\rho_i = \sum_i p_i \ket{\psi_i}\bra{\psi_i}.
\ee
 \sloppy Immediately after\footnote{Free evolution of the system is neglected here.}, ${\cal S}$ is coupled to the apparatus measuring observable $\hat M$. The trajectory $\gamma$ is determined by the initial state $\ket{\psi_i}$ and the single measurement outcome $r[\gamma]$ of this measurement. Outcome $r[\gamma]$ is obtained with the conditional probability $p(r[\gamma]\vert i) = \left\vert\langle m_{r[\gamma]}\vert\psi_i\rangle\right\vert^2$. The state after the measurement is denoted \linebreak \mbox{$\ket{\psi_f} = \ket{m_{r[\gamma]}}$}. Taking into account the selection of the initial state, the probability of the trajectory $\gamma = \{\ket{\psi_i},\ket{\psi_f}\}$ is:
\bb
p[\gamma] = p({r[\gamma]}\vert i)p_i \label{d:pgamma}.
\ee \fussy\\

In analogy with classical stochastic thermodynamics, the entropy production $\Delta_\text{i} s[\gamma]$ during a single realization of a protocol corresponding to the stochastic trajectory $\gamma$ is defined by the following formula \cite{Seifert05}:
\bb
\Delta_\text{i} s[\gamma] \underset{\text{def.}}{=} k_\text{B}\log\dfrac{p[\gamma]}{\tilde p[\tilde\gamma]}\label{d:Disgamma},
\ee
\noindent where $p[\gamma]$ is the probability of the trajectory $\gamma$ in this protocol, and $\tilde p[\tilde\gamma]$ is the probability that the time-reversed trajectory $\tilde\gamma$ corresponding to $\gamma$ occurs when the time-reversed protocol is performed.


In this section, the time-reversed protocol is simple. The initial state of the system is drawn from the mixture of the different states obtained at the end of the direct protocol, i.e. 
\bb
\rho_f = \sum_f p_f\ket{\psi_f}\bra{\psi_f} = \sum_\gamma  p[\gamma]\ket{m_{r[\gamma]}}\bra{m_{r[\gamma]}}.
\ee 
Then a measurement of the observable $\hat A$ is performed, projecting the system onto the eigenstates associated with the obtained outcome.\\

Note that when the set of initial states $\{\ket{\psi_i}\}$ does \emph{not} correspond to the entire set of eigenstates of $\hat A$, there is a finite probability, that we denote $\bar p$, that after the time-reversed process, the system will be in a state which is not one of the possible initial states. This probability plays a role in the values that the entropy production associated with the measurement process can take as we explain below.

The entropy production fulfills a fluctuation theorem. To prove it, we first compute: 
\bb
\moy{e^{-\Delta_\text{i}s[\gamma]/k_\text{B}}}_\gamma &=& \sum_\gamma p(\gamma)\dfrac{\tilde p[\tilde\gamma]}{p(\gamma)}\nonumber\\
&=& \sum_\gamma \tilde p[\tilde\gamma]\\
&=& 1 - \bar p.\label{eq:AbsIFT0}
\ee
To compute the last line, we have used that the quantity $\sum_\gamma \tilde p[\tilde\gamma]$ represents the total probability that the time-reversed trajectories end in any of the states $\{\ket{\psi_i}\}$ of the initial mixture. It therefore corresponds to $1-\bar p$ according to our definition of $\bar p$.

In the case where the initial mixture exhibits a finite probability $p_i >0$ for every eigenstate of $\hat A$, the probability $\bar p$ vanishes and Eq.\eqref{eq:AbsIFT0} takes the form of the so-called Integral Fluctuation Theorem (IFT)\cite{Seifert05}:

\bb
\moy{e^{-\Delta_\text{i}s[\gamma]/k_\text{B}}}_\gamma = 1.
 \label{d:IFT}
\ee
This is for example the case when the system is initially in a thermal distribution, such that $\hat A$ is the Hamiltonian of the system. The Second Law of thermodynamics automatically follows from the IFT owing to the convexity of the function exponential:
\bb
\moy{\Delta_\text{i} s[\gamma]}_\gamma \geq 0 \label{d:2ndLaw1}.
\ee\\

Conversely, when $\bar p$ is strictly positive, the Fluctuation Theorem reads:
\bb
\moy{e^{-\Delta_\text{i}s[\gamma]/k_\text{B}}}_\gamma = 1 - \bar p.
\ee 
leading to an entropy production fulfilling:
\bb
\moy{\Delta_\text{i} s[\gamma]}_\gamma \geq  - k_\text{B}\log(1-\bar p) > 0\label{d:2ndLawAbs}.
\ee
In this case, the average entropy production is strictly positive and the transformation cannot be reversible. This phenomenon has been called ``absolute irreversibility'' and studied extensively in Refs.\cite{Murashita14,MurashitaThesis,Funo15}.\\

We now identify the time-reversed trajectory $\tilde\gamma$ associated with $\gamma$ as the trajectory starting in $\ket{\psi_f}$ and ending in $\ket{\psi_i}$. The probability of this trajectory is \linebreak\mbox{$\tilde p(\tilde\gamma) = p_f \tilde p(\tilde r[\tilde\gamma]\vert f)$}, where $p_f = \bra{\psi_f}\rho_f\ket{\psi_f}$ is the probability to start in state $\ket{\psi_f}$ and $p(\tilde r[\tilde\gamma]\vert f) = \vert\bra{\psi_i}\psi_f\rangle\vert^2$ is the conditional probability to get the corresponding measurement outcome. As the $\ket{\psi_i} = \ket{m_{r[\gamma]}}$, we have $p(\tilde r[\tilde\gamma]\vert f) = p(r[\gamma]\vert i)$.

Finally, the entropy produced during $\gamma$ reads:
\bb
\Delta_\text{i} s[\gamma] &=& k_\text{B}\log\dfrac{p_i}{p_f} + k_\text{B}\log\dfrac{p(r[\gamma]\vert i)}{p(r[\tilde\gamma]\vert f)}\noindent \\
&=& k_\text{B}\log\dfrac{p_i}{p_f}.\label{eq:DSVN}
\ee

Using Eq.\eqref{eq:DSVN}, we compute the average of $\Delta_\text{i} s[\gamma]$ over the trajectories $\gamma$:
\bb
\moy{\Delta_\text{i} s[\gamma]}_\gamma &=& k_\text{B}\sum_\gamma p[\gamma]\log\dfrac{p_i}{p_f}\nonumber\\
&=& k_\text{B}\sum_i p_i \sum_{r[\gamma]} p(r[\gamma]\vert i)\log p_i - k_\text{B}\sum_f p_f \sum_{r[\gamma]} p(r[\gamma]\vert f)\log p_f\nonumber\\
&=& k_\text{B}\sum_i p_i \log p_i - k_\text{B}\sum_f p_f \log p_f\nonumber\\
&=& k_\text{B}\left(S_\text{VN}[\rho_f] - S_\text{VN}[\rho_i]\right). \label{eq:DiSVN}
\ee
where $S_\text{VN}[\rho] = -\text{Tr}\{\rho\log\rho\}$ is the Von Neumann entropy of the mixed state $\rho$.

 Eq.\eqref{eq:DiSVN} relates the thermodynamic entropy $\Delta_i s[\gamma]$ to the information-theoretic entropy $S_\text{VN}$: in the situation under study, the irreversibility of the transformation is directly caused by the uncertainty about the measurement result and therefore the output state of the system, which is quantified by the Von Neumann entropy of the mixture of the possible final states. 
 
 The second law for a quantum system undergoing quantum measurement finally reads:
\bb
\Delta S_\text{VN} = S_\text{VN}[\rho_f] - S_\text{VN}[\rho_i] \geq 0 \label{d:2ndLaw2}
\ee

This 2nd law can be interpreted in several ways:
\begin{itemize}
\item The entropy produced during the measurement reflects the fact that prior to the measurement, the operator has uncertainty about what will be the system's state after the measurement (in contrast with a unitary evolution). We emphasize that once the measurement is performed and the measurement outcome is read, the state of the system is a perfectly known pure state of zero Von Neumann entropy. Nonetheless, any thermodynamic protocol including a measurement process yields a random trajectory of the system, and therefore some uncertainty limiting the thermodynamic efficiency of the protocol (e.g. the amount of work that can be extracted). This uncertainty is quantified by Eq.\eqref{eq:DiSVN}.
\item When starting in a pure known state $\ket{\psi_i}$ ($S_\text{VN}[\rho_i] = 0$), quantum measurement induces irreversibility quantified by $S_\text{VN}[\rho_f] \geq 0$. The only situation in which measurement is reversible in that case is when $\ket{\psi_i}$ is an eigenstate of the measured observable $\hat M$, such that the measurement outcome is certain.
\item When the initial state is drawn from a mixture $\rho_i$, quantum measurement produces entropy except if $\rho_i$ is diagonal in the eigenbasis of the measured observable (e.g. if $\hat A = \hat M$).   
\item The mixture of maximal entropy is $\rho_i = \idop/d_{\cal S}$. This state carries the maximum uncertainty about the future measurement result. A measurement performed on such a mixture is reversible in the sense that it does not increase the uncertainty about the future state of the system.
\end{itemize} 

We can now wonder what is the link between the entropy produced in a quantum measurement and the quantum heat $Q_\text{q}$ introduced in previous section. The expression of the average entropy production Eq.\eqref{eq:DiSVN} differs from the case of an isothermal transformation. There seems there is not in general a term proportional to the quantum heat appearing in the entropy produced during a measurement\footnote{Furthermore, there is no temperature involved.}. However, we note that a non-zero histogram of quantum heat corresponds to a non-zero average entropy production. For instance, in the example of the readout of a Qubit initially in a superposition $\ket{+_\theta}$ of its energy eigenstates (cf section \ref{s:ExQED1}), the average entropy production is $k_\text{B} H_\text{Sh}[\cos^2(\theta/2)]$ where we have defined the Shannon entropy of the distribution $\{p,1-p\}$:
\bb
H_\text{Sh}[p] \underset{\text{def.}}{=} - p\log p - (1-p)\log(1-p) \label{d:Shannonp}
\ee
It is therefore positive for every value of $\theta$ except for $0$ or $\pi$ where it vanishes. These two values of $\theta$ are the only ones for which the probability distribution of the quantum heat is delta-peaked at $0$.

Note that the measurement process in this case is ``absolutely irreversible'' as there is a finite probability $\bar p = \sin(\theta)/2$ that the Qubit is projected onto state $\ket{-_\theta}$ after applying the time-reversed process.



\section{Application: a measurement-powered engine}
\label{s:MPE}

As a proof of principle for the framework introduced above, we now propose an engine in which work is extracted from a measuring apparatus instead of a hot thermal reservoir. This engine is presented in \cite{MPE}. The idea of measurement-based engine cycle is also exploited in \cite{Yi17}.

In this Measurement Powered Engine (MPE), the role of the working agent is played by a Qubit whose ground and excited levels are denoted $\ket{e}$ and $\ket{g}$ respectively. This engine belongs to the class of engines assisted by a Maxwell's demon, i.e. it involves a measurement of the state of the working substance, followed by a feedback step optimizing work extraction.

We first recall some properties about Maxwell's demons and provide an example of such a heat engine. Then we present the principle of the MPE and characterize its performances. Eventually, we propose a concrete implementation based on circuit Quantum ElectroDynamics (QED).

\subsection{Engine assisted by a Maxwell's demon}
\label{s:ThermalDemon}
The name ``Maxwell's demon'' refers to a celebrated thought experiment proposed by J. Maxwell in 1871, originally as a paradox apparently violating the second law of thermodynamics. The main idea is that an intelligent entity (the demon) which would be able to measure the position and the velocity of the particles of a gas in a box, could sort them and decrease their entropy without paying work \cite{Maxwell}. 
If such task could be performed cyclically, it would constitute a violation of the second law of thermodynamics. In particular, it would allow to build an engine extracting work from a single thermal bath. The solution to the paradox emerged after many decades of controversy, in particular owing to articles by L. Szilard \cite{Szilard29}, L. Brillouin, R. Landauer \cite{Landauer61} \& C. Bennett \cite{Bennett82}. 

It is nowadays well-known that a key quantity that has to be included in the physical description of the problem is the entropy of the memory in which the demon stores its measurement results. In order to repeat many times this process, this memory has to be reset periodically, which means decreasing its entropy. This operation is therefore associated to a work cost. The famous result known as Landauer's principle sets that if one has access to a thermal bath at temperature $T$, resetting a memory costs at least an amount of work $k_\text{B} T \log(2)$, where $k_\text{B}$ is Boltzmann's constant. This minimum value is reached in the case of a reset operation being performed quasi-statically, i.e. reversibly \cite{Berut12,DelRio11,Landauer61}. If reset of the memory of the demon is indeed taken into account, the cyclic work extraction is not anymore in contradiction with the second law of thermodynamics. 

This idea can be generalized to a variety of protocols involving a measurement process to optimize work extraction, providing a class of engines that have been studied \cite{Lloyd97,Ruschhaupt06,Strasberg13} and implemented in various systems \cite{Toyabe10,Koski14,KoskiMutual14,Vidrighin16}.


\subsection{Thermally driven Maxwell's demon engine}

Before detailing our proposal, we recall a possible operating mode for an elementary engine extracting work from a hot thermal reservoir owing to a Maxwell's demon. This engine uses a Qubit as working fluid and is presented in \cite{Lloyd97}.

\begin{figure}[h!]
\begin{center}
\includegraphics[width=0.8\textwidth]{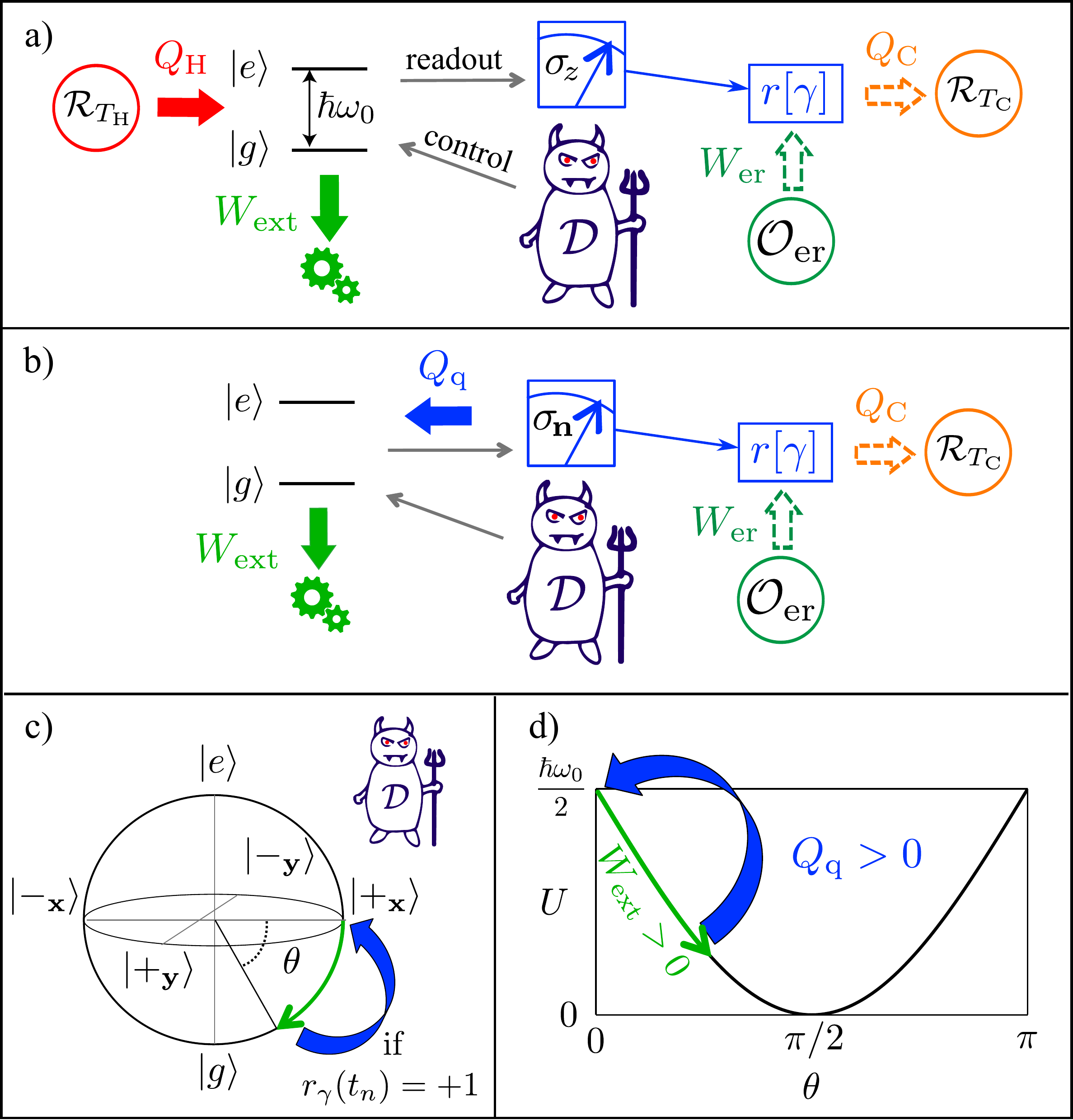}
\end{center}
\caption[Measurement Powered Engine]{Measurement Powered Engine. a) Thermally-driven engine Maxwell's demon engine. The demon ${\cal D}$ assists the extraction of heat $Q_\text{H}$ from the hot thermal reservoir ${\cal R}_{T_H}$ owing to measurements of $\sigma_z$ (readout) and feedback (control). The measurement record $r[\gamma]$ is stored in a classical memory which is erased at the end of every cycle. The erasure process involves an energy source ${\cal O}_\text{er}$ and a cold bath ${\cal R}_{T_C}$. b) Measurement Powered Engine (MPE): The demon's apparatus now performs measurement of $\sigma_\text{n}$. This measurement provides quantum heat $Q_\text{q}$ to the Qubit which is then extracted under the form of work $W_\text{ext}$. c) Cycle of the MPE in the Bloch sphere. Between two measurements the evolution due to a resonant driving field is a rotation around the $y$ axis at Rabi frequency $\Omega$. The measurement projects the Qubit onto states $\ket{\pm_\textbf{x}}$. d) Evolution of the Qubit's internal energy at the end of the work extraction step $U(\tau_\text{W})=\bra{\psi_\text{q}(\tau_\text{W})}H_0\ket{\psi_\text{q}(\tau_\text{W})}$ as a function of the Rabi angle $\theta = \Omega \tau_w$ characterizing the Bloch vector rotation for a duration $\tau_w$ between two successive measurements. During the Hamiltonian evolution the work $W_\text{ext} = \hbar\omega_0/2- U(\tau_\text{W})$ is extracted into the resonant driving field. The measurement step provides the amount of quantum heat $Q_\text{q} =W_\text{ext}$ back to the Qubit whatever the measurement outcome.\label{f:MPE}}
\end{figure}

The working agent is a Qubit whose energy eigenstates are $\ket{g}$ and $\ket{e}$, and transition frequency $\omega_0$. The bare system Hamiltonian is $H_0 = \hbar \omega_0\ket{e}\bra{e}$. Before the cycle starts, the Qubit is in the pure state $\ket{g}$ of zero internal energy and Von Neumann entropy. The cycle contains four steps:

\begin{enumerate}
\item {\it Thermalization}: At initial time $t_i=0$ the Qubit is coupled to a hot bath at temperature $T_\text{H}$ and thermalized in the Gibbs state $\rho_i=Z_\text{i}^{-1} \exp(- H_0/k_\text{B}T_\text{H} )$. $Z_\text{i}$ is the partition function, $k_\text{B}$ is Boltzmann's constant. The temperature is chosen sufficiently large $k_\text{B} T_\text{H}\gg \hbar \omega_0$ such that $\rho_i=\idop/2$. As a consequence, during this step the Von Neumann entropy of the Qubit is $S_\text{VN}[\rho_i] = \log 2$, and the mean internal energy is $U_i=\hbar \omega_0/2=Q_\text{H}$, where $Q_\text{H}$ is the average heat extracted from the hot bath in this step.\\
\item {\it Readout}: The Qubit is then decoupled from the bath and a demon ${\cal D}$ performs a projective measurement of the observable $\sigma_z$ and stores the outcome in its memory. As a first step, we shall not focus on the physical implementation of the demon and treat it as a device extracting and storing information on the Qubit onto some classical memory (readout), and exerting some action on the Qubit, conditioned to the readout (control). The physical durations of the readout, feedback and erasure steps are neglected, and additional energetic costs related to amplification of measurement outcomes are not considered. After the readout step, the states of the demon's memory are classically correlated with the Qubit's states $\ket{g}$ and $\ket{e}$, such that the memory's entropy $S_{\cal D}$ satisfies $S_{\cal D}=S_\text{VN}(\rho_i)=\log(2)$. \\
\item{\it Work extraction}: The work extraction step is triggered if the memory indicates that the Qubit is in the state $\ket{e}$, which happens half of the times since we have considered large temperatures $T_\text{H}$. A convenient way to extract work consists in resonantly coupling the Qubit to a resonant classical drive inducing Hamiltonian $H_y(t) = i(\hbar g/2) (\sigma_- e^{i\omega_0 t} - \sigma^\dagger_- e^{-i\omega_0 t})$, which induces a rotation around the $y$-axis in the Bloch sphere. 
The coupling time $\tau_{\pi}$ is tuned such that the qubit undergoes a $\pi$ pulse that coherently brings the system from state $\ket{1}$ to $\ket{0}$. 
The energy change induced by the $\pi$ pulse is equal to $-\hbar \omega_0$ if the Qubit is found in $\ket{e}$ and zero otherwise, such that the average is $\hbar\omega_0/2$. This decrease of the Qubit's energy quantifies the extracted work, that is used to coherently amplify the propagating driving field of the $\pi$ pulse by one extra photon. In practice, the Qubit could then provide work to power up another processing unit of a quantum machine in photonic or microwave circuits~\cite{Ikonen16}. Note that we consider large enough drive amplitudes, so that the extra photon has negligible effect on the coupling Hamiltonian. Finally, the mean extracted work equals $W_\text{ext} =\hbar \omega_0/2 = Q_\text{H}$. \\
\item {\it Reset of the memory}: The cycle is closed with the erasure of the demon's classical memory. This step is realized by performing a standard Landauer protocol \cite{Landauer61, Berut12,DelRio11}, which requires some work source ${\cal O}_\text{er}$ and cold bath ${\cal R}_{T_\text{C}}$ of temperature $T_\text{C}$ as additional resources (se Fig.\ref{f:MPE}a). When it is performed quasi-statically, erasure costs a minimal work (according to Landauer's principle) $W_\text{er}=-Q_\text{C}=S_{\cal D} T_\text{C}$, where $-Q_\text{C}$ is the heat dissipated in the cold bath during the process. 
\end{enumerate}

We characterize this thermally driven engine with its efficiency. We use the definition proposed in \cite{Lloyd97} which is inspired from standard heat engines like Carnot engine\footnote{According to our definitions, $W_\text{ext}$, $W_\text{er}$ and $Q_\text{H}$ are all positive when the cycles work as an engine.}:
\bb
\eta_\text{th} &\underset{\text{def.}}{=}& \dfrac{W_\text{ext}-W_\text{er}}{Q_\text{H}}\nonumber\\
&=& 1-\dfrac{2k_\text{B}T_\text{C}\log(2)}{\hbar\omega_0}\label{d:etath}
\ee

\noindent This efficiency does not depend on the hot bath temperature $T_\text{H}$ because of the saturation of the Qubit which cannot store more energy than $\hbar\omega_0$. We can check that $\eta_\text{th}$ is lower than the Carnot efficiency $\eta_\text{C} = 1- T_\text{C}/T_\text{H}$ as $k_\text{B}T_\text{H} \gg \hbar\omega_0$. 
As in the classical case, a maximum value of $\eta_\text{cl}=1$ can be reached in the limit $T_\text{C}\to 0$. However the condition of reversibility would then impose a vanishing engine power, since the duration of erasure diverges when it involves a zero temperature reservoir as a consequence of the third law of thermodynamics \cite{Wilks}.

%
%

\subsection{Measurement Powered Engine (MPE)}
\label{s:MPEfeedback}

\subsubsection{Cycle and performances}

We now introduce a new protocol to operate the engine, for which the demon is allowed to perform projective measurements of some arbitrary observable $\sigma_\textbf{n}$ of the Qubit. The eigenstates $\{\ket{+_\textbf{n}},\ket{-_\textbf{n}}\}$ of $\sigma_\textbf{n}$ are the two states of the Bloch sphere associated with the unit vector:
\bb
\textbf{n} = (x_\textbf{n},y_\textbf{n},z_\textbf{n}) = \bigg(\sin(\theta_\textbf{n})\cos(\phi_\textbf{n}),\sin(\theta_\textbf{n})\sin(\phi_\textbf{n}),\cos(\theta_\textbf{n})\bigg).
\ee  
They fulfil (using definition \ref{d:PsiThetaPhi}):
\bb
\ket{+_\textbf{n}} &=& \ket{\psi_\text{q}(\theta_\textbf{n},\phi_\textbf{n})}\\
\ket{-_\textbf{n}} &=& \ket{\psi_\text{q}(\pi-\theta_\textbf{n},\pi - \phi_\textbf{n})},
\ee

As evidenced in Section \ref{s:ExQED1}, this measurement is associated to some exchange of quantum heat which is here the only fuel for our engine. No hot thermal bath is thus required (Fig.~\ref{f:MPE}b). To simplify the calculation, we first present the engine cycle in the case $\textbf{n}=\textbf{x} = (1,0,0)$ and thus $(\theta_\textbf{n},\phi_\textbf{n}) = (\pi/2,0)$. As proven later, it corresponds to the choice which maximizes the efficiency. The measured observable is therefore  $\sigma_x$ of eigenstates $\ket{\pm_\textbf{x}} = (\pm\ket{1}+\ket{0})/\sqrt{2}$. The role of the direction of $\textbf{n}$ is discussed later on. The demon initializes the Qubit in state $\ket{+_\textbf{x}}$ and its memory is initially empty. The initial internal energy of the Qubit is therefore $U_0 = \hbar\omega_0/2$.  A cycle consists in the four following steps:

\begin{enumerate}
\item  {\it Work extraction} The Qubit is coupled to the drive (Hamiltonian $H_y(t)$) during a time $\tau_\text{w}$. Introducing the Rabi angle $\theta = g \tau_\text{w}$ (Fig.~\ref{f:MPE}c), the extracted work reads $W_\text{ext} = \hbar\omega_0/2- U(t_\text{w}) =   \hbar \omega_0 \sin(\theta)/2$. We have introduced the internal energy $U(t_\text{w})$ of the Qubit at the end of the rotation which satisfies $U(t_\text{w}) = \bra{\psi_\text{q}(\pi/2+\theta,0)}H_0\ket{\psi_\text{q}(\pi/2+\theta,0)}$. The extracted work $W_\text{ext}$ is strictly positive as long as $\theta \leq \pi$ (Fig.\ref{f:MPE}d). Each cycle gives rise to the same amount of work, which is extracted from the coherence of the $\ket{+_\textbf{x}}$ state \cite{Cottet17,Uzdin15}. Reciprocally, starting from the state $\ket{-_\textbf{x}}$ would give rise to energy absorption from the drive and negative work extraction. 
\item {\it Readout}. The demon measures the Qubit's observable $\sigma_x$, yielding state $\ket{+_\textbf{x}}$ (resp. $\ket{-_\textbf{x}}$) with probability $\cos^2(\theta/2)$ (resp. $\sin^2(\theta/2)$), which corresponds to a mixture of entropy $k_\text{B}H_\text{Sh}[\cos^2(\theta/2)]$, where the Shannon entropy is defined by Eq.\eqref{d:Shannonp}. The two possible Qubit's states are classically correlated with the states of the demon's memory: Therefore the entropy of the demon's memory satisfies $S_{\cal D} = k_\text{B}H_\text{Sh}[\cos^2(\theta/2)]$. On the other hand, as a result of the particular choice $\textbf{n} = \textbf{x}$, whatever its outcome, the measurement deterministically restores the Qubit's internal energy to its initial value. Indeed, $\bra{\pm_\textbf{x}} H_0 \ket{\pm_\textbf{x}} = \hbar\omega_0/2$, such that the measurement provides an amount of quantum heat $Q_\text{q} = W_\text{ext}$.
Importantly here, measurement plays three roles. First, as in classical Maxwell's demon engines, it allows extracting information on the Qubit's state. Second, and this is a quantum property of the measurement, it erases coherences in the measurement basis and therefore increases the Qubit's entropy.  Finally, it provides energy to the Qubit since the measurement basis does not commute with the bare energy basis. These two last characteristics make the connection between the measurement process and the action of a thermal reservoir. The present proposal relies on this analogy.
\item{\it Feedback}. If the Qubit is found in state $\ket{-_\textbf{x}}$ (outcome $r_\gamma(t_n) = -1$), a feedback pulse prepares the Qubit back in state $\ket{+_\textbf{x}}$. This step has no energy cost, e.g. it can be realized by letting the Qubit freely evolve (Rotation around the $z$-axis of the Bloch sphere) during some appropriate time. At the end of this step the Qubit is prepared back in the pure state $\ket{+_\textbf{x}}$.
\item{\it Erasure}. The classical demon's memory is finally erased to close the cycle. Just like in Eq.\eqref{d:etath}, we consider the minimal bound for the erasure work, which is reached for quasi-static processes. 
\end{enumerate}

\begin{figure}
\begin{center}
\includegraphics[width=0.8\textwidth]{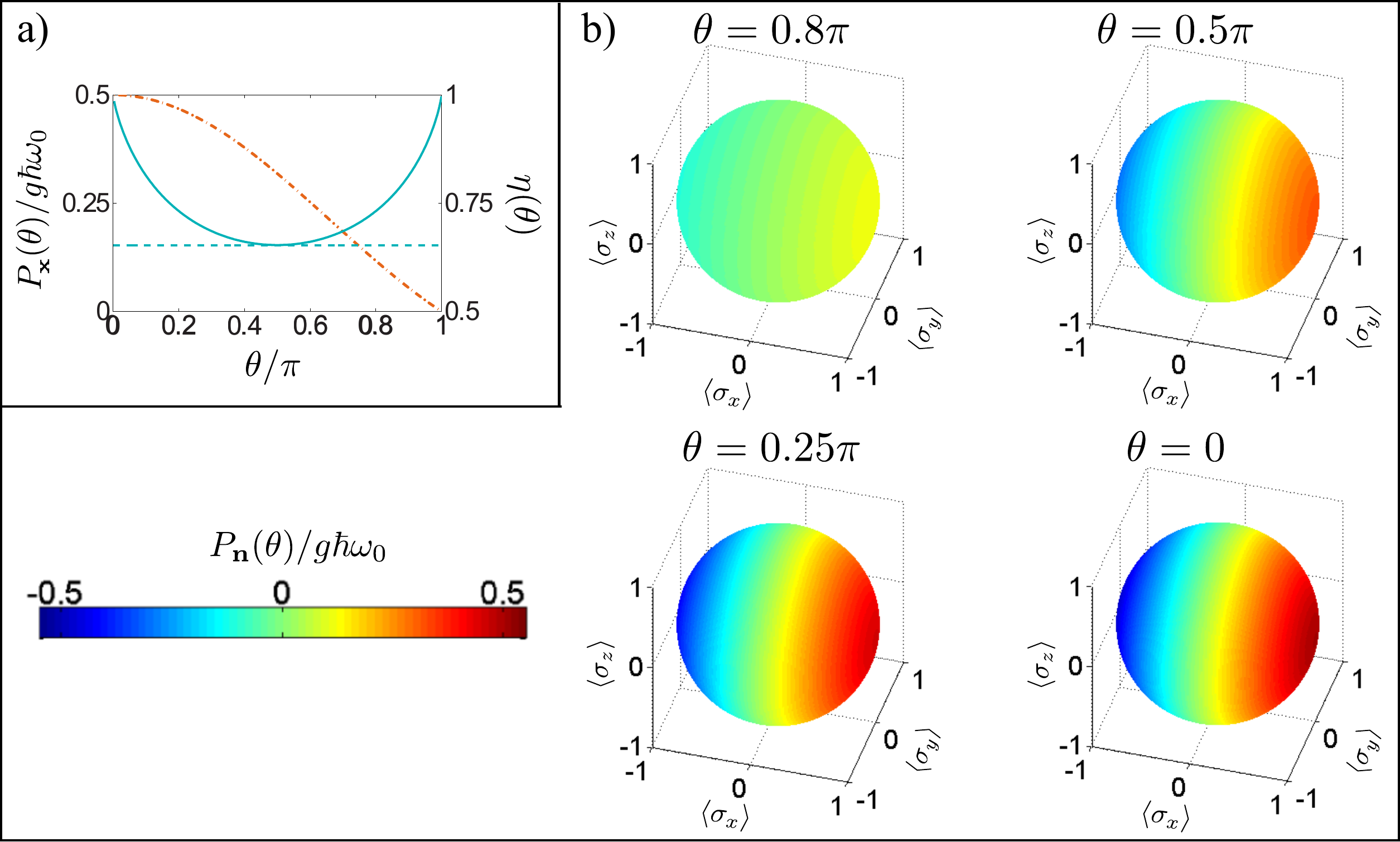}
\end{center}
\caption[Performances of the MPE]{ Performances of the MPE. a) Efficiency $\eta(\theta)$ (solid blue) and normalized power $P_{\textbf{x}}(\theta)/g\hbar\omega_0$ when the working point of the engine is $\ket{+_\textbf{x}}$, as a function of the Rabi angle $\theta$. b) Normalized power $P_\textbf{n}(0)/g\hbar\omega_0$ as a function of the working point in the Bloch sphere.\label{f:MPEPerf}}
\end{figure}

The efficiency of the engine reads:
\begin{equation}
\eta(\theta) = 1-\frac{2 k_\text{B} T_\text{C}}{\hbar \omega_0}\dfrac{H_\text{Sh}[\cos^2(\theta/2)]}{ \sin(\theta) },
\end{equation}

\noindent and is plotted Fig.\ref{f:MPEPerf}a. The minimum value is reached for $\theta=\pi/2$ where $\eta(\pi/2)=\eta_\text{th}$: Here both the thermally driven engine and the MPE provide the same amount of mean extracted work $\hbar\omega_0/2$ and require the same amount of work $W_\text{er} = k_\text{B} T_\text{C} \log(2)$ to erase the demon memory. When $\theta \neq \pi/2$, erasure cost decreases faster than the work extracted per cycle, leading to a larger efficiency than in the thermal case.

We now focus on the extracted power. We first neglect the duration of the feedback and erasure steps, such that the duration of a cycle is $\tau_w = \theta/\Omega$. This allows to compute the extracted power $P_{\textbf{x}}(\theta)$ for an engine involving a measurement of $\sigma_x$:

\begin{equation}
P_{\textbf{x}}(\theta) =g\dfrac{\hbar\omega_0}2 \frac{\sin(\theta)}{\theta}.
\end{equation}

\subsubsection{Zeno limit}

As it can be seen on Fig.\ref{f:MPEPerf}a, the efficiency is maximized in the limit $\theta\rightarrow 0$, where $\eta \rightarrow 1$. This corresponds to the Zeno regime where stroboscopic readouts are performed at a rate faster than the Rabi frequency ($\tau_\text{w} \ll g^{-1}$) such that the Qubit is frozen in state $\ket{+_\textbf{x}}$.

In contrast with thermally-driven engine, the power is finite and even maximal in the Zeno limit. Indeed, although the work extracted per cycle goes to zero in that limit, it behaves $O(\theta)$ like the duration of the cycle, leading to a finite extracted power.

In this regime, the dynamics induced by the measuring apparatus is not stochastic anymore. As a consequence the transfer of energy from the measuring apparatus is not anymore associated with a production of entropy: In this particular case, the energy provided by the measuring apparatus is more similar to work than heat, and can therefore be extracted with perfect efficiency. When operated in the Zeno regime, the cycle under study implements a measurement-controlled transducer.

As the probability of finding $\ket{-_\textbf{x}}$ is vanishingly small in the Zeno regime, the feedback loop may seem redundant: However, it has recently be argued that the measuring apparatus would still need to be reset in its initial reference state in order to be available for next measurement, which in the case of the Zeno regime corresponds to a diverging cost \cite{Abdelkhalek16}. The divergence originates in the linear increase in the number of performed projective measurements the probability of error (i.e. of finding $\ket{-_\textbf{x}}$ instead of $\ket{+_\textbf{x}}$). Maintaining the feedback step in the protocol allows to avoid such increase and therefore to maintain the unity efficiency of the engine.  

Conversely, as soon as a finite angle $\theta$ is considered, the stochasticity of the measurement and the associated increase of the memory's entropy forbid reaching unity efficiency just like for a classical heat engine. 

\subsubsection{Influence of the measurement basis}

We now study the engine's power $P_{\textbf{n}}(\theta)$, as a function of the measurement orientation $\textbf{n}$ in the Bloch sphere. The general expression of the work extracted in the first step is:
\bb
W_\text{ext} = \dfrac{\hbar\omega_0}{2}\big((1-\cos(\theta))\cos(\theta_\textbf{n})+\sin(\theta)\sin(\theta_\textbf{n})\cos(\phi_\textbf{n})\big).
\ee

In addition, for vectors $\textbf{n}$ pointing outside the equator of the Bloch sphere ($\theta_\textbf{n} \neq \pi/2$), the feedback step has a non zero work cost $W_\text{fb}$ which must be subtracted from $W_\text{ext}$:
\bb
W_\text{fb}=\hbar\omega_0 \sin^2(\theta/2) \cos(\theta_\textbf{n})(1-\sin^2(\theta_\textbf{n})\sin^2(\phi_\textbf{n})).
\ee

Finally, the average extracted power reads:

\bb
P_\textbf{n}(\theta) = {g}\dfrac{\hbar\omega_0}{2} \left(\dfrac{\sin(\theta)}{\theta}x_\textbf{n} + \dfrac{1-\cos(\theta)}{\theta}z_\textbf{n} y_\textbf{n}^2\right).
\ee

In Fig.\ref{f:MPEPerf}, the extracted power is plotted as a function of the position of working point in Bloch sphere, for different values of $\theta$. As expected, for any value of $\theta$, measuring the Qubit in its bare energy basis, i.e. $\textbf{n} = (0,0,\pm1)$, leads to zero power extraction since the measurement channel does not provide any energy. The engine also switches off if the demon measures the Qubit in the coupling Hamiltonian eigenbasis $\ket{\pm_\textbf{y}}$, since these states do not give rise to any work exchange between the Qubit and the drive. As mentioned previously, for a given value of $\theta$, maximal power is obtained if $\textbf{n} = \textbf{x} = (1,0,0)$: this corresponds to the point of the Rabi oscillation where the slope is maximal. Therefore in this operating point, the Qubit coherently provides energy to the drive in the fastest way. Reciprocally, using the state $\ket{-_\textbf{x}}$ as the operating point of the engine triggers the reverse mode where the engine coherently extracts maximal power from the drive. Finally, the Zeno limit $\theta\to 0$ provides the higher extracted power.

\subsection{Implementation in circuit QED} 
\label{s:ImplcQED}

\subsubsection{Finite measurement time correction}

In order to pave the road towards realistic implementation, we must examine a practical limitation: In any practical setup, the measurement actually takes some finite time $\tau_\text{m}$. This delay originates both from the time required for the measurement to be strong\footnote{The transfer of information occurs at a typical rate $\Gamma_\text{meas}$. The measurement is projective if the apparatus is coupled to the system during a time much larger than $1/\Gamma_\text{meas}$. Otherwise, the measurement is said to be weak: it does not perfectly project the system onto an eigenstate of the observable (see Chapter \ref{Chapter2} and \cite{Steck06}.}, and from the response time of the device used to perform the measurement. In addition, the feedback operation takes some finite time $\tau_\text{fb}$. The actual average duration of a cycle is then $\tau_\text{cyc} = \tau_\text{w} + \tau_\text{m} + p(-1)\tau_\text{fb}$,  where $p(-1) = \sin^2(\theta/2)(1-y_\textbf{n}^2)$ is the probability to obtain outcome $r_\gamma(t_n) = -1$. Therefore, the power has to be rescaled by a factor $\tau_\text{w}/\tau_\text{cyc}$. 

The rescaling factor has an impact on the engine performances as it forbids to reduce the duration $\tau_\text{w}$ of the work extraction step below $\tau_\text{m}$. As a consequence, the optimal power does not correspond anymore to the  Zeno regime which can be reached solely for a low Rabi frequency $g$ (see Fig.\ref{f:MPE_Impl}a).

\subsubsection{Setup}

\begin{figure}
\begin{center}
\includegraphics[width=0.85\textwidth]{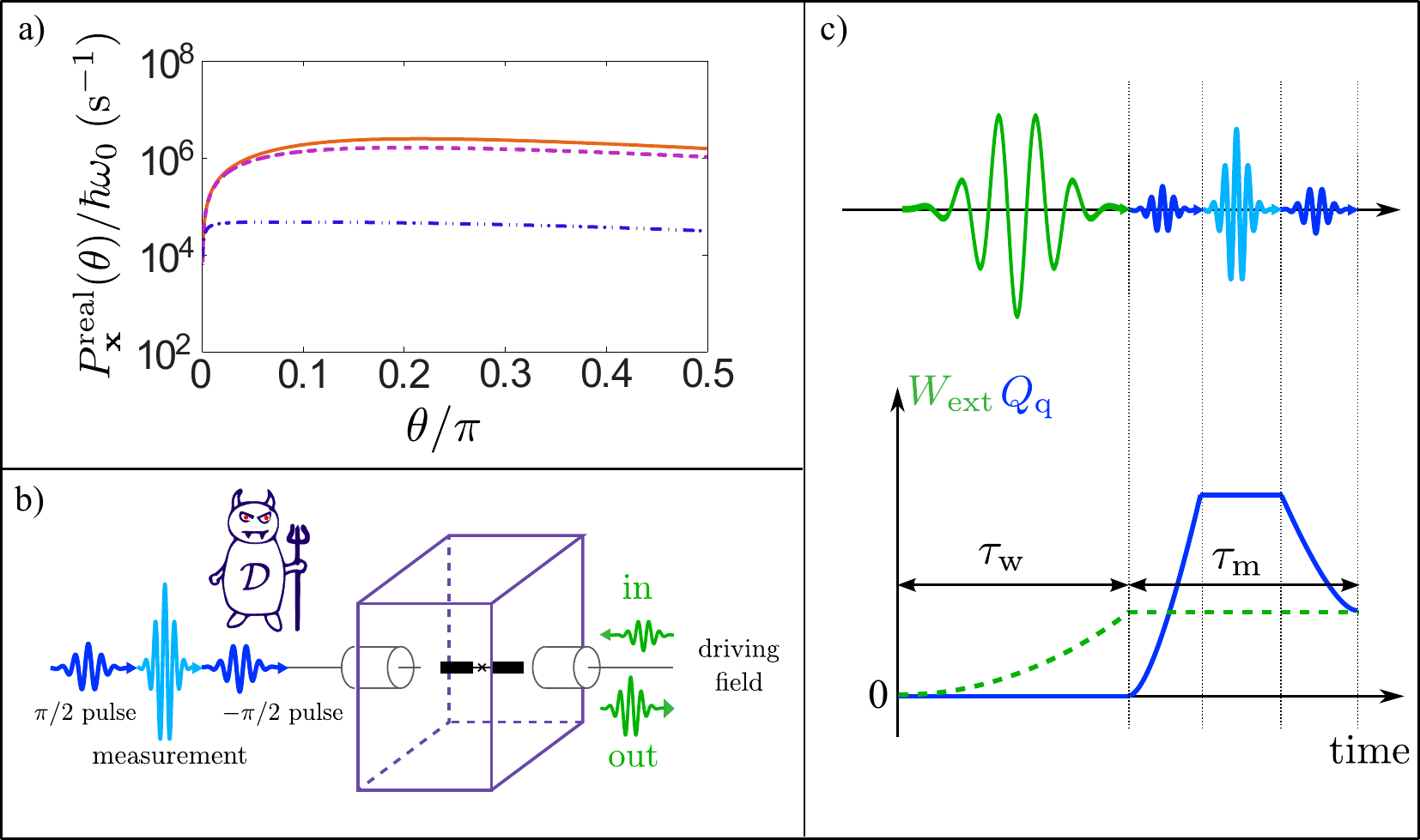}
\end{center}
\caption[Implementation of the MPE in circuit QED]{\label{f:MPE_Impl} Implementation of the MPE in circuit QED. a) Average extracted power $P^\text{real}_\textbf{x}(\theta)$ for the direction $\textbf{n} = \textbf{x}$ when the finite durations of the measurement and feedback steps are taken into account, as a function of the angle $\theta$.  We have considered three values of $g$: $1$ GHZ (orange solid), $10$ MHz (purple dashed) and $100$ kHz (blue dotted-dashed). {\it Parameters}: $\tau_\text{m} = 70$ns, $\tau_\text{fb} = 500$ns. b) Scheme of the experiment. The Qubit is a transmon in a 3D cavity. The thermodynamic cycle is realized by a sequence of pulses at various frequencies, and work is extracted as microwave propagating photons. c) Sequence of pulses corresponding to one engine cycle. Green: Drive. Dark blue: $\pi/2$ pulses. Light blue: readout tone. Bottom: Work extracted (green dashed) and quantum heat (dark blue) as a function of time.}
\end{figure}

We now study an implementation of the MPE based on a circuit QED setup pictured in Fig.\ref{f:MPE_Impl}b. Such setups use superconducting circuits to build on-chip quantum systems, like Qubits and harmonic oscillators interacting together \cite{Girvin08,Wallraff04}. The Qubit is a ``3D transmon'' involving a single Josephson Junction \cite{Koch07,Paik11}, of typical frequency of the order of $1$~GHz. In order to perform the measurement of $\sigma_x$, the Qubit is embedded in a 3D cavity of Hamiltonian $\hbar\omega_\text{c}a^\dagger a$. In the so-called dispersive limit of large Qubit-cavity detuning, one can engineer a Qubit-cavity coupling, such that a light field sent through the cavity evolves in a way sensitive to the Qubit state \cite{Wallraff04}. By measuring the field going out of the cavity, a Quantum Non Demolition projective measurement of $\sigma_z$ can be performed in about $50$~ns for optimized geometries \cite{Walter17,Kerman13,Billangeon15,Didier15,Richer16}. It is then possible to realize a periodic measurement of $\sigma_x$, and control feedback as demonstrated in Ref.~\cite{Campagne13} using two short $\pi/2$ pulses applied before and after the projective measurement of $\sigma_z$ (see section \ref{s:ExQED2} and \ref{f:MPE_Impl}c). The $\pi/2$ pulses take of the order of $10$~ns, such that the effective projective readout of $\sigma_x$ can be completed within a realistic time $\tau_\text{m} = 70$~ns. The feedback time is typically $\tau_\text{fb} = 500$~ns \cite{Campagne13}. The regime of small angle $\theta$ requires ${g}^{-1}\gg \tau_\text{w} \gg \tau_\text{m}$: These conditions can be fulfilled with more than one order of magnitude between each timescale, as evidenced by the observations of the Zeno regime in circuit-QED setups \cite{Bretheau15,Slichter16}. 

\section{Summary}

We have investigated the thermodynamics of a driven quantum system undergoing projective quantum measurements. Due to the randomness of the measurement-induced evolution, the quantum system under study follows a stochastic trajectory $\gamma$ conditioned on the measurement record. 

At the level of each trajectory, we have identified a stochastic and uncontrolled change of the system's internal energy induced by the measurement process that we have called quantum heat. We have also defined the entropy production induced by the measurement at the level of a single trajectory, and shown that it fulfills a fluctuation theorem. Finally, we have analyzed an example of genuinely quantum engine in which heat is extracted from a measurement process instead of a thermal reservoir and converted into work. 


\chapter{Stochastic thermodynamics of quantum open systems}
\label{Chapter2}

\minitoc

\lettrine{M}{any} situations of interest in thermodynamics involve open quantum systems and therefore go beyond the scope of Chapter \ref{Chapter1}. For instance, we want to describe a quantum system $\cal S$ driven by an operator and also be weakly coupled to a reservoir $\cal R$. We are particularly interested in two cases:
\begin{enumerate}
\item $\cal R$ is a thermal reservoir.
\item $\cal R$ induces pure dephasing in the eigenbasis of an observable of the system.
\end{enumerate}

Situation 1. is well-known from the thermodynamic point of view. Pioneer studies have already proven that quantum Fluctuation Theorems (FTs) like the quantum Jarzynski Equality (JE) \cite{Talkner07,Esposito09,Campisi11} are valid in this case. Yet, these seminal works have triggered numerous investigations, motivated by two major challenges. First, testing experimentally these FTs for quantum open systems is difficult. For instance, testing the JE requires to measure the energy change of the reservoir during the transformation, which represents an extremely small relative variation. The second challenge is of fundamental nature, and consists in identifying quantum signatures in quantum FTs. For instance, the coupling to a thermal bath erases quantum coherences during the system's evolution. This is an irreversible process, expected to induce some genuinely quantum entropy production. 

Situation 2. is much easier to investigate experimentally and has been extensively studied from the quantum open system point of view \cite{Skinner86,Carmichael1}. However, its thermodynamic interpretation is elusive.\\

In order to describe the thermodynamics of these situations, we present an extension of the formalism presented in Chapter \ref{Chapter1} based on the notion of quantum trajectories \cite{Molmer93}. The underlying idea is to formulate the non-unitary dynamics induced by the reservoir as a quantum measurement process. The thermodynamic formalism we present in this chapter is exposed in \cite{Elouard17}.
We then investigate applications in which the quantum system $\cal S$ is a Qubit.

\section{Formalism}

\subsection{Dynamics of quantum open systems}

\begin{figure}[h!]
\begin{center}
\includegraphics[width=0.57\textwidth]{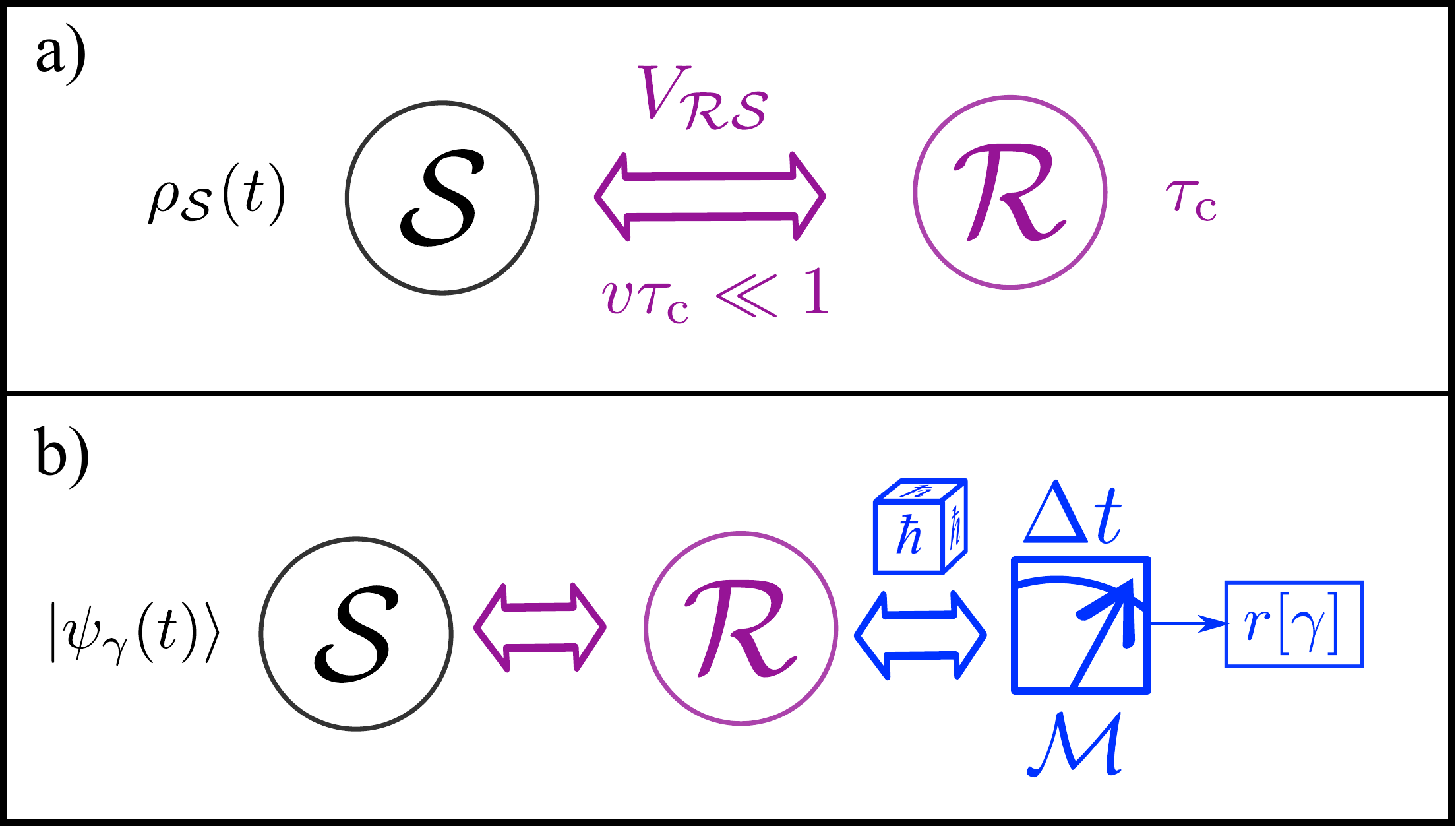}
\end{center}
\caption[Quantum open systems]{Quantum open systems. a) The system ${\cal S}$ is coupled to a reservoir ${\cal R}$ via coupling Hamiltonian $V_{\cal RS}$. The reservoir has a short correlation time $\tau_\text{c}$ such that $v\tau_\text{c}\ll 1$, where $v$ is the magnitude of $V_{\cal RS}$. b) Corresponding Kraus representation: The reservoir $\cal R$ is monitored by the measuring apparatus $\cal M$ which performs a projective measurement every $\Delta t$. If the measurement outcomes are read, the evolution of the system can be described as a quantum trajectory $\gamma$ of pure states $\ket{\psi_\gamma(t)}$, conditioned on the measurement record $r[\gamma]$. This trajectory is stochastic as it involves a quantum measurement process, which is symbolized by the $\hbar$-dice.
\label{f2:SceneryTraj}}
\end{figure}

We consider a quantum system ${\cal S}$ of Hamiltonian $H_{\cal S}$ coupled via the interaction Hamiltonian $V_{\cal SR}$ to a reservoir $\cal R$ (not necessarily thermal) of Hamiltonian $H_{\cal R}$ (see Fig.\ref{f2:SceneryTraj}a). The total density matrix $\rho_{\cal SR}(t)$ of the system ${\cal SR}$ evolves according to the Liouville - Von Neumann equation:
\bb
\dot\rho_{\cal SR}(t) = - \dfrac{i}{\hbar} [H_{\cal S}+V_{\cal SR}+H_{\cal R},\rho_{\cal SR}(t)] \label{d2:VNE}.
\ee

Because of the coupling Hamiltonian $V_{\cal RS}$ which acts on both $\cal S$ and $\cal R$, the two systems get entangled through their joint dynamics. However, the reservoir ${\cal R}$ is characterized by a very short correlation time $\tau_\text{c}$. According to Bloch-Redfield theory \cite{Bloch57,Redfield57}, a Markovian master equation for the density operator $\rho_{\cal S}(t)$ of the system alone
can be derived from Eq.\eqref{d2:VNE}. Such derivation requires that \cite{CCT,Breuer}: (i) $\tau_\text{c}$ is the shortest time-scale of the problem, (ii) the coupling between the system and the reservoir is weak, i.e. $v\tau_\text{c}/\hbar \ll 1$ with $v$ the typical magnitude of $V_{\cal RS}$. 
Finally, the master equation can be formulated as a Lindblad master equation \cite{Lindblad76}:
\bb
\dot\rho_{\cal S}(t) = - \dfrac{i}{\hbar} [H_{\cal S},\rho_{\cal S}(t)] + {\cal L}[\rho_{\cal S}(t)]\label{d2:LE},
\ee
\noindent where the Lindbladian superoperator reads:
\bb
{\cal L}[\rho] \underset{\text{def.}}{=} \sum_{k=1}^{\cal N} D[L_k] \rho_{\cal S}(t).\label{d2:Lindbladian}
\ee
\noindent We have introduced:
\bb
D[X]\rho \underset{\text{def.}}{=} X\rho X^\dagger - \tfrac{1}{2}\{X^\dagger X,\rho\}\label{d2:D}
\ee
\noindent with $\{A,B\} = AB+ BA$ the anticommutator. The operators $L_k$ are called jump operators. The expression and the number ${\cal N}$ of relevant jump operators depend on the coupling between ${\cal S}$ and ${\cal R}$.

The derivation of Eq.\eqref{d2:LE} from the fully quantum mechanical description Eq.\eqref{d2:VNE} is called microscopic derivation of the Lindblad equation and may be difficult for complex systems. In this Chapter, we focus on textbook situations for which the Lindblad equation is known. In Chapter \ref{Chapter3} and \ref{Chapter4}, non-textbook examples of such derivation are carried out. The key feature of the derivation is a coarse-graining in time of Eq.\eqref{d2:VNE} such that the new time step $\Delta t$ is much larger than $\tau_\text{c}$.

\subsection{Quantum trajectories}
\label{s2:DefQtraj}
The Lindblad equation \eqref{d2:LE} can be interpreted as resulting from a generalized measurement performed on the system (see Fig.\ref{f2:SceneryTraj}b). More precisely, Eq.\eqref{d2:LE} can be cast under the so-called Kraus representation \cite{Kraus83}:
\bb
\rho_{\cal S}(t+\Delta t) = \rho_{\cal S}(t) + {\cal L}[\rho_{\cal S}(t)]\Delta t \underset{\text{def.}}{=} \sum_r M_r \rho_{\cal S}(t)M_r^\dagger.\label{eq3:KrausLE}
\ee
\noindent where the operators $\{M_r\}$ are called Kraus operators and satisfy:
\bb
\sum_r M_r^\dagger M_r = \idop. \label{eq2:KrausSum}
\ee

Now, we interpret the action of operator $M_r$ on the system's state as the effect of a projective measurement performed on the reservoir ${\cal R}$, when a particular measurement outcome $r$ is obtained. In this description, the number of operators involved in decomposition \eqref{eq3:KrausLE} is the number of possible outcomes of the measurement performed on $\cal R$. As it is not a direct measurement of an observable of the Qubit, this measurement is called generalized measurement \cite{WisemanBook,Haroche}. 
If the outcome of this measurement is not read, the best description of the evolution of $\rho_{\cal S}$ is Eq.\eqref{eq3:KrausLE}, i.e. the Lindblad equation. \\

Conversely, if the measurement outcome is read, the system's state evolution is obtained by applying the corresponding Kraus operator \cite{Wiseman96}. Because of the stochasticity of the measurement process, the dynamics of the system takes the form of a stochastic quantum trajectory $\gamma$ conditioned on the sequence of measurement outcomes.

The quantum trajectory representation of the quantum open system dynamics has first be introduced as an efficient numerical tool to simulate large systems dynamics \cite{Molmer93}. However, the situation depicted in Fig.\ref{f2:SceneryTraj}b can nowadays be implemented in various setups. In those case, the quantum trajectories have a physical meaning: they correspond to the evolution of the system in a single realization of the studied protocol.  

In thermodynamics, using the quantum trajectory formalism to describe the thermodynamics of quantum open systems has a double interest. The first interest is conceptual: formulating the evolution of the open quantum system as a measurement process allows to interpret the thermodynamics in term of the quantities introduced in Chapter \ref{Chapter1}. The second interest is practical: Any experimental investigation requires to perform some measurement. An unraveling of the Lindblad equation is a particular measurement scheme which allows to extract information without perturbing the average evolution of the system's density operator. Quantum trajectories therefore provide a natural way of testing experimentally theoretical thermodynamic predictions while minimally disturbing the phenomena of interest.\\

As in Chapter \ref{Chapter1}, we assume that for every single realization of the protocol under study, the system is initially in a known pure state. Therefore, a trajectory $\gamma$ corresponds to a sequence of pure states $\{\ket{\psi_\gamma(t)}\}$. We discretize time using the time-step $\Delta t$ of the Lindblad equation, which corresponds to the typical period at which the apparatus $\cal M$ performs measurements on the reservoir. We introduce time $t_n = t_i + n\Delta t$ for $n \in \llbracket 0,N\rrbracket$, with $t_N = t_f$. The continuous measurement record is identified with the discrete sequence of measurement outcomes: $r[\gamma] = \{r_\gamma(t_n)\}_{1\leq n \leq N}$. Denoting $\ket{\psi_\gamma(t_n)}$ the state of the system at time $t_n$ along trajectory $\gamma$, we can write the stochastic evolution between $t_n$ and $t_{n+1}$ as:
\bb
\ket{\psi_\gamma(t_{n+1})}  = \dfrac{M_{r_\gamma(t_{n})}}{\sqrt{\moy{M_{r_\gamma(t_{n})}^\dagger M_{r_\gamma(t_{n})}}_{\psi_\gamma(t_{n})}}}\ket{\psi_\gamma(t_{n})}.  \label{d2:dPsi}
\ee

The precise expression of $M_{r}$ is depends on the way the environment is monitored. Contrary to the case of projective measurements performed directly on $\cal S$, $M_r$ is in general different from a projector on a eigenstate of the measured observable, and might even be non-hermitian. In addition, the Kraus representation (and therefore the set of Kraus operators) corresponding to a given Lindblad equation is not unique. It is fixed, though, when a particular monitoring scheme performed on the environment is precised. The choice of a given set of Kraus operators is called unraveling of the Lindblad equation.

\subsection{Unravelings}

In this dissertation, we focus on two types of unravelings: one is called Quantum Jump (QJ), and the other Quantum State Diffusion (QSD).

\subsubsection{Quantum jump unraveling}

In the QJ unraveling, there is a discrete set of possible measurement results: $r_\gamma(t_n) \in \llbracket 0,{\cal N}\rrbracket$. The measurement outcomes satisfying $r_\gamma(t_n) \geq 1$ correspond to a strong effect on the system's state called ``jump''. The corresponding Kraus operator is proportional to one of the jump operators:
\bb
M_{k} = \sqrt{dt}L_k, \quad k\geq 1.
\ee
\sloppy
\noindent Between times $t_n$ and $t_{n+1}$, a jump occurs with an infinitesimal probability \mbox{$p_k(t_n) = \bra{\psi_\gamma(t_n)}M_{k}^\dagger M_{k}\ket{\psi_\gamma(t_n)}=O(\Delta t)$}. Most of the time, no jump occur at time $t_n$, which corresponds to outcome $0$. The associated system's evolution is encoded in $M_{0}$. This evolution is infinitesimal (very close to the identity operator) and fulfils:
\bb
M_{0} = 1-\dfrac{i\Delta t}{\hbar} H_{\cal S}(t) -\dfrac{1}{2}\sum_{k=1}^{\cal N} M_{k}^\dagger M_{k},
\ee
such that the set of Kraus operators satisfies Eq.\eqref{eq2:KrausSum}.
\fussy

An example of QJ unraveling for a two-level atom in contact with a zero temperature electromagnetic reservoir is obtained by detecting the emitted photons with a time-resolved detector \cite{Haroche}. 


\subsubsection{Quantum state diffusion}
\label{s2:defQSD}

Another unraveling which is frequently used is called Quantum State Diffusion (QSD). In QSD trajectories, the measurement outcome $r_\gamma(t_n)$ takes continuous values and the effect on the system is weak whatever the measurement outcome. The Kraus operator applied at time $t_n$ has the general form:
\bb
M_{r_\gamma(t_n)}(t_n) = \prod_k \sqrt{p(dw_\gamma^{(k)}(t_n))}\left(\idop -  \dfrac{i\Delta t}{\hbar} H_{\cal S}(t_n) -\dfrac{1}{2}\sum_{k=1}^{d_{\cal S}^2-1} M_{r_k}^\dagger M_{r_k} + \sum_{k=1}^{d_{\cal S}^2-1} dw_\gamma^{(k)}(t_n)L_k\right)\nonumber\\\label{d2:MQSD}
\ee
\noindent where the $dw_\gamma^{(k)}(t_n)$ are $k$ Wiener increments, i.e.  (potentially complex) stochastic Gaussian variables of zero mean and variance $\Delta t$. We denote $p(dw_\gamma^{(k)}(t_n))$ the probability distribution of $dw_\gamma^{(k)}(t_n)$. The Wiener increments quantify the (normalized) fluctuations of the measurement outcome around their expected values.

Famous examples are the homodyne and heterodyne measurements of the light emitted by a two-level atom \cite{WisemanBook}.

\subsection{First law of thermodynamics}
\label{s2:1stLaw}
As in the first chapter, we consider thermodynamic protocols defined by:
\begin{itemize}
\item An initial pure state $\ket{\psi_i}$ of the system at time $t_i$ which can be the same for every trajectory or drawn from a mixture with probability $p_i$.
\item A drive of the system between time $t_i$ and $t_f$ corresponding to a time-dependent contribution $H_\text{d}(t)$ to its Hamiltonian.
\end{itemize}

The internal energy of the system along quantum trajectory $\gamma$ reads:
\bb
U_\gamma(t_n) \underset{\text{def.}}{=}\bra{\psi_\gamma(t_n)}H_{\cal S}(t_n)\ket{\psi_\gamma(t_n)}. \label{d2:U}
\ee

This quantity varies along trajectory $\gamma$ because of the energy injected by the drive, and the energy exchanged with the monitored reservoir. As in Chapter \ref{Chapter1}, we first identify the increment of work $\delta W_\gamma(t_n)$ injected by the drive between $t_n$ and $t_{n+1}$ :
\bb
\delta W_\gamma(t_n) \underset{\text{def.}}{=} \bra{\psi_\gamma(t_n)}\dot H_\text{d}(t_n)\ket{\psi_\gamma(t_n)}\label{d2:dW}.
\ee
Then, the heat increment $\delta Q_\gamma(t_n)$ is defined as:
\bb
\delta Q_\gamma(t_n) &\underset{\text{def.}}{=}& U_\gamma(t_{n+1})- U_\gamma(t_n) - \delta W_\gamma(t_n)\nonumber\\
&=& dU_\gamma(t_n) - \delta W_\gamma(t_n).\label{d2:dQ}
\ee
In the absence of reservoir and measuring apparatus, $\delta Q_\gamma(t_n)$ vanishes. 

The total work $W[\gamma]$ and heat $Q[\gamma]$ exchanged along trajectory $\gamma$ are obtained by summing up all the increments from $n$ equal $1$ to $N$. The first law can be written either between $t_n$ and $t_{n+1}$, or for a complete trajectory $\gamma$:
\bb
dU_\gamma(t_n) &=& \delta W_\gamma(t_n) + \delta Q_\gamma(t_n)\\
\Delta U[\gamma] &=& W[\gamma] + Q[\gamma]\label{d2:FirstLaw}
\ee

Finally, we also define the average power $\dot W(t)$ and heat flow $\dot Q(t)$ using the average over the trajectories $\gamma$ denoted $\moy{\cdot}_\gamma$:
\bb
\dot W(t_n) &=& \left\langle\dfrac{\delta W_\gamma(t_n)}{\Delta t}\right\rangle_\gamma = \text{Tr}\{\dot H_\text{d}(t_n) \rho_{\cal S}(t_n)\} \label{d2:AvPower}\\
\dot Q(t_n) &=& \left\langle\dfrac{\delta Q_\gamma(t_n)}{\Delta t}\right\rangle_\gamma = \text{Tr}\{ H_\text{d}(t_n) \dot \rho_{\cal S}(t_n)\} \label{d2:AvHeatFlow}.
\ee
We have used that the average over the measurement outcomes of the system's (pure state) stochastic density operator $\rho_\gamma(t_n) = \ket{\psi_\gamma(t_n)}\bra{\psi_\gamma(t_n)}$ corresponds to the solution of the Lindblad equation, i.e. $\moy{\rho_\gamma(t_n)}_\gamma = \rho_{\cal S}(t_n)$.

\subsection{Entropy production}

\subsubsection{Definition and IFT}

As in Chapter \ref{Chapter1}, we use the following definition for the entropy produced along trajectory $\gamma$, directly imported from the classical stochastic thermodynamics framework \cite{Seifert08,Sekimoto10}:
\bb
\Delta_\text{i} s[\gamma] \underset{\text{def.}}{=}k_\text{B}\log\left(\dfrac{p(\gamma)}{\tilde p(\tilde\gamma)}\right), \label{d2:Dis}
\ee
\noindent where $p(\gamma)$ is probability that $\cal S$ follows trajectory $\gamma$ in a single realization of the protocol under study. $\tilde p(\tilde\gamma)$ is the probability that the system follows the time-reversed trajectory $\tilde\gamma$ corresponding to $\gamma$ when a time-reversed protocol is applied. This probabilities are assumed to fulfil\footnote{For certain situations, e.g. when some of the possible initial states of the system are forbidden, this equality is not verified and the sum \eqref{eq2:GamGamTilde} is lower than one. This phenomenon which has been called ``absolute irreversibility'' in \cite{Murashita14} leads to a strictly positive entropy production (see also Section \ref{s:EntropyMeas}). In this chapter, we do not discuss such cases and rather focus on system prepared in initial mixtures, such as thermal distributions, for which such phenomenon does not occur.}:
\bb
\sum_\gamma \tilde p(\tilde\gamma) =1\label{eq2:GamGamTilde}
\ee
\noindent  The value of this entropy production is not constrained to be positive, however its satisfies the IFT:
\bb
\moy{e^{-\Delta_\text{i}s[\gamma]/k_\text{B}}}_\gamma &=& \sum_\gamma p(\gamma)\dfrac{\tilde p(\tilde\gamma)}{p(\gamma)}\nonumber\\
&=& \sum_\gamma \tilde p(\tilde\gamma)\nonumber\\
&=& 1. \label{d2:IFT}
\ee

The IFT (Eq.\eqref{d2:IFT}) leads to the second Law of thermodynamics: Using the convexity of the exponential function $\moy{e^{-\Delta_\text{i}s[\gamma]/k_\text{B}}}_\gamma \geq e^{-\moy{\Delta_\text{i}s[\gamma]/k_\text{B}}_\gamma}$, we find:
\bb
\moy{\Delta_\text{i}s[\gamma]}_\gamma \geq 0 \label{d2:SecondLaw}.
\ee
The IFT also leads to other famous FTs such that Jarzynski Equality when particular assumptions about the protocol and properties of the thermal reservoir are made \cite{Seifert08}.\\

The probability $p(\gamma)$ of a trajectory $\gamma$ can be split in two factors: (i) the probability $p_i$ for the system to start in the pure state $\ket{\psi_i}$ and (ii) the conditional probability $p(r[\gamma]\vert i)$ to obtain the measurement record $r[\gamma]$ given the initial state:
\bb
p(\gamma) =  p(r[\gamma]\vert i)p_i \label{d2:pgamma}.
\ee

The Kraus operators are normalized such that the probability of obtaining the outcome $r_\gamma(t_n)$ when the system is in state $\ket{\psi_\gamma(t_n)}$ is given by:
\bb
p(r_\gamma(t_{n})) = \bra{\psi_\gamma(t_n)}M_{r}^\dagger M_{r}\ket{\psi_\gamma(t_n)}.
\ee
Injecting the form of the evolution Eq.\eqref{d2:dPsi}, we find that the conditional probability $p(r[\gamma]\vert i)$ satisfies:
\bb
p(r[\gamma]\vert i) = \left\vert\bra{\psi_f} \overleftarrow{\prod_{n=1}^N} M_{r_\gamma(t_n)} \ket{\psi_i} \right\vert^2 \label{eq2:pcond}
\ee
\noindent where the arrow on the product sign indicates that the operators are ordered from the right to the left, and $\ket{\psi_f}$ is the final state of the stochastic trajectory. First derivations of quantum FTs have systematically involved protocols including some final projective measurement of an observable \cite{Esposito09,Manzano15} where $\ket{\psi_f}$ is the corresponding eigenstate. In the present framework, the considered trajectories do not necessarily end up with such a projective measurement: Then $\ket{\psi_f}$ is the final state of the direct stochastic evolution. \\

The probability of the time-reversed trajectory can be expressed in analogy to the direct case:
\bb
\tilde p(\tilde\gamma) = p_f \tilde p(\tilde r[\tilde \gamma \vert f),
\ee
\noindent where $p_f$ is the probability for $\tilde\gamma$ to start in $\ket{\psi_f}$ and $\tilde p(\tilde r[\tilde \gamma]\vert f)$ is the conditional probability of the time-reversed measurement record $\tilde r[\tilde \gamma]$ given the initial state $\ket{\psi_f}$ of $\tilde\gamma$, which reads:

\bb
\tilde p(\tilde r[\tilde\gamma]\vert f) &=& \left\vert\bra{\psi_i} \overleftarrow{\prod_{n=1}^N} \tilde M_{\tilde r_{\tilde\gamma(t_n)}}\ket{\psi_f}\right\vert^2.\label{eq2:ptildecond}
\ee
$\{\tilde M_{r}\}$ is the set of time-reversed Kraus operators built in the next section.

Finally, we rewrite the entropy produced along trajectory $\gamma$:

\bb
\Delta_\text{i}s[\gamma] &=& - k_\text{B}\log \dfrac{p_f}{p_i} - k_\text{B}\log \dfrac{\tilde p(\tilde r[\tilde \gamma \vert f)}{p(r[\gamma]\vert i)} \nonumber\\
&=& \Delta_\text{i}^\text{b}s[\gamma] + \Delta_\text{i}^\text{c}s[\gamma],
\ee
where we have defined the boundary and conditional entropy productions respectively denoted $\Delta_\text{i}^\text{b}s[\gamma]$ and $\Delta_\text{i}^\text{c}s[\gamma]$. 

\subsubsection{Time reversal}
\label{s2:TimeReversal}

Up to now, we did not explain how to find the time-reversed protocol and compute the probability $\tilde p(\tilde\gamma)$. Actually, there is a certain freedom in defining such a time-reversed process. Indeed, the IFT is a very general mathematical property of probabilities, and Eq.\eqref{d2:IFT} is valid for any probability law $\tilde p(\tilde\gamma)$ fulfilling Eq.\eqref{eq2:GamGamTilde}. As a consequence, there are in general different relevant choices for the time-reversed process leading to an expression of the IFT involving physically meaningful quantities. \\

First of all, there are several possible choices about the initialization of the system for the time-reversed protocol \cite{Seifert08}. The sole constraint is that the system must be initialized in a mixture $\rho_f = \sum_f p_f\ket{\psi_f}\bra{\psi_f}$ of the possible final states $\{\ket{\psi_f}\}$ of the direct trajectories. However, different prescriptions can be applied to choose the probabilities $p_f$. One natural choice is to take the final mixture of the direct process as an input for the reversed process. In this case $p_f$ is the sum of the probabilities of the direct trajectory ending in $\ket{\psi_f}$. Another possibility is to impose that the system starts the reversed process in equilibrium with the reservoir. In that case, $\rho_f$ is the equilibrium density operator and $p_f$ is the equilibrium population of $\ket{\psi_f}$ corresponding to the final Hamiltonian $H_{\cal S}(t_f)$. These two possible choices lead to different FTs. \\

Second, there is a freedom in defining the set of time-reversed operators ${\tilde M_r}$. In order to impose that the time-reversed transformation is a physical process, they must satisfy:
\bb
\sum_r \tilde M_r^\dagger \tilde M_r = \idop
\ee
Moreover, we need to relate each direct Kraus operator $M_r$ to one of the time-reversed Kraus operator $\tilde M_{\tilde r}$ and as a consequence it is  required that the numbers of reversed and direct Kraus operators are the same. However, additional constraints are needed to precise the expressions of the $\{\tilde M_r\}$ and also which is the label $\tilde r_{\tilde\gamma}(t_n)$ applied in the $n$th step of the time-reversed protocol. 

In this dissertation we use the prescription proposed in \cite{Crooks08} which consists in imposing that no entropy is produced if the system remains in equilibrium. More precisely, we define the equilibrium density operator $\pi_\text{eq}$ with the condition ${\cal L}[\pi_\text{eq}] = 0$\footnote{There may be several equilibrium states. Then a set of time-reversed operators can be defined from each of these equilibrium states \cite{Manzano15}.}, with ${\cal L}$ the Linbdladian defined in Eq.\eqref{d2:Lindbladian}. We then impose that the probabilities of a two-point measurement record $r[\gamma] = \{r_1,r_2\}$ and of its time-reversed $\tilde r[\tilde \gamma] = \{\tilde r_2,\tilde r_1\}$ are the same when the initial state of both the direct and reversed protocols is drawn from distribution $\pi_\text{eq}$. Explicitly, the condition reads:
\bb
\text{Tr}\{M_{r_2}M_{r_1}\pi_\text{eq}M_{r_1}^\dagger M_{r_2}^\dagger\} = \text{Tr}\{\tilde M_{\tilde r_1}\tilde M_{\tilde r_2}\pi_\text{eq}\tilde M_{\tilde r_2}^\dagger \tilde M_{\tilde r_1}^\dagger\}.\label{eq2:CondCrooks}
\ee
This allows to identify the time-reversed operator applied at the $n$th time-step of the reversed-process as:
\bb
\tilde M_{\tilde r_{\tilde \gamma(t_{n})}} = \pi_\text{eq}^{1/2} M_{r_\gamma(t_{N-n+1})}^\dagger\pi_\text{eq}^{-1/2}.\label{eq2:TimeRevCrooks}
\ee

Such a condition correctly inverse the direction of the jumps, and link the evolution at time $t_n$ in the time-reversed trajectory to to the evolution at time $t_{N-n+1}$ of the direct trajectory, as expected. However, in order to ensure that the Hamiltonian part of the evolution indeed propagates backwards, one must in addition transform certain observables of the system \cite{Messiah,Campisi11,Manzano15} (e.g. the sgn of the ``odd'' variables like momentum or spin operators must inverted). Such transformation (which does not affect the probability computed in Eq.\eqref{eq2:CondCrooks}) is summarized by the so-called time-reversal operator denoted $\Theta$ \cite{Messiah}, such that the most general form of the time-reversed operator is $\tilde M_{\tilde r_{\tilde \gamma(t_{n})}} = \Theta\pi_\text{eq}^{1/2} M_{r_\gamma(t_{N-n+1})}^\dagger\pi_\text{eq}^{-1/2}\Theta^{-1}$. In this thesis, we focus on applications in which such operation boils down to considering the Hamiltonian $H_{\cal S}(t_f-t)$, and we therefore do not write it explicitly.

Note that with this definition of the time-reversed operators, the reversed trajectory $\tilde\gamma$ does not necessarily follows the same sequence of quantum states as the direct trajectory. The condition rather imposes the reversed \emph{sequence} of measurement outcomes. Consequently, ensuring that the time-reversed trajectory $\tilde\gamma$ ends in state $\ket{\psi_i}$ requires to perform a projective measurement in the basis $\ket{\psi_i}$ at the beginning (resp. at the end) of the direct (resp. reversed) protocol. In the examples analyzed in this dissertation, we either explicitly include this initial measurement, or consider a known pure initial state for each trajectory such that this initial measurement has no consequences on the dynamics.

Another possible time-reversal condition has been introduced in \cite{Dressel16} in the context of QSD unraveling. In this case the system is imposed to follow the very same sequence of states backwards, which leads to relate the reversed Kraus operator to the inverse of the direct Kraus operators.\\

In the remaining of the chapter, we study more specific examples and their implementations: First, the case of a Qubit in contact with a thermal reservoir. We present an engineered environment allowing to implement the corresponding QJ unraveling. Then, we consider the case of the weak continuous monitoring of an observable of a quantum system, which implements a QSD unraveling.  

\section{Probing Fluctuation theorems in a engineered environment}

\subsection{System}

The system under study in this section is a Qubit weakly coupled to a thermal reservoir ${\cal R}_T$ at temperature $T$. The Qubit is also driven; such that its total Hamiltonian $H_\text{q}(t) = H_0 + H_\text{d}(t)$ is time-dependent. $H_0 = \hbar\omega_0\ket{e}\bra{e}$ is the bare Qubit Hamiltonian and $H_\text{d}(t)$ is the time-dependent contribution to the Qubit's Hamiltonian induced by the drive. The dynamics of the density operator $\rho_\text{q}(t)$ of the Qubit is captured by the Lindblad equation \cite{Carmichael1}:
\bb
\dot\rho_\text{q}(t) = -\dfrac{i}{\hbar}[H_\text{q}(t),\rho_\text{q}(t)] + {\cal L}_T[\rho_\text{q}(t)] \label{eq2:LEQubitTh}
\ee
where the thermal Lindbladian reads:
\bb
{\cal L}_T = \gamma_\text{q}(n_\text{q}+1)D[\sigma_-] + \gamma_\text{q}n_\text{q}D[\sigma_+] \label{eq2:LindbladianTh}.
\ee
We have introduced the spontaneous emission rate $\gamma_\text{q}$ of the Qubit and the population of the reservoir mode resonant with the effective Qubit frequency $\omega_\text{q}$ (which is in general different from $\omega_0$ because of the drive): 
\bb
n_\text{q} = \dfrac{1}{e^{\hbar\omega_q/k_\text{B}T}-1}.
\ee
Note that due to the drive, the effective Qubit frequency and therefore $n_\text{q}$ may depend on time.

\subsection{Measuring the energy of the reservoir}
\label{s2:SetupEngineered}

In order to define an unraveling of Eq.\eqref{eq2:LEQubitTh}, we have to choose a way to monitor the reservoir. A option particularly well-suited for thermodynamics consists in measuring the change in the reservoir's energy so as to have access to the heat exchanged with the reservoir. 
In pioneer papers, the IFT under the form of the JE \cite{Jarzynski97,Jarzynski11} has been demonstrated to be valid for quantum systems. Such equality was first demonstrated for a closed quantum system driven out-of-equilibrium in an adiabatic way\footnote{Adiabatic has to be understood in the thermodynamic sense here: the quantum system does not exchange heat with its environment. Practically, it generally corresponds to a ``quench'', i.e. a fast transformation during which the system does not have time to thermalize.}\cite{Talkner07}. Denoting $\Delta U[\gamma]$ the change in internal energy obtained by two projective energy measurements, and $\Delta F$ the difference between the initial and final equilibrium free energy of the system, JE reads in that case:
\bb
\left\langle e^{-\Delta U[\gamma]/k_\text{B}T} \right\rangle_\gamma = e^{-\Delta F}.\label{eq2:JEClosed}
\ee

Then it was extended to the open quantum system case by considering the reservoir and the system as an isolated system \cite{Campisi09,Esposito09,Campisi11}. The total energy change of this global system is $\Delta U_{\cal SR}[\gamma] =  \Delta U[\gamma]- Q_\text{cl}[\gamma]$, where we define the classical heat $Q_\text{cl}$ as the negative change in the thermal reservoir's energy. This leads to:
\bb
\left\langle e^{-(\Delta U[\gamma]-Q_\text{cl}[\gamma])/k_\text{B}T} \right\rangle_\gamma = e^{-\Delta F},\label{eq2:JETPMP}
\ee
Where $\Delta F$ is again the variation of the system's equilibrium free energy.

Probing such equality theoretically requires to measure the change of the internal energy of the reservoir from projective energy measurements performed at the beginning and at the end of the transformation, which is very challenging.

One possible strategy to access $Q_\text{cl}$ relies on a finite size reservoir whose temperature is sensitive to heat exchanges with the system \cite{Pekola13, Hekking13,Campisi15}, such that both emission and absorption events can be detected using fine calorimetry. But such reservoirs are by essence non Markovian, therefore affecting the dynamics of the system and the resulting work distribution \cite{Suomela16}.\\

\begin{figure}[h!]
\vspace{0.1cm}
\begin{center}
\includegraphics[width=\textwidth]{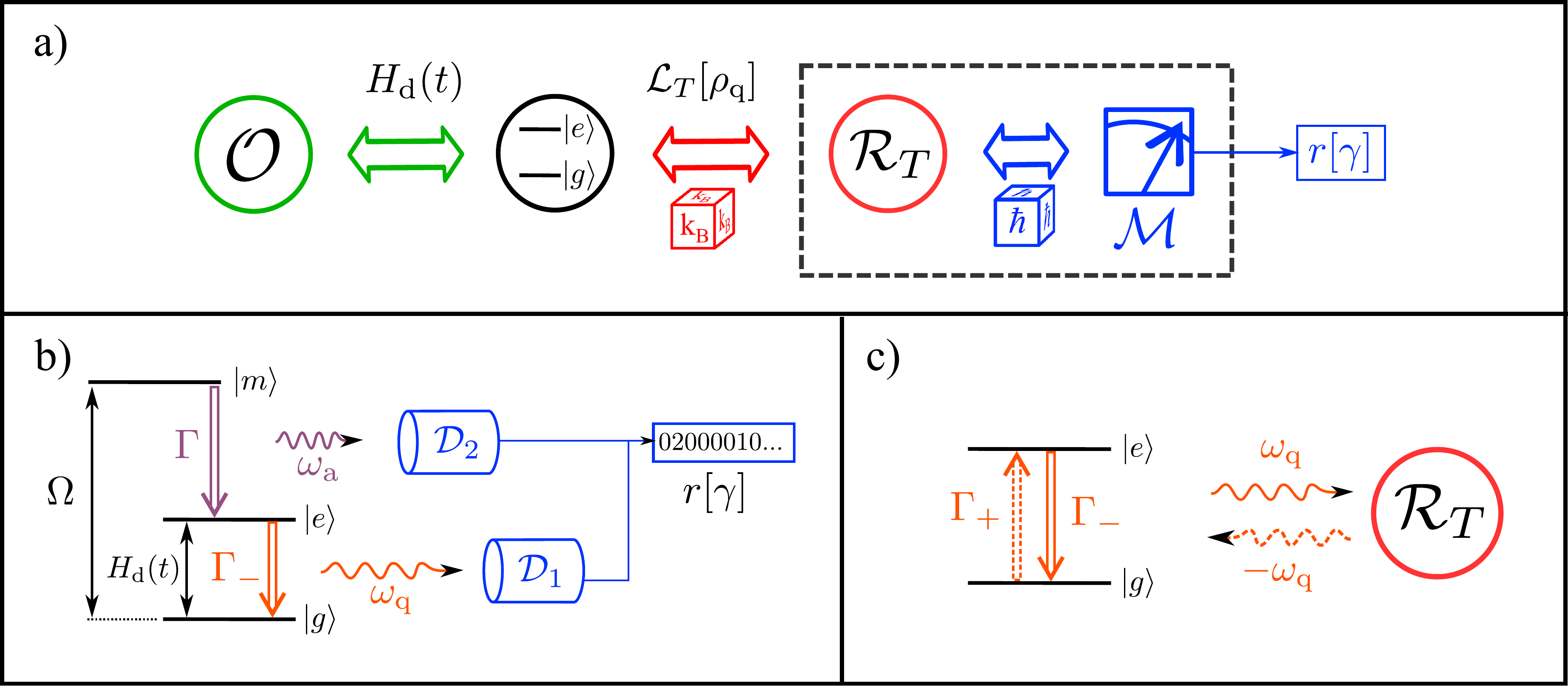}
\end{center}
\caption[Setup to test Qubit's Fluctuation Theorems]{Setup to test Qubit's Fluctuation Theorems. a)~Situation under study: a Qubit is driven by an operator $\cal O$ and in contact with a thermal reservoir ${\cal R}_T$. In presence of a measuring apparatus monitoring the reservoir, the system follows a quantum trajectory $\gamma$ of pure states $\ket{\psi_\gamma(t)}$, conditioned on the measurement record $r[\gamma]$.
b)~Engineered environment based on a three-level atom. The state $\ket{m}$ has a very short lifetime due to spontaneous emission in vacuum characterized by the rate $\Gamma$. The transition $\ket{e}$ to $\ket{g}$ (resp. $\ket{m}$ to $\ket{e}$) is associated with emission of a photon of energy $\omega_\text{q}$ (resp. $\omega_\text{a}$) detected by ${\cal D}_1$ (resp. ${\cal D}_2$). The transition $\ket{e} \rightarrow \ket{g}$ is driven by the Hamiltonian $H_\text{d}(t)$.
c) Equivalent description in terms of an effective thermal bath. In the presence of a weak classical drive of Rabi frequency $\Omega$ resonant with the transition $\ket{m} \rightarrow \ket{e}$, level $\ket{m}$ can be adiabatically eliminated. This result in an effective incoherent rate $\Gamma_+ = 4\Omega^2/\Gamma$ associated with transition $\ket{g} \rightarrow \ket{e}$ (dotted arrow), simulating a Qubit coupled to an effective thermal bath of temperature $T_\text{eff} = \hbar\omega_\text{q}/k_\text{B}\log(\Gamma_-/\Gamma_+)$. \label{f2:Engineered}}
\end{figure}

Another strategy is based on a time-resolved monitoring of the environment which corresponds to the situation of Fig.\ref{f2:Engineered}a. The environment is monitored in such a way that the photons exchanged with the Qubit can be tracked. 
In convenient platforms like superconducting circuits and semi-conducting quantum photonics, it is possible to detect the energy emitted by the system in the reservoir. In both cases, the radiation produced by quantum emitters (superconducting Qubits or quantum dots) is efficiently funneled into well-designed waveguides, and is thus recorded with high efficiency. This recent experimental ability has lead to the development of bright single photon sources \cite{Claudon10, Somaschi16} and to the monitoring of quantum trajectories of superconducting quantum bits \cite{Riste13,Murch13,Weber14}. However, standard schemes based on photo-counters do not allow for the recording of photons absorbed by the system, a major drawback for quantum thermodynamics purposes.\\

In this section, we present a proposal of setup, that we have exposed in \cite{FTeng}, allowing to test the JE for a Qubit in contact with a thermal reservoir. It is based on a scheme to detect the quantum jumps of a Qubit first introduced in \cite{Carvalho11,Santos11}. The key idea is to use a three-level atom instead of a Qubit and to engineer its electro-magnetic environment to keep track of the absorptions of photons.

\subsubsection*{Proposal}

We consider an atom whose three levels are denoted by $\ket{m}$, $\ket{e}$ and $\ket{g}$ of respective energy $E_m>E_e>E_g$ (see Fig.\ref{f2:Engineered}b). This three-level atom is coupled to an electro-magnetic reservoir at zero temperature, such that $\ket{m}$ decays to state $\ket{e}$ (resp. $\ket{e}$ decays to $\ket{g}$) with a spontaneous emission rate $\Gamma$ (resp. $\Gamma_-$). We assume that $\ket{m}$ is a metastable level, such that $\Gamma \gg \Gamma_-$. We denote $\omega_0 = (E_e-E_g)/\hbar$ and $\omega_\text{a} = (E_m-E_e)/\hbar$. In addition, the atom is weakly driven by a laser resonant with transition $\omega_0+\omega_\text{a}$ and of Rabi frequency $\Omega$, satisfying $\Omega \ll \Gamma$. Due to its short lifetime, level $\ket{m}$ can be adiabatically eliminated, resulting in an effective incoherent transition rate $\Gamma_+ = 4\Omega^2/\Gamma$ from state $\ket{g}$ to state $\ket{e}$ \cite{Carvalho11}. The states $\ket{e}$ and $\ket{g}$ define our effective Qubit of interest (Fig.\ref{f2:Engineered}c). Without external drive, this Qubit relaxes towards some effective thermal equilibrium characterized by the effective temperature $T$, satisfying $e^{-\hbar \omega_\text{q}/k_\text{B}T} = \frac{\Gamma_+}{\Gamma_-}$.

The dynamics of the Qubit density matrix $\rho_\text{q}(t)$ is ruled by the master equation Eq.\eqref{eq2:LEQubitTh}. Assuming that a photo-counter ${\cal D}_1$ (resp. ${\cal D}_2$) detects the photons emitted at frequency $\omega_\text{q}$ (resp. $\omega_\text{a}$), one can formulate the evolution of the Qubit state conditioned on the measurement records of the detectors within the QJ unraveling.

Neglecting the probability that both detectors ${\cal D}_1$ and ${\cal D}_2$ detect a photon between $t_n$ and $t_{n+1}$, we sum up the measurement records of both detectors under the form: $r[\gamma] = \{r_\gamma(t_n)\}_{1\leq n \leq N}$, with $r_\gamma(t_n) \in \{1,2,0\}$. The three possible outcomes correspond to detector ${\cal D}_1$ clicking, detector ${\cal D}_2$ clicking, and none of the detector clicking respectively. Detecting a photon on detector ${\cal D}_1$ (resp. ${\cal D}_2$) at time $t_n$ corresponds to applying the operator $M_1 = \sqrt{\Gamma_-\Delta t}\sigma_-$ (resp. $M_2 = \sqrt{\Gamma_+\Delta t}\sigma_+$) on the Qubit's state $\ket{\psi_\gamma(t_n)}$. Remarkably in this scheme, the absorption of a photon from the effective thermal reservoir (Fig.\ref{f2:Engineered}c) is detectable and actually corresponds to the emission of a photon of frequency $\omega_\text{a}$ (Fig.\ref{f2:Engineered}b). If no photon is detected, the no-jump operator $M_0 = \idop -  \dfrac{i\Delta t}{\hbar} H_\text{q}(t_n) - \tfrac{1}{2} M_1^\dagger M_1^{}- \tfrac{1}{2} M_2^\dagger M_2^{}$ is applied.\\ 

So as to ensures that the scheme is valid in presence of the drive, some conditions must be satisfied:
\begin{itemize}
\item $\Gamma$ has to be much larger than any frequency involved in the driven dynamics.
\item If the drive induces some time-dependence in the effective Qubit frequency, the rates $\Gamma_\pm$ has to be tuned to coincide at any time with the rates involved in the Lindbladian ${\cal L}_T$, namely: 
\bb
\Gamma_-(t) &=& \gamma_ \text{q}(n_\text{q}(t)+1)\label{eq2:GMin_t}\\
\Gamma_+(t) &=& \gamma_ \text{q}n_\text{q}(t).\label{eq2:GPl_t}
\ee 
Such constraints can be fulfilled owing to two control parameters: the Rabi frequency $\Omega$ and the rate $\Gamma_-(t)$ of the transition $\ket{e}\to\ket{g}$. The latter be varied by embedding the three level atom in a quasi-resonant cavity.
\end{itemize}

Finally, note that the engineered reservoir can be used to simulate a large class of Markovian reservoirs which is not restricted to thermal baths. The effective temperature $T$ can be tuned e.g. to be negative, or can be varied independently of the effective Qubit frequency. 

\subsection{Thermodynamics along quantum jump trajectories}

\subsubsection{Classical and quantum heat}
\label{s2:QqQcl}

The scheme presented above allows to count the photons emitted and absorbed by the Qubit in a time-resolved manner. As a consequence, the variation of the reservoir's energy can be inferred along each trajectory, considering that a photon exchanged at time $t_n$ carries a quantum of energy $\hbar\omega_\text{q}(t_n)$.

We underline that assuming a well-defined energy for the detected photons is not in contradiction with the time-energy Heisenberg uncertainty as the integration time of the photo-detection, which corresponds to the coarse-graining time-step $\Delta t$ used to derived the Lindblad equation \eqref{eq2:LEQubitTh}, fulfills $\omega_\text{q} \Delta t \gg 1$ \cite{CCT}.

Therefore, the increment of classical heat (i.e. the energy provided by the reservoir) at time $t_n$ reads:
\bb
\delta Q_{\text{cl},\gamma}(t_n) = \left\{\begin{array}{cl}
-\hbar\omega_\text{q}, &r_\gamma(t_n) = 1 \\ 
\hbar\omega_\text{q}, &r_\gamma(t_n) = 2 \\ 
0, &r_\gamma(t_n) = 0.
\end{array} \right. \label{d2:dQcl}
\ee\\

This contribution does not in general correspond to the total heat increment which is given by Eq.\eqref{d2:dQ}. If the Qubit's state at time $t_n$ is in a superposition of states $\ket{e}$ and $\ket{g}$, the total energy change contains another contribution that corresponds to what we called quantum heat in Chapter \ref{Chapter1}. Indeed, the generalized measurement affects the coherences in the $\{\ket{e},\ket{g}\}$ basis, which has a footprint in the energy balance. In particular, when a jump occurs (outcomes $1$ or $2$), the all the coherences in the $\{\ket{e},\ket{g}\}$ basis are erased, just like a projective measurement of $\sigma_z$ would do. 

The elementary evolution of the Qubit between times $t_n$ and $t_{n+1}$ corresponds to a generalized measurement process with three outcomes. Obtaining outcome $r_\gamma(t_n) = 1$ (resp. $r_\gamma(t_n) =2$) is equivalent to a finding the Qubit in state $\ket{e}$ (resp. $\ket{g}$). Using the results of Section \ref{s:ExQED1} in which the case of a projective measurement of $\sigma_z$ is studied, we show that the increment of quantum heat during these two events satisfies:
\bb
\delta Q_{\text{q},\gamma}(t_n) = \left\{\begin{array}{ll}
\hbar\omega_\text{q} P_{g,\gamma}(t_n), &r_\gamma(t_n) = 1 \\ 
-\hbar\omega_\text{q}P_{e,\gamma}(t_n), &r_\gamma(t_n) = 2,
\end{array} \right. \label{d2:dQq12}
\ee
where we have introduced the populations of state $\ket{e}$ and $\ket{g}$ respectively $P_{e,\gamma}(t_n) = \vert\bra{e}\psi_\gamma(t_n)\rangle\vert^2$ and $P_{g,\gamma}(t_n) = \vert\bra{g}\psi_\gamma(t_n)\rangle\vert^2$. Note that for these two outcomes, the post-measurement state is different from the case of a plain $\sigma_z$ measurement: For outcome $r_\gamma(t_n) =1$ which corresponds to finding the Qubit in the excited level, the measurement projects it onto $\ket{g}$. Reciprocally, for outcome $r_\gamma(t_n) =2$ the Qubit is projected onto $\ket{e}$. This means that this measurement process is a destructive measurement of $\sigma_z$.

When outcome $r_\gamma(t_n) =0$ is obtained, the ``no-jump'' operator $M_{0}$ is applied on $\ket{\psi_\gamma(t_n)}$. This evolution can be seen as a weak measurement of the observable $\sigma_z$ in the sense that it weakly affects the relative weights of states $\ket{e}$ and $\ket{g}$ in the Qubit's state. This induces a quantum heat contribution proportional to $\Delta t$: 
\bb
\delta Q_{\text{q},\gamma}(t_n) &=& -\dfrac{1}{2}\bra{\psi_\gamma(t_n)}\{M_1^\dagger M_1 - p_1(t_n),H_\text{q}(t)\}\ket{\psi_\gamma(t_n)}\nonumber\\
&&-\dfrac{1}{2}\bra{\psi_\gamma(t_n)}\{M_2^\dagger M_2 - p_2(t_n),H_\text{q}(t)\}\ket{\psi_\gamma(t_n)}, \quad r_\gamma(t_n) = 0.\quad \label{d2:dQq0}
\ee
Whatever the measurement outcome, the quantum heat increment vanishes if the Qubit state does not carries coherences in the $\{\ket{e},\ket{g}\}$ basis.

Eventually, the total heat increment at time $t_n$ introduced for a general unraveling in Section \ref{s2:1stLaw} can be split into:
\bb
\delta Q_\gamma(t_n) = \delta Q_{\text{cl},\gamma}(t_n) + \delta Q_{\text{q},\gamma}(t_n).
\ee
The two contributions to the heat flow echoes the double role played by the thermal reservoir: As a classical reservoir, it randomly exchanges excitations with the Qubit; it also extracts information about the system and destroys its coherences just like a measuring apparatus.

\subsubsection{Two-point measurement protocol and Second law}
\label{s2:TPMP}

The IFT can be probed owing to the platform presented above. In this section, we consider the protocol in which the JE is satisfied, i.e.:
\begin{itemize}
\item The Qubit is initially in equilibrium with the thermal reservoir (the drive is switched off $H_\text{q}(t_i) = H_0$). Its initial internal energy is measured, yielding outcome $U_i$. The corresponding energy eigenstate state $\ket{\psi_i} = \ket{\psi_\gamma(t_i)}$ is obtained with probability $p_i = e^{-U_i/k_\text{B}T}/Z_i$. 
\item The drive is switched on. The driven Qubit evolves in contact with the reservoir up to time $t_f$ where the drive is switched off.
\item A final energy measurement is performed. Its outcome is denoted $U_f$ and the final state of the trajectory is $\ket{\psi_f} = \ket{\psi_\gamma(t_f)}$.
\end{itemize}

If we moreover define the time-reversed protocol as starting in equilibrium with the reservoir such that $p_f = e^{-U_f/k_\text{B}T}/Z_f$, the IFT takes the form of the JE. This can be proved noting that the boundary entropy production is:
\bb
\Delta_\text{i}^\text{b} s[\gamma] = \dfrac{1}{T}(\Delta U[\gamma]- \Delta F),
\ee
and computing the conditional entropy production. For this we first determine the equilibrium state: 
\bb
\pi_\text{eq} = \exp(-\omega_\text{q}(t)\sigma_z)/\text{Tr}\{\exp(-\omega_\text{q}(t)\sigma_z)\} = P_e^{(eq)}(t)\ket{e}\bra{e} + P_g^{(eq)}(t)\ket{g}\bra{g},
\ee
with $P_e^{(eq)}(t) = 1- P_g^{(eq)}(t) = e^{-\hbar\omega_\text{q}(t)/k_\text{B}T}/(1+e^{-\hbar\omega_\text{q}(t)/k_\text{B}T})$. This allows to rewrite the time-reversal condition \eqref{eq2:TimeRevCrooks} in that case as:
\bb
\tilde M_{\tilde r_{\tilde \gamma}(t_n)} = e^{\delta Q_{\text{cl},\gamma}(t_{N-n+1})/k_\text{B}T}M_{ r_{\gamma}(t_{N-n+1})}^\dagger\label{eq2:DBCop}
\ee
Using Eq.\eqref{eq2:pcond}, we can write the conditional probability of the time-reversed trajectory $\tilde\gamma$ as:
\bb
\tilde p(\tilde r[\tilde\gamma]\vert f)&=& e^{Q_\text{cl}[\gamma]/k_\text{B}T} \left\vert \bra{\psi_i}\overleftarrow{\prod_{n=1}^N} M^\dagger_{r_\gamma(t_{N-n+1})}\ket{\psi_f}\right\vert^2\nonumber\\
&=& e^{Q_\text{cl}[\gamma]/k_\text{B}T} \left\vert \bra{\psi_f}\overrightarrow{\prod_{n=1}^N} M^\dagger_{r_\gamma(t_{N-n+1})}\ket{\psi_i}\right\vert^2\nonumber\\
&=&  e^{Q_\text{cl}[\gamma]/k_\text{B}T} \left\vert \bra{\psi_f}\overleftarrow{\prod_{n=1}^N}  M^\dagger_{r_\gamma(t_{n})}\ket{\psi_f}\right\vert^2\nonumber\\
&=& e^{Q_\text{cl}[\gamma]/k_\text{B}T} p(r[\gamma]\vert i).
\ee
such that $\Delta_\text{i}^\text{c}s[\gamma] =e^{Q_\text{cl}[\gamma]/k_\text{B}T}$. Finally we get:
\bb
\Delta_\text{i}s[\gamma] = \dfrac{1}{T}(\Delta U[\gamma] - Q_\text{cl}[\gamma] - \Delta F) \label{eq2:DisQubitThTPMP},
\ee
which proves Eq.\eqref{eq2:JETPMP}.\\

As mentioned in Section \ref{s2:TimeReversal}, initializing the reversed protocol in a thermal equilibrium distribution is not the only relevant choice. The IFT can also be computed when the reversed protocol is assumed to start from the final state of the direct protocol. The allowed class of direct protocols can also be extended (while staying within the two-point measurement framework): we now consider that the initial state $\ket{\psi_i}$ (resp. final state $\ket{\psi_f}$) of trajectory $\gamma$ is obtained from a projective measurement of an arbitrary observable of the Qubit, performed on the initial (arbitrary) mixed state $\rho_i = \rho_\gamma(t_i)$ (resp. the final mixture $\rho_f = \rho_\text{q}(t_f)$). This leads to a more general version of the IFT proved and the associated form of the 2nd law \cite{Horowitz13}:
\bb
\moy{\Delta_\text{i}s[\gamma]}_\gamma = k_\text{B}\Delta S_\text{VN} - \dfrac{\moy{Q_\text{cl}[\gamma]}_\gamma}{T} \label{eq2:2ndLawTPMP}
\ee
where we have introduced the variation of Von Neumann entropy of the Qubit's density operator:
\bb
\Delta S_\text{VN} = S_\text{VN}[\rho_f]-S_\text{VN}[\rho_i].
\ee

\subsection{Application to a time-dependent Stark shift}
\label{s2:Stark}
In order to illustrate the JE can be tested in the platform proposed in Section \ref{s2:SetupEngineered}, we consider the case of a drive of the form:
\bb
H_\text{d}(t) = \hbar \omega_0 \mu(t-t_i)\ket{e}\bra{e}, \quad t\in [t_i,t_f] \label{d2:StarkDrive}.
\ee
where $\mu \ll \Gamma$ is the rate at which the Qubit frequency is varied. Such drive can be obtained owing to a linearly time-dependent Stark shift.

We consider the protocol leading to JE detailed in Section \ref{s2:TPMP}.  As the drive $H_\text{d}$ is diagonal in the bare Qubit eigenbasis $\{\ket{e},\ket{g}\}$ and the Qubit is initialized either in $\ket{e}$ or $\ket{g}$, the Qubit's state $\ket{\psi_i\gamma(t_n)}$ is at any time either $\ket{e}$ or $\ket{g}$. As a consequence, the quantum heat increment is always zero in this example.

The increment of work between $t$ and $t+\Delta t$ verifies:
\bb
\delta W_\gamma(t) = \hbar\omega_0 \mu P_{e,\gamma}(t),
\ee
\noindent where $P_{e,\gamma}(t) = \bra{\psi_\gamma(t)}\Pi_e\ket{\psi_\gamma(t)}$. The increment of classical heat (which corresponds to the total increment of heat) is given by:
\bb
\delta Q_{\text{cl},\gamma}(t) &=& \left\{ \begin{array}{c}
-\hbar\omega_\text{q}(t), \quad r_\gamma(t) = 1 \\ 
\hbar\omega_\text{q}(t), \quad r_\gamma(t) = 2 \\ 
0, \quad r_\gamma(t) = 0.
\end{array}\right.  \label{eq2:dQclHdiag}
\ee
\noindent with $\omega_\text{q}(t) = \omega_0(1+\mu(t-t_i))$.

\begin{figure}
\begin{center}
\includegraphics[width=0.8\textwidth]{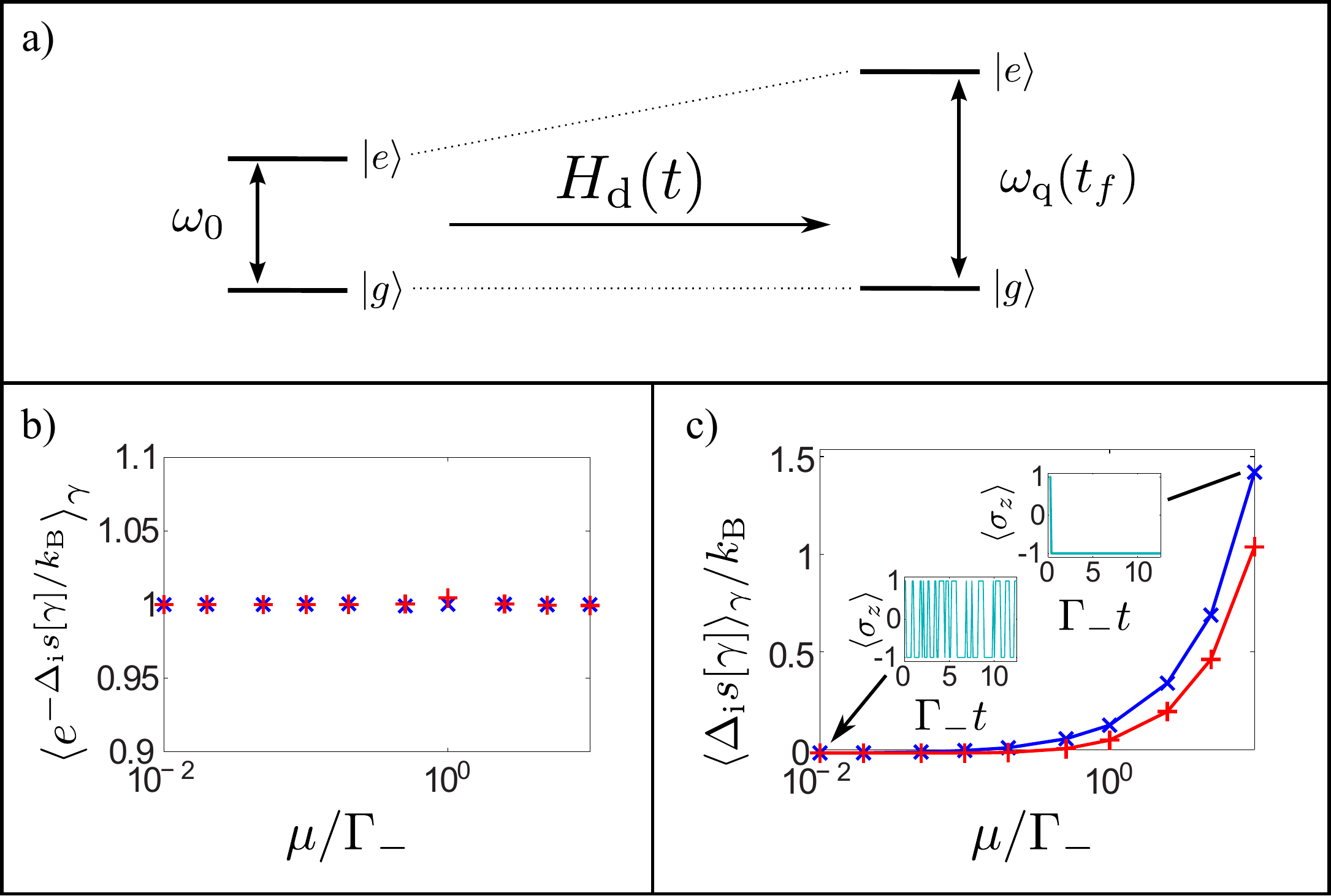}
\end{center}
\caption[Application: Time-dependent Stark shift of a Qubit]{Application: Time-dependent Stark shift of a Qubit. a) Transformation performed on the Qubit. The frequency is shifted from $\omega_\text{q}(t_i) = \omega_0$ to $\omega_\text{q}(t_f)$ with a constant slope $\mu$. The Qubit is in contact with the thermal reservoir at temperature $T$ at any time. b) Test of JE for different value of the slope $\mu$ of the frequency shift. c) Entropy production averaged over the trajectories as a function of the slope $\mu$. Insets: Expectation value of $\sigma_z$ along a single trajectory $\gamma$. For b) and c), we used two different temperatures fulfilling $\hbar\omega_0/k_\text{B}T = 0.3$ (red '+') and $\hbar\omega_0/k_\text{B}T = 3$ (blue 'x'). Number of trajectories: $10^5$.  \label{f2:StarkShift}}
\end{figure}

When there is no quantum heat, Eq.\eqref{eq2:DisQubitThTPMP} can be simplified to:
\bb
\Delta_\text{i}s[\gamma] = \dfrac{1}{T}(W[\gamma] -\Delta F),
\ee
We simulated the a sample of $10^5$ quantum trajectories for a  high and a low temperature, for different values of the slope $\mu$ of the frequency shift. We fixed the duration of the trajectory to $t_f - t_i = 10/\Gamma_-$. In Fig.\ref{f2:StarkShift}b, we show that the JE is satisfied in all the considered cases. In Fig.\ref{f2:StarkShift}c, we plot the average entropy production as a function of $\mu$. For $\mu$ much smaller than the rate of emission of photons $\Gamma_-$, the entropy production is very small: this corresponds to a quasi-static transformation in which the Qubit has enough time to reach the equilibrium state associated with every value of the effective frequency $\omega_\text{q}(t)$. In such a transformation, a very large number of photons is exchanged with the reservoir (see inset). In the opposite limit of a very fast transformation $\mu \gg \Gamma_-$ (a ``quench''), the Qubit is drawn out of equilibrium and the entropy production diverges. In a single trajectory in this limit, there is typically a few photons exchanged with the reservoir (see inset). We have run the simulation for two different temperatures. The higher temperature leads to a smaller entropy production which reflects the fact that the thermalization time typically scales like $1/n_\text{q}$.

\subsection{Influence of the finite detection efficiency}

Up to now, we have assumed that the environment is monitored with $100\%$ efficient measuring apparatuses, i.e. that the detected ${\cal D}_1$ and ${\cal D}_2$ were able to detect all the photons emitted by the three-level atom. This is a necessary condition to generate trajectories of pure states as we have studied above. In practice, the monitoring of the reservoir has a finite detection efficiency: e.g. state-of-the art photon-detectors only allow for a fraction $\eta$ of the emitted photons to be indeed detected \cite{Natarajan12,Goltsman01}. 

In this section, we investigate how the QJ unraveling is modified when only a fraction $\eta$ of the photons are detected and what are the consequences on the test of JE. 

\subsubsection{Modified unraveling}

The finite detection efficiency adds some classical uncertainty on the evolution of the Qubit: when no photon is detected between $t_n$ and $t_{n+1}$, the operator does not know if (i) there was indeed no photon emitted, (ii) a photon was emitted by the Qubit and detector ${\cal D}_1$ missed it, or (iii) a photon was absorbed by the Qubit and detector ${\cal D}_2$ missed it. Consequently, there exist several different possible (or fictitious) trajectories $\gamma_\text{F}$ of pure states, which cannot be distinguished by the measurement record: Each of these trajectories corresponds to a particular sequence of undetected emissions and absorptions. 

As a consequence, in a single realization of the protocol, the state of the Qubit conditioned on the detected photons only must be described by a stochastic density matrix $\rho_\gamma$, rather than a wavevector. In the limit where $\eta = 0$ (no detector), the evolution of this density matrix is captured by the Lindblad equation Eq.\eqref{eq2:LEQubitTh}, such that $\rho_\gamma(t_n) = \rho_\text{q}(t_n)$. For $0<\eta<1$, the evolution of $\rho_\gamma(t_n)$ is still conditioned on the stochastic measurement outcome $r_\gamma(t_n) \in \{1,2,0\}$ of the detector at time $t_n$. However, each measurement outcome is now associated to a super-operator ${\cal E}^{(\eta)}_{k}[\rho_\gamma]$ acting on $\rho_\gamma(t_n)$. Detecting one photon in detector ${\cal D}_1$ (resp. ${\cal D}_2$) between time $t_n$ and $t_{n+1}$ corresponds to applying the super-operator ${\cal E}^{(\eta)}_{1}[\rho_\gamma] = \eta M_1 \rho_\gamma M_1^\dagger$ (resp. ${\cal E}^{(\eta)}_{2}[\rho_\gamma] = \eta M_2 \rho_\gamma M_2^\dagger$), which occurs with probability $p_1^{(\eta)}(t) = \eta \text{Tr}\{M_1^\dagger M_1\rho_\gamma(t)\}$ (resp. $p_2^{(\eta)}(t) = \eta \text{Tr}\{M_2^\dagger M_2\rho_\gamma(t)\}$). When no photon is detected, which occurs with probability $p_0^{(\eta)}(t) = 1-p_1^{(\eta)}(t)-p_2^{(\eta)}(t)$, a super-operator decreasing the state purity is applied\footnote{${\cal E}_0^{(\eta)}$ is a linear interpolation between ${\cal E}_0^{(1)}[\rho] = M_0 \rho_\gamma M_0^\dagger$ applied in the perfect efficiency quantum jump formalism and ${\cal E}_0^{(0)}[\rho] = {\cal L}_T[\rho]\Delta t$ where ${\cal L}_T$ is the Lindbladian defined in Eq.\eqref{eq2:LindbladianTh}}: 

\bb
{\cal E}^{(\eta)}_{0}[\rho_\gamma] &=& M_0\rho_\gamma M_0^\dagger + (1-\eta)(M_1 \rho_\gamma M_1^\dagger + M_2 \rho_\gamma M_2^\dagger)\nonumber\\
&=& \idop -idt[H_\text{q}(t_n),\rho_\gamma(t_n)] - \dfrac{\eta \Delta t}{2}(\Gamma_-(t_n)\left\{\Pi_e,\rho_\gamma\right\} + \Gamma_+(t_n)\left\{\Pi_g,\rho_\gamma\right\})\nonumber\\
&&+(1-\eta)\left(\Gamma_-(t_n)\Delta t D[\sigma_-]+  \Gamma_+\Delta t D[\sigma_+]\right)\rho_\gamma.
\label{E0eta}\ee

\noindent where $\Pi_e$ (resp. $\Pi_g$) is the projector onto state $\ket{e}$ (resp. $\ket{g}$). After applying the super-operator ${\cal E}^{(\eta)}_{k}$ at time $t_n$, the density matrix has to be divided by $p_k^{(\eta)}(t_n)$ in order to be renormalized. \\

\subsubsection{Generalized Jarzynski equality.}

\noindent We now extend the definitions of thermodynamic quantities into the regime $\eta <1$. For the sake of simplicity, we restrict the study to the time-dependent Stark shift driving Hamiltonian defined in Eq.\eqref{d2:StarkDrive}, such that no quantum heat is exchanged. Moreover, we suppose that the protocol leading to JE is applied.

\begin{figure}[h!]
\begin{center}
\includegraphics[width=\textwidth]{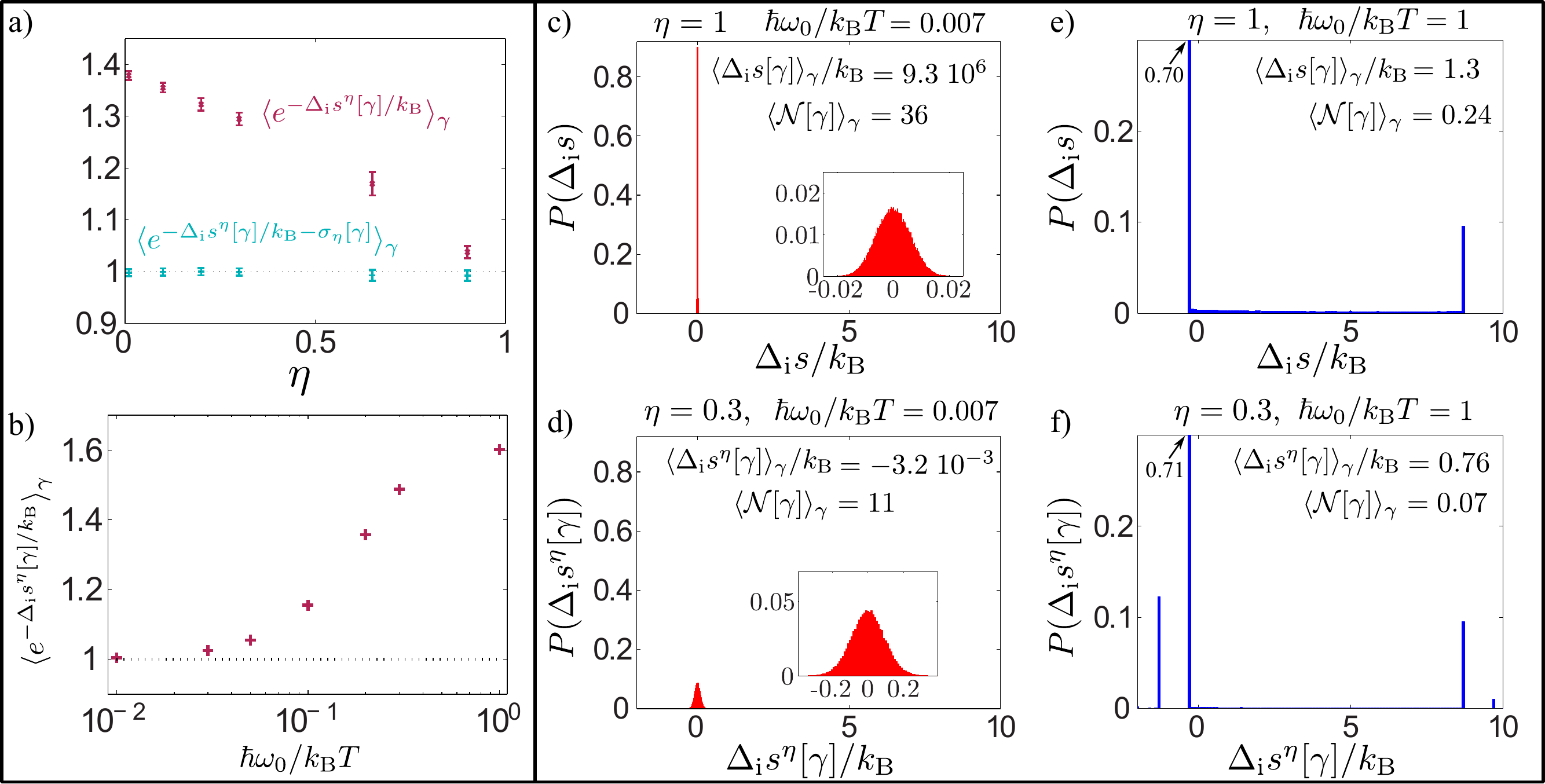}
\end{center}
\caption[Thermodynamic consequences of finite detection efficiency]{Thermodynamic consequences of finite detection efficiency. a) Test of Jarzynski equality at finite efficiency, for a drive $H_\text{d}(t) = \hbar \omega_\text{q}\mu (t-t_i)\Pi_e$. Purple `x': Measured parameter $\langle e^{-\Delta_\text{i}s^\eta[\gamma]/k_text{B}}\rangle_\gamma$. The magnitude of the deviation (positive for this transformation) increases monotonically when $\eta$ decreases. Blue `+': The plotted parameter included the correction term $\sigma[\gamma]$. We have used $N_\text{traj} = 10^4$ trajectories $\gamma$ and $10^4$ ficitious trajectories $\gamma_F$ for each trajectory. The error bars stand for the standard deviation divided by $\sqrt{N_\text{traj}}$. b) Jarzynski's parameter without correction for $\eta = 0.1$ as a function of the temperature. The error bars are smaller than the marker size. c)-f) Distributions of the entropy production during the transformation for $\eta = 1$ in c),e) and $\eta =  0.3$ in d),f). c) and d) correspond to high temperature ($\hbar\omega_0/k_\text{B}T = 7\; 10^{-3}$) and therefore a large mean number of photons exchanged per trajectory $\moy{{\cal N}[\gamma]}_\gamma = N - \moy{{\cal N}_0[\gamma]}_\gamma$. The inserts show the same distribution with rescaled axes to evidence the similarity of shape. e) and f) correspond to a lower temperature ($\beta\omega_1 = 1$) and a few photon exchanged. {\it Parameters}: For a) $\hbar\omega_0/k_\text{B}T  = 0.1$, $\gamma_\text{q} t_f = 0.5$, $\epsilon/\gamma = 600$.  For b) and c): $\Gamma_- t_f = 1$, $\mu/\Gamma_- = 9$, $N_\text{traj} = 10^5$. \label{f2:FiniteEff}
}
\end{figure}

With the definitions proposed above, the increment of measured classical heat reads:
\bb
\delta Q_\text{cl}^\eta(t_n) =  \left\{\begin{array}{c}
-\hbar\omega_\text{q}(t_n),\quad r_\gamma(t_n)=1 \\ 
\hbar\omega_\text{q}(t_n),\quad r_\gamma(t_n)=2 \\
0, \quad r_\gamma(t_n)=0 
\end{array} \right.
\label{dQR}.
\ee
Despite its apparent similarity with the definition Eq.\eqref{d2:dQcl} of the classical heat increment in the perfect efficiency regime, this energy flow does not capture the entire classical heat flow dissipated in the heat bath, as it does not take into account the energy carried by the undetected photons. Therefore, the measured entropy production 
\bb
\Delta_\text{i}s^\eta[\gamma] = \dfrac{1}{T}(\Delta U[\gamma]-Q_\text{cl}^\eta[\gamma]-\Delta F),
\ee
\noindent does not check JE as illustrated in Fig.\ref{f2:FiniteEff}a and b. The difference with $1$ is all the larger as the efficiency $\eta$ is smaller, and JE is recovered in the limit $\eta\to 1$, showing that our definitions for the finite-efficiency case still hold in the perfect efficiency limit. Remarkably, however, it is possible to formulate another equality taking into account the finite efficiency, converging towards the standard JE when $\eta\rightarrow 1$:
\bb
\moy{e^{-\Delta_\text{i}s^\eta[\gamma]/k_\text{B} - \sigma_\eta[\gamma]}}_\gamma = 1. \label{eq2:JEeta}
\ee

This equality is reminiscent of JE, but it includes a trajectory-dependent correction term $\sigma_\eta[\gamma]$. We now show that $\sigma_\eta[\gamma]$ can be computed from the measurement record only.\\

The probability if a trajectory  $\gamma$ can still be written under the form $p(\gamma) = p_ip(r[\gamma]\vert i)$, but now $p(r[\gamma]\vert i)$ reads: 
\bb
p(r[\gamma]\vert i) = \text{Tr}\left\{\ket{f}\bra{f}\left(\overleftarrow{\prod_{n=1}^N} {\cal E}_{r_\gamma(t_n)}^{(\eta)}\right)\big[\ket{i}\bra{i}\big]\right\},\label{eq2:Pdeta}
\ee
\noindent As in section \ref{s2:Stark}, the probability for trajectory $\gamma$ to start in state $\rho_\gamma(t_i) = \ket{\psi_i}\bra{\psi_i}$ fulfils thermal distribution $p_i = Z^{-1}_ie^{-U_i/k_\text{B}T}$. Similarly, we introduce $\tilde p(\tilde\gamma) = p_f \tilde p(\tilde r [\tilde \gamma]\vert f)$, with $p_f = Z^{-1}_f e^{-U_f/k_\text{B}T}$, the probability of the time-reversed trajectory $\tilde\gamma$ corresponding to $\gamma$. It is generated by the time-reversed drive $H_\text{d}(t_f-t)$ and the sequence of time-reversed super-operators $\{\tilde{\cal E}^{(\eta)}_{\tilde r_{\tilde \gamma}(t_n)}\}$ such that:

\bb
\tilde p(\tilde r[\tilde \gamma]\vert f) = \text{Tr}\left\{\ket{i}\bra{i}\left(\overleftarrow{\prod_{n=1}^N} \tilde{\cal E}_{{\tilde r}_{\tilde\gamma}(t_n)}^{(\eta)}\right)\big[\ket{f}\bra{f}\big]\right\},\label{eq2:Preta}
\ee

When $\eta=1$, the the time-reversed operators at time $t_n$ can be deduced from Eq.\eqref{eq2:DBCop}:
\bb 
\tilde {\cal E}^{(1)}_1[\rho_\gamma](t_n) &=& e^{-\hbar \omega_\text{q}(t_n)/k_\text{B}T}M_1^\dagger \rho_\gamma M_1 \label{Er1}\\
\tilde{\cal E}^{(1)}_2[\rho_\gamma] (t_n)&=& e^{\hbar \omega_\text{q}(t_n)/k_\text{B}T}M_2^\dagger \rho_\gamma M_2 \label{Er2}\\
\tilde {\cal E}^{(1)}_0[\rho_\gamma](t_n) &=& M_0^\dagger \rho_\gamma M_0.\label{Er0}
\ee
\noindent which implies that $\Delta_\text{i}^\text{c} s[\gamma] = e^{Q_\text{cl}[\gamma]/k_\text{B}T}$. When $\eta < 1$, the ratio $\tilde p(\tilde r[\tilde\gamma]\vert f)/p(r[\gamma]\vert i)$ is modified. \\

It is useful to decompose ${\cal E}^{(\eta)}_0$ into three super-operators conserving the purity of the Qubit's state: 
\bb
{\cal E}^{(\eta)}_0 = {\cal E}^{(\eta)}_{00}+{\cal E}^{(\eta)}_{01}+{\cal E}^{(\eta)}_{02},\label{decomposition}
\ee 
with 
\bb
{\cal E}^{(\eta)}_{01}[\rho_\gamma] &=& (1-\eta){\cal E}^{(1)}_1[\rho_\gamma], \nonumber\\
{\cal E}^{(\eta)}_{02}[\rho_\gamma] &=& (1-\eta){\cal E}^{(1)}_2[\rho_\gamma], \nonumber\\
{\cal E}^{(\eta)}_{00}[\rho_\gamma] &=& {\cal E}^{(1)}_0[\rho_\gamma].\label{eq2:E0i}
\ee

\noindent From this decomposition, we generate the set ${\cal F}[\gamma]$ of the fictitious trajectories $\gamma_F$ of pure states, compatible with a given real trajectory $\gamma$: a trajectory $\gamma_F \in  {\cal F}[\gamma]$ is built by applying a sequence of super-operators $\{{\cal E}_{\displaystyle r_{\gamma_F}(t_n)}\}_{1\leq n \leq N}$, according to the following rules:
\begin{itemize}
\item $r_{\gamma_F}(t_n) = r_{\gamma}(t_n)$ if $r_{\gamma}(t_n)$ belongs to $\{1,2\}$ (i.e. when a photon has been detected).
\item $r_{\gamma_F}(t_n)$ is drawn from $\{00,01,02\}$ with probability $p_{0k} = \text{Tr}\{{\cal E}_{0k}[\rho_\gamma(t_n)\}/p_0^{(\eta)}$, $k\in\{1,2,0\}$, when $r_{\gamma}(t_n) = 0$ (no photon detected).
\end{itemize}
\noindent  The set ${\cal F}[\gamma]$ thus contains $3^{{\cal N}_0[\gamma]}$ fictitious trajectories, where ${\cal N}_0[\gamma]$ is the number of time-steps during which no photon is detected in $\gamma$. \\

The perfect-efficiency fictitious trajectories are interesting because they can be time-reversed using the rules Eq.\eqref{Er1}-\eqref{Er0} and as a consequence fulfil $\tilde p(\tilde \gamma_F) = \exp(-(\Delta U[\gamma_F]- Q_\text{cl}[\gamma_F]-\Delta F))\tilde p(\gamma_F)$. Here, because of the initial and final projective measurements of the system's energy (which are assumed to be of perfect efficiency), we have $\Delta U[\gamma_F] = \Delta U[\gamma]$ if $\gamma_F\in {\cal F}[\gamma]$. The classical heat $Q_\text{cl}[\gamma_F]$ along trajectory $\gamma_F$ can be computed from Eq.\eqref{eq2:dQclHdiag}. Besides, the probabilities of the fictitious trajectories in ${\cal F}[\gamma]$ are related to the probability of $\gamma$ according to:
\bb
  p(\gamma) = \sum_{\gamma_F\in \mathcal{F}[\gamma] } p(\gamma_F).
\ee

Exploiting these properties, the following equality can be derived:
\bb
1 &=& \sum_{\gamma}\tilde p(\tilde\gamma)\nonumber\\
&=&  \sum_{\gamma} \sum_{{\cal F [\gamma]}}  \tilde p(\tilde\gamma_F)\nonumber\\
&=& \sum_{\gamma} \sum_{{\cal F [\gamma]}} p(\gamma_F) e^{-(\Delta U[\gamma_F]-\Delta F - Q_\text{cl}[\gamma_F])/k_\text{B}T}\nonumber\\
&=& \sum_{\gamma} p(\gamma) e^{-(\Delta U[\gamma]-\Delta F - Q_\text{cl}^{\eta}[\gamma])/k_\text{B}T}\nonumber\\
&&\quad\times\sum_{{\cal F [\gamma]}} p(\gamma_F\vert \gamma) e^{-( Q_\text{cl}^{\eta}[\gamma]- Q_\text{cl}[\gamma_F])/k_\text{B}T}.\nonumber\\\label{Pdeta2}
\ee
\noindent We have introduced the conditional probability $p(\gamma_F\vert \gamma) = p(\gamma_F)/p(\gamma)$ of the fictitious trajectory $\gamma_F\in {\cal F}[\gamma]$, given the detected trajectory $\gamma$. In order to find Eq.\eqref{eq2:JEeta}, we define: 
\bb
\sigma_\eta[\gamma] = -\log \sum_{{\cal F}[\gamma]} p(\gamma_F\vert\gamma) e^{(Q_\text{cl}[\gamma_F]-Q_\text{cl}^\eta[\gamma])/k_\text{B}T_\text{eff}}.
\ee\\

The correction $\sigma_\eta[\gamma]$ can be numerically computed, based solely on the (imperfect) measurement record $r[\gamma]$. The computation involves the simulation of a sample of the trajectories $\gamma_F \in {\cal F}[\gamma]$ for each trajectory $\gamma$. The corrected expression $\langle e^{-\Delta_\text{i}s[\gamma]/k_\text{B} -\sigma_\eta[\gamma]} \rangle_\gamma$ is plotted in Fig.\ref{f2:FiniteEff}a: JE is verified, showing that the experimental demonstration of a FT in a realistic setup is within reach.

\subsubsection{A quantum to classical boundary}

The correction term $\sigma_\eta[\gamma]$ can be seen as a correction to the detected entropy production taking into account the information loss due to missed photons. Interestingly, the information carried by a single photon is not always the same, depending on the temperature or the speed of the drive. 

\sloppy
In Fig.\ref{f2:FiniteEff}b, the deviation from JE is studied as a function of the reservoir temperature for $\eta = 0.1$. We see that at high temperature such that $\hbar\omega_0/k_\text{B}T \ll 1$, JE is verified even without taking into account the correction. This effect can be seen as a quantum-to-classical border in the thermodynamic framework, which is illustrated in Fig.\ref{f2:FiniteEff}c to f. In the quasi-static limit characterized by \mbox{$\mu \ll \Gamma_- $} (high temperature case), many photons are exchanged and the system is allowed to thermalize before any significant work is done: the transformation is reversible. We see from Fig.\ref{f2:FiniteEff}c and d that in this limit, the effect of detection inefficiency in the entropy production is merely to widen the spread of the distribution but conserving its Gaussian shape around zero.  The statistical properties of the distribution are not affected, even if many photons are missed, and JE is verified (Fig.\ref{f2:FiniteEff}b). This corresponds to a semi-classical situation where, even though the exchange of heat is quantized, the overall thermodynamic behavior of the system is equivalent to a classical ensemble. In this limit, there is no significant information loss due to the undetected photons.

On the other hand, as soon as work and heat are exchanged at similar rates \mbox{$\epsilon \sim \Gamma_-$}, the system is brought out of equilibrium and the transformation is irreversible. In this limit, few photons are exchanged before a non negligible amount of work is done on the system. Each missed photon represents, then, a significant information loss,  such that the measured distribution of entropy production is severely affected (see Fig.\ref{f2:FiniteEff}e and f). Such information loss is quantifiable by the violation of the non-corrected JE, and is a direct consequence of the quantization of heat exchanges.
\fussy

\subsection{Beyond the two point measurement protocol: case of the spontaneous emission.}

The formalism presented in this chapter allows to go beyond the two-point measurement approach and look at transformations in which no projective measurement is performed on the Qubit at the end of the transformation. In such a case, the computation of the conditional entropy production presented in Section \ref{s2:TPMP} is still valid. However, the average boundary entropy production is not any more related to the variation of Von Neumann entropy because the set of the possible final states $\{\ket{\psi_f}\}$ of the trajectories is not necessarily orthogonal. 
The average boundary entropy production reads:
\bb
\moy{\Delta_\text{i}^\text{b}s[\gamma]}_\gamma = -\sum_f p_f\log p_f + \sum_i p_i\log p_i.
\ee
where $-\sum_\text{f} p_f\log p_f$ is in general different from $S_\text{VN}[\rho_f]$ because $\rho_\text{f}$ cannot always be written as a diagonal matrix with diagonal coefficients $\{p_f\}$. This phenomenon reflects the fact that the measurement record $r_\gamma(t)$ allows to distinguish without ambiguity between two different trajectories, while their final states have an overlap and therefore could not be perfectly distinguished owing to a single projective measurement of an observable. In this situation, the system's state and the memory of measuring apparatus $\cal M$ have non trivial correlations, the Von Neumann entropy of the Qubit is not anymore the good thermodynamic entropy. \\

As an example, we consider that the Qubit is prepared in state $\ket{\psi_i} = \ket{+} = (\ket{e} + \ket{g})/\sqrt{2}$ at time $t_i$ and coupled to a thermal reservoir at temperature $T = 0K$ in which it can spontaneously emit a photon. We consider trajectories between $t_i = 0$ and $t_f$. The Qubit is not driven such that $H_\text{q}(t) = H_0$. For the sake of simplicity, we assume a perfect detection efficiency $\eta = 1$.

\begin{figure}[h!]
\begin{center}
\includegraphics[width=0.65\textwidth]{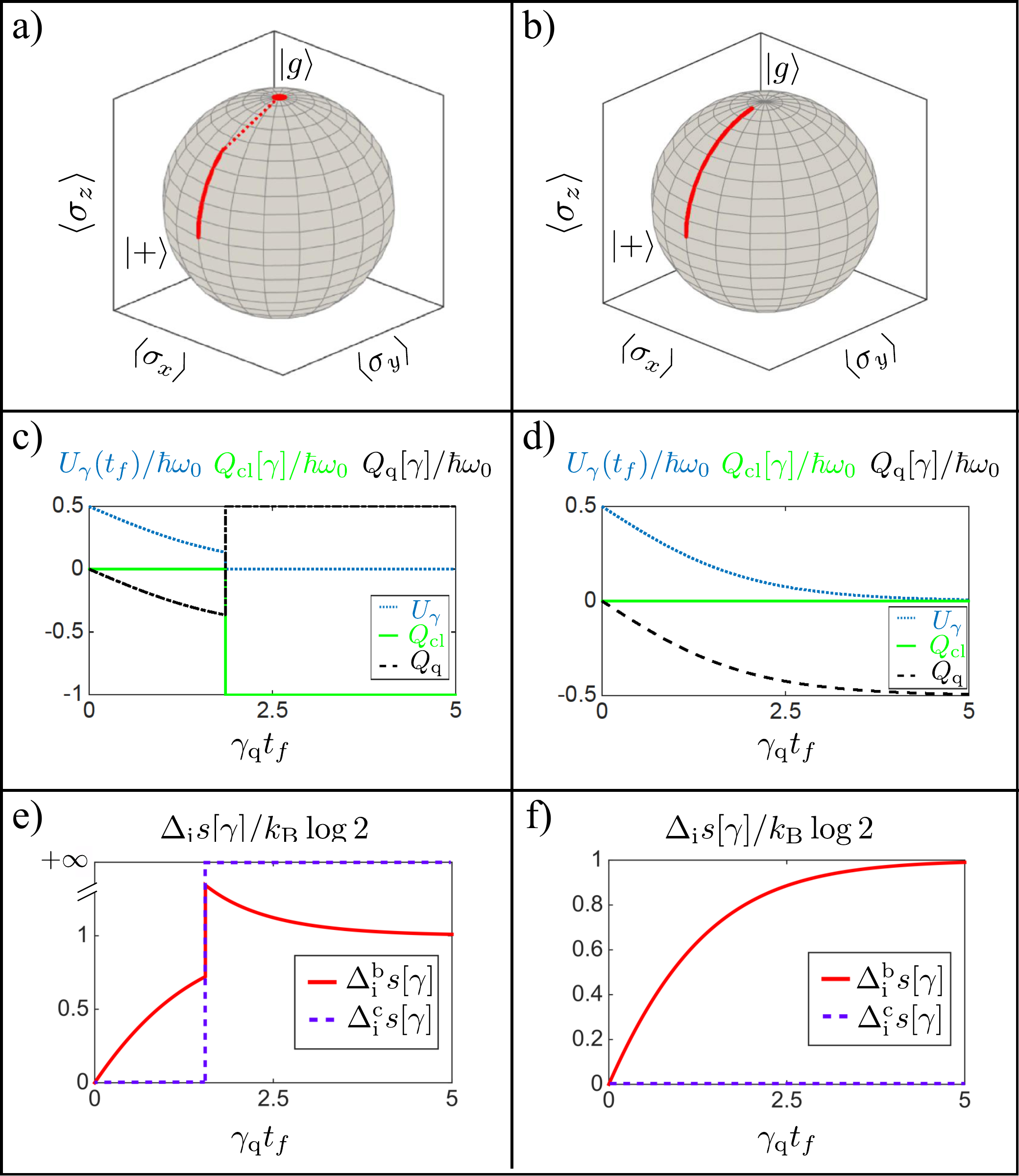}
\end{center}
\caption[Thermodynamics of spontaneous emission]{Thermodynamics of spontaneous emission. A Qubit is prepared in the state $\ket{+} = (\ket{e}+\ket{g})/\sqrt{2}$ and coupled to a zero-temperature reservoir monitored with a photo-counter. a), b): Quantum trajectories of the Qubit's state in the Bloch sphere, in the case of a ``jump'' trajectory (a) and ``no-jump'' trajectory (b). In order to enhance the visibility, the sphere has been oriented with the south pole (state $\ket{g}$) at the top. c), d): Internal energy, classical and quantum heat contributions in units of $\hbar \omega_0$ as a function of time. e),f): Two contributions to entropy production: $\Delta_\text{i}^\text{c}s[\gamma]$ and $\Delta_\text{i}^\text{b}s[\gamma]$, in unit of bits.}
\label{f2:SpontEm} \end{figure}

There are two classes of trajectories giving rise to two possible final states $\ket{\psi_\gamma(t_f)}$. In the ``jump" trajectories (see Fig.\ref{f2:SpontEm}a), the Qubit relaxes to the ground state by emitting a photon, such that $\ket{\psi_\gamma(t_f)} = \ket{g}$, which happens with a probability $p_\text{j}(t_f) = (1-e^{-\Gamma t_f})/2$. Reciprocally the ``no-jump" trajectory (see Fig.\ref{f2:SpontEm}b) occurs with the probability $p_\text{nj}(t_f) = (1+e^{-\Gamma t_f})/2$: The Qubit deterministically evolves under the effective non-hermitian Hamiltonian $H_\text{eff} = H_0 -(i\gamma_\text{q}/2)\ket{e}\bra{e}$ until $t_f$, such that the final state reads: 
\bb
\ket{\psi_\gamma(t_f)}  = \dfrac{e^{-(\gamma_\text{q}/2+i\omega_0)t_f}}{\sqrt{1+e^{-\gamma_\text{q}t_f}}}\ket{e}+\dfrac{1}{\sqrt{1+e^{-\gamma_\text{q}t_f}}}\ket{g}.
\ee

Remarkably for $t_f\gg \gamma_\text{q}^{-1}$, the Qubit also ends up in the ground state, while no photon has been emitted: Just like recording a click, not detecting a photon increases our knowledge on the system's state. Eventually, the whole process of spontaneous emission at large times $t_f\gg \gamma_\text{q}^{-1}$ can be seen as a measurement of the Qubit observable $\sigma_z$, recording  (resp. not recording) a click boiling down to finding the Qubit in $\ket{e}$ (resp. $\ket{g}$). However, such measurement differs from a projective measurement of $\sigma_z$ because the Qubit always ends up in $\ket{g}$ \footnote{According to the classification of Ref.\cite{WisemanBook}, this measurement is \emph{not} minimally disturbing contrary to a projective measurement, i.e. it involves Kraus operators which are not self-adjoint.}.\\ 

We first analyse the energy exchanges along the two kind of trajectories. We first look at the no-jump trajectories: At time $t_f$, the internal energy of the Qubit is
\bb
U_\gamma(t_n) =  \hbar\omega_0\dfrac{e^{-\gamma_\text{q}t_f}}{{1+e^{-\gamma_\text{q}t_f}}} \leq U_\gamma(0) = \dfrac{\hbar\omega_0}{2}.
\ee
\noindent As the Qubit's Hamiltonian is time-independent, no work has been performed. Moreover, no photon has been emitted such that no classical heat has been dissipated in the reservoir: this internal energy variation is only due to quantum heat. If $t_f\gg \gamma_\text{q}^{-1}$, the total quantum heat exchanged is $Q_\text{q}[\gamma] = -\hbar\omega_0/2$, which is consistent with a projective measurement of $\sigma_z$ with outcome $\ket{g}$ (see Section \ref{s:ExQED1}). In the case of a ``jump'' trajectory, a photon is emitted at time $t_j \in [0,t_f]$, which corresponds to an exchange of classical heat $Q_\text{cl} = -\hbar \omega_0$. After the jump, the variation of internal energy is $\Delta U[\gamma] = -\hbar\omega_0/2$, and the total quantum heat exchanged is $Q_\text{q}[\gamma] = \hbar\omega_0/2$: this is consistent with a measurement of $\sigma_z$ with outcome $\ket{e}$. 
The stochastic quantities $U_\gamma(t)$, $Q_\text{cl}[\gamma]$ and $Q_\text{q}[\gamma]$ are plotted on Fig.\ref{f2:SpontEm}c and d for a ``jump'' and ``no-jump'' trajectory respectively.\\

We then analyze the entropy production. As long as no jump takes place, the conditional term $\Delta_\text{i}^\text{c}s[\gamma](t)$ equals $0$ (the evolution is deterministic). It diverges as soon as a click is recorded, which is typical of spontaneous emission. 

Interestingly, the boundary term remains finite in any case, verifying $\Delta_\text{i}^\text{b}s[\gamma] = -k_\text{B} \log[p_\text{nj}(t_f)]$ (resp $-k_\text{B}\log[p_\text{j}(t_f)]$) for the no-jump trajectory (resp. for any jump trajectory). The average boundary entropy production therefore reads $\langle \Delta_\text{i}^\text{b}s[\gamma] \rangle_\gamma  = k_\text{B}H_\text{Sh}[p_\text{nj}(t_f)]$, where $H_\text{Sh}[p] = -p\log(p ) - (1-p)\log(1-p)$ stands for the Shannon's entropy. This entropy production quantifies the information acquired up to time $t_f$ owing to the measurement record. This information allows to distinguish two non-orthogonal states $\ket{e}$ and $\ket{\psi_\text{nj}(t_f)}$. For $t_f\gg \gamma_\text{q}^{-1}$, $\ket{\psi_\text{nj}(t)}$ converges towards $\ket{g}$ and the two states distinguished by the measurement become orthogonal, such that $\langle \Delta_\text{i}^\text{b}s[\gamma] \rangle_\gamma$ converges towards $k_\text{B}\log(2)$ which is not equal to the final Von Neumann entropy $S_\text{VN}[\ket{g}\bra{g}] = 0$ of the Qubit.  \\

\section{Thermodynamics of weak continuous measurement}

\label{s2:ContMeas}
\subsection{Continuous monitoring of an observable}

In this section we present another situation which can be treated within the framework presented in this chapter: the continuous monitoring of an observable of the system. 
This corresponds situations in which the measurement record $r[\gamma]$ is related to an observable $X = \sum_k x_k \ket{x_k}\bra{x_k}$ of the system. Such continuous measurement process can be modelled as a sequence of short measurements of duration $\Delta t$ \cite{Steck06}. These short measurements are weak, in the sense that they do not extract enough information to completely assign a value of the observable, and therefore do not project the system $\cal S$ onto one of the eigenstates of $X$. More precisely, the effect of each of these weak measurements can be modelled by a Kraus operator:
\bb
M_r &=& \lambda e^{-\Gamma_\text{meas}\Delta t(r-X)^2}\nonumber\\
&=& \lambda\sum_k e^{-\Gamma_\text{meas}\Delta t(r-x_k)^2}\ket{x_k}\bra{x_k} \label{eq2:Mweak},
\ee
where $\lambda$ is a normalization factor ensuring that $\sum_r M_r^\dagger M_r = \idop$. Here $\Gamma_\text{meas}$ quantifies the rate at which information about the measured observable is acquired. The strong (projective) measurement is recovered when $\Gamma_\text{meas}\Delta t \gg 1$. In this limit, the only measurement outcomes obtained with a non-zero probability fulfil $r\in\{x_k\}$ and each Kraus operator is a projector onto one of the states $\{\ket{x_k}\}$. 

Here we focus on the limit $\Gamma_\text{meas}\Delta t \ll 1$ in which the effect of the measurement on the system's state is to infinitesimally change the weight of each of the eigenstates of $X$. In this case, the density operator of the system obeys a Lindblad equation with only one jump operator which is the observable $X$ that is measured \cite{Steck06}:
\bb
\dot\rho_{\cal S}(t) = -\dfrac{i}{\hbar}[H_{\cal S}(t),\rho_{\cal S}] + \Gamma_\text{meas} D[X]\rho_{\cal S}(t).
\ee
\noindent Therefore, the effect of the measurement process on $\rho_{\cal S}$ is to induce pure dephasing at a typical rate $\Gamma_\text{meas}$ in the eigenbasis $\{\ket{x_k}\}$ of $X$. Reciprocally, a Lindbladian corresponding to a pure-dephasing channel can be unraveled using the continuous measurement of an observable of the system.\\

The evolution for each single realization of the continuous measurement process can be written as a QSD trajectory conditioned on the stochastic measurement record 
\bb
r_\gamma(t) = \bra{\psi_\gamma(t)}X\ket{\psi_\gamma(t)} + \dfrac{dw_\gamma(t)}{2 \sqrt{\Gamma_\text{meas}}\Delta t},
\ee
where $dw_\gamma(t)$ is a Wigner increment with real values. The Kraus operator that has to be applied on the system's state at time $t_n$ is obtained by expending Eq.\eqref{eq2:Mweak} to first order in $\Delta t$, using I\^o's stochastic calculus conventions\footnote{In particular, $dw_\gamma(t)$ is of order $\sqrt{\Delta t}$ and satisfies $(dw_\gamma(t))^2 = \Delta t$.}. It takes the form of Eq.\eqref{d2:MQSD} with only one jump operator (and therefore one Wigner increment) \cite{Steck06}:
\bb
M_{r_\gamma(t)} = \sqrt{p(dw_\gamma(t))}\left(1-i\Delta t H_{\cal S}(t)- \tfrac{1}{2} \Gamma_\text{meas} \Delta t X^2 + dw_\gamma(t)\sqrt{\Gamma_\text{meas}}X\right). \label{eq2:MX}
\ee

An example of implementation is provided by the hybrid optomechanical system studied in Chapter \ref{Chapter4}, in which monitoring the light emitted by a Qubit allows to monitor continuously the position of a mechanical oscillator.

\subsection{Thermodynamics}
\subsubsection{Nature of the heat}

In this situation, there is no exchange of excitations with the reservoir: the heat increment entirely corresponds to the quantum heat contribution. It reads:
%
\bb
\delta Q_{\text{q},\gamma}(t) &=& dU_\gamma(t)- \delta W_\gamma(t)\nonumber\\
&=& \sqrt{\Gamma_\text{meas}}dw_\gamma(t)\left(\moy{\{X,H_{\cal S}(t)\}}_{\psi_\gamma(t)}- 2U_\gamma(t) \moy{X}_{\psi_\gamma(t)}\right)\nonumber\\
&& + \Gamma_\text{meas}\Delta t \left\langle X H_{\cal S}(t)X -\dfrac{1}{2}\{X^2,H_{\cal S}(t)\}\right\rangle_{\psi_\gamma(t)}\label{eq2:QqContMeas}
 \ee
This expression is the direct generalization of the quantum heat exchanged during a projective measurement identified in Chapter \ref{Chapter1}. It vanishes if $\ket{\psi_\gamma(t)}$ is an eigenstate of $X$ such that measurement has no effect on the system.

Eq.\eqref{eq2:QqContMeas} can be averaged over the trajectories (or equivalently over the stochastic measurement record) to yield the average quantum heat flow. Noting that the term proportional to $dw_\gamma(t)$ has a zero average \cite{Steck06}, we find:

\bb
\dfrac{d}{dt}\moy{Q_{\text{q}}[\gamma]}_\gamma &=&  \Gamma_\text{meas}\text{Tr}\left\{ \left(X H_{\cal S}(t)X -\dfrac{1}{2}\{X^2,H_{\cal S}(t)\}\right)\rho_\text{S}(t)\right\},
 \ee
\noindent which vanishes if $X$ and $H_{\cal S}(t)$ commute.

Eventually, the First law of thermodynamics reads:
\bb
\Delta U[\gamma] = Q_\text{q}[\gamma] + W[\gamma],
\ee
where the work increment is given by Eq.\eqref{d2:dW}.

\subsubsection{Entropy production}

In order to compute the entropy production along such QSD trajectories, we need to extend the definition of the time-reversed Kraus operators. The rule \eqref{eq2:TimeRevCrooks} can still be applied in that case. We first identify the equilibrium state. Any classical mixture of the eigenstates of $X$ is an equilibrium state of the Lindbladian ${\cal L}_X = \Gamma_\text{meas}D[X]$. However, only the mixtures involving non-zero populations for every eigenstates of $X$ can be inverted. All these invertible mixtures lead to the same reversed Kraus operators, such that we can take the simplest one which is the maximally mixed state. We therefore use:
\bb
\pi_\text{eq} = \dfrac{\idop}{d_{\cal S}},
\ee
which leads to define:
\bb
\tilde M_{r_\gamma(t_n)} = M_{r_\gamma(t_n)}^\dagger = \sqrt{p(dw_\gamma(t))}\left(1+i\Delta tH_{\cal S}(t)- \tfrac{1}{2} \Gamma_\text{meas} \Delta t X^2 + dw_\gamma(t)\sqrt{\Gamma_\text{meas}}X\right).\nonumber\\
\ee
\noindent This relation is very similar to Eq.\eqref{eq2:DBCop} with $\delta Q_{\text{cl},\gamma}(t) = 0$. We can therefore derive the CFT and the second law:

\bb
\left\langle e^{-\Delta_\text{i}^\text{b}s[\gamma]/k_\text{B}}\right\rangle_\gamma &=& 1 \label{d2:CFTcont}\\
\moy{\Delta_\text{i}^\text{b}s[\gamma]}_\gamma &\geq & 0.
\ee

The boundary entropy production reads:
\bb
\moy{\Delta_\text{i}^\text{b}s[\gamma]}_\gamma = -\sum_f p_f\log p_f + \sum_i p_i\log p_i,
\ee 
\noindent which is proportional to the variation of Von Neumann entropy within the two-point measurement framework. Eq.\eqref{d2:CFTcont} generalizes the second law in presence of a measuring apparatus derived in Chapter \ref{Chapter1} to the case of weak continuous measurement.\\

\subsection{Application to a feedback stabilizing protocol}
\label{s2:FeedbackQq}

As an illustration we consider a Qubit of Hamiltonian $H_0$ prepared at time $t_i$ in state $\ket{\psi_i} = \ket{+_\theta} =\cos(\theta/2)\ket{e}+\sin(\theta/2)\ket{g}$. Then the observable $\sigma_z$ is continuously monitored at a rate $\Gamma_\text{meas}$.

 Using the expression of the Kraus operator \eqref{eq2:MX} at time $t_n$ with $X = \sigma_z$, we write the elementary evolution of the Qubit between times $t_n$ and $t_{n+1}$\footnote{The terms proportional to $\moy{\sigma_z}_{\psi_\gamma(t)}$ come from the normalization of the state at each time step.}: 

\begin{align}
\ket{\psi_\gamma(t+\Delta t)} = \bigg[ -i\dfrac{\omega_0 \Delta t}{2}\sigma_z &- \dfrac{\Gamma_\text{meas} \Delta t}{2}\left(\sigma_z-\left\langle\sigma_z\right\rangle_{\psi_\gamma(t)}\right)^2 \nonumber\\
&+\sqrt{\Gamma_\text{meas}}dw_\gamma(t)(\sigma_z-\left\langle\sigma_z\right\rangle_{\psi_\gamma(t)})\bigg]\ket{\psi_\gamma(t)}.\label{eq:SSEdecoh}
\end{align} 


\begin{figure}[h!]
\begin{center}
\includegraphics[width=\textwidth]{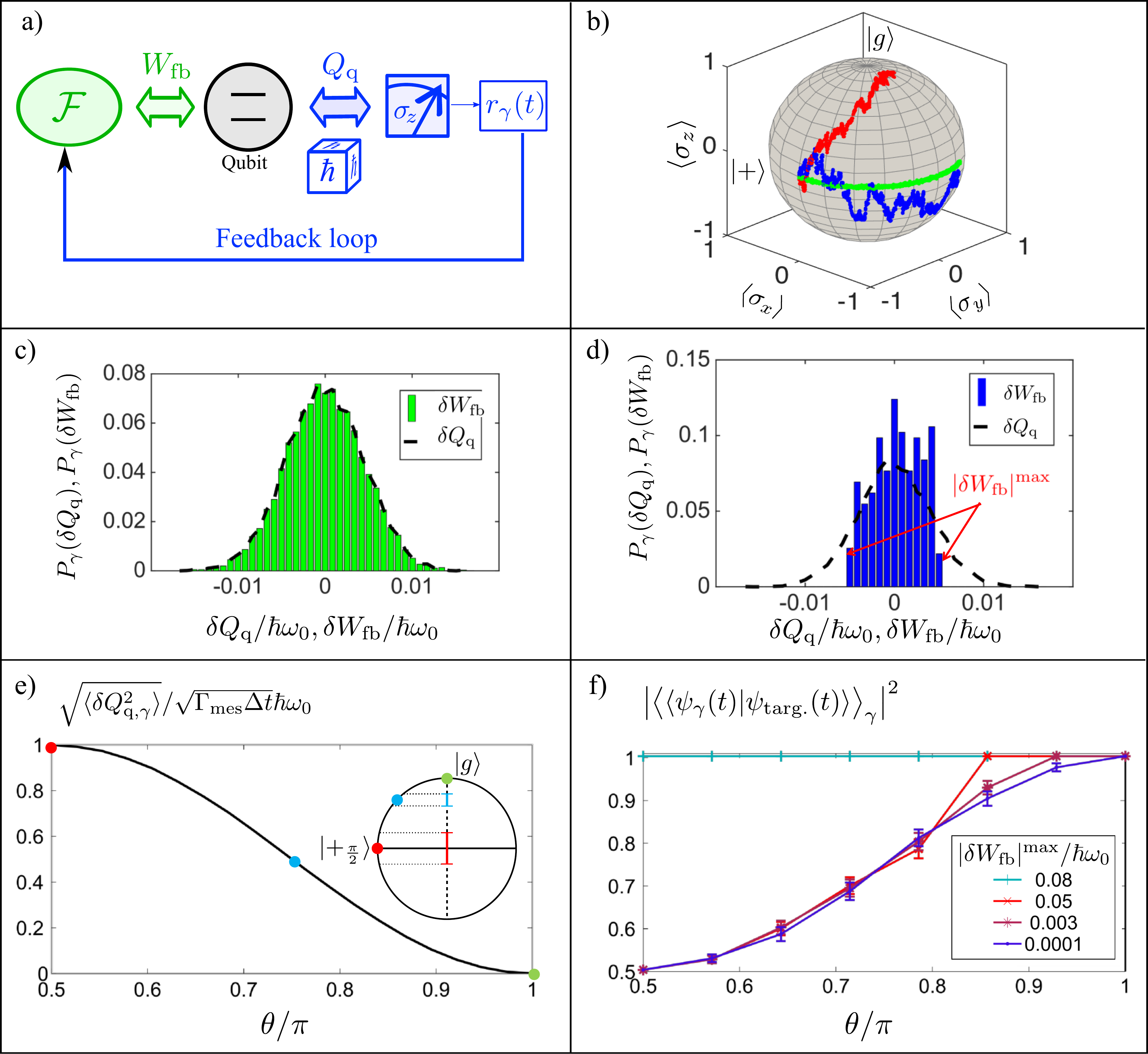}
\end{center}
\caption[Analysis of a feedback protocol]{Analysis of a feedback protocol stabilizing the state $\ket{\psi_\text{target}(t)} = \exp(-i\omega_0 \sigma_z t/2)\ket{+}$. a): The Qubit ${\cal S}$ is coupled to a measuring apparatus continuously measuring the observable $\sigma_z$. The measurement result $r_\gamma(t_n)$ obtained between times $t_n$ and $t_{n+1}$ is sent to a classical apparatus ${\cal F}$ implementing a feedback on the Qubit. b): Trajectory in the Bloch sphere in the case of perfect feedback (green), imperfect feedback (blue) and without feedback (red). c),d): Normalized distributions of quantum heat increments $P_\gamma(\delta Q_\text{q})$ (dashed black) and feedback work $P_\gamma(\delta W_\text{fb})$ (bars) performed by the feedback source ${\cal F}$. 
c): The two distributions match, such that the state is perfectly stabilized. d): The distribution of $P_\gamma(\delta W_\text{fb})$ is bounded and the feedback is not perfect.  {\it Parameters:} Evolution time $T= 1.5 / \omega_0$, pure dephasing rate $\Gamma_\text{meas} = 0.1 \omega_0$, feedback work cutoff: $\vert\delta W_\text{fb}\vert^\text{max} = 0.05 \hbar \omega_0$.  e): Standard deviation of the quantum heat increment $\delta Q_{\text{q},\gamma}(t)$ depending on the colatitude $\theta$ on the Bloch sphere of the target state $\ket{\psi_\text{targ.}(t)}$. The insets shows the position in the Bloch sphere of three different target states states corresponding to $\theta = \pi/2$ (red), $3\pi/4$ (blue) and $\pi$ (green), and the corresponding magnitude of the energy fluctuations in the target state $\sqrt{\moy{H_0^2}_{\psi_\text{targ.}(t)} - \moy{H_0}_{\psi_\text{targ.}(t)}^2}$ is depicted by the vertical segments. The equator state features the largest energy fluctuation, and requires the largest amount of energy. f) : Average fidelity of the final state $\ket{\psi_\gamma(t_N)}$ to the target state $\ket{\psi_\text{targ.}(t)}$ as a function of the colatitude $\theta$, for different values of the feedback work cutoff $|\delta W_\text{fb}|^\text{max}$. The error bars stand for the $99\%$ confidence interval. The trajectories last for $t= 20/\Gamma_\text{meas}$.}
\label{f2:Fb1}\end{figure}

The Hamiltonian of the Qubit being time-independent, the stochastic variation of the internal energy during an elementary measurement step of duration $\Delta t$ is only due to quantum heat, i.e. $U_\gamma(t+\Delta t) - U_\gamma(t) =\delta Q_{\text{q},\gamma}(t)$. We compute this variation using the rules of It\^o's calculus:

\bb \label{eq:fluc}
dU_\gamma(t)= \delta Q_\mathrm{q}[\gamma](t) &=& d\bra{\psi_\gamma(t)}H_0\ket{\psi_\gamma(t)}+\bra{\psi_\gamma(t)}H_0 d\ket{\psi_\gamma(t)}+d\bra{\psi_\gamma(t)}H_0 d\ket{\psi_\gamma(t)} \nonumber\\
&=& \hbar  \omega_0\sqrt{\Gamma_\text{meas}}dw_\gamma(t)\left(1-(\moy{\sigma_z}_{\psi_\gamma(t)})^2\right)\nonumber\\
 &=& 4\sqrt{\Gamma_\text{meas}}\hbar \omega_0 dw_\gamma(t)\vert\left\langle\sigma_- \right\rangle_{\psi_\gamma(t)}\vert^2,
\ee

\noindent where to go to the last line we have used the identity:
\bb
1-(\langle\sigma_z(t)\rangle_{\psi_\gamma(t)})^2 = 4 \bra{\psi_\gamma(t)} e\rangle\bra{e}\psi_\gamma(t)\rangle\bra{\psi_\gamma(t)} g\rangle  \bra{g}\psi_\gamma(t)\rangle = 4\vert\langle\sigma_-\rangle_{\psi_\gamma(t)}\vert^2.
\ee

 The stochastic energy variations are all the larger as the measurement strength $\Gamma_\text{meas}$ and the Qubit's coherences $\vert\left\langle\sigma_- \right\rangle\vert$ are large. They vanish when the Qubit's state has converged into one of the stable points $\ket{e}$ or $\ket{g}$ where measurement is completed. Once integrated between $0$ and $t \gg 1/{\Gamma^{*}}$, the quantum heat and internal energy change converge towards $\Delta U[\gamma] = Q_\text{q}[\gamma] = \hbar \omega_0 \sin(\theta/2)^2$ (resp. $-\hbar \omega_0 \cos(\theta/2)^2$) as the Qubit's state is approaching $\ket{e}$ (resp. $\ket{g}$). Note that the continuous measurement during a time  $t_f \gg \Gamma_\text{meas}^{-1}$ is equivalent to a strong projective measurement, and we therefore retrieve the same amount of quantum heat as in the example of Section \eqref{s:ExQED1}.\\

In order to assign an operational meaning to the quantum heat in this situation, we study a protocol in which a feedback loop is used to counteract the decoherence induced by the continuous measurement. We assume that the measurement record is continuously sent to a classical apparatus performing feedback on the Qubit, denoted ${\cal F}$ (see Fig. \ref{f2:Fb1}a). This device uses the measurement outcome $r_\gamma(t_n)$ obtained between at time $t_n$ to stabilize the state $\ket{\psi_\text{targ.}(t)} = e^{-i\omega_0 t \sigma_z/2} \ket{+_\theta}$ owing to a classical drive. 

\sloppy
The feedback drive has to induce rotation around the $y$-axis in the frame rotating at the frequency $\omega_0$, which corresponds to Hamiltonian $H_\text{fb}(t_n)=g(r_\gamma(t_n))\left(\sigma_-e^{i\omega_0t_n}+\sigma_+e^{-i\omega_0t_n}\right)$. The amplitude of the laser (and therefore the Rabi frequency) is tuned depending on $r_\gamma(t_n)$.  For the sake of simplicity, we assume that the time needed to transfer the measurement outcome to ${\cal F}$, as well as the time $\tau_\mathrm{fb}$ required to perform the actuation of the feedback is much shorter than $\Delta t$.  As a consequence, we neglect the decoherence during the feedback process, and we approximate the evolution of the Qubit by as a repeated cycle of two steps: (i) A measurement-induced evolution during $\Delta t$ described by Eq.\eqref{eq:SSEdecoh} and rotating the Qubit's state of an angle $\delta \theta_\gamma(t_n)$ around the $y$-axis and (ii) an instantaneous change $\delta \theta_{\text{fb},\gamma}(t_n)$ of the colatitude induced by the feedback. 

If the feedback is perfect, the two angles compensate each other $\delta \theta_{\text{fb},\gamma}(t_n) = -\delta \theta_\gamma(t_n)$. In this case, the feedback loop provides an amount of work compensating for the quantum heat. This gives an operational meaning to the quantum heat: it is (minus) the energy required to contrast the effect of the measurement. In the following we analyze energy exchanges around single trajectories and look at the influence of a possible limitation of the feedback loop. 
\fussy

Under monitoring and feedback, the Qubit follows trajectories between the initial time $t_i = 0$ and $t_f = t_N = N\Delta t$. Each elementary measurement at time $t_n = n\Delta t$, $n\in \llbracket 0,N\rrbracket$ gives rise to a quantum heat increment $\delta Q_{\text{q},\gamma}(t_n)$, which must be compensated by some work increment $\delta W_{\text{fb},\gamma}(t_n)$ provided by the drive for the state to be stabilized. The trajectory $\gamma$ thus gives rise to two normalized distributions 

\bb \label{eq2:distQq}
P_\gamma(\delta Q_\mathrm{q}) = \sum_{n=0}^{N} \delta^{\cal D} (\delta Q_\mathrm{q} - \delta Q_{\mathrm{q},\gamma}(t_n) ) / Q_\mathrm{q}[\gamma],
\ee
\noindent and
\bb \label{eq2:distW}
P_\gamma(\delta W_\mathrm{fb}) = \sum_{n=0}^{N} \delta^{\cal D} (\delta W_\mathrm{fb} - \delta W_{\mathrm{fb},\gamma}(t_n) ) / W_\mathrm{fb}[\gamma],
\ee
\noindent which match each other if the feedback is perfect (see Fig.\ref{f2:Fb1}c). Just like the energy fluctuations (Eq.\ref{eq:fluc}), the quantum heat distribution quantifies the strength of the measurement performed by the reservoir: Its typical support $|\delta Q|^\mathrm{max}$ is all the larger as the rate of decoherence $\Gamma_\text{meas}$ or the magnitude of the Qubit's coherences $\vert \langle \sigma_-\rangle_{\psi_\gamma(t)}\vert^2$ are increased. This puts physical constraints on the feedback source, which must be able to provide a power ${\cal P} = |\delta Q|^\text{max}/\tau_\mathrm{fb}$ to stabilize the state. A limitation of the feedback loop could come with a finite maximum source power ${\cal P}^\text{max}$ leading to a cutoff in the work distribution's support $|\delta W_\text{fb}|^\text{max} = {\cal P}^\text{max} \tau_\mathrm{fb}$. More precisely, if we assume that the loop only corrects angle variations $\delta\theta_\gamma$ such that the required amount of work $\vert\delta W_\text{fb}\vert$ is lower than $|\delta W_\text{fb}|^\text{max}$, we found that the stabilization might be altered (See Fig.\ref{f2:Fb1}b-d).\\


 We have also studied the dependence of the feedback's performances, as a function of the state to stabilize $\ket{\psi_\text{targ.}(t)}$. As expected, the support of the quantum heat distribution is all the smaller as the state approaches the poles of the Bloch sphere, which are stable under the monitoring process. Therefore the feedback requires less and less power (See Fig.\ref{f2:Fb1}e,f). Just like the response time (here taken as infinitely short), the quantum heat distribution appears as an essential tool to evaluate the quality of a feedback loop.

\section{Summary}

In this chapter, we have proposed a framework to describe the stochastic thermodynamics of open quantum systems. The framework is based on a quantum trajectory description, corresponding to the dynamics followed by the system when a measuring apparatus continuously monitors the reservoir. Such trajectories can be observed experimentally: In particular, we have presented a setup allowing to access the quantum jump trajectories for a Qubit in contact with a thermal reservoir. We have defined the thermodynamic quantities, work, heat and entropy production, at the level of the single trajectory.

In the case of a Qubit interacting with a thermal reservoir, we have identified two contribution to the heat flow: (i) the classical heat corresponding to the change of the reservoir's energy and responsible for an increase of the reservoir's entropy, and (ii) the quantum heat related to coherence erasure. We have identified several forms of the Integral fluctuation theorem depending on the applied protocol, among them the Jarzynski Equality. In addition, in order to bridge the gap towards realistic implementations, we have investigated the influence of detection inefficiency, which prevents in practice from observing pure states. We have derived a correction that has to be taken into account for Jarzynski equality to be valid despite imperfect detection.

In the case of the continuous measurement of observable (or of a reservoir inducing pure dephasing), we have identified the form of the IFT, generalizing results of Chapter \ref{Chapter1}. We have studied a protocol stabilizing a state against measurement-induced pure dephasing, demonstrating that the quantum heat in this case represents the energy required to perform a perfect feedback.

\chapter{Thermodynamics of fluorescence}
\label{Chapter3}

\minitoc

\lettrine{I}{n} Chapter \ref{Chapter1}, we have seen that the level of description of the quantum system under and the processes performed on it have an impact on its thermodynamic behavior. In Chapter \ref{Chapter2}, we have stressed that for a quantum open system, the unraveling, i.e. the way the environment is monitored, also influences the stochastic dynamics and therefore the fluctuation of thermodynamic quantities. In this chapter, we show that another parameter plays a paramount role in the thermodynamics description: the time step used to describe the dynamics of the system.
 
The issue of the relevant time step to describe the dynamics of a system is particularly crucial for quantum open systems. Indeed, when deriving the Lindlad equation capturing the evolution of a quantum system in contact with a Markovian reservoir, a coarse-graining in time is needed. This step is essential to reduce the exact coupled dynamics of the system and the reservoir to a Markovian evolution for the system alone. After the coarse-graining, the dynamics is described with a new time step, that we call coarse-graining time step in the following. It must be chosen much larger than the correlation time of the reservoir.

\begin{figure}[h!]
\begin{center}
\includegraphics[width=0.75\textwidth]{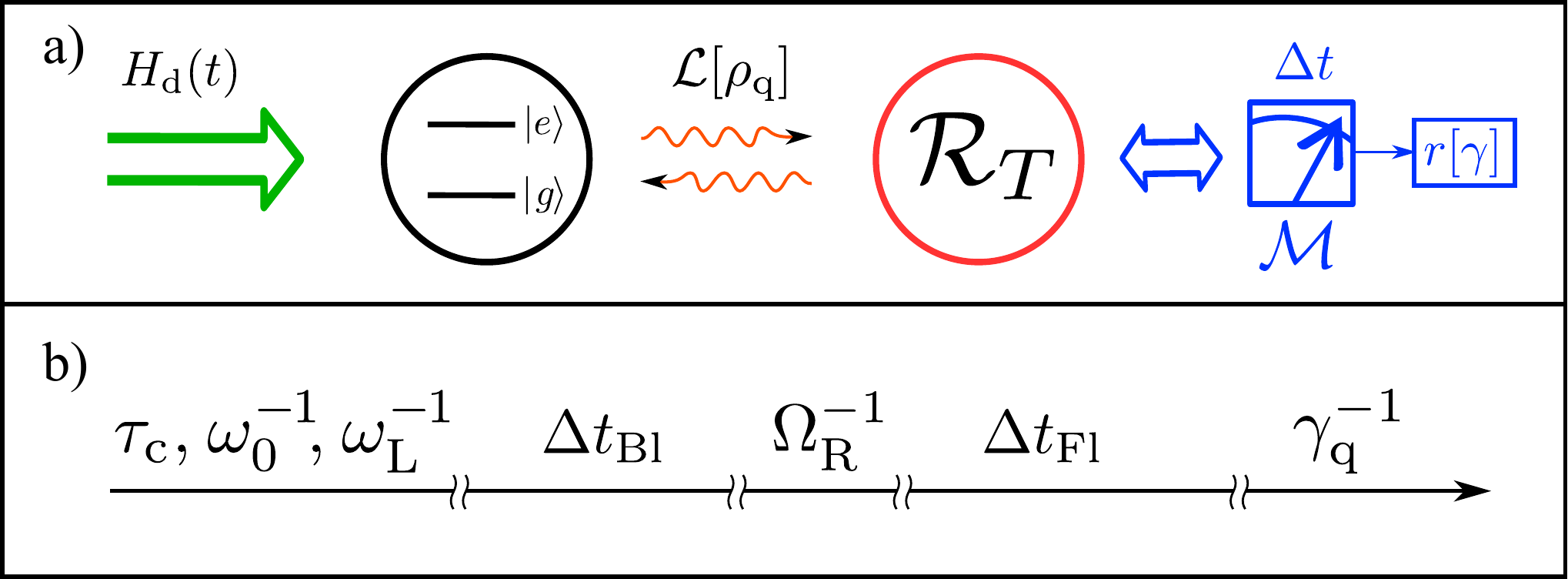}
\end{center}
\caption[Timescales of the problem]{Timescales of the problem. a) Situation under study: a Qubit is driven by a coherent drive generating the time-dependent term $H_\text{d}(t)$ in its Hamiltonian. The Qubit is coupled to an electromagnetic reservoir assimilated to a thermal reservoir at temperature $T$. b) Relative magnitude of the frequency parameters and of the two possible coarse-graining time steps leading to the FME ($\Delta t_\text{Fl}$) and to the OBEs ($\Delta t_\text{Bl}$).\label{f3:Scenery}}
\end{figure}

In certain situations, several relevant coarse-graining time steps can be chosen, allowing for several different master equation descriptions. To illustrate the influence of this parameter, we focus on a very fundamental situation for quantum optics: the case of a dissipative Qubit driven by a classical field (see Fig.\ref{f3:Scenery}a). As it involves a coherently driven system coupled to a thermal reservoir, it is also a fundamental situation of quantum thermodynamics. 

So far, quantum optical and thermodynamic studies of this problem have relied on two different descriptions, based on two different coarse-graining time steps: 
\begin{itemize}
\item In thermodynamics, the case of the driven Qubit has solely been analyzed \cite{Alicki13,Langemeyer14,Bulnes15} using the so-called Floquet Master Equation (FME) which involves a coarse-graining time step longer than a Rabi period.
\item In quantum optics, the Qubit's dynamics is described by the Optical Bloch Equations (OBEs) \cite{CCT}. This description involves a coarse-graining time shorter than a Rabi period and has lacked a thermodynamic analysis so far.
\end{itemize} 

Here, we complete this picture with (i) a stochastic thermodynamics analysis of the FME which allows to retrieve the average thermodynamic quantities found in the literature, (ii) a derivation of the OBEs in presence of non-zero temperature thermal reservoir which allows to precise their domain of validity, (iii) a stochastic thermodynamic analysis of the OBEs, in which we evidence the role of the coherences in the bare Qubit eigenbasis, and (iv) a comparison of the two thermodynamic descriptions.

\section{Definitions}

\subsection{Microscopic model}

We consider a Qubit of frequency $\omega_0$, driven by a quasi-resonant classical light source of frequency $\omega_\text{L} = \omega_0 - \delta$. The coupling strength is denoted $g$ such that the total Qubit Hamiltonian reads $H_\text{q}(t) = H_0 + H_\text{d}(t)$, where:

\bb
H_0 &=& \dfrac{\hbar\omega_0}{2}\sigma_z\\
H_\text{d}(t) &=& \dfrac{\hbar g}{2}\left(e^{i\omega_\text{L}t}\sigma_- + e^{-i\omega_\text{L}t} \sigma_+\right)
\ee

We have introduced the Pauli matrix $\sigma_z = \ket{e}\bra{e}-\ket{g}\bra{g}$ and the lowering operator $\sigma_- = \ket{g}\bra{e} = \sigma_+$, where $\ket{e}$ and $\ket{g}$ are the excited and ground states of the Qubit, respectively, and the Rabi frequency $g$. 

In the absence of reservoir, the drive induces so-called Rabi oscillations, i.e. coherent reversible oscillations of the state of the Qubit in the Bloch sphere around an axis tilted by an angle $2\theta = \arctan(g/\delta)$ with respect to the $z$-axis. This axis is rotating at frequency $\omega_\text{L}$ around the $z$-axis and have a frequency:
\bb
\Omega_\text{R} = \sqrt{g^2+\delta^2}, \label{d3:RabiFreq}
\ee
\noindent called the generalized Rabi frequency and which corresponds to $g$ at resonance $\delta =0$.\\

The Qubit is also coupled to a thermal reservoir ${\cal R}_T$ of temperature $T$. The reservoir is modeled as a collection of harmonic oscillators, such that the coupling Hamiltonian $V$ and the reservoir's Hamiltonian $H_{\cal R}$ read:
\bb
H_{\cal R} &=&\sum_k \hbar\omega_k \left(a^\dagger_k a_k + \dfrac{1}{2}\right)\nonumber\\
V &=&\sum_k \hbar g_k \left(a_k\sigma_+ + a_k^\dagger \sigma_-\right) = \sum_{l=\pm} R_l \sigma_l.
\ee
\noindent In these definitions, $a_k$ is the lowering operator in the bosonic mode $k$ of the reservoir that has frequency $\omega_k$. The coupling strengths  $g_k$ are taken real without loss of generality. We also introduce the two operators acting on the reservoir:
\bb
 R_- = \sum_k \hbar g_k a_k^\dagger = R_+^\dagger.
\ee

\subsection{Dynamics}

The density operator $\rho_\text{tot}$ of the global system obeys the Von Neumann equation:
\bb
\dot\rho_\text{tot} = -\dfrac{i}{\hbar}\left[H_\text{q}(t) + V + H_{\cal R},\rho_\text{tot}\right].\label{d3:VNE}
\ee
\noindent We consider the limit of weak coupling to the reservoir, defined by the condition $v\tau_\text{c}/\hbar \ll 1$, where $\tau_\text{c}$ is the correlation time of the reservoir and $v$ the typical magnitude of the coupling  Hamiltonian $V$.  In this limit, a Lindblad master equation capturing the effective Markovian dynamics of the sole Qubit can be derived \cite{CCT,Breuer,Carmichael1}:
\bb
\dot\rho_\text{q}(t) = -\dfrac{i}{\hbar}[H_\text{q}(t),\rho_\text{q}(t)] + {\cal L}[\rho_\text{q}(t)] \label{d3:LE},
\ee
\noindent where ${\cal L}$ is a superoperator encoding the non-unitary part of the dynamics. The derivation of its precise form is called microscopic derivation of the Lindblad equation. It requires to coarse-grain Eq.\eqref{d3:VNE} in time, such that the new time step $\Delta t$ is much larger than the reservoir's correlation time, but still much shorter than the characteristic time of the Qubit's dynamics generated by $\cal L$, i.e. the spontaneous emission time $1/\gamma_\text{q}$, with:
\bb
\gamma_\text{q} = \sum_k g_k^2 \delta_\text{D}(\omega_k-\omega_0) \label{d3:gammaq}.
\ee
\noindent Here $\delta_\text{D}$ is the Dirac distribution.

In the situation under study, two choices are possible for the magnitude of the coarse-graining time step $\Delta t$, leading to two different superoperators $\cal L$: (i) If we choose $\Delta t = \Delta t_\text{Fl}$ much longer than one Rabi period, we find the FME description associated with Lindbladian ${\cal L}_\text{Fl}$, (ii) if we take $\Delta t = \Delta t_\text{Bl}$ much shorter than one Rabi period we obtain  Lindbladian ${\cal L}_\text{Bl}$ which characterize the OBEs\footnote{The subscripts ``Fl'' and ``Bl'' respectively stand for Floquet and Bloch.}. 

In the following, we focus on the regime characterized by:
\bb
\tau_\text{c}, \omega_\text{L}^{-1},\omega_0^{-1} \ll \Omega_\text{R}^{-1} \ll \gamma_\text{q}^{-1} \label{d3:Regime}.
\ee
This condition forbids the ultra-strong driving regime $\Omega_\text{R} \gtrsim \omega_0, \omega_\text{L}$ and the weak driving regime $\Omega_\text{R} \ll \gamma_\text{q}$. In this regime, both FME and OBEs approaches are valid, and the choice of the coarse-graining time step is a turning knob allowing to switch from a description to the other (see Fig.\ref{f3:Scenery}b).

\subsection{Useful notations}

In order to simplify further calculations, it is useful to introduce the frame rotating at the drive frequency. This frame is related to the lab frame via the following transformation acting on the Qubit's subspace only: 
\bb
U_\text{R}(t) = e^{i(\omega_L t/2) \sigma_z}\label{eq3:UR}
\ee
Operators and states in the rotating frame are denoted with a tilde. In particular:
\bb
\tilde\rho_\text{tot}(t) = U_\text{R}(t)\rho_\text{tot}(t)U_\text{R}(t)^\dagger.
\ee

It is also useful to introduce the effective Qubit's Hamiltonian in the rotating frame:
\bb
\tilde H_\text{eff} &=& \tilde H_\text{q} - \dfrac{\hbar\omega_L}{2}\sigma_z\nonumber\\
&=& \dfrac{\hbar\delta}{2}\sigma_z + \dfrac{\hbar g}{2}\sigma_x \label{d3:Heff}
\ee
\noindent in which the frequency shift due to the rotation is taken into account. The Von Neumann equation in the rotating frame reads:
\bb
\dot{\tilde \rho}_\text{tot}(t) = -\dfrac{i}{\hbar}\left[\tilde H_\text{eff} + \tilde V(t) + H_{\cal R},\tilde\rho_\text{tot}(t)\right] \label{eq3:VNErot},
\ee
\noindent where the interaction Hamiltonian in the rotating frame is now time-dependent:
\bb
\tilde V(t) = \sum_{l=\pm} R_l \sigma_l e^{-i\omega_L t}.
\ee\\



Finally, we introduce the  dressed states $\ket{+}$ and $\ket{-}$  which are the two eigenstates of $\tilde H_\text{eff}$ of respective energies $\epsilon_+$ and $\epsilon_-$ \cite{CCT}, which fulfil:
\bb
\ket{+} &=& \cos(\theta)\ket{e} + \sin(\theta)\ket{g}\label{d3:plus}\\
\ket{-} &=& -\sin(\theta)\ket{e} + \cos(\theta)\ket{g}\label{d3:minus}\\
\epsilon_\pm &=& \pm \dfrac{\hbar\Omega_\text{R}}{2}\label{d3:epsilonpm}
\ee
\noindent where $\theta$ is defined by:

\bb
\cos(\theta) &=& \sqrt{\dfrac{\Omega_\text{R} + \delta}{\Omega_\text{R}}}\nonumber\\
\sin(\theta) &=& \sqrt{\dfrac{\Omega_\text{R} - \delta}{\Omega_\text{R}}}. \label{d3:Theta}
\ee

When written in the fixed frame, these two states are referred to as the Floquet states \cite{Langemeyer14,Bulnes15}, satisfying $\ket{\pm(t)} = U_\text{R}(t)\ket{ \pm}$.

\section{Long coarse-graining time step: Floquet Master Equation}

We first focus on the case (i) defined above, i.e. a coarse-graining time step $\Delta t = \Delta t_\text{Fl}$, where $\Delta t_\text{Fl}$ satisfies:
\bb
\tau_\text{c},\omega_0^{-1},\omega_\text{L}^{-1},g^{-1} \ll \Delta t_\text{Fl} \ll \gamma_\text{q}^{-1} \label{d3:DtFl}.
\ee

We first derive the explicit form of the FME Lindbladian ${\cal L}_\text{Fl}$. Although this derivation is not new, we present it for the sake of completeness, and because it highlights the role of the coarse-graining time step. We then analyze the resulting dynamics and study the stochastic thermodynamics within a quantum jump unraveling. Finally, we recover the expression of the average thermodynamic quantities already derived in the literature.

\subsection{Microscopic derivation of the Floquet Master Equation}
\label{s3:MicroscopicDerivation}
\label{s3:DerivationFME}

The derivation of the FME follows the standard form of the microscopic derivation of the Lindblad equation \cite{Breuer}.

The first step to derive ${\cal L}_\text{Fl}$ is to write Eq.\eqref{eq3:VNErot} in the interaction picture with respect to Hamiltonians $\tilde H_\text{eff}$ and $H_{\cal R}$. In the following, operators and states in the interaction picture will be denoted with a subscript $I$, e.g.:
\bb
V^I(t) = e^{-\tfrac{it}{\hbar}(\tilde H_\text{eff} + H_{\cal R})}\tilde V e^{\tfrac{it}{\hbar}(\tilde H_\text{eff} + H_{\cal R})}.
\ee

\noindent We obtain: 
\bb
\dot \rho^I_\text{tot}(t) = -\dfrac{i}{\hbar} [V^I(t), \rho^I_\text{tot}(t)].\label{eq3:VNI}
\ee

We now perform the coarse-graining: we integrate Eq.\eqref{eq3:VNI} between $t$ and $t+\Delta t_\text{Fl}$:
\bb
\rho_\text{tot}^I(t+\Delta t_\text{Fl}) - \rho^I_\text{tot}(t) = -\dfrac{i}{\hbar}\int_t^{t+\Delta t_\text{Fl}} dt' [V^I(t'),\rho^I_\text{tot}(t')].\label{eq3:pre1}
\ee

Then, we rewrite Eq.\eqref{eq3:pre1} as an expansion of the evolution of $\rho^I_\text{tot}(t)$ to second order in $V^I(t)$. This can be done by iterating Eq. \eqref{eq3:pre1}:
\bb
\rho_\text{tot}^I(t+\Delta t_\text{Fl}) - \rho^I_\text{tot}(t) &=& -\dfrac{i}{\hbar}\int_t^{t+\Delta t_\text{Fl}} dt' [V^I(t'),\rho^I_\text{tot}(t)]\nonumber\\
&&-\dfrac{1}{\hbar^2}\int_t^{t+\Delta t_\text{Fl}} dt'\int_{t}^{t'} dt'' [V^I(t'),[V^I(t''),\rho^I_\text{tot}(t'')]].
\ee
A master equation for the Qubit alone can be found by tracing over the reservoir degrees of freedom, and applying the following assumptions (valid in the weak coupling regime $v\Delta t_\text{Fl} \ll 1$ and when Eq.\eqref{d3:DtFl} is satisfied\footnote{see section IV-D of \cite{CCT} for a detailed discussions of these approximations.}):
\begin{itemize}
\item {\it Born approximation}: The total density operator at time $t''$ is assumed to have a factorized form $\rho_\text{tot}^I(t'') = \rho_\text{q}(t'')\otimes \rho_{\cal R}^I$, where $\rho_{\cal R}^I$ is the density operator of the reservoir in the interaction picture, which is time-independent \cite{CCT}.
\item {\it Markov approximation}: The density operator of the Qubit $\rho_\text{q}(t'')$ in the integrand of the double-integral is replaced by $\rho_\text{q}(t)$. In other words, we neglect the evolution of the Qubit's density operator in the interaction picture during $\Delta t_\text{Fl}$. This is legitimate as the evolution generated by the coupling Hamiltonian $V$ occurs on time-scales large with respect to $\Delta t_\text{Fl}$.
\end{itemize} 

In addition, we assume that the average value of $V$ in the reservoir's density operator $\rho_{\cal R}$ verifies $\moy{a_k} = 0$. This is in particular true for a thermal state. Consequently, the integrand of the single integral vanishes when taking the trace over the reservoir subspace. Denoting $\Delta \rho_\text{q}^I(t) = \rho_\text{tot}^I(t+\Delta t_\text{Fl}) - \rho^I_\text{tot}(t)$, we find:
\bb
\Delta \rho_\text{q}^I(t) = -\dfrac{1}{\hbar^2}\int_t^{t+\Delta t_\text{Fl}} dt'\int_{t}^{t'} dt'' [V^I(t'),[V^I(t''),\rho^I_\text{q}(t)\otimes\rho^I_{\cal R}]]. \label{eq3:pre2}
\ee

We now use that the coupling has the form $V^I(t) = \sum_{l=\pm} R_l^I(t) \sigma_l^I(t)$. When expanding the double commutator in Eq.\eqref{eq3:pre2}, we thus find terms proportional to two-time correlation functions $C_{ll'}(t',t'')$ of the reservoir, defined by:
\bb
C_{ll'}(t',t'') = \dfrac{1}{\hbar^2}\text{Tr}\{(R_l^I(t'))^\dagger R_{l'}^I(t'')\rho_{\cal R}^I\}, \quad l,l'\in \{+,-\}.\label{d3:Cllp}
\ee
Due to the short correlation time $\tau_\text{c}$ of the reservoir, these correlations functions are non-zero only if $\tau = t'-t'' \lesssim \tau_\text{c}$. Changing the integration variables in Eq.\eqref{eq3:pre2} to $\tau$ and $t'$ and extending the integration over $\tau$ to $[0,+\infty]$, we find the usual precursor of the Lindblad equation:

\bb
\Delta \rho_\text{q}^I(t) &=&  - \int_t^{t+\Delta t_\text{Fl}} dt' \int_0^\infty d\tau \sum_{ll'} C_{ll'}(t',t'-\tau)\nonumber\\
&& \times \left[ \left(\sigma_l^I(t')\right)^\dagger  \sigma_{l'}^I(t'-\tau)\rho_\text{q}^I(t) - \sigma_{l'}^I(t'-\tau)\rho_\text{q}^I(t)\left(\sigma_l^I(t')\right)^\dagger\right]    + \text{h.c.},\label{eq3:pre3}
\ee

Up to now the derivation was very general and we did not inject the precise form of the coupling. For this we need to compute the evolution generated by $\tilde H_\text{eff}$. To this end, we decompose $\sigma_l^I$ in the dressed basis $\{\ket{+},\ket{-}\}$. We write:
\bb
\sigma_l^I(t) = \sum_{\omega\in \{0,\pm\Omega_\text{R}\}} \tilde \sigma_l(\omega)e^{-i\omega t} \label{eq3:slomega}.
\ee
Identifying
\bb
\tilde \sigma_\pm(0) &=& \dfrac{g}{2\Omega_\text{R}}\big(\ket{+}\bra{+} - \ket{-}\bra{-}\big)\\
\tilde \sigma_\pm(\Omega_\text{R}) &=& \mp \dfrac{\Omega_\text{R}\mp \delta}{2\Omega_\text{R}}\ket{-}\bra{+}\\
\tilde \sigma_\pm(-\Omega_\text{R}) &=& \pm \dfrac{\Omega_\text{R}\pm \delta}{2\Omega_\text{R}}\ket{+}\bra{-},
\ee
\noindent we finally obtain:
\bb
V^I(t) = \sum_{l=\pm}\sum_{\;\;\omega\in\{0,\pm\Omega_\text{R}\}} R_l\tilde\sigma_l(\omega) e^{-il\omega_\text{L}t}e^{-i\omega t}.
\ee
We can then rewrite the master equation as:
\bb
\Delta \rho^I_\text{q}(t) &=&  - \int_t^{t+\Delta t_\text{Fl}} dt' \sum_{ll'} \sum_{\omega,\omega'} e^{i(\omega - \omega')t'+i(l-l')\omega_\text{L}t} \alpha_{ll}(\omega' + l'\omega_\text{L})\quad\quad\quad\quad\nonumber\\
&&\quad\quad\quad\quad\quad\quad\quad\times \left[ \tilde\sigma_l^\dagger (\omega) \tilde\sigma_{l'}(\omega')\rho_\text{q}^I(t) - \tilde\sigma_{l'}(\omega')\rho_\text{q}^I(t) \tilde\sigma_l^\dagger (\omega) \right] + \text{h.c.},\nonumber\\\label{eq3:pre4}
\ee
We have introduced:
\bb
\alpha_{ll'}(\omega)  = \int_0^\infty d\tau  C_{ll'}(t',t'-\tau'), \quad l,l'\in \{+,-\}.
\ee
which is assumed to be independent of $t'$ (the reservoir is at any time in equilibrium). Moreover, we use the specific form of $R_\pm$ tom compute the value of $\alpha_{ll'}(\omega)$. The terms $\alpha_{ll'}(\omega)$ with $l\neq l'$ are proportional to expectations values of the form $\moy{a_k^2}$ and $\moy{(a_k^\dagger)^2}$ which vanish for a reservoir at thermal equilibrium. The only non-zero terms correspond to $l = l'$ and fulfill:
\bb
\alpha_{++}(\omega) &=& \dfrac{1}{2}\Gamma(\omega)N(\omega)\\
\alpha_{--}(\omega) &=& \dfrac{1}{2}\Gamma(-\omega)(N(-\omega)+1),
\ee
with:
\bb
\Gamma(\omega) &\underset{\text{def.}}{=}& \sum_k g_k^2 \delta_\text{D}(\omega-\omega_k)\\
N(\omega) &\underset{\text{def.}}{=}& \dfrac{1}{e^{\hbar\omega/k_\text{B}T}-1}\\
\ee

As a consequence, only the terms satisfying $l=l'$ have to be kept in Eq.\eqref{eq3:pre4}, leading to:
\bb
\Delta \rho^I_\text{q}(t) &=&  - \int_t^{t+\Delta t_\text{Fl}} dt' \sum_l \sum_{\omega,\omega'} e^{i(\omega - \omega')t'} \alpha_{ll}(\omega' + l\omega_\text{L})\nonumber\\   &&\quad\quad\quad\quad\times \left[ \tilde\sigma_l^\dagger (\omega) \tilde\sigma_l(\omega')\rho_\text{q}^I(t) - \tilde\sigma_l(\omega')\rho_\text{q}^I(t) \tilde\sigma_l^\dagger (\omega) \right] + \text{h.c.},\label{eq3:precursor}
\ee

At this point, the master equation is not yet under the Lindblad form $\dot\rho_\text{q}^I(t) = \sum_k D[L_k]\rho_\text{q}^I(t)$ which ensures that the master equation is a valid CPTP map. In the standard approach in quantum open systems \cite{Breuer}, the Lindblad form is obtained by applying the secular approximation \cite{CCT}. This crucial step consists in neglecting off-diagonal terms involving $\omega \neq \omega'$ in the integrand. Indeed, such terms oscillate over $t'$ at a frequency $\vert\omega-\omega'\vert \geq \Omega_\text{R}$ larger than $1/\Delta t_\text{Fl}$ and would give a very small contribution to the master equation once averaged between $t$ and $t + \Delta t_\text{Fl}$. 

Performing such approximation finally leads to the FME:
\bb
\Delta \rho_\text{q}^I(t) &=& -\Delta t_\text{Fl} \sum_{l,\omega}\alpha_{ll}(\omega)(\omega + l\omega_L)\left[\tilde\sigma_l^\dagger (\omega) \tilde\sigma_l(\omega) \rho_\text{q}^I(t) - \tilde\sigma_l(\omega)\rho_\text{q}^I(t) \tilde\sigma_l^\dagger (\omega)\right] + \text{h.c.} \nonumber\\
&=& \Delta t_\text{Fl} \sum_{l,\omega} 2 \alpha_{ll}(\omega)(\omega + l\omega_L)D[\tilde\sigma_l(\omega)]\rho_\text{q}^I(t).\label{eq3:FMEI}
\ee

\subsection{Properties of the FME dynamics}
\label{s3:FMEinterpretation}

In order to find a more intuitive form of the FME, we write back Eq.\eqref{eq3:FMEI} in the Schr\"odinger picture (but still in the rotating frame). We find:
\bb
\dot {\tilde \rho}_\text{q}(t) = -\dfrac{i}{\hbar}[\tilde H_\text{eff},\tilde\rho_\text{q}(t)] + {\tilde{\cal L}}_\text{Fl}[\tilde\rho_\text{q}(t)]\label{eq3:FME}.
\ee

The effective Hamiltonian of the Qubit in the rotating frame satisfies 
\bb
\tilde H_\text{eff} = \dfrac{\hbar\Omega_\text{R}}{2}\Sigma_z
\ee
and the Lindbladian $\tilde {\cal L}_\text{Fl}$ splits into three parts: $\tilde{\cal L}_\text{Fl} = \tilde{\cal L}_0 + \tilde{\cal L}_1 + \tilde{\cal L}_2$, with:
\bb
\tilde{\cal L}_0 &=& \dfrac{\gamma_{0,\downarrow}+\gamma_{0,\uparrow}}{2}D[\Sigma_z]\\
\tilde{\cal L}_1 &=& \gamma_{1,\downarrow}D[\Sigma_-]+\gamma_{1,\uparrow}D[\Sigma_+]\\
\tilde{\cal L}_2 &=& \gamma_{2,\downarrow}D[\Sigma_-(t)]+\gamma_{2,\uparrow}D[\Sigma_+].
\ee
We have introduced the Pauli matrices in the dressed basis:
\bb
\Sigma_z &=&\ket{+}\bra{+}-\ket{-}\bra{-}\\
\Sigma_- &=&\ket{-}\bra{+} = \Sigma_+^\dagger.
\ee

$\tilde{\cal L}_0$ corresponds to the Lindbladian of a pure-dephasing channel (in the dressed basis), involving the rates:
\bb
\gamma_{0,\downarrow} &=& \dfrac{1}{4}\sin^2(2\theta)\Gamma(\omega_L)(N(\omega_L)+1)\\
\gamma_{0,\uparrow} &=& \dfrac{1}{4}\sin^2(2\theta)\Gamma(\omega_L)N(\omega_L).
\ee

$\tilde{\cal L}_1$ and $\tilde{\cal L}_2$ correspond to two thermal relaxation channels in the dressed basis involving the rates:
\bb
\gamma_{1,\downarrow} &=&\cos^4(\theta)\Gamma(\omega_\text{L}+\Omega_\text{R})(N(\omega_\text{L}+\Omega_\text{R})+1)\\
\gamma_{1,\uparrow} &=&\cos^4(\theta)\Gamma(\omega_\text{L}+\Omega_\text{R})N(\omega_\text{L}+\Omega_\text{R})\\
\gamma_{2,\downarrow} &=& \sin^4(\theta)\Gamma(\omega_\text{L}-\Omega_\text{R})N(\omega_\text{L}-\Omega_\text{R})\\
\gamma_{2,\uparrow} &=& \sin^4(\theta)\Gamma(\omega_\text{L}-\Omega_\text{R})(N(\omega_\text{L}-\Omega_\text{R})+1).
\ee\\



\begin{figure}[h!]
 \begin{center}
 \includegraphics[width=0.8\textwidth]{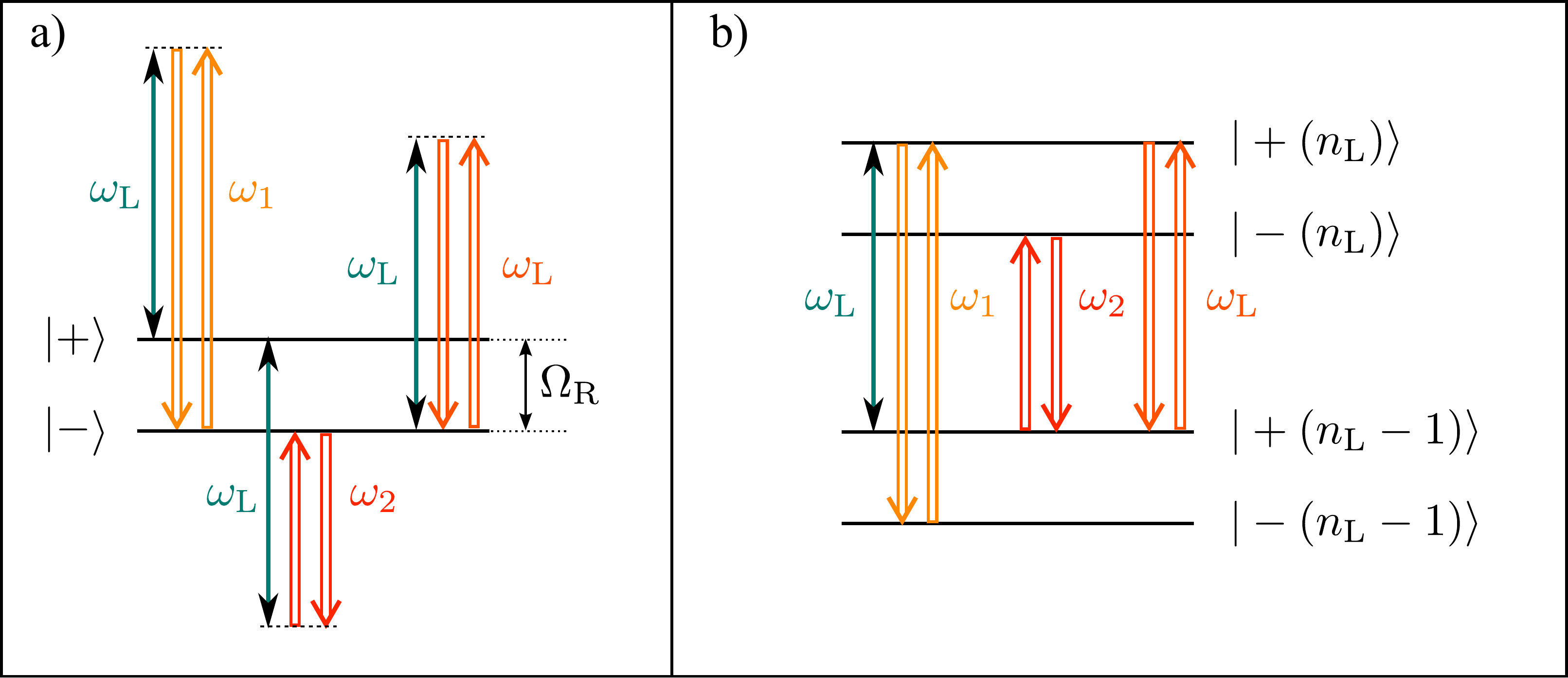}
 \end{center}
 \caption[Interpretation of Floquet dynamics]{Interpretation of Floquet dynamics. a) The three frequencies involved in the FME. b) The ladder of the energy eigenstates of the joint field-Qubit system when the field is quantized.\label{f3:RadiativeCasc}}
 \end{figure}

The three Lindbladians $\tilde{\cal L}_0$, $\tilde{\cal L}_1$ and $\tilde{\cal L}_2$ are associated with exchanges of photons at three different frequencies: respectively $\omega_L$ and $\omega_1 = \omega_L + \Omega_\text{R}$ and $\omega_2 = -\omega_L + \Omega_\text{R}$. These exchanges of photons correspond to transitions in the dressed basis $\{\ket{+},\ket{-}\}$ assisted by the monochromatic laser drive which provides or takes a quantum of energy $\hbar\omega_L$ (see Fig.\ref{f3:RadiativeCasc}a). For instance, the emission (resp. absorption) of photon of frequency $\omega_1$ corresponds to a transition from $\ket{+}$ to $\ket{-}$ (resp. $\ket{-}$ to $\ket{+}$), during which the drive has provided (resp. taken away) a photon of frequency $\omega_L$.\\
 
 These transitions can be understood in the radiative cascade picture: the monochromatic classical drive can be equivalently described as a quantized field coupled to the Qubit \cite{CCT}. The eigenstates $\{\ket{+(n_\text{L}),\ket{-(n_\text{L})}}\}$ of the joint field-Qubit Hamiltonian form a Jaynes-Cummings ladder indexed by the number $n_L$ of photons in the field (see Fig.\ref{f3:RadiativeCasc}b). This energy diagram is non-linear: the transition frequency between the energy levels $\ket{+(n_\text{L})}$ and $\ket{-(n_\text{L})}$ scales like $\sqrt{n_\text{L}}$. 

 The classical limit corresponds to a field in a coherent state of large amplitude $\ket{\alpha}$ such that $\moy{n_\text{L}} = \vert \alpha \vert^2 \gg 1$. As $\sqrt{\moy{n_\text{L}}+1} \simeq \sqrt{\moy{n_\text{L}}}$, the ladder is almost periodic (the non-linearity can be neglected) and the study of the dynamics can be restricted to the four states $\ket{+(n_\text{L})}$, $\ket{-(n_\text{L})}$, $\ket{+(n_\text{L}-1)}$ and $\ket{-(n_\text{L}-1)}$ between which the reservoir induces transitions. When tracing over the field states, one recovers the picture of the FME where the two reduced Qubit states $\ket{+}$ and $\ket{-}$ are connected by the three types of transitions\footnote{These three types of transitions are responsible for the Mollow triplet.}~. When engineered, these transitions can be e.g. exploited to cool down the Qubit in an arbitrary state of the Bloch sphere \cite{Murch12}.

 \subsection{Stochastic thermodynamics}

 \subsubsection{FME Quantum Jump trajectory}
 \label{s3:QJtraj}
 
In order to study the stochastic thermodynamics of the Qubit, we now introduce the Quantum Jump (QJ) unraveling of Eq.\eqref{eq3:FME}. The trajectories generated by this unraveling correspond to single realizations of the Qubit's dynamics if the reservoir is continuously monitored by a photocounter of time-resolution $\Delta t_\text{Fl}$ (see section \ref{s2:DefQtraj}). More precisely, this device is able to detect whether there was an emission or an absorption of photon between times $t$ and  $t+\Delta t_\text{Fl}$, and to distinguish the three frequencies $\omega_\text{L}$, $\omega_\text{L} + \Omega_\text{R}$ and $\omega_\text{L} - \Omega_\text{R}$. 

Note that such detection resolved both in time and in frequency does not violate the time-frequency uncertainty as the detector has an integration time $\Delta t_\text{Fl} \gg 1/\Omega_\text{R}$ (see condition \eqref{d3:DtFl}).

The measurement record $r[\gamma] = \{r_\gamma(t_n)\}_{1\leq n \leq N}$ obtained by the photocounter between times $t_i$ and $t_f$ conditions a stochastic pure-state quantum trajectory $\gamma$ followed by the system. Here we have discretized the time introducing $t_n = t_i + n\Delta t_\text{Fl}$ for $n\in\{1,N\}$ with $t_N = t_f$. 

The considered monitoring scheme has $7$ different measurement outcomes at time $t_n$ denoted $r_\gamma(t_n)\in \{k\}_{0\leq k\leq 6}$ associated with the following events: 
\begin{itemize}
\item Outcomes $r_\gamma(t_n)=1$ and $r_\gamma(t_n)=2$ correspond to the emission and the absorption of a photon of frequency $\omega_1$, respectively.
\item Outcomes $r_\gamma(t_n)=3$ and $r_\gamma(t_n)=4$ correspond to the emission and the absorption of a photon of frequency $\omega_2$, respectively.
\item Outcomes $r_\gamma(t_n)=5$ and $r_\gamma(t_n)=6$ correspond to the emission and the absorption of a photon of frequency $\omega_L$, respectively.
\item Outcome $r_\gamma(t_n)=0$ corresponds to the absence of photon exchange at time $t$.
\end{itemize}

Each measurement outcomes is associated with a Kraus operators $M_{r_\gamma(t_n)} \in \{M_{k}\}_{1\leq k \leq 7}$. When written in the rotating frame, these Kraus operators are simply related to the dressed states:
 \bb
\tilde M_1 &=& \sqrt{\gamma_{1\downarrow} \Delta t}\Sigma_-\\
\tilde M_2 &=& \sqrt{\gamma_{1\uparrow} \Delta t}\Sigma_+\\
\tilde M_3 &=& \sqrt{\gamma_{2\downarrow} \Delta t}\Sigma_-\\
\tilde M_4 &=& \sqrt{\gamma_{1\uparrow} \Delta t}\Sigma_+\\
\tilde M_5 &=& \sqrt{\gamma_{0\downarrow} \Delta t}\Sigma_z\\
\tilde M_6 &=& \sqrt{\gamma_{0\uparrow} \Delta t}\Sigma_z\\
\tilde M_0 &=& 1-i\Delta t_\text{Fl} \tilde H_\text{eff} - \dfrac{\Delta t_\text{Fl}}{2}\sum_{k=1}^6 \tilde  M_k^\dagger \tilde M_k.
\ee

We recall that the evolution of the Qubit's state $\ket{\tilde\psi_\gamma(t_n)}$ between $t_n$ and $t_{n+1}$ reads:
\bb
\ket{\tilde \psi_\gamma(t_{n+1})} = \dfrac{\tilde M_{r_\gamma(t_n)}\ket{\tilde \psi_\gamma(t_n)}}{\sqrt{\langle \tilde M_{r_\gamma(t_n)}^\dagger \tilde M_{r_\gamma(t_n)}\rangle_{\tilde \psi_\gamma(t_n)}}}.
\ee \\
Note that as $\tilde H_\text{eff} = \frac{\Omega_\text{R}}{2}\Sigma_z$ is diagonal in the dressed basis, it does not build coherences between states $\ket{+}$ and $\ket{-}$. As a consequence, the amplitude of the coherence 
\bb
\tilde\rho_{+-,\gamma}(t) = \bra{\tilde \psi_\gamma(t_n)} - \rangle\langle + \ket{\tilde \psi_\gamma(t_n)},
\ee
along trajectory $\gamma$ takes non-zero values solely if the initial state of is a superposition of $\ket{+}$ and $\ket{-}$, and only before the first jump. In other words, after a typical time of $1/\gamma_\text{q}$, the Qubit's state $\ket{\tilde \psi_\gamma(t_n)}$ is always either $\ket{+}$ or $\ket{-}$ and jumps between these two states when a photon of frequency $\omega_1$ or $\omega_2$ is detected. As a consequence, the populations of the dressed states $\tilde P_{\pm,\gamma}(t) = \vert \bra{\pm} \tilde \psi_\gamma(t_n)\rangle\vert^2$ along trajectory $\gamma$ belongs to $\{0,1\}$ after the transient evolution, i.e. for times $t_n \gg 1/\gamma_\text{q}$.

\subsubsection{Thermodynamic quantities}

As in the previous chapter, the internal energy of the Qubit along trajectory $\gamma$ is given by the expectation value of the Hamiltonian:
\bb
U_\gamma(t_n) &=& \bra{\tilde \psi_\gamma(t_n)}\tilde H_\text{eff}\ket{\tilde\psi_\gamma(t_n)}\nonumber\\
&=& \dfrac{\Omega_\text{R}}{2}\left(\tilde P_{+,\gamma}(t_n)-\tilde P_{+,\gamma}(t_n)\right).
\ee

 The expression of the elementary work provided by the drive between $t_n$ and $t_{n+1}$ is obtained from the time-derivative of the Qubit's Hamiltonian. As mentioned earlier, the time-derivative of the Hamiltonian has to be computed in the fixed frame. However, in the case under study, it is interesting to express the work in term of the Qubit's state in the rotating frame. To do so, we define the operator $\delta \hat w= \Delta t_\text{Fl}\dot H_\text{q}(t)$ such that $\delta W_\gamma(t) = \bra{\psi_\gamma(t)}\delta \hat w \ket{\psi_\gamma(t)}$. It satisfies:
 \bb
\delta \hat w =  i\omega_\text{L}\Delta t_\text{Fl}\hbar g (e^{i\omega_\text{L}t}\sigma_- - e^{-i\omega_\text{L}t}\sigma_+).
 \ee
Note that the infinitesimal increment of work can be written as an observable, while \emph{no} observable can be associated with the work performed during a finite thermodynamic transformation as this quantitity does not corresponds to a state function \cite{Talkner07}. The operator $\delta \hat w$ can be written in the rotating frame, leading to:
\bb
\delta \tilde{\hat w} = \hbar\omega_\text{L} g\Delta t_\text{Fl} \sigma_y= \hbar\omega_\text{L} g\Delta t_\text{Fl} \Sigma_y.\label{d3:PowerOp}
\ee
where we have used the identity $\sigma_y = \Sigma_y$.
 
Finally, the increment of work between times $t_n$ and $t_{n+1}$ reads:
\bb
\delta W_{\gamma}(t_n) &=&  \bra{\tilde\psi_\gamma(t_n)}\delta \tilde{\hat w}\ket{\tilde\psi_\gamma(t_n)}\\
&=& -\hbar\omega_L g \Delta t_\text{Bl} \text{Im} \tilde \rho_{+-,\gamma}(t). \label{s3:dW}
\ee
As it is proportional to the coherences in the dressed basis, the increment of work vanishes when $t_n \gg 1/\gamma_\text{q}$.\\

 We now compute the increment of classical heat. As the unraveling under study corresponds to a photon counting scheme, it directly allows to know what is the energy variation of the reservoir every $\Delta t_\text{Fl}$, and therefore what it the increment of classical heat (i.e. the energy provided by the thermal reservoir). From the analysis of section \ref{s3:FMEinterpretation}, we find (see also \cite{Alicki13,Langemeyer14}):

\bb
\delta Q_{\text{cl},\gamma}(t_n) = \left\{\begin{array}{ll}
-(\hbar\omega_\text{L} + \hbar\Omega), &r_\gamma(t_n) = 1 \\ 
\hbar\omega_\text{L} + \hbar\Omega, &r_\gamma(t_n) = 2 \\ 
\hbar\omega_\text{L} - \hbar\Omega, &r_\gamma(t_n) = 3 \\ 
-(\hbar\omega_\text{L} - \hbar\Omega), &r_\gamma(t_n) = 4 \\ 
-\hbar\omega_\text{L}, &r_\gamma(t_n) = 5 \\ 
\hbar\omega_\text{L}, &r_\gamma(t_n) = 6 \\ 
0, &r_\gamma(t_n) = 0.
\end{array} \right. \label{d3:dQclFME}
\ee
 
Note that the jump $M_3$ (resp. $M_4$) corresponds to an absorption (resp. emission) of photon of frequency $\omega_\text{L}-\Omega_\text{R}$ while the Qubit is projected onto $\ket{-}$ (resp. $\ket{+}$).\\

The change of  the internal energy contains in general two other contributions:

\begin{enumerate}
\item When the Qubit's state carries coherences in the basis $\{\ket{+},\ket{-}\}$, there is a quantum heat exchange associated with the variations of these coherences. This contribution can be computed in analogy with Eq.\eqref{d2:dQq12} (see Section \ref{s2:QqQcl}), noting that measurement outcomes $r_\gamma(t_n) \in \{1,3\}$ (resp. $r_\gamma(t_n) \in \{2,4\}$) are equivalent to measure the Qubit in state $\ket{+}$ (resp. $\ket{-}$). Such projection is associated with a suppression of the coherences between states $\ket{+}$ and $\ket{-}$. The quantum heat increment accounts for the corresponding variation of the internal energy $U_\gamma$. Outcomes $r_\gamma(t_n) \in \{5,6\}$ have no effect on the magnitude of the coherence and is not associated to a quantum heat increment. Finally $M_0$ is a weighted sum of projectors onto states $\ket{+}$ and $\ket{-}$ and outcome $r_\gamma(t_n) =0$ therefore has an effect analogous to a weak measurement of $\Sigma_z$, which is associated with an infinitesimal quantum heat increment. Eventually, the quantum heat increment at time $t_n$ reads:
\bb
\delta Q_{\text{q},\gamma}(t_n) = \left\{\begin{array}{ll}
\hbar\Omega_\text{R}\tilde P_{-,\gamma}(t), &r_\gamma(t_n) \in\{1,3\} \\ 
-\hbar\Omega_\text{R}\tilde P_{+,\gamma}(t), &r_\gamma(t_n) \in\{2,4\} \\ 
0, &r_\gamma(t_n) \in \{5,6\}\\ 
-\hbar\Omega_\text{R}(\gamma_{1,\downarrow}-\gamma_{1,\uparrow}+\gamma_{2,\downarrow}-\gamma_{2,\uparrow})\Delta t_\text{Fl}\vert\tilde \rho_{+-,\gamma}(t_n)\vert^2, &r_\gamma(t_n)=0.
\end{array} \right.\quad \label{d3:dQqFME}
\ee

This increment is non-zero at time $t_n$ only if $\tilde\rho_{+-,\gamma}(t_n)$ is non-zero. Consequently, the quantum heat increment always vanishes for $t_n \gg \gamma_\text{q}$.\\

\item When a jump occurs, the drive assists the transition by providing or taking a quantum $\hbar\omega_\text{L}$. This energy exchange is stochastic as it depends on the measurement outcome. However, it does not correspond to what we have called quantum heat in the previous sections: in particular, it is not associated with the erasure of coherences due to a measurement. This energy variation is an artefact induced by the trace over the field's degrees of freedom which allows to switch from the exact picture of the Jaynes-Cummings ladder (see Fig.\ref{f3:RadiativeCasc}b) to the reduced description of Qubit driven by a classical time-dependent field (see Fig.\ref{f3:RadiativeCasc}a). In the reduced description, the laser drive becomes an external incoherent source of energy for the Qubit. We denote $\delta Q_{\text{L},\gamma}(t_n)$ the increment of energy provided this way by the drive. It reads:
\bb
\delta Q_{\text{L},\gamma}(t_n) = \left\{\begin{array}{ll}
\hbar\omega_\text{L}, &r_\gamma(t_n) \in \{1,4,5\} \\ 
-\hbar\omega_\text{L}, &r_\gamma(t_n) \in \{2,3,6\} \\ 
0, &r_\gamma(t_n) = 0.
\end{array} \right. \label{d3:dQL}
\ee
\end{enumerate}

\vspace{2cm}

We finally obtain the first law:
\bb
\Delta U[\gamma] = W[\gamma] + Q_\text{cl}[\gamma] + Q_\text{L}[\gamma] + Q_\text{q}[\gamma].
\ee\\

In order to compute the entropy production, we need to specify more the applied protocol. For instance, we can consider a two-point measurement protocol (an observable of the Qubit is projectively measured at the initial and final times of the trajectory, see Section \ref{s2:TPMP}). This leads to the following expression for the entropy produced along trajectory $\gamma$ (see section \ref{s2:TPMP}):
\bb
\Delta_\text{i} s[\gamma]  = k_\text{B}\log\left(\dfrac{p_i}{p_f}\right) - \dfrac{Q_\text{cl}[\gamma]}{T}.
\ee

This entropy production fulfills the IFT Eq.\eqref{d2:IFT}.

\subsection{Average thermodynamics}

The analysis along stochastic trajectories allows to find the average thermodynamic quantities. We compute the averages of the stochastic expressions above over the measurement outcomes using that the probability of a jump of kind $k$ is \linebreak$p_k(t_n) = \bra{\psi_\gamma(t_n)}\tilde M_k^\dagger \tilde M_k \ket{\psi_\gamma(t_n)}$. 


We find the following expressions for the average work derivative \linebreak\mbox{$\dot W(t) = \moy{\delta W_\gamma(t)/\Delta t_\text{Fl}}_\gamma$}, the uncontrolled heat flow provided by the laser drive \linebreak\mbox{$\dot Q_\text{L}(t) = \moy{\delta Q_{\text{L},\gamma}(t)/\Delta t_\text{Fl}}_\gamma$} and the heat flow provided by the reservoir \linebreak\mbox{$\dot Q_\text{cl}(t) = \moy{\delta Q_{\text{cl},\gamma}(t)/\Delta t_\text{Fl}}_\gamma$}:
\bb
\dot W(t) &=& \hbar\omega_\text{L}g \text{Im}\rho_\text{+-}(t),\label{eq3:W}\\
\dot Q_\text{L}(t) &=& \hbar\omega_\text{L}\Big[(\gamma_{1,\downarrow}-\gamma_{2,\downarrow})P_+(t)+(\gamma_{2,\uparrow}-\gamma_{1,\uparrow})P_-(t)+ (\gamma_{0,\downarrow}-\gamma_{0,\uparrow})\Big],\label{eq3:QL}\\
\dot Q_\text{cl}(t) &=& - \hbar\omega_1(\gamma_{1,\downarrow}P_+(t)-\gamma_{1,\uparrow}P_-(t))- \hbar\omega_2(\gamma_{2,\downarrow}P_+(t)-\gamma_{2,\uparrow}P_-(t))\nonumber\\
&& - \hbar\omega_\text{L} (\gamma_{0,\downarrow}-\gamma_{0,\uparrow}).\label{eq3:Pres}
\ee  

The average quantum heat is zero at any time. At steady state (i.e. after the transient decay of the coherences in the dressed basis), the average work derivative vanishes, and the energy balance reads:
\bb
{\cal P}_\text{L}^\infty + {\cal P}_\text{res}^\infty = 0, 
\ee
where we have defined ${\cal P}_\text{L}^\infty = \dot Q_\text{L}(t)$ and ${\cal P}_\text{res}^\infty = \dot Q_\text{cl}(t)$ for large times $t \gg 1/\gamma_\text{q}$.
This energy balance is in agreement with expressions derived in the literature \cite{Alicki13,Langemeyer14,Bulnes15}. In \cite{Bulnes15}, the classical heat flow is derived using the so-called full counting statistics method \cite{Esposito09}. This method boils down to keeping track of the classical heat exchanges occurring along each quantum jump trajectories defined in section \ref{s3:QJtraj}, and storing the result in a generating function, from which the statistics of classical heat exchanges can be retrieved. 

In the FME description, as it provides an uncontrolled energy flow, the drive has some similarities with a reservoir. It differs from a thermal reservoir though because it has a ``colored'' spectrum: indeed for large detunings $\delta \gg g$, the laser drive decouples from the Qubit and ${\cal P}_\text{L}^\infty$ vanishes.\\


We also compute the average rate of entropy production:
\bb
\sigma_\text{i}(t) = \dfrac{d}{dt}\moy{\Delta_\text{i}s[\gamma]}_\gamma = \dot S_\text{VN}[\rho_\text{q}(t)] - \dfrac{\dot Q_\text{cl}(t)}{T} \geq 0.\label{eq3:Sirr}
\ee
 At steady-state, the rate of entropy production takes the form $\sigma_\text{i}^\infty = - {\cal P}_\text{res}^\infty/T$. This continuous production of entropy is the hallmark of a out-of-equilibrium steady state \cite{Oono98,Esposito07}: the drive continuously draw the Qubit out of its equilibrium with the reservoir. Eq.\eqref{eq3:Sirr} is also in agreement with results published in \cite{Alicki13,Bulnes15}. This average entropy production rate can be recovered from the Spohn inequality \cite{Spohn78}:
\bb
\text{Tr}\{{\cal L}[\rho](\log \rho - \log \pi^\infty)\} \leq 0,
\ee
valid for any Lindbladian ${\cal L}$ and its stationary state $\pi^\infty$. 

\section{Short coarse-graining time step: Optical Bloch Equations}
\label{s3:OBE}
\subsection{Approximation of independent rates of variation}

In quantum optics, the dynamics of the driven Qubit is generally described by the Optical Bloch Equations (OBEs), which correspond to a Lindblad equation involving the total Hamiltonian $H_\text{q}(t)$ and the same Lindbladian as derived for the bare Qubit, i.e. \cite{CCT}:

\bb
\dot \rho_\text{q}(t) = -\dfrac{i}{\hbar}[H_\text{q}(t),\rho_\text{q}(t)] + {\cal L}_\text{Bl}[\rho_\text{q}(t)] \label{eq3:OBE},
\ee
where 
\bb
{\cal L}_\text{Bl} = \gamma_\text{q}n_\text{q} D[\sigma_+] + \gamma_\text{q}(n_\text{q}+1) D[\sigma_-] \label{eq3:LindbladianOBE}.
\ee

The approximation corresponding to use this unperturbed Lindbladian is called the approximation of independent rates of variation in \cite{CCT}. While it is natural in the limit of weak coupling to the drive $g \ll \gamma_\text{q}$, proving the validity of such approximation in the case of strong drive requires further checks. 

In this section, we show that Eq.\eqref{eq3:OBE} is obtained instead of the FME when considering a coarse-graining time step $\Delta t_\text{Bl}$ much shorter than $\Delta t_\text{Fl}$ and satisfying:
\bb
\tau_\text{c},\omega_0^{-1},\omega_\text{L}^{-1},g^{-1} \ll \Delta t_\text{Bl} \ll \gamma_\text{q}^{-1} \label{d3:DtBl},
\ee
Moreover, the OBEs require the following additional constraints that are fulfilled under standard conditions in quantum optics:
\bb
\Omega_\text{R} &\ll& \omega_0,\omega_\text{L} \label{d3:NotUltraStr}\\
\Gamma(\omega + l\omega_\text{L}) &\simeq & \Gamma(\omega_0) = \gamma_\text{q}, \quad\forall \omega \in \{0,\pm\Omega_\text{R}\} \label{d3:FlatSpec}\\
N(\omega + l\omega_\text{L}) &\simeq & N(\omega_0)= n_\text{q}, \quad\forall \omega \in \{0,\pm\Omega_\text{R}\} \label{d3:NpmN0}
\ee

Condition \eqref{d3:NotUltraStr} forbids ultra-strong coupling between the Qubit and its drive and the dispersive regime of very large detuning. This condition is consistent with the regime under study as defined by Eq.\eqref{d3:Regime}. Condition \eqref{d3:FlatSpec} imposes that the spectrum of the electromagnetic reservoir is flat around $\omega_0$. This condition is generally fulfilled, but may be wrong when the electromagnetic environment of the Qubit is structured e.g. if the Qubit is embedded in a nearly resonant a cavity \cite{ Lewenstein87,Murch12,Carmichael1}. Finally, Eq.\eqref{d3:NpmN0} constrains the regime of temperature that can be studied with the OBEs.  This condition is fulfilled for $\hbar\omega_0/k_\text{B}T \gtrsim 10$ which is naturally true at room temperature for optical frequencies.

Equation Eq.\eqref{eq3:OBE} can be expressed as a function of the average population of the excited state $P_e(t) = \bra{e}\rho_\text{q}(t)\ket{e}$\footnote{The population of state $\ket{e}$ is the same in the rotating and fixed frames: $\tilde P_e(t) = \bra{e}\tilde \rho_\text{q}(t)\ket{e} = P_e(t)$.} and the average amplitude of the coherences in the $\{\ket{e},\ket{g}\}$ basis in the rotating frame $\tilde s(t) = \bra{e}\tilde\rho_\text{q}(t)\ket{g} = e^{i\omega_L t}\bra{e}\rho_\text{q}(t)\ket{g}$, leading to the usual form of the Optical Bloch Equations (OBEs):
\bb
\dot P_e(t) &=& -\gamma_\text{q}(2n_\text{q}+1) P_e(t) + \gamma_\text{q}n_\text{q} + \dfrac{ig}{2}(\tilde s(t)-\tilde s(t)^*)\label{eq3:OBEPe}\\
\dot{\tilde s} &=& -\left(i\delta + \dfrac{\gamma_\text{q}(2n_\text{q}+1)}{2}\right)\tilde s(t)  + ig\left(P_e(t)-\dfrac{1}{2}\right)\label{eq3:OBEs}.
\ee

\subsection{Microscopic derivation of the OBEs}

In order to derive Eq.\eqref{eq3:OBE}, we use the same procedure as in Section \ref{s3:MicroscopicDerivation}, except that the coarse-graining time step is set to be $\Delta t_\text{Bl}$ satisfying $\Omega_\text{R}^{-1} \gg \Delta t_\text{Bl} \gg \tau_\text{c}$.  Repeating the same steps up to Eq.\eqref{eq3:precursor}, we find:

\bb
\Delta \rho^I_\text{q}(t) &=&  - \int_t^{t+\Delta t_\text{Bl}} dt' \sum_l \sum_{\omega,\omega'} e^{i(\omega - \omega')t'} \alpha_{ll}(\omega' + l\omega_\text{L}) \left[ \tilde\sigma_l^\dagger (\omega) \tilde\sigma_l(\omega') - \tilde\sigma_l(\omega')\rho_\text{q}^I(t) \tilde\sigma_l^\dagger (\omega) \right] \nonumber\\   &&+ \text{h.c.}.\label{eq3:precursorOBE}
\ee

Because of conditions \eqref{d3:FlatSpec}-\eqref{d3:NpmN0}, the coefficients $\alpha_{ll}(\omega + l\omega_\text{L})$ do not depend on $\omega$:
\bb
\alpha_{ll}(\omega + l\omega_\text{L}) \simeq \alpha_{ll}(l\omega_0),
\ee 
such that the sums over $\omega$ and $\omega'$ can be expressed as a function of the operators $\sigma_l$ and $\sigma_{l'}$ in the interaction picture:
\bb
\sum_m e^{-i\omega t'}\tilde \sigma_l(\omega) = \sigma_l^I(t').
\ee
Moreover, as $\Delta t_\text{Bl}$ is much smaller than $g^{-1}$, we can neglect the evolution induced by the drive in $\sigma_l^I(t')$:
\bb
\sigma_l^I(t') \simeq e^{il\omega_0 t'}\sigma_l
\ee

As a consequence, the exponential factor vanishes in Eq.\eqref{eq3:precursorOBE} and the integration over $t'$ can be replaced by a factor $\Delta t_\text{Bl}$. It yields:
\bb
\Delta \rho_\text{q}^I(t) &=& -\Delta t_\text{Bl}\sum_l \alpha_{ll}(l\omega_0)[\sigma_l^\dagger \sigma_l \rho_\text{q}^I(t) - \sigma_l \rho_\text{q}^I(t)\sigma_l^\dagger] + \text{h.c.}\nonumber\\
&=& 2\Delta t_\text{Bl} \sum_l \alpha_{ll}(l\omega_0) D[\sigma_l] \rho_\text{q}^I(t).
\ee
Finally, by identifying the coarse-grained derivative $\Delta \rho_\text{q}^I(t)/\Delta t_\text{Bl} = \dot \rho_\text{q}^I(t)$ and going back to Schr\"odinger picture (and the lab frame), we find Eq.\eqref{eq3:OBE}.

\subsection{Thermodynamic analysis}

\subsubsection{Unraveling the OBEs}

We now introduce a QJ unraveling of Lindblad equation \eqref{eq3:OBE}. Such unraveling is implemented if the photons exchanged with the reservoir are monitored by a photocounter of time resolution $\Delta t_\text{Bl}$. Such a detection scheme can be realized owing to the engineered environment presented in section \ref{s2:SetupEngineered}. 

The measurement outcome at time $t_n$ has three possible values $r_\gamma(t_n)\in\{1,2,0\}$ corresponding to the detection of an emission, of an absorption and no exchange of photon, respectively. As above, the evolution of the Qubit's state conditioned on the measurement record $r[\gamma] = \{r_\gamma(t_n)\}_{1\leq n \leq N}$ reads:
\bb
\ket{\psi_\gamma(t_n)} = \dfrac{M_{r_\gamma(t_n)}\ket{\psi_\gamma(t_n)}}{\sqrt{\langle M_{r_\gamma(t_n)}^\dagger M_{r_\gamma(t_n)}\rangle_{\psi_\gamma(t_n)}}}
\ee
where the three measurement operators are the same as in section \ref{s2:SetupEngineered}, namely:
\bb
M_1 &=& \sqrt{\gamma_\text{q}(n_\text{q}+1)\Delta t_\text{Bl}}\sigma_-\\
M_2 &=& \sqrt{\gamma_\text{q}n_\text{q}\Delta t_\text{Bl}}\sigma_+\\
M_0 &=& 1-\dfrac{i\Delta t_\text{Bl}}{\hbar}H_\text{q}(t)-\dfrac{\gamma_\text{q}(n_\text{q}+1)\Delta t_\text{Bl}}{2}\Pi_e-\dfrac{\gamma_\text{q}n_\text{q}\Delta t_\text{Bl}}{2}\Pi_g .
\ee

In these trajectories, the jumps induced by the reservoir link states $\ket{e}$ and $\ket{g}$ and not states $\ket{+}$ and $\ket{-}$. As the Hamiltonian $H_\text{q}(t)$ which appears in the no-jump measurement operator $M_0$ is not diagonal in the $\{\ket{e},\ket{g}\}$ basis, it builds coherences between the states $\ket{e}$ and $\ket{g}$. This coherences are erased each time a jump occurs, leading to a rich dynamics (see inset of Fig.\ref{f3:JE}b).

\subsubsection{Stochastic Thermodynamic quantities}

The internal energy of the Qubit is:
\bb
U_\gamma(t) &=& \bra{\psi_\gamma(t)}H_\text{q}(t)\ket{\psi_\gamma(t)}\nonumber\\
&=& \dfrac{\hbar\omega_0}{2}(2P_{e,\gamma}(t)-1) + \hbar g \text{Re}\tilde s_\gamma(t)
\ee
where $P_{e,\gamma}(t) = \vert\langle e\ket{\psi_\gamma(t)}\vert^2$ and $\tilde s_\gamma(t) = \bra{\psi_\gamma(t)} e^{i\omega_\text{L}t} \sigma_- \ket{\psi_\gamma(t)}= \bra{\tilde \psi_\gamma(t)} \sigma_- \ket{\tilde \psi_\gamma(t)}$ are respectively the population of state $\ket{e}$ and the amplitude of the coherences in the $\{\ket{e},\ket{g}\}$ basis in the rotating frame along trajectory $\gamma$.

The expression of the elementary work provided by the drive between $t_n$ and $t_{n+1}$ is obtained from the time-derivative of the Hamiltonian $H_\text{q}(t)$:
\bb
\delta W_{\gamma}(t_n) &=&  -\hbar\omega_\text{L} g \Delta t_\text{Bl} \bra{\psi_\gamma(t_n)}\dot H_\text{q}(t)\ket{\psi_\gamma(t_n)}\\
&=& -\hbar\omega_\text{L} g \Delta t_\text{Bl} \text{Im} \tilde s_\gamma(t).
\ee

Once again, we use that the photon counting scheme allows to know the variation of the reservoir's energy so as to compute the increment of classical heat between $t_n$ and $t_{n+1}$:

\bb
\delta Q_{\text{cl},\gamma}(t_n) = \left\{\begin{array}{cl}
-\hbar\omega_0, &r_\gamma(t_n) = 1 \\ 
\hbar\omega_0, &r_\gamma(t_n) = 2 \\ 
0, &r_\gamma(t_n) = 0.
\end{array} \right. \label{d3:dQcl}
\ee
We moreover assumes that the integration time of the detector, which corresponds to the coarse-graining time step $\Delta t_\text{Fl}$ fulfills $\omega_0\Delta t_\text{Bl} \gg 1$, which ensures that the detected photons have a well-defined frequency $\omega_0$. Such condition is compatible with the constraints previously imposed for $\Delta t_\text{Fl}$ and is generally assumed in standard microscopic derivations of the Lindblad equation \cite{CCT}.

Finally, some quantum heat is provided to the Qubit because of the presence of coherences in the $\{\ket{e},\ket{g}\}$. This increment can be computed by evaluating the change of the internal energy $U_\gamma$ associated with the variations of this coherences induced by each type of Kraus operator (see Section \ref{s2:QqQcl}):
\bb
\delta Q_{\text{q},\gamma}(t_n) = \left\{\begin{array}{ll}
\hbar\omega_0 P_{g,\gamma}(t_n), &r_\gamma(t_n) = 1 \\ 
-\hbar\omega_0P_{e,\gamma}(t_n), &r_\gamma(t_n) = 2 \\ 
-\hbar g\dfrac{\gamma_\text{q}(2n_\text{q}+1)}{2}\Delta t_\text{Bl} \text{Re}\tilde s_{\gamma}(t_n) - \hbar\omega_0 \gamma_\text{q}\Delta t_\text{Bl}\vert \tilde s_{\gamma}(t_n) \vert^2 , &r_\gamma(t_n) = 0.
\end{array} \right. \quad \label{d3:dQq}
\ee

Note that contrary to the case of the long coarse-graining time step, the work and quantum heat increments are non zero all along the trajectory even at long times as coherences in the $\{\ket{e},\ket{g}\}$ are continuously created by the drive.

Eventually, the first law reads:
\bb
\Delta U[\gamma] = W[\gamma] + Q_\text{q}[\gamma] + Q_\text{cl}[\gamma].
\ee

As above, we compute the entropy produced during a two-point measurement protocol. We find a rate production of entropy:
\bb
\Delta_\text{i}s[\gamma] = \log\left(\dfrac{p_i}{p_f}\right) - \dfrac{Q_\text{cl}[\gamma]}{T}.
\ee

\subsubsection{Average thermodynamic quantities}

We deduce the average thermodynamic quantities corresponding to the stochastic work, classical and quantum heats and entropy production, using the probabilities of the two types of jump:
\bb
p_1 = \Delta t_\text{Bl}\gamma_\text{q}(n_\text{q}+1)P_{e,\gamma}(t)\\
p_2 = \Delta t_\text{Bl}\gamma_\text{q}n_\text{q}(1-P_{e,\gamma}(t)).
\ee 

We find:
\bb
\dot W(t) &=& -\hbar \omega_\text{L} g \text{Im}\tilde s(t) \\
\dot Q_\text{cl}(t) &=&  -\hbar\omega_0\left(\gamma_\text{q}(n_\text{q}+1)P_e(t) - \gamma_\text{q}n_\text{q}P_g(t)\right)\\
\dot Q_\text{q}(t) &=& -\dfrac{\gamma_\text{q}(2n_\text{q}+1)}{2}\text{Re}\tilde s(t).
\ee

As before, we want to introduce the flow of energy provided by the drive to the Qubit ${\cal P}_\text{L}$ and the heat flow provided by the reservoir ${\cal P}_\text{res}$. To this end, it has to be noted that the steady-state coherences in the $\{\ket{e},\ket{g}\}$ basis are solely due to the drive. We can therefore use the arguments of Section \ref{s:ExQED1}, in which the readout of a Qubit initially in a superposition of $\ket{e}$ and $\ket{g}$ is analyzed, to evidence that this quantum heat actually corresponds to energy provided by the drive. We therefore have:
\bb
{\cal P}_\text{L} &=& \dot W(t) + \dot Q_\text{q}(t)\\
{\cal P}_\text{res} &=& \dot Q_\text{cl}(t)
\ee\\

At steady state, the population $P_e(t)$ and the coherences $\tilde s(t)$ verify:
\bb
P_e^{\infty} &=& \dfrac{1}{2n_\text{q}+1}\left(n_\text{q} + \dfrac{1/2}{1+ 2\dfrac{\delta^2}{g^2}+ \dfrac{\gamma_\text{q}^2(2n_\text{q}+1)^2}{2g^2}}\right)\nonumber\\
\tilde s^{\infty} &=& -\dfrac{\dfrac{\delta}{g(2n_\text{q}+1)} + i \dfrac{\gamma_\text{q}}{g}}{1+ 2\dfrac{\delta^2}{g^2}+ \dfrac{\gamma_\text{q}^2(2n_\text{q}+1)^2}{2g^2}},
\ee
leading to the following expressions for the steady state energy exchanges:
\bb
\dot W^\infty &=& \hbar\omega_\text{L} \dfrac{ \gamma_\text{q}}{1+ 2\dfrac{\delta^2}{g^2}+ \dfrac{\gamma_\text{q}^2(2n_\text{q}+1)^2}{2g^2}}\label{eq3:Wss}\\
\dot Q_\text{cl}^\infty &=&  -\hbar\omega_0 \dfrac{\gamma_\text{q}/2}{1+ 2\dfrac{\delta^2}{g^2}+ \dfrac{\gamma_\text{q}^2(2n_\text{q}+1)^2}{2g^2}}\label{eq3:Qclss}\\
\dot Q_\text{q}^\infty &=& \dfrac{\gamma_\text{q}\delta /2g}{1+ 2\dfrac{\delta^2}{g^2}+ \dfrac{\gamma_\text{q}^2(2n_\text{q}+1)^2}{2g^2}}\label{eq3:Qqss}.
\ee

Just as in the FME approach, the steady state energy balance reads ${\cal P}_\text{L}^\infty + {\cal P}_\text{res}^\infty = 0$. However, as a direct consequence of the short coarse-graining time step, part of the energy provided by the laser drive is here described as a controlled work contribution (and the remaining as quantum heat). In the case of a resonant driving, the average quantum heat vanishes and the laser only provides work. 

The controlled nature of the energy exchange with the drive is therefore well captured by the optical Bloch description owing to its short time step able to describe the Rabi oscillations. Conversely, because it requires a time-averaging of theses oscillations, the Floquet approach only describes an effective incoherent energy transfer between the drive and the Qubit.

 
Finally, we also compute the average entropy production rate:
\bb
\sigma_\text{i}(t) = \dot S_\text{VN}(t) - \dfrac{{\cal P}_\text{res}}{T} \geq 0.
\ee
which takes non-zero value at steady state, reflecting the out-of-equilibrium nature of the Qubit's state in presence of the drive:
\bb
\sigma_\text{i}^\infty =  - \dfrac{\dot Q_\text{cl}^\infty}{T} \geq 0.
\ee
Eq.\eqref{eq3:Qclss} allows to check that this stationary entropy production vanishes for large detuning $\delta \gg g$ for which the Qubit decoupled from the drive.

\subsection{Connexion between quantum heat and entropy production}

In this section, we aim at illustrating the link between entropy production and quantum heat in the situation under study. So as to broaden the discussion, we extend the analysis to allow both strong and weak drives which can be addressed by the OBE approach.

We consider Jarzynski's protocol \cite{Jarzynski97,Seifert05}: the Qubit is initially in thermal equilibrium with the thermal reservoir at temperature $T$ (the drive is switched off) and its energy is measured at time $t_i$ with outcome $U_i \in \{-\hbar\omega_0/2,\hbar\omega_0/2\}$. The initial state of each trajectory is $\ket{\psi_i} \in \{\ket{e},\ket{g}\}$. Then the drive is switched on up to time $t_f$ where a second projective measurement of the Qubit's energy is performed, with outcome $U_f\in \{-\hbar\omega_0/2,\hbar\omega_0/2\}$. The final state of the trajectory is $\ket{\psi_f} \in \{\ket{e},\ket{g}\}$. As the initial and final Hamiltonian are both equal to $H_0$, the variation of free energy $\Delta F$ vanishes. 

In this context, the Central Fluctuation Theorem is equivalent to Jarzynski Equality (see section \ref{s2:TPMP}):
\bb
\moy{e^{-\Delta_\text{i} s[\gamma]/k_\text{B}}}_\gamma = \moy{e^{- \left(\Delta U[\gamma] - Q_\text{cl}[\gamma] - \Delta F\right)/T}}_\gamma = 1.
\ee
and the entropy produced in a single trajectory reads:
\bb
\Delta_\text{i}s[\gamma] &=&  \dfrac{1}{T}\left(\Delta U[\gamma] - Q_\text{cl}[\gamma] - \Delta F\right)\nonumber\\
&=& \dfrac{1}{T}\left(W[\gamma] - \Delta F\right) + \dfrac{Q_\text{q}[\gamma]}{T}. \label{eq3:disgamma}
\ee

The last term is a contribution to the entropy production which is proportional to the quantum heat. It corresponds to a genuinely quantum irreversibility associated to the presence of coherences in the $\{\ket{e},\ket{g}\}$ basis, which are continuously built up by the drive and erased by the reservoir.

\begin{figure}[h!]
\vspace{0.1cm}
\begin{center}
\includegraphics[width=\textwidth]{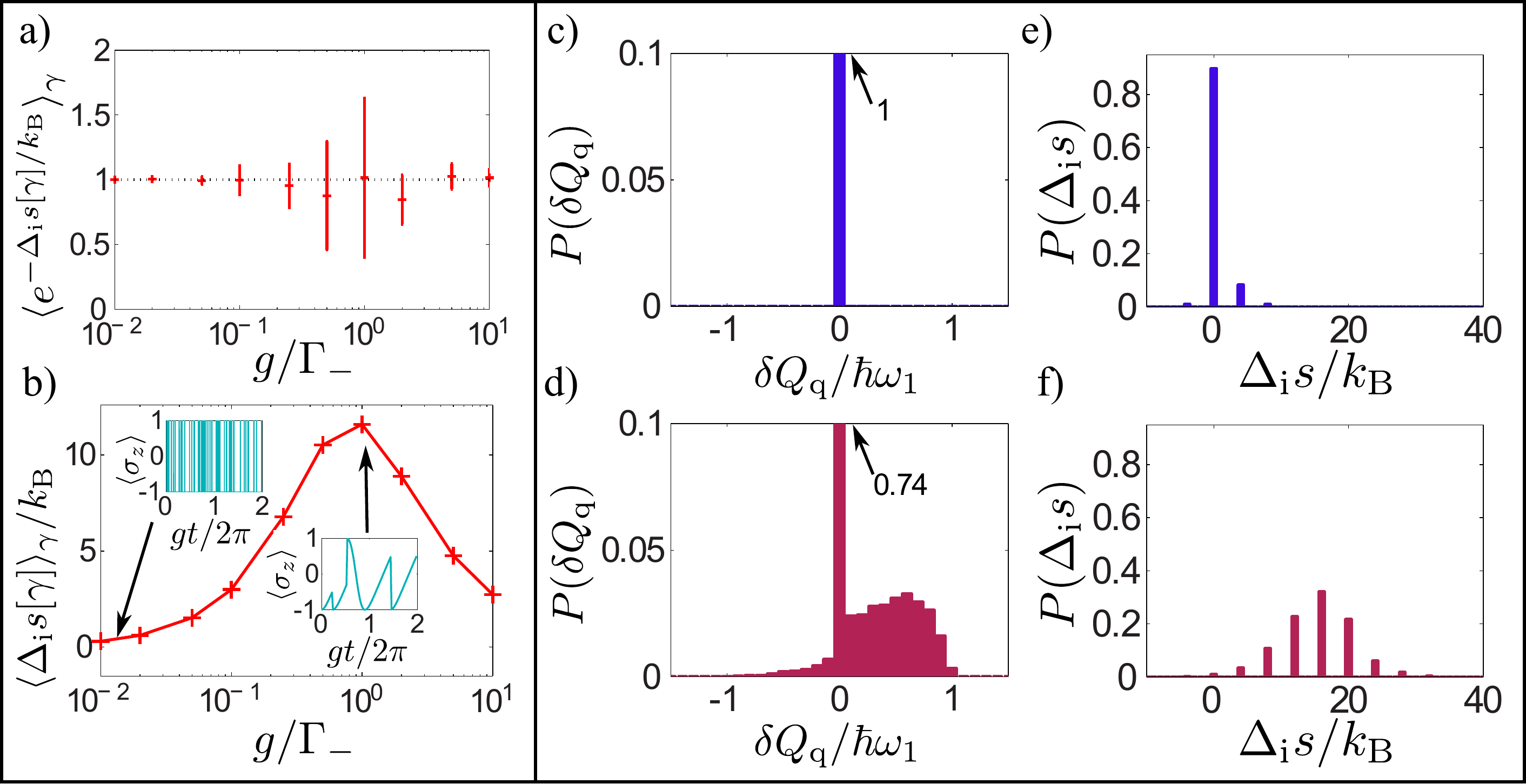}
\end{center}
\caption[Connection between quantum heat and entropy production]{Connection between quantum heat and entropy production. a): Numerical test of Jarzynski Equality as a function of the ratio $g/\Gamma_-$. b): Average entropy production as a function of $g/\Gamma_-$. Insets: expectation value of $\sigma_z$ along a single trajectory $\gamma$. c),d): Distribution of the quantum heat increment $\delta Q_{\text{q},\gamma}$ for $g/\Gamma_-= 0.01$ (c) and for $g/\Gamma_-= 1$ (d). e),f): Distribution of the entropy produced along a single trajectory for $g/\Gamma_-= 0.01$ (e) and for $g/\Gamma_-= 1$ (f). {\it Parameters}: $\hbar\omega_0/k_\text{B}T = 3$ for all plots, number of trajetories $N_\text{traj}  = \numprint{2e6}$ for a),b), $N_\text{traj}  = \numprint{1e3}$ for c) to f).\label{f3:JE}}
\end{figure}

Simulating a fair sample of this Quantum Jump trajectories allows testing such a Fluctuation Theorem (see Fig.\ref{f3:JE}a) and compute the entropy production (see Fig.\ref{f3:JE}b). For this simulation, we took a resonant drive $\delta = 0$ and set the duration of the transformation to $t_f - t_i = 4\pi/g$ (two complete Rabi oscillations). For a weak drive verifying $g/\Gamma_- \ll 1$ (with $\Gamma_- = \gamma_\text{q}(n_\text{q}+1)$ the emission rate), the Qubit is always in a thermal equilibrium state. This regime corresponds to reversible quasi-static isothermal transformations. In the other extreme, i.e. when $g \gg \Gamma_-$, the drive is strong enough to induce almost unperturbed Rabi oscillations. In this case, the transformation is too fast to allow for a stochastic event to take place and heat to be exchanged, i.e. it is adiabatic in the thermodynamic sense and once again reversible.

The intermediate scenario where $g\sim \Gamma_-$ gives rise to a maximal average entropy production. In this situation, the thermodynamic time arrow has a completely different nature from the case of the time-dependent Stark shift studied in Section \ref{s2:Stark}. Here, entropy is mostly produced by the frequent interruptions of the Qubit Rabi oscillations caused by the stochastic exchanges of photons with the bath (Inset of Fig\ref{f3:JE}d). During such quantum jumps, the quantum coherences induced by the driving process are erased, giving rise to the exchange of quantum heat, and to entropy production of quantum nature. The connection between quantum heat and entropy production is captured in Eq.\eqref{eq3:disgamma} where they are quantitatively related, but also in the corresponding histograms (see Fig\ref{f3:JE}c to f). Indeed the histograms of entropy production and quantum heat both concentrate around zero in the reversible regime, and take finite values in the irreversible regime. This observation is another evidence of the connection between quantum irreversibility and energy fluctuations that has been highlighted in Chapter \ref{Chapter1} in the context of measurement induced irreversibility. In both cases, quantum heat exchanges and entropy production appear as two thermodynamic signatures of the same phenomenon, i.e. coherence erasure.

\section{Discussion}

We have analyzed two approaches to describe the driven Qubit dynamics, both valid in the regime \eqref{d3:Regime}, but corresponding to different coarse-graining time steps. We have found two types of trajectories and stochastic thermodynamic descriptions.

In order to check the consistency of the two methods, we may compare the values of quantities that have a physical meaning in both approaches. As the FME dynamics is not defined for time-scales shorter than a Rabi period, we focus on steady state properties.

\begin{figure}[h!]
\begin{center}
\includegraphics[width=0.8\textwidth]{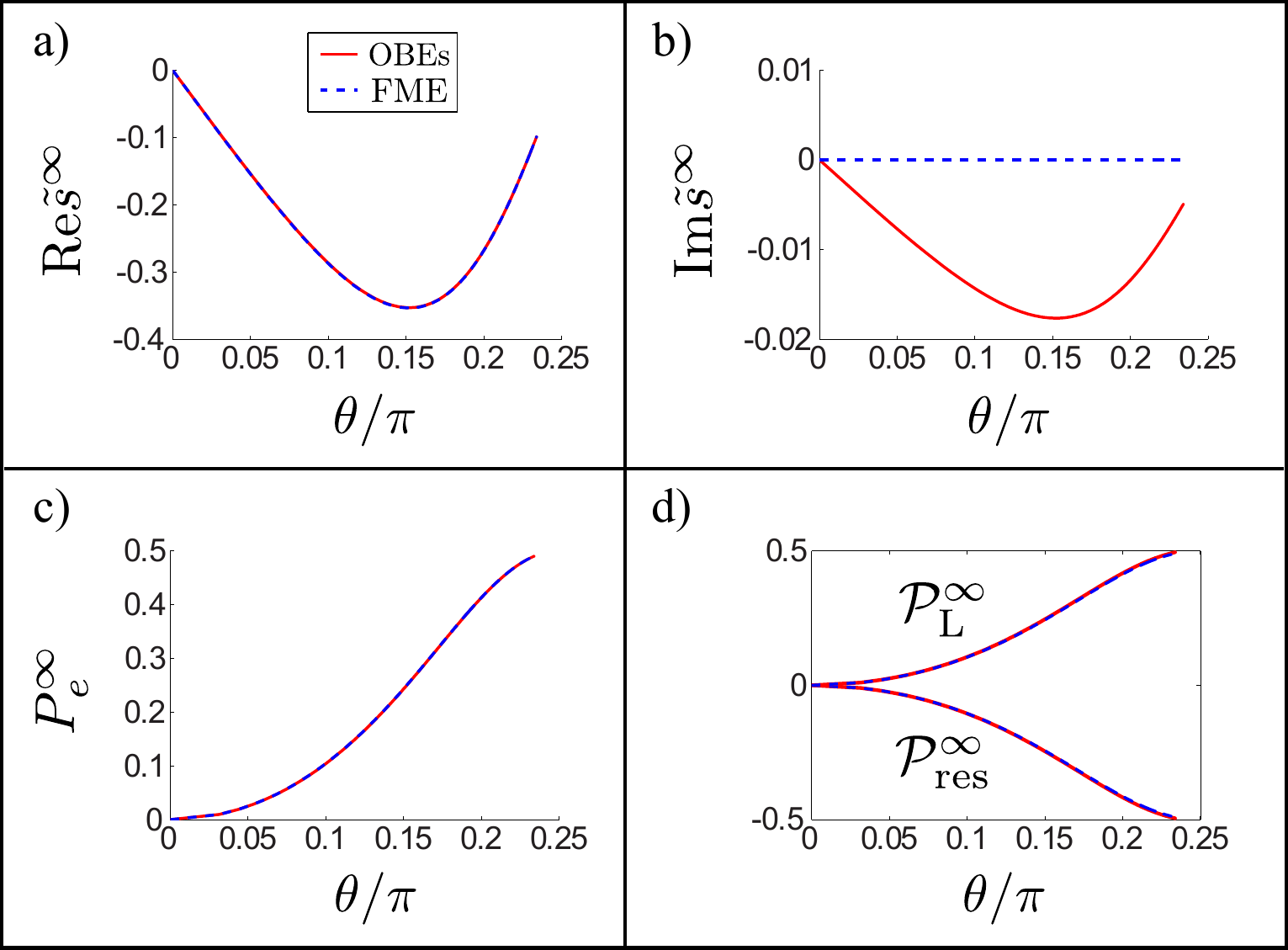}
\end{center}
\caption[Comparison of the approaches]{Comparison of the approaches. Steady-state values of (a) the real and (b) the imaginary parts of the amplitude of the coherences $\tilde s^\infty$, (c) the steady-state value of the excited state population $P_\text{e}^\infty$ and (d) the steady-state energy flows provided by the laser ${\cal P}_L$ and by the reservoir ${\cal P}_\text{res}$, predicted by the FME (blue dotted) and the OBEs (solid red), as a function of the angle $\theta$ defining the dressed basis (see Eqs.\eqref{d3:Theta}). {\it Parameters}: $\gamma/\omega_\text{L} = \numprint{1e-3}$, $\delta/\omega_\text{L} = \numprint{1e-2}$, $\hbar\omega_\text{L}/k_\text{B}T = 10$, $g/\omega_\text{L} \in [0;0.1]$.\label{f3:Comparison}}
\end{figure}

In Fig.\ref{f3:Comparison}a to c, we plot the steady state amplitude of the coherences in the $\{\ket{e},\ket{g}\}$ basis (in the rotating frame) $\tilde s^\infty$ and the steady state population of state $\ket{e}$ as a function of the angle $\theta$ defining the dressed basis. We fix the detuning to $\delta= \omega_\text{L}/100$ and vary the vacuum Rabi frequency $g$ from $0$ to $\omega_\text{L}/10 \gg \delta$ in order to explore all the values of $\theta$ from $0$ (the dressed basis and the bare basis coincide) to the maximum value $\theta = \pi/4$ (i.e. the resonant case in which $\ket{\pm} = (\pm\ket{e}+\ket{g})/\sqrt{2}$). We find that both approaches are in very good agreement except that the FME predicts a purely imaginary steady state amplitude of the coherences.

From a thermodynamic point of view, we note that the total stationary power provided by the drive ${\cal P}_\text{L}^\infty$ and classical heat flow provided by the reservoir  ${\cal P}_\text{res}^\infty$ predicted by both approaches coincide. Consequently, the change of coarse-graining only affects the way the energy exchanges are shared between work and heat: the coherent nature of the energy provided by the laser drive can solely be seen on time-scales allowing the Rabi oscillations to be observed. Note that as they are proportional to the classical heat flows, the average entropy productions predicted by both approaches also coincide. However, the link to coherence erasure which is the evidence of the genuinely quantum nature of this steady state entropy production can only be observed with the short coarse-graining time step. Eventually, once Rabi oscillations are coarse-grained, the laser drive behaves as an incoherent pump which solely differs from a thermal reservoir because of its ``colored'' spectrum: the incoherent energy flow provided by the drive vanishes for detuning much larger than $g$.

\section{Summary}

We have studied the fluorescence of a Qubit in contact with a thermal reservoir. We have shown that two different descriptions of the dynamics found in the literature correspond to two different choices of coarse-graining time steps when deriving the Lindblad equation. Using the formalism presented in previous chapters, we have studied the stochastic thermodynamics associated with both descriptions.

While the total power provided by the drive to the Qubit predicted with both approaches coincide, we have shown that the splitting between work and heat is affected by the choice of the time step. This study highlights that the time step impacts the thermodynamic descriptions.
 
\chapter{Hybrid Optomechanical systems: platforms for quantum thermodynamics}
\label{Chapter4}

\minitoc



\begin{figure}
\begin{center}
\includegraphics[width=\textwidth]{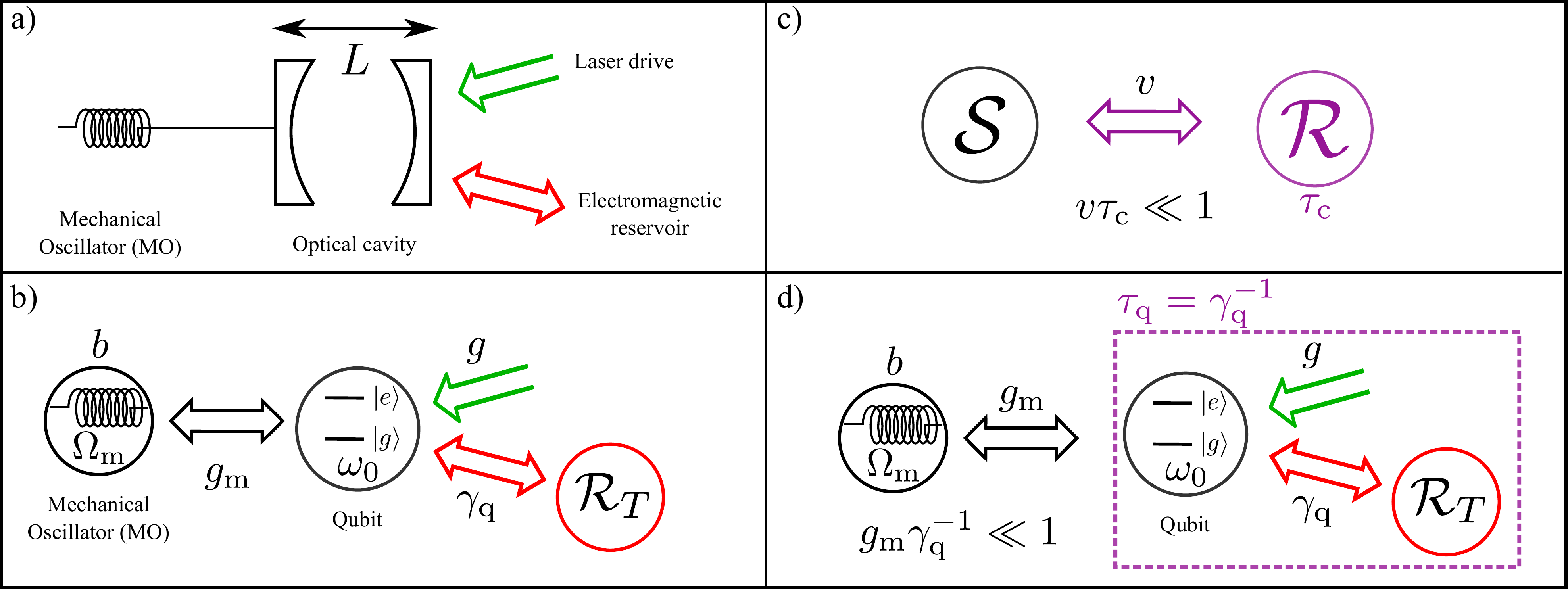}
\end{center}
\caption[Hybrid optomechanical system]{Hybrid optomechanical system. a) Cavity optomechanics: the light in an optical cavity is coupled to the position of a mechanical oscillator. b) Hybrid optomechanical setup under study: the role of the cavity is played by a two-level quantum emitter (Qubit) of frequency $\omega_0$. It is coupled with a strength $g_\text{m}$ to the position of a Mechanical Oscillator (MO) of frequency $\Omega_\text{m}$ and lowering operator $b$. The Qubit (resp. MO) is coupled to  an electromagnetic (resp. phonon) reservoir of temperature ${T}$ (resp. $T_\text{m}$), such that it has a decay rate $\gamma_\text{q}$ (resp. $\Gamma_\text{m}$). c) Canonical situation of open quantum systems: a system $\cal S$ is weakly coupled  to a reservoir ${\cal R}$ whose correlation time is denoted $\tau_\text{c}$. Denoting $v$ the strength of the coupling, the weak-coupling condition $v\tau_\text{c}\ll 1$ allows to coarse-grain the joint dynamics and derive an effective Markovian relaxation channel for ${\cal S}$ characterized by a rate of mangitude $v^2\tau_\text{c}$. d) The Qubit coupled to its electromagnetic environment is equivalent to a reservoir of correlation time $\tau_\text{q} = 1/\gamma_\text{q}$, such that the weak coupling condition $g_\text{m}/\gamma \ll 1$ is statisfied in the regime under study. \label{f4:CavityOM}}
\end{figure}

\lettrine{O}{ptomechanics} is a blooming field, made possible by the discovery of radiation pressure \cite{Lebedew1901,Nichols1901}. The canonical optomechanical system (see Fig.\ref{f4:CavityOM}a) involves an optical cavity with one of the two end-mirrors fixed on a Mechanical Oscillator (MO) \cite{Marquardt09}. 
The coupling between the light in the cavity and the oscillator acts in two ways: the light in the cavity affects the motion of the mirror through radiation pressure, and the motion of the mirror changes the cavity frequency which affects the amplitude of the light in the field. Cavity optomechanics is a research area which involves various setups from the microscopic \cite{Metzger04,Gigan06,Arcizet06,Favero07} to the kilometer scale \cite{Corbitt04,Corbitt07}. Applications are multiple: e.g. sensing \cite{Krause12,Abadie11}, generation of non-classical mechanical states \cite{OConnel10}, microwave-to-optical transducers \cite{Bochmann13}.

Later on, hybrid optomechanical setups with various architectures were proposed \cite{Treutlein}. The one studied in this thesis couples a MO with a two-level quantum emitter (a Qubit) instead of a cavity (see Fig.\ref{f4:CavityOM}b). The equivalent of the radiation pressure is a force on the MO depending on the Qubit population. Reciprocally, the transition frequency of the Qubit is modulated during mechanical oscillations. This situation can be realized in several platforms. For instance:
\begin{itemize}
\item In Ref.\cite{Arcizet11} the Qubit is a nitrogen-vacancy (NV) center, i.e. a defect in a nano-diamond which bears a spin $\frac{1}2$. The spin is fixed at the end of a vibrating SiC nanowire, and immersed in a strong magnetic field gradient which enables the coupling: the spin experiences a Zeeman shift of its transition frequency which varies strongly with the position of the mechanical oscillator. 
\item In Ref.\cite{Yeo13} the Qubit is a semiconduting quantum dot embedded at the bottom of a trumpet-shaped semiconducting nanowire whose top can oscillate. The coupling is mediated by the mechanical strain field around the quantum dot, which strongly depends on the nanowire's deflection, in particular when the quantum dot is eccentric.
\item In Refs.\cite{Pirkkalainen13,Lahaye09}, the mechanical resonator is integrated in a superconducting circuit. The Qubit is based on a Josephson junction, and the oscillations of the resonator modulate a capacitance, which affects the Qubit frequency. 
\end{itemize}

In contrast with cavity optomechanical setups, hybrid systems are non-linear: while the cavity can be filled with a large number of photons, leading to a larger effect on the oscillator, the Qubit is saturated with a single excitation. Interestingly, several hybrid setups \cite{Arcizet11,Yeo13,Pirkkalainen13} are about to reach the so-called ultra-strong coupling regime, in which a single event of absorption or emission of a photon by the Qubit has a visible effect on the MO motion.

In this chapter, we study the hybrid setup from two complementary perspectives. First, we show that it provides an natural platform to test thermodynamics of a Qubit: the MO behaves as a nano-battery providing work to the Qubit while the electromagnetic environment of the Qubit behaves as a thermal reservoir. As a great interest of this platform, the work can be directly measured in the battery which is providing it. This idea is published in \cite{Elouard15}.

In a second part, we look at the hybrid system from the point of view of quantum open systems. We show that when it is driven by a laser, the Qubit behaves like a reservoir inducing non-thermal relaxation of the MO. This reservoir is unusual because it is of finite size and therefore is affected by the MO dynamics. We show that a master equation can be derived and we use it to investigate the long-time evolution of the hybrid system in the ultra-strong coupling regime. This study is presented in \cite{OpticalNoise}.

\section{The hybrid optomechanical system}

\subsection{Setup}
The hybrid optomechanical system we consider is pictured on Fig.\ref{f4:CavityOM}b. It is composed of a Qubit of frequency $\omega_0$, coupled to a Mechanical Oscillator (MO) of frequency $\Omega_\text{m}$. The strength of the opto-mechanical coupling is characterized by the frequency $g_\text{m}$. Besides, the Qubit interacts with its electromagnetic reservoir ${\cal R}_T$, i.e. a thermal reservoir of temperature ${T}$ and may be driven by a laser field of Rabi frequency $g$. We denote $\gamma_\text{q}$ the spontaneous emission rate of the Qubit. 

We focus on a regime with the following characteristics:

\begin{itemize}
\item The interaction is dispersive: $\omega_0 \gg \Omega_\text{m}$. The two systems are very far from resonance.
\item The Qubit adiabatically follows the oscillator: $\omega_0,\gamma_\text{q} \gg \Omega_\text{m}$. The dynamics of the Qubit is very fast with respect to that of the MO, and it therefore adapts its state to the instantaneous mechanical position.
\item The coupling is ``ultra-strong'' for the MO: $g_\text{m} \gtrsim \Omega_\text{m}$. As a consequence, the absorption or the emission of a photon induces a displacement of the MO larger than its zero-point position fluctuation $x_0$ (namely, than the width of its ground state wavefunction).    
\end{itemize} 

Eventually, we identify a last condition that is crucial to describe the MO as a classical battery generating a time-dependant Hamiltonian on the Qubit: 
\bb 
g_\text{m} \ll \gamma.\label{d4:semiclcond}
\ee
\noindent In the following, we call semi-classical regime the regime in which condition \eqref{d4:semiclcond} is satisfied. In summary, the parameters involved in the hybrid system dynamics respect the following inequalities:
\bb
\omega_0,\gamma \gg g_\text{m} \gtrsim \Omega_\text{m}  
\label{d4:regime}.
\ee

This regime is relevant for the implementations cited before (see table \ref{t:OMparams}). The only exception is reference \cite{Pirkkalainen13} in which the adiabatic condition is not satisfied. \\

Note that in any practical implementation, the MO is also coupled to some phononic thermal reservoir inducing some damping $\Gamma_\text{m}$. In order to characterize the non-thermal relaxation channel induced by the Qubit on the MO, we neglect in the present chapter this thermal reservoir, which is justified by the large quality factors reached by the implementations (see table \ref{t:OMparams}).

\begin{table}[h!]
\vspace{0.2cm}
\hspace*{-0.85cm}
{\tabulinesep=1.2mm
\begin{tabu}{|l|c|c|c|c|c|}
\hline 
Variable & Symbol & \cite{Yeo13} & \cite{Arcizet11} & \cite{Pirkkalainen13} & \cite{Lahaye09} \\ 
\hline 
\hline 
MO frequency & $\Omega_\text{m}/2\pi$ & $530$ kHz & $1$ MHz & $72$ MHz & $58$ MHz\\ 
\hline 
MO quality factor & $\Omega_\text{m}/\Gamma_\text{m}$ & \numprint{3e3} & $>10^4$& $5500$& $\numprint{3e4}$ - $\numprint{6e4}$\\ 
\hline
MO zero point fluctuation & $x_{0}$ & \numprint[fm]{11} & $0.7$ pm& $4$ fm & $13$ fm \\ 
\hline
MO thermal population & $n_\text{m}$ & $2\; 10^5$ & $2\; 10^3$ & $\sim 6$ & $34$\\
\hline
Qubit frequency & $\omega_0/2\pi$  & $333$ THz & $18$ GHz & $\sim 1$ GHz& $\sim 10$ GHz\\ 
\hline
Qubit decay rate & $\gamma_\text{q}/2\pi$ & $157$ GHz & $7$ MHz & $3$ MHz & $170$ - $800$ MHz\\ 
\hline 

Optomechanical coupling & $g_\text{m}/2\pi$ & $450$ kHz & $100$ kHz & $4.5$ - $25$ MHz & $0.3$ - $2.3$ MHz\\ 
\hline 
Semi-classical quality factor & $Q_\text{sc}$ & \numprint{4e5} & \numprint{700} & \numprint{0.3} & \numprint{2e3}-\numprint{5e5}\\ 
\hline  
\end{tabu}}
\caption{Typical parameters for state-of-the-art hybrid optomechanical system.} 
\label{t:OMparams}
\end{table}

\subsection{The Semi-classical behavior}
\label{s4:semiCl}

In this chapter we evidence that the regime allows for a semi-classical treatment of the hybrid system, in which the joint system is always described by a factorized density operator $\rho_\text{qm}(t) = \rho_\text{q}(t)\otimes\rho_\text{m}(t)$, in spite of the large coupling strength $g_\text{m}$.  

This factorized form is valid on two time-scales we shall consider to describe the hybrid system's dynamics: On the fast time-scale corresponding to the Qubit dynamics, the MO can be considered as frozen. In the line of the textbook situation of molecular dynamics, we can use the Born-Oppenheimer approximation \cite{Born27} and look for solutions of the Qubit dynamics in which the position of the MO is a fixed parameter.

On the long time-scale, which is relevant for the MO dynamics, the Qubit can be seen as a reservoir weakly coupled to the MO. Indeed, because of its electromagnetic environment, the Qubit dynamics has a typical correlation time $\tau_\text{q} = 1/\gamma_\text{q}$. In the situation of a system weakly coupled to a reservoir of correlation time $\tau_\text{c}$ (see Fig.\ref{f4:CavityOM}c), Bloch-Redfield theory allows to derive a Markovian master equation for the system alone. The weak-coupling condition reads $v\tau_\text{c} \ll 1$, where $v$ is the typical strength of the coupling between the system and the Qubit. The derivation involves a coarse-graining of the exact system-reservoir dynamics with a time-step $\Delta t_\text{cg}$ much larger than $\tau_\text{c}$ (see Section \ref{s3:MicroscopicDerivation} and the fourth chapter of Ref.\cite{CCT}). Owing to this process, the effect of the reservoir is encompassed in an effective relaxation channel. 

For the hybrid system under study, such the weak-coupling condition reads $g_m/\gamma_\text{q} \ll 1$ which is satisfied in the regime under study: a master equation for the MO can therefore be derived by coarse-graining the hybrid system dynamics using a new time-step $\Delta t_\text{q} \gg 1/\gamma_\text{q}$. \\

The case of the hybrid system still differs from the textbook open quantum system situation for two reasons that will be explored in the two next sections of this chapter: 
\begin{itemize}
\item The optomechanical coupling has non zero effect at first order in $g_\text{m}$. This first order is responsible for effective time-dependent Hamiltonians of the Qubit and the MO, which are studied in this chapter with the thermodynamic point of view.
\item At second order in $g_\text{m}$, the usual derivation of a master equation has to be adapted because the Qubit is of finite size, unlike standard reservoirs. As a consequence, the properties of the dissipation channel we obtain are affected by the MO dynamics, and therefore time-dependent. 
\end{itemize}

\subsection{Master equation for the coupled hybrid system}

The total Hamiltonian of the hybrid optomechanical system reads:
\bb
H_\text{qm} = \hbar\omega_0 \Pi_e + H_\text{L}(t) + V + \hbar\Omega_\text{m} b^\dagger b. 
\ee

The Qubit is modelled as in previous chapters, while the MO is identified with a harmonic oscillator of mechanical frequency $\Omega_\text{m}$ and lowering operator $b$. We have neglected the vacuum energy contribution which plays no role in the dynamics. The Qubit is driven by a laser of Rabi frequency $g$, which adds a term $H_\text{L}(t) = \hbar g (e^{-i\omega_L t}\sigma_+ + e^{i\omega_L t}\sigma_-)$ in its Hamiltonian. The resonator and the Qubit are coupled via the Hamiltonian:
\bb
V = \hbar g_\text{m} (b+b^\dagger) \Pi_e = \hbar g_\text{m} \hat{X} \Pi_e,
\ee 

\noindent where $g_\text{m}$ is the strength of the optomechanical coupling. Such parametric coupling induces a shift of the Qubit frequency proportional to the position of the MO. In the following, we use the reduced position operator $\hat{X} = b + b^\dagger$, which is linked to the real displacement of the mechanical degree of freedom by multiplication with the zero-point motion\footnote{That is to say the width of the ground state wavefunction.} $x_0 = \sqrt{2\hbar/m_\text{eff}\Omega_\text{m}}$, with $m_\text{eff}$ the effective mass of the oscillator. $x_0$ is typically in the fm range.


In the setups cited in the introduction of the chapter, the Qubit is weakly coupled to a thermal reservoir at temperature ${T}$ which is modeled like in Section \ref{s3:MicroscopicDerivation} as a collection of Harmonic oscillator of frequencies $\{\omega_k\}$. The coupling Hamiltonian $V_{\cal R}$ between the Qubit and the reservoir and the reservoir Hamiltonian read:

\bb
V_{\cal R} &=& \sum_k \hbar g_k\left(a_k \sigma_+ + a_k^\dagger \sigma_-\right)\\
H_{\cal R} &=& \sum_k \hbar\omega_k a^\dagger_k a_k.
\ee

The total master equation $\rho_\text{tot}$ of the hybrid system and the reservoir obeys the Von Neumann equation:

\bb
\dot\rho_\text{tot} = -i [H_\text{qm}(t) + V_{\cal R} + H_{\cal R},\rho_\text{tot}].
\ee

As in section \eqref{s3:MicroscopicDerivation}, the exact dynamics can be coarse-grained in order to eliminate the reservoir from the description. The new time-step $\Delta t_\text{q}$ is chosen to be much larger than the reservoir correlation time, but smaller than $v_{\cal R}^2\tau_{\cal R}$, where $v_{\cal R}$ is the typical magnitude of $V_{\cal R}$. This procedure allows to derive a Lindbladian capturing the relaxation channel induced by the reservoir on the Qubit, whose typical rate is the spontaneous emission rate of the Qubit:
\bb
\gamma_\text{q} = \sum_k g_k^2\delta_\text{D} (\omega_0 - \omega_k).
\ee
 In this chapter, when we consider a coherent drive, we neglect any deviation from the optical Bloch equations picture due e.g. to the structure of the electro-magnetic reservoir (see Chapter \ref{Chapter3}).

Contrary to Section \ref{s3:MicroscopicDerivation}, we have to take into account the presence of the coupling to the MO which affects the Qubit frequency: $H_0 + V = \hbar(\omega_0 + g_\text{m}\hat X)\Pi_e$. Owing to the adiabatic condition $\gamma_\text{q} \gg \Omega_\text{m}$, the MO can be considered as frozen during one coarse-graining time-step $\Delta t_\text{q} \ll 1/\Omega_\text{m},1/g_\text{m}$. The MO position can therefore be seen as a fixed parameter. As long as we restrict our analysis to MO states which have a quantum variance in position small enough around the average $x(t)$, we can consider that the Qubit has an effective frequency:
\bb
\omega_\text{q}(t) = \omega_0 + g_\text{m}x(t).\label{d4:omq}
\ee
More precisely, introducing the MO's reduced density operator $\rho_\text{m}(t) = \text{Tr}_\text{q}\{\rho_\text{qm}(t)\}$, where $\rho_\text{qm}(t)$ is the density operator of the total hybrid system, and defining the average MO position and position variance:
\bb 
x(t) &=& \text{Tr}_\text{m}\{\rho_\text{m}(t)\hat X\}\\
 V_\text{X} &=& \text{Tr}\{\rho_\text{m}(t)\hat X^2\}-x(t)^2,
\ee
we require:
\bb
g_m V_X \ll \gamma_\text{q}, \omega_0.\label{d4:VXcond}
\ee 
\noindent This condition is typically verified for a coherent state such that $V_X = 1$.\\

Finally, assuming that we are in the conditions such that the optical Bloch equations are valid (see Section \ref{s3:OBE}), we find the following Lindbladian:
\bb
{\cal L}_\text{q}^{x(t)} = \gamma_\text{q}\left(({n}[x(t)]+1){\cal D}[\sigma_-] + {n}[x(t)]{\cal D}[\sigma_+]\right).\nonumber\\ \label{d4:LqX}.
\ee

\noindent with 
\bb
n[x(t)] = 1/(e^{\hbar\omega_\text{q}(t)/k_\text{B}{T}}-1) = 1/(e^{\hbar(\omega_0 + g_\text{m}x(t))/k_\text{B}{T}}-1).
\ee

The dynamics of the density operator of the Hybrid system $\rho_\text{qm}(t)$ is therefore ruled by the master equations:
\bb
\dot \rho_\text{qm}(t) = -i \left[H_\text{qm}(t),\rho_\text{qm}(t)\right] + {\cal L}_\text{q}^{x(t)}[\rho_\text{qm}(t)] \label{eq4:MEHybrid},
\ee
 
\noindent Typical values of the parameters involved in the model are provided in table \ref{t:OMparams} for the implementations mentioned in the introduction.

\section{A platform for information thermodynamics}

In this section, we restrict our analysis to the case where the Qubit is not driven by a laser ($g=0$), and we focus on the dynamics at first order in $g_\text{m}$, which contains the average effect of the MO on the Qubit and of the Qubit on the MO. Our goal is to evidence that the hybrid optomechanical system naturally provides an implementation of a generic situation of thermodynamics involving a system (the Qubit) coupled to a thermal reservoir (the electromagnetic reservoir) and an operator driving it (the MO). Indeed, as we show below, the dynamics of the MO generates a time-dependent term in the Qubit Hamiltonian, associated to work provided the MO. 

Note that the total Hamiltonian of the hybrid system is time-independent: this setup is therefore an example of autonomous quantum machine \cite{Brask15} in which the role of the operator is played by a finite-size system that we call a battery \footnote{Sometimes, in theoretical studies of autonomous systems, the role played by the MO here is shared between two devices, a battery solely providing the energy and a ``clock'' \cite{Buzek99,Woods16} whose evolution triggers the time-dependence in the effective system's Hamiltonian.}. Studying such a device therefore gives insights about the emergence of time-dependent Hamiltonians which plays a paramount role in quantum thermodynamics. Here, the factorized dynamics of the hybrid system (see Section \ref{s4:semiCl}) is a crucial condition for this classical behavior of the MO to emerge.

Besides, the MO provides an interesting battery because its energy is finite, in spite of its classical behavior. The work performed on the Qubit is associated with a decrease of the MO mechanical amplitude. In order to be measured, this decrease of amplitude has to be sufficiently large with respect to fluctuations. The ultra-strong coupling condition $g_\text{m} \gtrsim \Omega_\text{m}$ is useful to improve the signal to noise ratio of such a measurement.

 We first present the dynamics of the two systems, and then present the thermodynamic description of the Qubit dynamics. We finally provide an application in the framework of information thermodynamics.

\subsection{First order dynamics without laser}

As justified by the analysis of Section \ref{s4:semiCl}, we introduce a factorized ansatz for the joint hybrid system $\rho_\text{qm}(t) = \rho_\text{q}(t)\otimes\rho_\text{m}(t)$. We also consider that the MO is initialized in a pure coherent state of complex amplitude $\beta_i$, i.e. $\rho_\text{m}(t_i) = \ket{\beta_i}\bra{\beta_i}$. Owing to the form of the ansatz, the evolution equation Eq.\eqref{eq4:MEHybrid} can be recast as a system of two coupled master equations for the MO and Qubit reduced density matrices $\rho_\text{m}(t)$ and $\rho_\text{q}(t)= \text{Tr}_\text{m}\{\rho_\text{qm}(t)\}$:
\bb
\dot \rho_\text{q}(t) &=& -i[H_\text{q}(t),\rho_\text{q}(t)] + {\cal L}_\text{q}^{x(t)}[\rho_\text{q}]\label{eq4:SCEq}\\
\dot \rho_\text{m}(t) &=& -i[H_\text{m}(t),\rho_\text{m}(t)]\label{eq4:SCEm}
\ee
\noindent where the effective Qubit and MO Hamiltonians read:
\bb
H_\text{q}(t) &=& \hbar(\omega_0+ g_m x(t))\Pi_e = \hbar\omega_\text{q}(t)\Pi_e \\
H_\text{m}(t) &=& \hbar\Omega_\text{m}b^\dagger b + \hbar g_m (b+b^\dagger) P_e(t).
\ee
\noindent The Qubit effective Hamiltonian contains the effective frequency $\omega_\text{q}(t)$ introduced in Eq.\eqref{d4:omq}, which it is useful to rewrite:
\bb
\omega_\text{q}(t) = \omega_0 + \delta_\text{m}(t),
\ee
where $\delta_\text{m} =  g_\text{m} x(t)$ is a shift of the Qubit proportional to the average MO position. The MO effective Hamiltonian includes a force proportional to the average population of the Qubit's excited state 
\bb
P_e(t) = \text{Tr}\{\rho_\text{q}(t)\Pi_e\},
\ee
 which is the direct analog of the radiation pressure appearing in the context of cavity optomechanics.

It is interesting to notice that the evolution of the MO Eq.\eqref{eq4:SCEm} preserves the nature of pure coherent state of the MO. As a consequence, we can characterize the state of the MO at any time $t\geq t_i$ with its coherent amplitude $\beta(t)$ obeying the dynamical equation:
\bb
\dot \beta(t) = -i\Omega_\text{m}\beta(t) - ig_\text{m} P_e(t) \label{eq4:dbeta}.
\ee

\subsection{Average thermodynamic quantities} 

We now describe the thermodynamics of the Qubit. In this chapter we focus on average thermodynamic quantities. We consider that the Qubit is prepared in a state that which does not carry coherences in the $\{\ket{e},\ket{g}\}$ basis. As $H_\text{q}(t)$ and the steady state of ${\cal L}_\text{q}^{x(t)}$ are both diagonal in this basis, the Qubit never get coherences, and its thermodynamic behavior is very reminiscent with the situation studied in Section \ref{s2:Stark}, except that the time-dependent frequency of the Qubit is changed according to the dynamics of the MO ruled by Eq.\eqref{eq4:SCEm}.

As mentioned above, the hybrid system is an autonomous system, in the sense that its total Hamiltonian $H_\text{qm}$ is time-independent (in the absence of other drive), and therefore does not receive any work from the outside. However, energy exchange occurs between the MO and the Qubit, and in the regime of parameter we are interested in, this can be formulated in term of a time-dependent Hamiltonians for the Qubit and the MO. \\

The thermodynamic transformation of the Qubit we consider corresponds to variation of its effective frequency $\omega_\text{q}(t)$ controlled by the MO's dynamics between $t_i$ and $t_f$. We define the average internal energy $U(t)$ of the Qubit at time $t$ and the increment of average work $\delta W(t)$ performed on (resp. increment of average heat\footnote{Here this term only contains classical heat provided by the thermal reservoir.} $\delta Q(t)$ received by) the Qubit between $t$ and $t+\Delta t_\text{q}$:
\bb
U(t) &=& \text{Tr}\{H_\text{q}(t) \rho_{\cal S}(t)\} = \hbar\omega_\text{q}(t) P_e(t)\label{d4:U}\\
\delta W(t) &=& \text{Tr}\{ \dot H_\text{q}(t)\rho_\text{q}(t)\}\Delta t_\text{q} = \hbar g_\text{m} \Delta t_\text{q}\dot x(t) P_e(t)\label{d4:dW}\\
\delta Q(t) &=& \text{Tr}\{H_\text{q}(t) \dot \rho_\text{q}(t)\}= -\gamma_\text{q}\Delta t_\text{q}\hbar\omega_\text{q}(t)\left((2 n[x(t)] +1) P_e(t)-n[x(t)]\right)\label{d4:dQ},
\ee
\noindent The integrated work $W = \int_{t_i}^{t_f} \delta W(t)$ and heat $Q = \int_{t_i}^{t_f} \delta Q(t)$ along the transformation fulfil the first law of thermodynamics:
\bb
\Delta U = W + Q,
\ee\\

We also introduce the average entropy production $\Delta_\text{i}S$ in an isothermal transformation at temperature $T$ in which the Qubit starts in thermal equilibrium\footnote{To obtain such entropy production, we considered a process including initial and final projective measurements of the energy of the Qubit (which does not affect the density operator already diagonal in the $\{\ket{e},\ket{g}\}$ basis) and we set the initial state of the direct and time-reversed processes to be the thermal equilibrium state associated with the initial and final values of the Qubit's effective frequency $\omega_\text{q}(t)$ (see Section \ref{s2:TPMP}).}:
\bb
\Delta_\text{i}S &=& \Delta S_\text{VN}-\dfrac{1}{k_\text{B} T} Q = \dfrac{1}{k_\text{B}T}(W-\Delta F) \geq 0 \label{d4:DiS},
\ee

\noindent where the Qubit's free energy reads $F(t) = U(t) - T S_\text{VN}[\rho_\text{q}(t)]$ and the Von Neuman entropy of the Qubit's density operator reads:
\bb
S_\text{VN}[\rho_\text{q}(t)] &=& -k_\text{B} \text{Tr}\{\rho_{\text{q}}(t)\log\rho_\text{q}(t)\} \nonumber\\
&=& - k_\text{B}P_e(t)\log P_e(t)-k_\text{B}(1-P_e(t))\log(1-P_e(t))\nonumber\\
&=& k_\text{B}H_\text{Sh}[P_e(t)] \label{d4:S},
\ee
\noindent  $H_\text{Sh}[p]$ is the Shannon entropy of distribution $\{p,1-p\}$.

\subsection{The MO as a finite-size battery}\label{s4:MObattery}

Because it does not become entangled with the Qubit at any time of the transformation, the MO plays the role of an external operator controlling the variation of a parameter of the Qubit's Hamiltonian: the frequency $\omega_\text{q}(t)$. Despite its classical behavior from the Qubit's point of view, the MO is a finite size system. As a consequence, the work it performs on the Qubit results in a finite variation of its internal energy
\bb
U_\text{m}(t) = \hbar\Omega_\text{m}N_\text{m}(t),
\ee
 where the average number of phonons in the MO is denoted:
 \bb
 N_\text{m} = \text{Tr}\{b^\dagger b\rho_\text{m}(t)\}.
 \ee
 
  The derivative of $U_\text{m}(t)$ can be computed from Eq.\eqref{eq4:SCEm}:
\bb
\dot U_\text{m}(t) &=& \hbar\Omega_\text{m}\dot N_\text{m}(t)\nonumber\\
&=& -i g_\text{m} \hbar\Omega_\text{m} P_e(t)(\beta^*(t)-\beta(t))\nonumber\\
&=& - \hbar g_m(\dot{\beta}^*(t)+\dot{\beta}(t)) P_e(t)\nonumber\\
&=& - \dot \delta_\text{m} P_e(t)\nonumber\\
&=& - \dfrac{\delta W(t)}{\Delta t_\text{q}},
\ee
\noindent which leads to the following equality once integrated between $t_i$ and $t_f$:
\bb
W = - \Delta U_\text{m} = \hbar\Omega_\text{m}(N_\text{m}(t_i) - N_\text{m}(t_f)).\label{d4:WmUm}
\ee

Here is a great interest of using the MO to implement a transformation of the Qubit: the work $W$ performed on the Qubit can be measured directly from the change of the internal energy of the MO between the initial and the final times of the transformation. As the MO remains in a coherent state during the whole transformation, the measurement of $N_\text{m}(t_{i/f}) = \vert\beta_{i/f}\vert^2$ can be replaced with a measurement of the initial and final positions and velocities, which can be realized experimentally e.g. using light deflection techniques \cite{Yeo13}. In practice, such a measurement contains various sources of noise, including the Brownian motion of the MO, such that the measurement has to be performed many times and the result averaged over the realizations. Ultimately, a variation of the MO reduced position can be measured if it is larger than $1$ (i.e. if the position variation is larger than the zero point motion $x_0$).

Note that using a battery of finite size also implies that its dynamics is affected by the transformation. This is in particular true if the initial mechanical energy $U_\text{m}(t_i)$ is of the order of the work $W$ performed on the Qubit during the transformation. The regime of an ideal classical operator is retrieved when the initial mechanical amplitude is large enough such that the work exchanges does not affect significantly the MO's dynamics and therefore the time-variation of $\omega_\text{q}(t)$.

\subsection{Information-to-energy conversions in a hybrid system}

We now study an application of this thermodynamic study in the framework of information thermodynamics. Since the resolution of Maxwell's demon paradox, it is known that information and work extraction are closely related. When the operator has information about a system, he can exploit it to extract work from the system \cite{Toyabe10,Koski14}. Conversely, when the operator does not have information about the system, and want to put it in a known state, he has to pay some work related to a reduction of the system's entropy \cite{Berut12}. Therefore, it is possible to convert forth and back information about a system into energy. Landauer's principle  sets that if one is able to perform these information-to-energy conversions in a reversible (quasi-static) and isothermal (in contact with a reservoir at a temperature $T$) manner, the work that can be extracted (resp. that has to be paid) when the operator has (resp. in order to acquire) $1$ bit of information is $W_1 = k_\text{B}T\log2$  \cite{DelRio11,Berut12,Landauer61}. 

The simplest setup in which this principle of information-to-energy conversion can be observed is a two-level system like the Qubit involved in the hybrid optomechanical system. Such system carries at most one bit of information. In this part, we show that the dynamics of the hybrid system can be interpreted as cycles of reversible information-to-energy conversions, and that the modulation of the MO energy is therefore related to the Qubit entropy variation.

Let us first analyse the first quarter of a mechanical oscillation which corresponds to the so-called Landauer erasure process in which work is paid to put the Qubit in a known pure state. The MO is initialized in a coherent state of purely imaginary amplitude $\beta_i \in i\mathbb{R}$, such that the initial average position  is $x(t_i) = 0$, and the initial average momentum is $p(t_i)= \beta_i/2$. We define the average momentum: 
\bb
p(t) \underset{\text{def}}{=} i\text{Tr}\{(b^\dagger - b)\rho_\text{m}(t)\}.
\ee

 The Qubit has an initial effective frequency $\omega_\text{q}(t_i) = \omega_0$ and is in equilibrium with its thermal reservoir, such that is density operator is a mixture $\rho_{\text{q},i} = P_e^\text{th}[x(t_i)]\ket{e}\bra{e} + (1-P_e^\text{th}[x(t_i)])\ket{g}\bra{g}$, where the thermal population corresponding to a position $x(t)$ of the MO satisfies:

\bb
P_e^\text{th}[x(t)] = \dfrac{e^{-\hbar\omega_\text{q}(t)/k_\text{B}{T}}}{Z(t)} = \dfrac{e^{-\hbar(\omega_0+g_\text{m}x(t))/k_\text{B}{T}}}{1+e^{-\hbar(\omega_0+g_\text{m}x(t))/k_\text{B}{T}}}.
\ee

\noindent $Z(t) = 1+e^{-\hbar\omega_\text{q}(t)/k_\text{B}{T}}$ is the canonical partition function. Because it is a mixed state, $\rho_{\text{q},i}$ has a non-zero Von Neumann entropy $S_\text{VN}[\rho_\text{q}(t_i)] = H_\text{Sh}[P_e^\text{th}[x(t_i)]]$. In the limit of high temperature $\hbar\omega_0 \ll k_\text{B}{T}$, $\ket{e}$ and $\ket{g}$ are equiprobable in the mixture and the initial entropy is $k_\text{B}\log(2)$. 

 At time $t_f$ corresponding to a quarter of oscillation later, the MO momentum is $0$ and the position its $x(t_f)$. The frequency of the Qubit is $\omega_\text{q}(t_f) = \omega_0 + g_\text{m}x(t_f)$. Therefore, the Qubit frequency has increased of an amount $g_\text{m}x(t_f)$. We assume a reversible transformation, i.e. that the Qubit is at any time in equilibrium with its reservoir at temperature, and therefore that the probability to be in state $\ket{e}$ at time $t$ is $P_e^\text{th}[x(t)]$. At the end of the transformation, the entropy of the Qubit has decreased to $S_\text{VN}[\rho_\text{q}(t_f)] = k_\text{B}H_\text{Sh}[P_e^\text{th}[x(t_f)]]$: some information has been acquired on the Qubit's state.  This entropy would be zero for large frequency $\hbar\omega_\text{q}(t_f) \gg k_\text{B}{T}$.  Note however,  that in order to stay in the weak coupling regime for the Qubit, the frequency modulation $\delta_\text{m}$ has to be kept much smaller than $\omega_0$, which corresponds to a constraint on the MO amplitude $\vert\beta\vert\ll \omega_0/g_\text{m}$. This prevents from doing complete information-to-energy conversion in fulfilling $\hbar\omega_\text{q}(t_i) \ll k_\text{B}{T} \ll \hbar\omega_\text{q}(t_f)$.   Notwithstanding, we can perform conversions involving less than one bit of information.
 
We can compute the work performed on the Qubit during such a transformation:
\bb
W^\text{rev} &=& \hbar g_\text{m}\int_{t_i}^{t_f} dt  \dot x(t) P_e^\text{th}[x(t)]\nonumber\\
&=& \hbar g_\text{m}\int_{0}^{x(t_f)} dx P_e^\text{th}[x]\nonumber\\
&=& -k_\text{B}{T} \log\left(\dfrac{1+e^{-\hbar(\omega_0+g_\text{m}x(t_f))/k_\text{B}{T}}}{1+e^{-\hbar\omega_0/k_\text{B}{T}}}\right)\nonumber\\
&=& \Delta U - k_\text{B}{T}\Delta S_\text{VN}\geq 0 \label{d4:Wrev},
\ee

\noindent and the heat received by the Qubit:
\bb
Q^\text{rev} =  \Delta U_\text{q} - W^\text{rev}= k_\text{B}{T}\Delta S_\text{VN}\leq 0 \label{d4:Qrev},
\ee

\noindent The proportionality between heat and the variation of entropy is a signature of the reversibility of the transformation, which requires to increase $x(t)$ slowly with respect to the thermalization rate wich is typically $\gamma_\text{q}(2{n}[\omega_q]+1)$. This condition is verified in regime \eqref{d4:regime} as long as the mechanical amplitude is not too large such that $\Omega\vert \beta_i\vert \ll \gamma_\text{q}(2{n}[\omega_q]+1)$. When the reversibility condition is violated, the work cost is larger than $W^\text{rev}$.

\begin{figure}[h!]
\begin{center}
\includegraphics[width=0.8\textwidth]{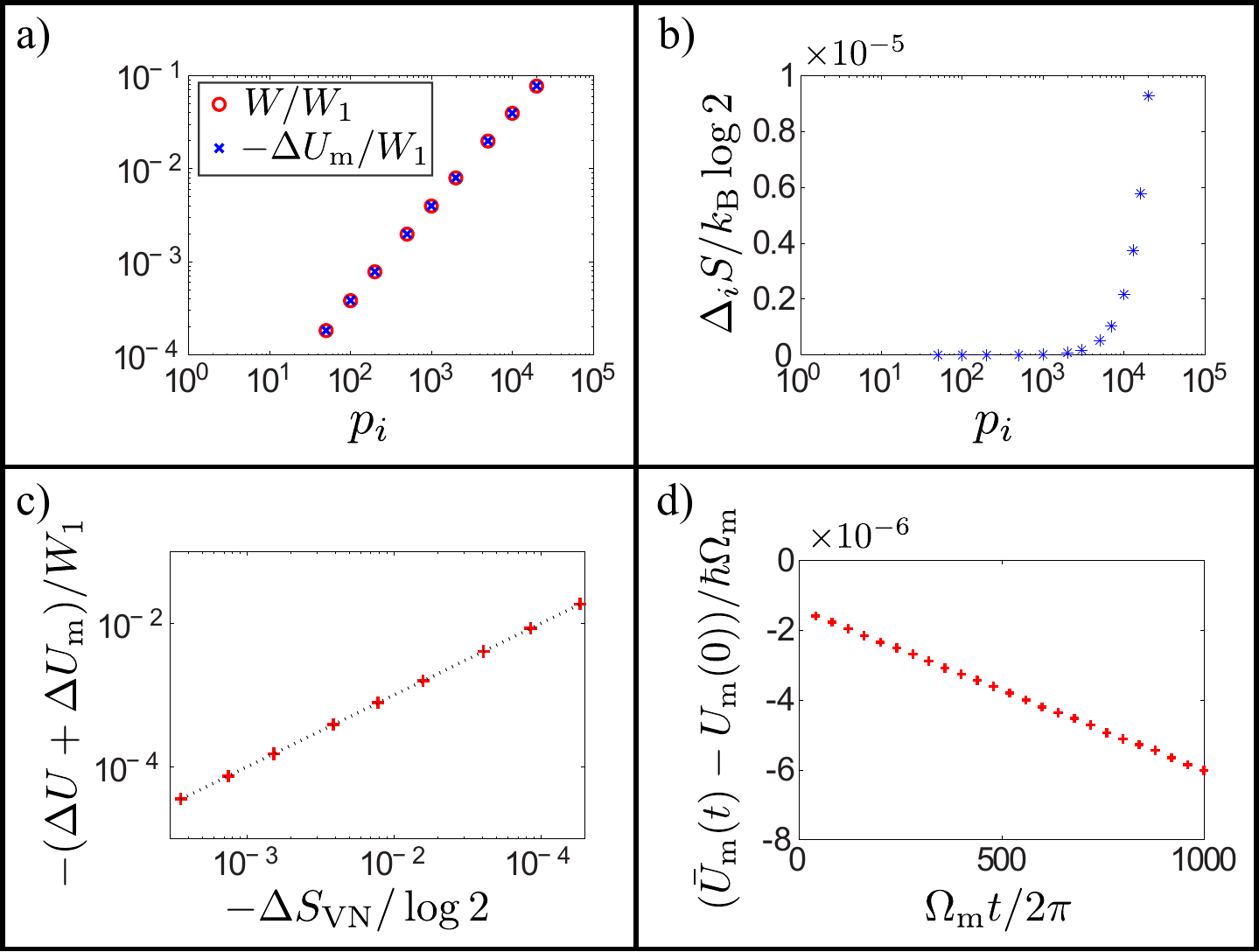}
\end{center}
\caption[Information-to-energy conversions]{Information-to-energy conversions. a,b,c) Thermodynamic of the first quarter of an oscillation (conversion of work into energy): Work $W$ and energy provided by the MO $\Delta U_\text{m}$ in unit of $W_1 = kT_\text{B}\log 2$ (a) and entropy produced in unit of $k_\text{B}\log 2$ (b) as a function of the initial momentum $p_i$ of the MO. c) Heat dissipated computed from the variation of the internal energy of the total system as a function of the variation of the Qubit's Von Neumann entropy. d) Decrease of the internal energy of the MO due to the irreversibility of the Qubit transformation. $\bar{U}_\text{m}(t)$ stands for the internal energy of the MO averaged over 20 mechanical oscillations around time $t$. {\it Parameters}: $\Omega_\text{m}/\gamma_\text{q} = 0.01$, $g_\text{m}/\gamma_\text{q} = 0.1$. a,b,c) $\omega_0/\gamma_\text{q} = \numprint{1e4}$ and $k_\text{B}T/\hbar\gamma_\text{q} = \numprint{3e4}$. d) $\omega_0/\gamma_\text{q} = \numprint{1e3}$, $k_\text{B}T/\hbar\gamma_\text{q} = 10$. \label{f4:Conversion}}
\end{figure}

Solving numerically Eq.\eqref{d4:WmUm} for different values of the initial mechanical amplitude $\beta_i = 2ip_i$, one can directly check that $W$ and $\Delta U_\text{m}$ compensate (see Fig.\ref{f4:Conversion}a). As the initial momentum $p_i$ increases, the transformation of the Qubit gets faster and more entropy is produced (see Fig.\ref{f4:Conversion}b). However, in the regime under study, the transformation is very close to be quasi-static, hence the proportionality between the heat $Q$ (that can be measured as the change of the total hybrid system's internal energy) and the variation of the Qubit entropy (see \ref{f4:Conversion}c).  The variation of mechanical amplitude $\vert\delta \beta\vert$ associated with the work $W$ can be estimated using $W = - \Delta U_\text{m} \sim 2\Omega\vert\beta_i\vert \vert\delta \beta\vert$, and therefore $\vert\delta\beta\vert \sim W/ 2\Omega\vert\beta_i\vert$ which corresponds to $\vert\delta\beta\vert \sim 4$ for the parameters or Fig.\ref{f4:Conversion}. This variation of amplitude is larger than the zero point motion and could therefore be measured.

During the second quarter of an oscillation, the frequency of the Qubit is decreased back to $\omega_0$, which corresponds to the reverse conversion: the information the operator has about the Qubit' state decrease, while work is extracted and stored in the MO. In the case of a reversible transformation, the work performed on the Qubit is $- W^\text{rev}$. When the transformation is too fast, less work can be extracted. The two remaining quarter of oscillations involve another cycle of information-to-energy conversion (the Qubit frequency is decreased and increased back). In the reversible limit, the total energy variation of the MO reduces to $0$. Conversely, irreversibility can be detected via a non-zero average energy change of the MO over a mechanical oscillations, corresponding to the fact that less energy is recovered when the Qubit's frequency is decreased that the energy invested when the Qubit's frequency is increased. This results in a damping of the MO (see Fig.\ref{f4:Conversion}d). This ``thermodynamic'' damping is initially linear in time, but the dissipated power actually dwindles while the mechanical amplitude and therefore the speed of erasure decrease.

\section{Optical noise on the MO}

Up to now we have used \eqref{eq4:SCEq}-\eqref{eq4:SCEm} which only include the first-order effect of the coupling. In this section, we go one step further and take into account the fluctuations of the Qubit's population in the MO dynamics. We also allows the Qubit to be driven by a laser. This study enables to precise the regime of validity of the semi-classical dynamics used in previous section in the ultra-strong coupling regime $g_\text{m}\gtrsim\Omega_\text{m}$. Note that similar studies have been performed in the case of weak coupling $g_\text{m}\ll \Omega_\text{m},\gamma_\text{m}$ but large mechanical amplitude $g_\text{m}\sqrt{N_\text{m}} \gg \gamma_\text{q}$ \cite{Rabl10}.

After presenting briefly the dynamics of the Qubit in presence of the laser drive, we derive the master equation ruling the MO dynamics. We study the relaxation channel induced by the Qubit on the MO and show that it results in a heating of the MO, causing the scattering of the MO position in phase space. Finally, we show that this scattering back-acts on the Qubit dynamics at long times scales. 
 
\subsection{Evolution of the Qubit}
\label{s4:EvolQubit}

In order to get more intuition about the behavior of the system in the presence of the laser, we use the semi-classical master equation \eqref{eq4:MEHybrid} to compute the evolution equation for the Qubit population $P_\text{e}(t)$ and the amplitude of coherences in the $\{\ket{e},\ket{g}\}$ basis (in the rotating frame) $\tilde s(t) = \text{Tr}\{\rho_\text{q}(t)\sigma_-e^{i\omega_L t}\}$.  
\bb
\dot P_e(t) &=& - \gamma_\text{q}(2 {n}[x(t)]+1)P_e(t) + \gamma_\text{q} {n}[x(t)] + \dfrac{ig}{2}({\tilde s}(t)-{\tilde s}^*(t))\label{eq4:SCPe}\\
\dot{\tilde s}(t) &=& - i\delta_\text{T}(t){\tilde s}(t) - \dfrac{\gamma_\text{q}}{2}(2 {n}[x(t)]+1))s(t) + ig(P_e(t)-1/2)\label{eq4:SCs},
\ee

\noindent  Where the total effective detuning between the Qubit and the laser is 
\bb
\delta_\text{T}(t) = \omega_0- \omega_L + \delta_\text{m}(t).
\ee

Eq. \eqref{eq4:SCPe}-\eqref{eq4:SCs} can be compared to the optical Bloch equations \eqref{eq3:OBEPe}-\eqref{eq3:OBEs} ruling the dynamics of the driven-dissipative Qubit in the absence of the MO. The only difference is that the detuning $\delta_\text{T}(t)$ is now time-dependent, and determined by the MO average position $x(t)$. 

\begin{figure}
\begin{center}
\includegraphics[width=\textwidth]{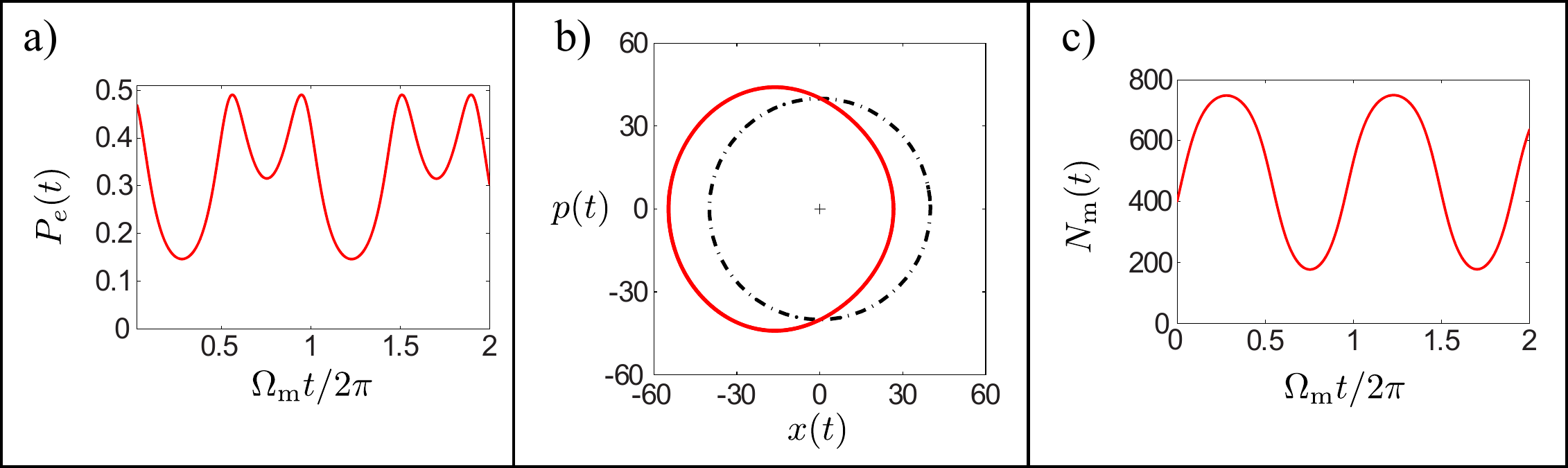}
\end{center}
\caption[Semi-classical dynamics]{Semi-classical dynamics. a) Average population $P_e(t)$ of the Qubit during two mechanical oscillations. b) Red: Average position in phase space of the MO $\beta(t) = (x(t)+ip(t))/2$. Black dot-dashed: evolution of the MO in the absence of coupling to the Qubit ($g_\text{m}=0$), starting in the same initial state. c) Relative variation of the average number of phonons in the MO $(N_\text{m}(t)-N_\text{m}(0))/N_\text{m}(0)$. {\it Parameters}: ${T} = 0$, $g/\gamma_\text{q} = 5$, $\Omega_\text{m}/\gamma_\text{q} = \numprint{5e-2}$, $g_\text{m}/\gamma_\text{q} = 0.1$, $\beta(0) = 20i$.\label{f4:SemiCl}}
\end{figure}

At first order in $g_\text{m}$, the MO always remains in a coherent state such that this position fulfils $x(t) = 2\text{Re}\beta(t)$ and the value of $\beta(t)$ is ruled by Eq.\eqref{eq4:dbeta}. Eq.\eqref{eq4:SCPe},\eqref{eq4:SCs} and \eqref{eq4:dbeta} can be easily be solved numerically. For instance, in Fig.\ref{f4:SemiCl} we have chosen for the Qubit a strong drive $g \gg \gamma_\text{q}$ and a zero temperature ${T} = 0$. We have set the laser-Qubit detuning to zero when the MO is at rest position, i.e. $\omega_0 = \omega_\text{L}$. The frequency $\omega_\text{q}(t)$ is slowly modulated at a frequency $\Omega_\text{m}$, such that the Qubit is brought periodically in and off resonance with the laser. Because of the adiabatic condition $\gamma_\text{q} \gg \Omega_\text{m}$, the Qubit population is modulated at the same frequency (see Fig.\ref{f4:SemiCl}a). As a consequence, the force on the MO is also modulated, leading to an asymmetrical motion in phase space (see Fig.\ref{f4:SemiCl}b), which can also be seen in the asymmetric extrema of the phonon number in Fig.\ref{f4:SemiCl}c), or the two local minima of $P_e(t)$ in Fig.\ref{f4:SemiCl}a which has different depth. Note that in order to emphasize the effect, the value of the ratio $g_\text{m}/\Omega_\text{m}$ in the simulation has been chosen larger than in the state-of-the-art implementations by a factor of about $10$.

\subsection{Master equation for the MO}

We now take into account the effect of coupling on the MO up to second order in $g_\text{m}$. Our starting point is the master equation Eq.\eqref{eq4:MEHybrid}. The derivation follows the same outline as the microscopic derivation of a master equation (see the fourth chapter of \cite{CCT}, and Section \ref{s3:MicroscopicDerivation} of the present dissertation), with the Qubit playing the role of the reservoir.

\subsubsection{Interaction picture with respect to the semi-classical evolution}

In order to recover a situation analogous to textbook open quantum system studies, we first remove the first order dynamics from master equation Eq.\eqref{eq4:MEHybrid}. To do that we introduce the interaction picture with respect to the semi-classical evolution. We first go to the frame rotating at the laser frequency. Then, we introduce the superoperator ${\cal K}(t_2,t_1)$ encoding the semi-classical evolution of the hybrid system (in the rotating frame), namely:

\bb
{\cal K}(t_2,t_1)\rho = {\cal T}\exp\int_{t_1}^{t_2} dt\left( {\cal L}_\text{q}^{x(t)}[\rho] -i [{\tilde H}_\text{q}(t) + H_\text{m}(t) ,\rho]\right),
\ee

\noindent where ${\tilde H}_\text{q}(t) = \delta_\text{T}(t)\Pi_e + (g/2) \sigma_x$ is the effective semi-classical Hamiltonian of the Qubit in the rotating frame and ${\cal T}\exp$ is a time-ordered exponential. We also introduce its inverse ${\cal K}^{-1}(t_2,t_1) = {\cal K}(t_1,t_2)$. ${\cal K}$ is defined such that if the hybrid system starts in a factorized state of the form $\rho_i = \rho_{\text{q},i}\otimes\ket{\beta_i}\bra{\beta_i}$ at $t_i = 0$, the evolved state ${\cal K}(t,0)\rho_i$ is still factorized and the average values of Qubit and MO observables are those given by Eq.\eqref{eq4:dbeta},\eqref{eq4:SCPe}  and \eqref{eq4:SCs}. This property formally means that we can write $K$ as a tensor product of superoperators acting solely on the Qubit and MO subspaces:
\bb
K(t_2,t_1) = K_\text{q}(t_2,t_1)\otimes K_\text{m}(t_2,t_1).\label{d4:Kfact}
\ee
\noindent with
\bb
K_\text{q}(t_2,t_1) &=& {\cal T}\exp\int_{t_1}^{t_2} dt\left( {\cal L}_\text{q}^{x(t)}[\rho] -i [{\tilde H}_\text{q}(t),\rho]\right),\label{d4:Kq}\\
K_\text{m}(t_2,t_1) &=& {\cal T}\exp\int_{t_1}^{t_2} dt\left( {\cal L}_\text{q}^{x(t)}[\rho] -i [H_\text{m}(t) ,\rho]\right),\label{d4:Km}
\ee

In analogy with usual (unitary) interaction picture, the density operator of the hybrid system in the interaction picture with respect to the semi-classical evolution is defined as:
\bb
{\cal K}(t,0){\rho}^I_\text{qm}(t) = \rho_\text{qm}(t).
\ee

Injecting this relation into Eq.\eqref{eq4:MEHybrid} allows to find the master equation ruling the dynamics of $\rho^I_\text{qm}(t)$:

\bb
\dot{\rho}^I_\text{qm}(t) = -i K^{-1}(t,0)[\delta V(t),K(t,0)\rho^I_\text{qm}(t)]\label{eq4:MEI}
\ee 
\noindent where:
\bb
\delta V(t) = \hbar g_m \delta \Pi_e(t)\delta \hat X(t) = \hbar g_m (\Pi_e-P_e(t))(\hat X-x(t))
\ee
is the optomechanical interaction from which have been substracted the average effect on the MO and on the Qubit. Note that $\delta V(t)$ depends explicitly on time through $P_e(t)$ and $x(t)$. For simplicity, we did not write $\delta V(t)$ itself in this generalized interaction picture, hence the presence of the operators $K(t,0)$ in Eq.\eqref{eq4:MEI}. Finally note that $\delta V(t)$ fulfils the usual property of the reservoir-system coupling to have a zero average: $\text{Tr}_\text{q}\{\rho_\text{q}(t)\delta V(t)\} = \text{Tr}_\text{m}\{\rho_\text{m}(t)\delta V(t)\} = 0$.

It is interesting to analyse the time-scales involved in Eq.\eqref{eq4:MEI}. The coupling operator $\delta V$ is of magnitude $g_m$ and depends on time through $P_e(t)$ and $x(t)$. The value of $P_e(t)$ evolves typically on two time-scales: a slow time-scale $\Omega_\text{m}^{-1}$, $g_\text{m}^{-1}$ corresponding to the variation of the MO position, and therefore of $\omega_\text{q}(t)$, and a fast time-scale given by the Qubit rates $g,\gamma$. As $\gamma \gg \Omega_\text{m}$, the Qubit follows adiabatically the position of the MO: $P_e(t)$ typically reaches very quickly the steady state corresponding to a given value of $x(t)$ before the MO position changes significantly. 
In summary:
\begin{itemize}
\item $\rho_\text{m}^I(t)$ has a typical evolution time-scale of $1/g_\text{m}$, $1/\Omega_\text{m}$.
\item $\rho_\text{q}^I(t)$ has evolves on both fast an slow time-scales. 
\end{itemize} .

\subsubsection{Lindbladian induced by the Qubit}

We now coarse-grain the evolution encoded in Eq.\eqref{eq4:MEI}. The new time-step $\Delta t_\text{m}$ is chosen such that:
\bb
\Omega_\text{m}^{-1},g_\text{m}^{-1},1/(g_\text{m}^2\tau_\text{q}) \gg \Delta t_\text{m} \gg \gamma_\text{q}^{-1} \label{d4:Dtm}
\ee
\noindent which is possible in the regime defined by Eq.\eqref{d4:regime}. Integrated between $t$ and $t+\Delta t_\text{m}$, the master equation \eqref{eq4:MEI} yields:
\bb
\Delta \rho^I_\text{qm}(t) &=& -i \int_t^{t+\Delta t_\text{m}} dt' K^{-1}(t',0)[\delta V(t'), K(t',0)\rho^I_\text{qm}(t)] \nonumber\\
&& - \int_t^{t+\Delta t_\text{m}} dt'  \int_t^{t'} dt''K^{-1}(t',0)[\delta V(t'), K(t',t'')[\delta V(t''), K(t'',0)\rho^I_\text{q}(t'')\otimes\rho^I_\text{m}(t)]]\label{eq4:precursor}\nonumber\\
\ee 
\noindent In the line of the microscopic derivation of the Lindblad master equation, we have replaced  $\rho^I_\text{qm}(t'')$ with a factorized state $\rho_\text{q}^I(t'')\otimes\rho_\text{m}^I(t)$ in the integrand of the two times integral. The MO state at time $t''$ is approximated by $\rho_\text{m}^I(t)$ (Markov approximation), which is justified by the fact that $\rho_\text{m}^I(t)$ does no vary significantly between $t$ and $t+\Delta t_\text{m}$ because of condition \eqref{d4:Dtm}. Contrary to the usual case however, we do not assume that the ``reservoir's state'' is time-independent: the Qubit's density operator evolves between $t$ and $t+\Delta t_\text{m}$. We now take the trace over the Qubit degrees of freedom. As the first order has already been removed, the first line right-hand term vanishes and we find:
\bb
\dot \rho_\text{m}^I(t) &\equiv& \dfrac{\text{Tr}_\text{q}\Delta \rho^I_\text{qm}(t)}{\Delta t_\text{m}} \nonumber\\
&=&  - \dfrac{1}{\Delta t_\text{m}}\int_t^{t+\Delta t_\text{m}} dt'  \int_t^{t'} dt''\text{Tr}_\text{q}\{[\delta V(t'), K(t',t'')[\delta V(t''), K(t'',0)\rho^I_\text{q}(t'')\otimes\rho^I_\text{m}(t)]]\}\label{eq4:precursor2} \nonumber\\
\ee 
\noindent In order to compute the integral, we expand the double commutator and use the form Eq.\eqref{d4:Kfact} of the operator ${\cal K}$. This yields terms of the form:
\bb
 - \dfrac{g_\text{m}^2}{\Delta t_\text{m}}\int_t^{t+\Delta t_\text{m}} dt'   \int_t^{t'} dt''&&\text{Tr}_\text{q}\big\{\delta \Pi_e(t')K_\text{q}(t',t'')\delta \Pi_e(t'')K_\text{q}(t'',0)\rho^I_\text{q}(t'')\big\} \nonumber\\
 &&\quad\quad\quad\quad\times \delta \hat X(t') K_\text{m}(t',t'')[\delta \hat X(t''), K_\text{m}(t'',0))\rho^I_\text{m}(t)]]\nonumber\\\label{eq4:term1}
\ee
\noindent In the integrand we identify the two-time correlation function of the Qubit operator $\delta \Pi_e(t)$:
\bb
C_\text{q}(t',t'') = \text{Tr}_\text{q}\big\{\delta \Pi_e(t')K_\text{q}(t',t'')\delta \Pi_e(t'')K_\text{q}(t'',0)\rho^I_\text{q}(t'')\big\}.\label{d4:correlatorq}
\ee
\noindent $C_\text{q}(t',t'')$ differs from the usual reservoir correlation function because it is not invariant by time translation $(t',t'')\to (t'+t,t''+t)$. This is a direct consequence of the fact that the Qubit is of finite size \cite{Whitney08}. However, in perfect analogy with the derivation made in section \eqref{s3:MicroscopicDerivation}, $C_\text{q}(t',t'')$ vanishes for time delay $\vert t'-t''\vert$ larger than $\tau_\text{q}$. We can therefore change the integration variable in the double integral of \eqref{eq4:term1} to make the delay $\tau = t'-t''$ appear and extend the integral over $\tau$ from zero to infinity, we find:

 \bb
 - \dfrac{g_\text{m}^2}{\Delta t_\text{m}}\int_t^{t+\Delta t_\text{m}} dt'   \int_0^{\infty} d\tau C_\text{q}(t',t'-\tau) \delta \hat X(t') K_\text{m}(t',t'-\tau)[\delta \hat X(t'-\tau), K_\text{m}(t'-\tau,0))\rho^I_\text{m}(t)]]\nonumber\\
\ee

\noindent The semi-classical evolution generated by $K_\text{m}$, the time-dependence in $t'$ of $C_\text{q}(t',t'-\tau)$ and of $\delta X(t')$ all occur on time-scales $\Omega_\text{m}^{-1}$, $g_\text{m}^{-1}$ and can therefore be neglected between $t$ and $t+\Delta t_\text{m}$. Term \eqref{eq4:term1} can  therefore be simplified into:
\bb
 - g_\text{m}^2 \int_0^{\infty} d\tau C_\text{q}(t,t-\tau) \delta \hat X(t) [\delta \hat X(t),\rho_\text{m}(t)]]\nonumber\\
\ee
Taking into account the other similar term coming from the expansion of the commutators in Eq.\eqref{eq4:precursor2}, we finally find the master equation (back in Schr\"odinger's picture):
\bb
\dot \rho_\text{m}(t) &=& -i[H_\text{m}(t),\rho_\text{m}(t)] -\dfrac{\Gamma_\text{opt}(t)}{2}[\hat X,[\hat X,\rho_\text{m}(t)]],\label{eq4:Mm}
\ee 

\noindent where
\bb
\Gamma_\text{opt}(t) = 2g_\text{m}^2\text{Re}\int_0^{\infty}d\tau C_\text{q}(t,t-\tau),
\ee

\noindent is the rate of the relaxation channel induced by the Qubit on the MO, whose typical magnitude is $g_\text{m}^2/\gamma_\text{q}$. Note that in deriving Eq.\eqref{eq4:Mm}, we have neglected a Hamiltonian term induced by the coupling which has the form:
\bb
-2i\text{Im}\int_0^{\infty}d\tau C_\text{q}(t,t-\tau) [\hat X^2,\rho_\text{m}(t)].
\ee
\noindent This term generates an anharmonicity of magnitude $g_\text{m}^2/\gamma_\text{q}$ much smaller than $\Omega_\text{m}$. \\

The form of the Lindbladian induced by the Qubit corresponds to the effect of a continuous weak measurement of the position $\hat X$ at a rate $\Gamma_\text{opt}(t)$ (see Section \ref{s2:ContMeas} and Ref.\cite{Steck06}). Indeed, some information about the mechanical position is extracted by the dissipative Qubit. This information could be recovered from the beam of photons exchanged between the Qubit and the thermal reservoir. For instance, the average intensity $I(t)$ of the beam is directly related to the MO population, which follows adiabatically $x(t)$ (see Fig.\ref{f4:SemiCl}). Similarly, the spectrum $S_\text{em}(t,\omega)$ of the emitted light, computed from the photons emitted between times $t$ and $t+\Delta t_\text{m}$ allows to known the Qubit effective frequency which is also related to $x(t)$. This information is not sufficient to completely determine the value of the mechanical position: Hence such a weak measurement does not project the oscillator onto an eigenstate of the position operator, but rather slightly affects the momentum variance (see Section \ref{s4:PropOpt}).\\

Because of the slow dependence of $P_e(t)$ on the average MO position $x(t)$, the measurement rate $\Gamma_\text{opt}(t)$ varies in time within a mechanical period. Indeed, depending on the position, the Qubit may be in or out of resonance with the laser. When the Qubit is out of resonance, its population vanishes and the coupling has no effect on the MO, such that $\Gamma_\text{opt}(t)$ is zero. Some properties of the Lindbladian induced by the Qubit are explored in next section.

\subsubsection{Expression of $\Gamma_\text{opt}$}

The heating rate induced by the Qubit is closely related to the noise spectrum $S_t(\omega)$ of the Qubit's population, defined as:
\bb
S_t(\omega) = \int_0^\infty d\tau C_\text{q}(t,t-\tau) e^{i\omega \tau}.
\ee

\noindent In the regime we are interested in, it is reasonable to identify $S_t(\omega)$ with the steady state noise spectrum corresponding to the value $\delta_\text{m}(t)$ of the detuning. In other words, we make the assumption that the dependence in $t$ of $C_\text{q}(t,t-\tau)$ only comes from the dependence on $\delta_\text{m}(t)$. The steady state noise spectrum can be computed using the Quantum Regression Theorem (QRT) \cite{Carmichael1}:
\begin{align}
S_t(\omega) &= -\dfrac{g_\text{m}^2}{4}(1,0,0).(i\omega\idop + A(t))^{-1}\vec R(t),
\end{align}

\noindent where we have introduced the matrix $A(t)$ encoding the optical Bloch equations:

\bb
A(t) = \left(\begin{array}{ccc}
-\gamma_\text{q}(2{n}[\omega_\text{q}(t)]+1) & -ig & ig \\ 
-\dfrac{ig}{2} & i\delta(t)-\dfrac{\gamma_\text{q}(2{n}[\omega_\text{q}(t)]+1)}{2} & 0 \\ 
\dfrac{ig}{2} & 0 & -i\delta_\text{T}(t)-\dfrac{\gamma_\text{q}(2{n}[\omega_\text{q}(t)]+1)}{2}
\end{array}\right),\nonumber\\
\ee

\noindent which is defined such that the optical Bloch equations can be written $\delta\dot{\vec{\sigma}} = A(t).\delta\vec{\sigma}$, where $\delta \vec\sigma = (\delta\sigma_z(t),\delta\sigma_-(t),\delta\sigma_+(t))$ and $\delta\sigma_i(t) = \sigma_i - \text{Tr}\{\rho_\text{q}(t)\sigma_i\}$, for $i\in\{+,-,z\}$. We have also introduced:
\bb
 \vec R(t) = \text{Tr}\{\delta\vec\sigma(t)\delta\sigma_z(t)\rho_\text{q}^{(\infty)}[\delta_\text{T}(t)]\},
 \ee

\noindent where $\rho_\text{q}^{(\infty)}[\delta_\text{T}(t)]$ is the steady state of Bloch equations corresponding to a detuning $\delta_\text{T}(t)$ (in the rotating frame), i.e. 

\begin{equation}
\rho_\text{q}^{\infty}[\delta_\text{T}(t)]= \left(\begin{array}{cc}
P_e^{(\infty)}[\delta_\text{T}(t)]& (s^{(\infty)}[\delta_\text{T}(t)])^* \\ 
s^{(\infty)}[\delta_\text{T}(t)] & 1-P_e^{(\infty)}[\delta_\text{T}(t)]
\end{array} \right)
\end{equation}
where
\bb
P_e^{(\infty)}[\delta_\text{T}(t)] &=& \dfrac{1}{2{n}[\omega_\text{q}(t)]+1}\left({n}[\omega_\text{q}(t)] + \dfrac{1/2}{1+2\delta(t)^2/g^2+\gamma_T^2/2g^2}\right)\nonumber\\
s^{(\infty)}[\delta_\text{T}(t)] &=& -\dfrac{1}{2{n}[\omega_\text{q}(t)]+1}\dfrac{\delta(t)/g + i \gamma_T/g}{1+ 2\delta^2(t)/g^2+ \gamma_T^2/2g^2}\label{steadystate}
\ee

\noindent We have denoted $\gamma_T = \gamma(2{n}[\omega_\text{q}(t)]+1)$ the total damping rate of the Qubit. Finally, the rate $\Gamma_\text{opt}(t)$ is twice the real part of $S_t(0)$, and can be computed analytically as a function of $\delta_\text{T}(t)$ by inverting the matrix $A(t)$.

It is interesting to consider two usual limits in which $\Gamma_\text{opt}$ has a relatively simple expression. First, the case of zero temperature for the Qubit at ${T} = 0$. We have:
\begin{equation}
\Gamma_\text{opt}(t) = 2\dfrac{g_\text{m}^2 g^2(4\delta(t)^2 + \gamma_\text{q}^2)(g^2 + 2\gamma_\text{q}^2)}{\gamma_\text{q}(4\delta(t)^2 + 2g^2 + \gamma_\text{q}^2)^3}.
\end{equation}

\noindent which takes a maximum value $\Gamma_\text{opt}^\text{max}\simeq 0.09*g_\text{m}^2/\gamma_\text{q}$ for $g \simeq \delta_\text{T} \simeq \gamma$.

Second, the case of no laser drive ($g=0$) and non-zero TLS temperature:
\begin{equation}
\Gamma_\text{opt}(t) = 2\dfrac{g_\text{m}^2}{\gamma}\dfrac{{n}[\omega_\text{q}(t)](1+{n}[\omega_\text{q}(t)])}{(2 {n}[\omega_\text{q}(t)]+1)^2},
\end{equation}
\noindent which is maximized $\Gamma_\text{opt}^\text{max}\simeq 0.5*g_\text{m}^2/\gamma_\text{q}$ for very high temperature $k_\text{B}T_\text{q}\gg \hbar\omega_\text{q}$ such that ${n}[\omega_\text{q}] \to \infty$. Note that $\Gamma_\text{opt}(t)$ vanishes if both ${T}$ and $g$ are zero as the Qubit never gets excited.

\begin{figure}
\begin{center}
\includegraphics[width=0.8\textwidth]{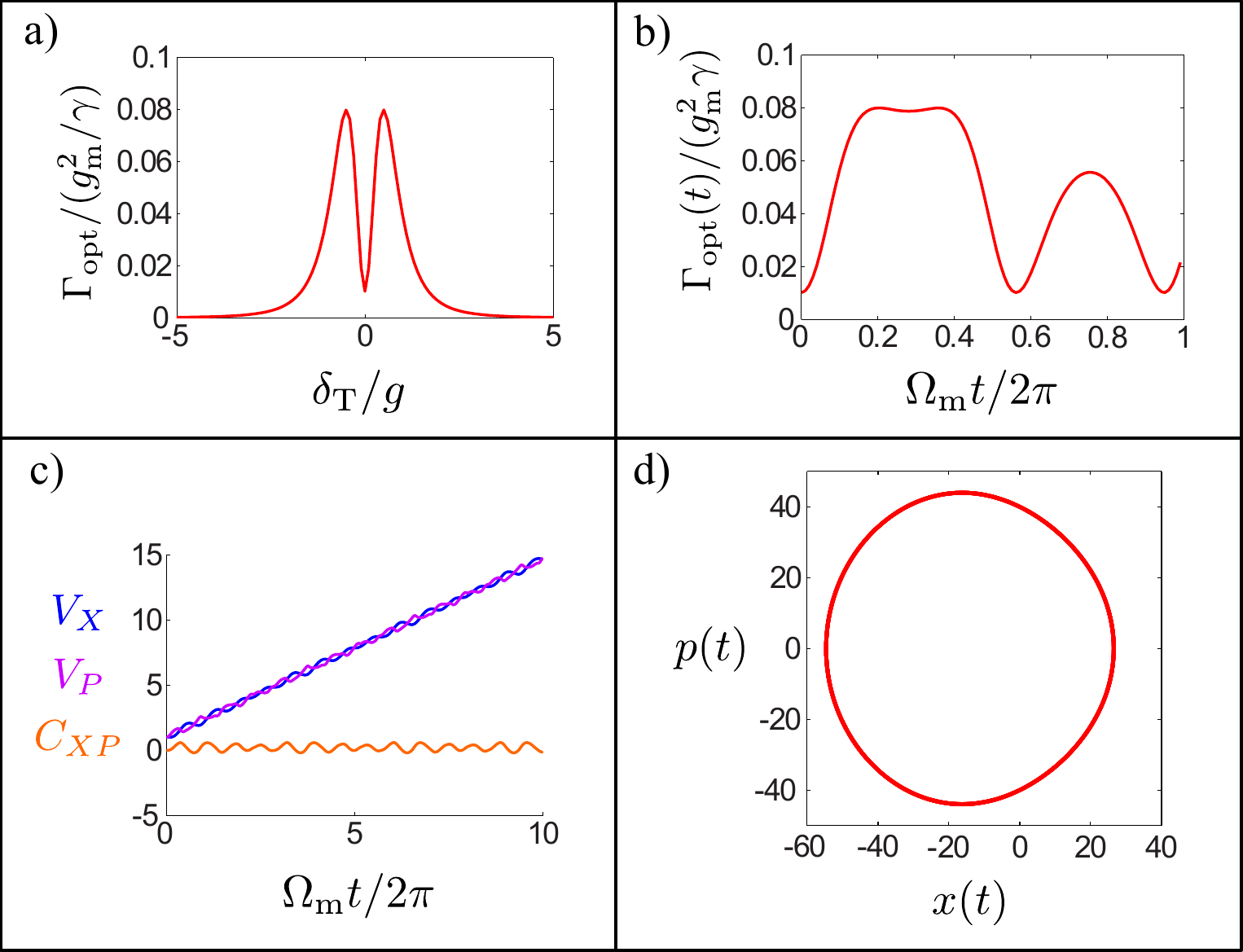}
\end{center}
\caption[Properties of the Qubit-induced Lindbladian]{Properties of the Qubit-induced Lindbladian. a) Real part of the qubit noise spectrum at zero frequency $\text{Re}S_t(0)$ as a function of the effective detuning $\delta_\text{m}$. b) Evolution of the heating rate $\Gamma_\text{opt}$ over one mechanical period. c) Evolution of $V_X$, $V_P$ and $C_{XP}$ during $10$ mechanical periods. d) Evolution of the position of the MO in phase space during $10$ mechanical periods. {\it Parameters}: ${T} = 0$, $g/\gamma_\text{q} = 5$, $\Omega_\text{m}/\gamma_\text{q} = \numprint{5e-2}$, $g_\text{m}/\gamma_\text{q} = 0.1$, $\beta(0) = 20i$.\label{f4:GamOptAv}}
\end{figure}

\subsubsection{Properties of the Qubit-induced Lindbladian}
\label{s4:PropOpt}

From Eq.\eqref{eq4:Mm}, we identify the Lindbladian induced by the Qubit on the MO:

\bb
{\cal L}_\text{opt}  = -\dfrac{\Gamma_\text{opt}(t)}{2}[X,[X,\rho_\text{m}(t)]]\label{d4:Lopt}
\ee

\noindent ${\cal L}_\text{opt}$ has no effect on the average position or momentum of the MO, but it increases the variance $V_P(t) = \text{Tr}\{\rho_\text{m}(t)\hat P^2\}-p(t)^2$ of the MO's momentum $\hat P = i(b^\dagger - b)$ at a rate $4\Gamma_\text{opt}$. The variance in position is not directly affected, however, it is coupled to $V_P$ due to free evolution, such that it increases as well with some delay. Because it affects differently $V_X$ and $V_P$, the master equation \eqref{eq4:Mm} does not transform a coherent state into a coherent state.\\

 However, this evolution preserves the Gaussian nature of a density operator. A Gaussian state is a state whose Wigner function ${\cal W}_\text{m}(x,p)$ is a Gaussian function of $x$ and $p$. This property is satisfied by coherent states but also squeezed coherent states. Gaussian states are simple to study as they are fully characterized by only five parameters $\{x(t),p(t),V_X(t),V_P(t),C_{XP}(t)\}$. We have defined $C_{XP}(t) = \text{Tr}\{\rho_\text{m}(t)(\hat X \hat P+ \hat P \hat X)\} - 2 x(t)p(t)$ related to the orientation in phase space of the Wigner function. The Wigner function of the MO at time $t$ is related to the five parameters according to:
\bb
{\cal W}_\text{m}(u,v) = \dfrac{1}{\pi}\sqrt{ac-b^2}\exp\left(-a (u-x(t))^2+ 2b (u-x(t))(v-p(t))-c (v-p(t))^2\right)\quad\label{eq4:WignerGaussian}
\ee
\noindent with
\bb
a &=& \dfrac{2V_P(t)}{4V_X(t)V_P(t) - C_{XP}(t)^2}\nonumber\\
b &=& \dfrac{C_{XP}(t)}{4V_X(t)V_P(t) - C_{XP}(t)^2}\nonumber\\
c &=& \dfrac{2V_X(t)}{4V_X(t)V_P(t) - C_{XP}(t)^2}\nonumber\\
\ee

The five parameters $\{x(t),p(t),V_X(t),V_P(t),C_{XP}(t)\}$ obey the following system of coupled differential equations: 
\bb
\dot x(t) &=& \Omega_\text{m}p(t) \label{eq4:dx}\\
\dot p(t) &=& -\Omega_\text{m}x(t)\\
\dot V_X(t) &=& \Omega_\text{m} C_{XP}(t)\\
\dot V_P(t) &=& -\Omega_\text{m} C_{XP}(t) + 4\Gamma_\text{opt}(t)\\
\dot C_{XP}(t) &=& \Omega_\text{m} (V_P(t)-V_X(t))\label{eq4:dcxp},
\ee

\noindent A numerical solution is plotted in Fig.\ref{f4:GamOptAv}, showing the increase of the variances. For long time-scales larger than $1/\Gamma_\text{opt} \geq \gamma/g_\text{m}^2$, the variance of the MO position becomes large with respect to $1$ and therefore condition \eqref{d4:VXcond} may be violated.

At this point, we can already conclude that the (first-order) semi-classical picture is a good approximation of the hybrid system dynamics as long as we look at the evolution on a time-scale short with respect to $\gamma/g_\text{m}^2$. Note that most state-of-the-art implementations are deep in the regime $g_\text{m} \ll \gamma$, such that this restriction is not too stringent in practice. We can define a ``Semi-classical quality factor'' $Q_\text{sc} = \Omega_\text{m}/\Gamma_\text{opt}\gtrsim \Omega_\text{m}/(g_m^2/\gamma)$ which gives an indication of the number of mechanical oscillations on which the semi-classical description is accurate. In the figure \ref{f4:GamOptAv}, this parameter is small (around $5$) and therefore the increase of the position and momentum variance can be seen solely after a few mechanical oscillations. For the different experimental setups, its value is given in Table \ref{t:OMparams} in the last line. $Q_\text{sc}$ is comparable to the mechanical quality factor for \cite{Yeo13} and  \cite{Lahaye09} which means that the semi-classical description is very good for those systems.  However, it is one hundred of magnitude smaller than the mechanical quality factor for Ref.\cite{Arcizet11}, which means that the noise induced by the Qubit may dominate against the Browian motion and therefore be observed in such systems. Setup \cite{Pirkkalainen13} is different from the others as it does not fulfils condition \eqref{d4:regime}: the Qubit's sponaneous emission time is smaller than the mechanical frequency, and therefore the adiabatic approximations are not valid and the semi-classical description is expected to fail.

\subsection{Evolution at longer time-scales: decoupling from the laser}

The description of the MO dynamics presented in previous section is limited to relatively short time-scales. Indeed, it predicts an increase of the position variance $V_X(t)$, bringing the MO's state out of the domain of validity of assumption \eqref{d4:VXcond}, which is essential to justify the Qubit semi-classical dynamics. In addition, this description does not include any mechanism limiting the growth of $V_X$ and $V_P$ (and therefore the growth of the mechanical energy) induced by the Qubit, which seems unphysical. In this section, we present a method based on quantum trajectories which fixes both of these limitations.

The Lindbladian ${\cal L}_\text{opt}$ decreases the purity of the MO's state $\rho_\text{m}$: for example, an initial pure coherent state is transformed into a mixed state during the evolution.  
Physically, this loss of purity originates from the noise induced on the MO by the Qubit. This noise reflects the stochastic variations of the force proportional to the population of the Qubit because of the emissions and absorptions of photons induced by the thermal reservoir. 

Alternatively, the evolution of the MO in presence of the Qubit-induced noise can be formulated as a stochastic sequence of pure-states owing to an unraveling of the Lindblad equation \eqref{eq4:Mm}. Such description exploits the record of a continuous measurement performed on the reservoir to describe the dynamics of the system under the form of a quantum trajectory (see Section \ref{s2:DefQtraj}). The master equation is recovered by averaging the system's state over all the possible trajectories, which boils down not reading the measurement record.

Here, a measurement scheme consistent with the master equation \eqref{eq4:Mm} could be the continuous monitoring of the beam of photons exchanged between $t$ and $t+\Delta t_\text{m}$. Because of the condition $\gamma_\text{q}\Delta t_\text{m} \gg 1$, this beam contains many photons that the coarse-grained description does not allow to order in time. However, collective properties of the beam averaged between $t$ and $t+\Delta t_\text{m}$ still carry information about the MO state: the intensity $I(t)$ of the beam averaged during $\Delta t_\text{m}$, or the spectrum $S_\text{em}(t,\omega)$ of the emitted light computed from the photons emitted between $t$ and $t+\Delta t_\text{m}$, contain information about $x(t)$. 

Because of their finite duration $\Delta t_\text{m}$, the variance of the outcome of such measurements of the mechanical position is very large: they therefore correspond to weak measurements of $\hat X$. 
When the record $r[\gamma]$ is read, such measurement scheme implements a quantum state diffusion unraveling of the master equation \eqref{eq4:Mm} (see Sections \ref{s2:defQSD} and \ref{s2:ContMeas}).

The evolution of the state of the MO conditioned to the continuous measurement record $r[\gamma] = \{r_\gamma(t_n)\}_{1\leq n \leq N}$ follows a quantum trajectory $\gamma$. We have discretized the time $t\to \{t_n\}_{1\leq n \leq N}$ and denoted $\ket{\phi_{\text{m},\gamma}(t_n)}$ the state of the MO at time $t_n$, with $\ket{\phi_{\text{m},\gamma}(0)}=\ket{\beta_i}$ the initial MO's state. The evolution between time $t_n$ and $t_{n+1} = t_n+\Delta t_\text{m}$ is encoded in a measurement operator $M_{r_\gamma(t_n)}$ conditioned to the outcome $r_\gamma(t_n)$:
\bb
\ket{\phi_{\text{m},\gamma}(t_{n+1})} = \dfrac{M_{r_\gamma(t_n)}\ket{\phi_{\text{m},\gamma}(t_n)}}{\moy{M_{r_\gamma(t_n)}^\dagger M_{r_\gamma(t_n)}}_{\phi_{\text{m},\gamma}(t_n)}}\label{d4:StochStep}
\ee
\noindent with
\bb
M_{r_\gamma(t_n)} = 1-idt H_\text{m}(t_n) - \dfrac{\Gamma_\text{opt}(t_n)\Delta t_\text{m}}{2}\hat X^2 - \sqrt{\Gamma_\text{opt}(t_n)}dw_\gamma(t_n)\hat X \label{d4:Mr}.
\ee
\noindent As in Chapter \ref{Chapter2}, we have introduced the Wiener increment $dw_\gamma(t)$, i.e. a random Gaussian variable of zero average and standard deviation $\Delta t_\text{m}$, which is related to the measurement result $r_\gamma(t_n)$ according to:
\bb
r_\gamma(t_n) = \moy{\hat X}_{\phi_{\text{m},\gamma}(t_n)} +  \dfrac{dw_\gamma(t_n)}{2\Delta t_\text{m}\sqrt{\Gamma_\text{opt}(t_n)}} \label{d4:rt}.
\ee
As for every unraveling, the master equation \eqref{eq4:Mm} can be recovered from Eq.\eqref{d4:StochStep} via an average over all the possible measurement results.\\

\begin{figure}
\begin{center}
\includegraphics[width=0.9\textwidth]{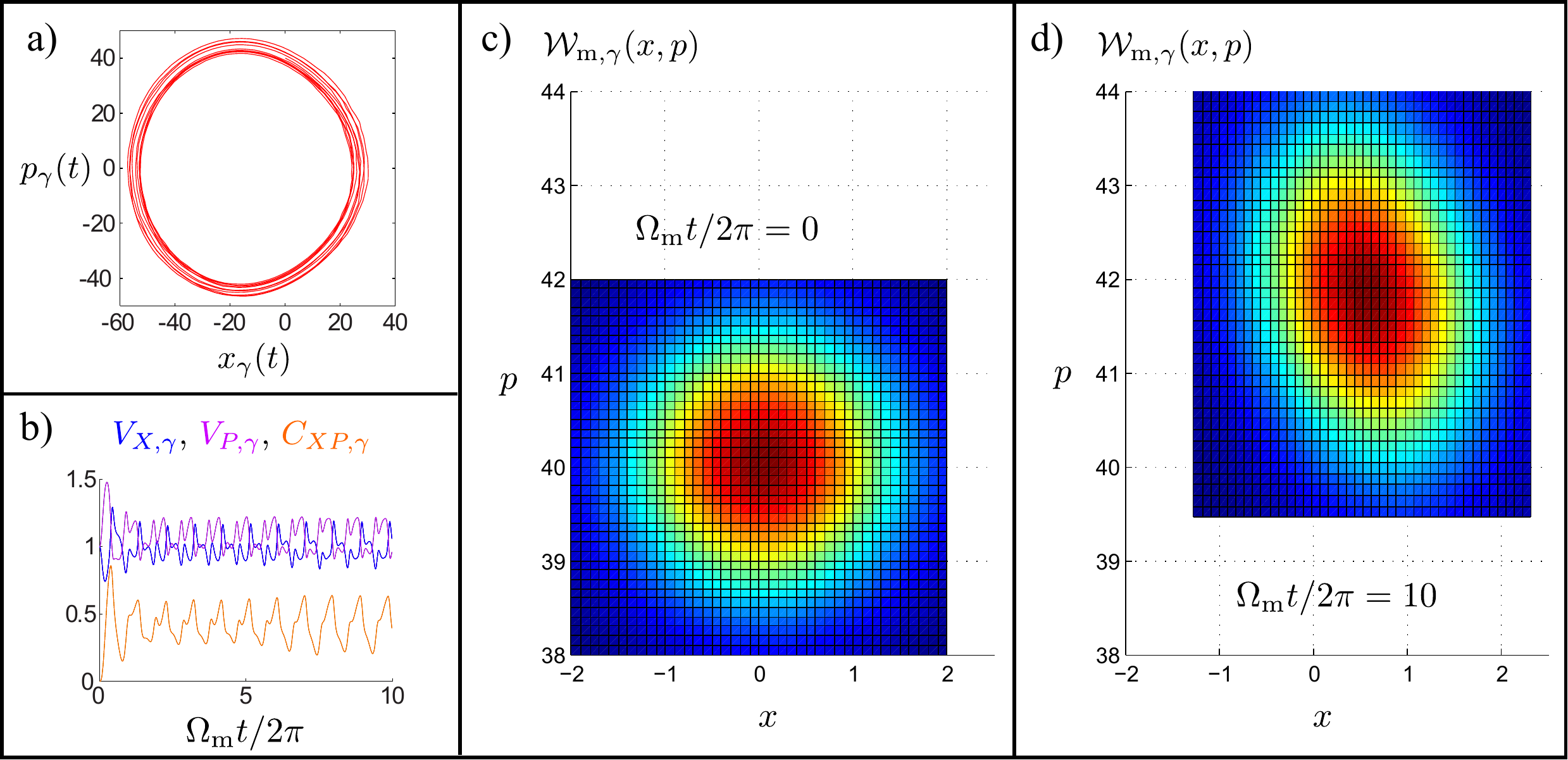}
\end{center}
\caption[Stochastic evolution of the MO]{Stochastic evolution of the MO. a)b): Evolution of $V_{X,\gamma}$, $V_{P,\gamma}$, $C_{XP,\gamma}$ (a), and if the position in phase space (b) along a single trajectory during $10$ mechanical periods. c),d): Wigner function ${\cal W}_{\text{m},\gamma}(x,p)$ of the MO at the initial time (c) and after $10$ mechanical periods (d). \label{f4:GamOptTraj}}
\end{figure}

Just like the solution of master equation \eqref{eq4:Mm}, the state of the MO along a given trajectory $\gamma$ is a Gaussian state characterized by five parameters. In order to distinguish them from the corresponding parameters computed from state $\rho_\text{m}(t)$ (containing also classical uncertainty), we introduce new notations indexed with $\gamma$:
\bb
x_\gamma(t) &=& \bra{\phi_{\text{m},\gamma}(t)}\hat X \ket{\phi_{\text{m},\gamma}(t)}\\
p_\gamma(t) &=& \bra{\phi_{\text{m},\gamma}(t)}\hat P \ket{\phi_{\text{m},\gamma}(t)}\\
V_{X,\gamma}(t) &=& \bra{\phi_{\text{m},\gamma}(t)}X^2(t)\ket{\phi_{\text{m},\gamma}(t)}-x_\gamma(t)^2\\
V_{P,\gamma}(t) &=& \bra{\phi_{\text{m},\gamma}(t)}P^2(t)\ket{\phi_{\text{m},\gamma}(t)}-p_\gamma(t)^2\\
C_{XP,\gamma}(t) &=& \bra{\phi_{\text{m},\gamma}(t)}(X(t)P(t)+P(t)X(t))\ket{\phi_{\text{m},\gamma}(t)}-2x_\gamma(t)p_\gamma(t),
\ee
\noindent and denote ${\cal W}_{\text{m},\gamma}(x,p)$ the Gaussian Wigner function corresponding to state $\ket{\phi_{\text{m},\gamma}(t)}$. The evolution of these five parameters can be computed from Eqs.\eqref{d4:StochStep}-\eqref{d4:rt}:
\bb
\dot x_\gamma(t) &=& \Omega_\text{m}p_\gamma(t) + 2\sqrt{\Gamma_\text{opt}(t)}dw_\gamma(t)V_{X,\gamma}(t)\label{eq4:xgamma}\\
\dot p_\gamma(t) &=& -\Omega_\text{m}x_\gamma(t) + \sqrt{\Gamma_\text{opt}(t)}dw_\gamma(t) C_{XP,\gamma}(t)\\
\dot V_{X,\gamma}(t) &=& \Omega_\text{m}C_{XP}(t) - 4\Gamma_\text{opt}(t) V_{X,\gamma}(t)^2\label{eq4:vxgamma}\\
\dot V_{P,\gamma}(t) &=& -\Omega_\text{m}C_{XP,\gamma}(t) + 4\Gamma_\text{opt}(t) - \Gamma_\text{opt}(t)C_{XP,\gamma}(t)^2\label{eq4:vpgamma}\\
\dot C_{XP,\gamma}(t) &=& \Omega_\text{m}(V_{P,\gamma}(t)-V_{X,\gamma}(t)) - 2 \Gamma_\text{opt}(t) V_{X,\gamma}(t) C_{XP,\gamma}(t)\label{eq4:cxpgamma}
\ee
Eqs.\eqref{eq4:xgamma}-\eqref{eq4:cxpgamma} contain new terms with respect to their counterparts Eqs.\eqref{eq4:dx}-\eqref{eq4:dcxp} derived from the master  equation. In the evolution of the average position and momentum, there is now a term proportional to $dw_\gamma(t)$ describing a stochastic scattering of the MO distribution in phase space induced by the Qubit. In the evolution of the variances $V_{X,\gamma}$ and $V_{P,\gamma}$, there are now negative terms limiting the increase, which represent the update of the knowledge on the MO's state due to information acquired during the measurement.

 Numerical integration of Eq.\eqref{eq4:xgamma}-\eqref{eq4:cxpgamma} allows to observe the scattering of the MO position in phase space (see Fig.\ref{f4:GamOptTraj}a). In Fig.\ref{f4:GamOptTraj}b we can see that $V_{X,\gamma}$ and $V_{P,\gamma}$ do not diverge and keep oscillating around $1$. This aspect is very important because it means that along each single trajectory $\gamma$ the condition \eqref{d4:VXcond} and the factorized description of the hybrid system remain valid, even on time-scales much larger than $\gamma_\text{q}/g_\text{m}^2$. Eventually Fig.\ref{f4:GamOptTraj}b shows that the parameter $C_{XP,\gamma}$ oscillates around a mean value of about $0.5$: the MO distribution gets slightly rotated in phase space as this can be seen in Fig.\ref{f4:GamOptTraj}c),d).\\

Interestingly, this description allows to see that the slow scattering of the MO position back-acts on the Qubit dynamics. Along a trajectory $\gamma$, the dynamics of the Qubit is captured by the effective Hamiltonian in which $x_\gamma(t)$ is used for the actual average MO position. Similarly, $\Gamma_\text{opt}(t)$ must be computed every $\Delta t_\text{m}$ using $\delta_T(t) = \omega_\text{L}-\omega_0 + g_\text{m}x_{\gamma}(t)$. These prescriptions together with stochastic equation \eqref{d4:StochStep} allow to compute the joint hybrid dynamics on arbitrary long time-scales. 

\begin{figure}
\vspace{0.1cm}
\begin{center}
\includegraphics[width=1.1\textwidth]{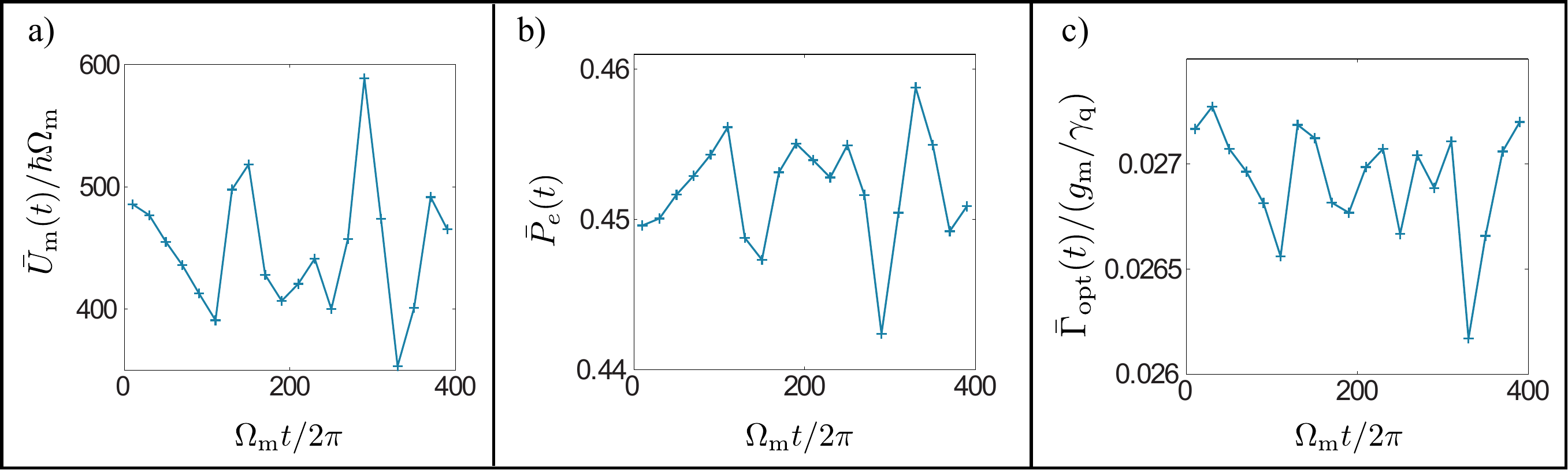}
\end{center}
\caption[Scattering at long time-scale]{Scattering at long time-scale. Internal energy $\bar{U}_\text{m}(t)$ of the MO (a), Qubit population $\bar{P}_e(t)$ (b) and heating rate $\bar{\Gamma}_\text{opt}(t)$ (c) averaged over $20$ mechanical periods. \label{f4:DriftMO}}
\end{figure}

After following a given trajectory during a time large with respect to $1/\Gamma_\text{opt}(t)$ the MO energy might have increased significantly. If the mechanical amplitude is so large such that it fulfills $\vert\beta\vert \gg g/g_\text{m}$, the average position of the MO spends most of each oscillation far from its rest position $x_\gamma(t) = 0$. Consequently, the Qubit is almost always off resonance with the laser and is almost never populated, such that both first and second order effect of the coupling on the MO vanish. This decoupling at large amplitude gives a natural bound to the noise induced by the Qubit. The slow drift of the MO energy, the Qubit population and the heating rate are plotted in Fig.\ref{f4:DriftMO}. For the sake of clarity, these quantities which oscillate at frequency $\Omega_\text{m}$ have been averaged over $20$ mechanical oscillations.


\section{Summary}

We studied a hybrid optomechanical system composed of a driven Qubit parametrically coupled to a Mechanical Oscillator (MO). Focusing on the dispersive, adiabatic, ultra-strong coupling regime which is relevant for several implementations, we proposed two different directions of study. 

We first exploited the average dynamics corresponding to the first order in $g_\text{m}$, and showed that provided it is initialized in a coherent state, the MO behaves as a battery generating a time-dependent term in the Qubit's Hamiltonian. This term induces a shift of the Qubit transition frequency proportional to the MO position. The whole evolution of the hybrid system can be analyzed as a thermodynamic transformation of the Qubit, driven by the MO. Because the battery is of finite size, the average work corresponds to an equal change in the mechanical energy. This energy change can be measured by looking at the variation of the MO coherent amplitude. As an application, we demonstrated that reversible information-to-energy conversions can be observed in state-of-the-art hybrid systems. 

In a second time, we showed that these device realize an unusual situation from the quantum open system point of view: the driven Qubit in contact with its thermal bath behaves as a reservoir for the MO responsible for a heating channel increasing the momentum and position variances. This heating channel acts on a typical time-scale which corresponds to many mechanical oscillations, such that the first order dynamics is a good approximation for most of the implementations. This reservoir induces a scattering of the MO distribution in phase space. Because of its finite size, this reservoir is affected by the MO position. The heating rate is modulated within each mechanical periods and follows the long-term MO position scattering. Eventually when the mechanical energy has increased enough the Qubit is most of the time off resonance with the driving laser and the heating process stops.

Several extensions of this study can be considered. First, in order to accurately model experiments, the thermal phononic reservoir connected to the MO has to be taken into account. This reservoir can be found to induce its own Lindbladian ${\cal L}_\text{m}$, which is just the Lindbladian of a free MO coupled to a thermal reservoir \cite{OpticalNoise}. ${\cal L}_\text{m}$ has to be added to the MO master equation and competes with the Qubit-induced heating. In the limit where the thermal MO reservoir dominates, the Qubit is responsible for a slight change of the effective reservoir temperature. The trajectory-based approach developed in the last section can also be extended to include thermal noise without specific complications.

A study of the fluctuations of the work performed on the Qubit show that they can be deduced from the fluctuations of the MO position and momentum variations, which can be measured. Such access to the work distribution would provide a new way to test experimentally fluctuation theorems such as Jarzynski equality for a quantum open system, which has remained elusive up to now \cite{Monsel}.

\addchap{Conclusion}

\lettrine{Q}{uantum} measurement is a key phenomenon to understand the thermodynamics of quantum systems. In this thesis, we have proposed a formalism to describe the thermodynamics of quantum systems in which measurement plays a central role: As it is the primary source of randomness in the system's dynamics, it plays a role analogous to the thermal reservoir of standard stochastic thermodynamics.\\

We have first analyzed the projective measurement itself as a thermodynamic process, and we have shown that it induces an uncontrolled change of the system's energy that we have called quantum heat, and an entropy production. As a proof of principle, we have proposed an engine powered by a quantum measurement instead of  a hot thermal reservoir. Then, we have used the decription of quantum open system dynamics in term of generalized measurement to build a formalism capturing the thermodynamics of driven quantum systems in contact with a reservoir. We have defined thermodynamic quantities like the work, the quantum and the classical heats, the entropy production, at the level of the single realization, and we have derived fluctuation theorems. We have studied various types of drives (frequency shift or Rabi oscillations), of reservoirs (thermal or pure-dephasing channel) and of monitoring schemes (quantum jump or quantum state diffusion unravelings).\\

An important purpose of our work is to bridge the gap between theoretical quantum thermodynamics and experimental quantum optics. With this aim, we have suggested a setup to test the thermodynamics of a driven qubit in contact with a reservoir. We have investigated the consequences of imperfections in the detection scheme, and showed how to correct them. 
Finally, we have focused on two situations drawn from quantum optics. First, we have investigated the thermodynamics of fluorescence. By applying our formalism to such fundamental example, we have provided a thermodynamic interpretation of the optical Bloch equation which had remained elusive so far. We have compared our analysis with the approach already present in the literature based on Floquet master equation.  Second, we have studied an optomechanical device involving a Qubit coupled to a mechanical resonator. We have shown that the oscillator behaves as a classical battery, providing work to the Qubit by shifting its transition frequency. Such device allows to implement and study canonical transformations of information thermodynamics such that Landauer's erasure. Looking at the long-term dynamics, we have shown that the Qubit behaves as finite-size reservoir inducing a scattering of the MO position.\\ 

The formalism presented in this thesis allows to address many of the questions which motivate the growing field of quantum thermodynamics. Indeed, it provides consistent definitions of the thermodynamic variables, and quantum versions of the laws of thermodynamics. Furthermore, it can be exploited to analyze various quantum heat engines and compare their performances with their classical counterparts. This formalism can be used to analyze the energy exchanges involved in quantum algorithms, and therefore bring new perspectives on their advantages with respect to classical computation, but also on their limitations. Finally, by describing quantum measurement as a thermodynamic process, it provides a large toolbox to design new types of thermodynamic cycles, and exploit more effectively the specificities of the quantum world \cite{Yi17}. \\

The applications studied in this thesis already opens new perspectives for quantum thermodynamics.

We have identified several examples of non-thermal reservoirs: First, the 
engineered environment of section \ref{s2:SetupEngineered} can be used to generate a negative temperature or a time-dependent temperature bath. Second the laser 
drive has been shown to behaves as a colored bath when described on a long 
enough time scale. Finally, the Qubit involved in the optomechanical system 
is seen as a finite size non-thermal reservoir by the oscillator. The next step 
is now to investigate the thermodynamic behaviour of these reservoirs: what 
kind of heat do they provide? What is the associated second law? Can we 
build engine to extract work from these genuinely quantum reservoirs?

The Zeno limit briefly investigated in the example of the Measurement 
Powered Engine in Chapter \ref{Chapter1} also brings interesting 
perspectives. Indeed, it allows measurement to have a deterministic effect on 
the working agent. The next steps are to study under which conditions measurement can be a source of work instead of heat, and discuss implementations.

Besides, the optomechanical platform presented in the last chapter can provide other applications. In this thesis, we have only focused on average thermodynamic quantities, but the study of the fluctuations induced by the Qubit on the oscillator opens interesting perspectives for out-of-equilibrium thermodynamics. In particular, the fluctuation of the mechanical position can be related to the fluctuations of work, providing another experimental check of fluctuation theorems for the Qubit  \cite{Monsel}.

The entropy production in presence of a measuring apparatus which does not 
perform projective measurements requires further investigations to be fully 
understood.  First steps are to quantify the correlations between the 
working agent and the measuring device, and to try to reformulate the second 
law in the absence of final projective measurement in term of information-
theoretic quantities like the mutual information between the apparatus and 
the system. This approach may allow to understand better the relations 
between the entropy production presented in this thesis and another 
approach recently presented in \cite{Dressel16}.

The contextual nature of quantum thermodynamics is also a fascinating line 
of research. We have shown that the implemented measurement scheme (or 
unraveling) had consequences on the fluctuations of thermodynamic 
variables like work or entropy production. Clearly understanding what are 
the phenomena which do not depend on the unraveling is a crucial 
stride in the understanding of out-of-equilibrium quantum thermodynamics. 
In particular, the fluctuation theorems are generally proven within a 
particular measurement scheme. A situation that may bring fruitful insights 
in such a problem is the case of the quantum-state diffusion unraveling of a 
thermal reservoir, as implemented e.g. with a homodyne or heterodyne setup. 
In this case, the lowering and raising operators of the systems are mixed 
together in the Kraus operators, preventing from giving a clear definition of the quantum and classical heat contributions.

The influence of the measurement scheme also urges to generalize the study of measurement imperfections and their thermodynamic consequences, so as to favourise experimental check of these concepts. In this thesis we have focused on a specific source of noise which corresponds to the case of photocounters, and we have shown that depending on the regime it may have a thermodynamic influence or not. Measurement imperfections of other origin, e.g. in the case of continuous measurement, have been extensively studied \cite{Wiseman02,Gambetta07,Steck06}, and their thermodynamic consequences must now be investigated.

Another immediate follow-up of the present thesis consists in studying 
multipartite quantum systems in the role of the working agent, in order to investigate how thermodynamics is affected by another fascinating quantum phenomenon: entanglement. This special type of quantum correlations has been shown to be a resource for many quantum tasks like superdense coding \cite{Wootters98}, quantum teleportation \cite{Bouwmeester97}, quantum cryptography \cite{Ekert91} or interferometry \cite{Pezze09}. As a consequence much effort has been devoted to design schemes allowing to generate entangled states. Schemes found in the literature involve e.g. unitary gates \cite{Filipp11} or measurement of a non-local observable like the parity of two Qubits \cite{Roch14,Riste13}. Applying the formalism of this thesis for entanglement generating protocols is a promising way to get insights about the fundamental properties of entanglement, but also to compare the costs involved in these various schemes \cite{MZI}. 

Finally, the versatility of the quantum trajectories approach also provides prospects. Indeed, such description can be used to investigate situation in which are not described by a Lindblad equation. In particular, trajectory-based schemes can be used to capture non-Markovian dynamics \cite{Breuer04}, which is another direction of research expected to bring new behaviors in thermodynamics \cite{Schmidt15,Whitney16}.



\appendix 




\printbibliography[heading=bibintoc]


\end{document}